%% file: main.tex
\documentclass[11pt,a4paper]{article}
\usepackage{jheppub}
\usepackage{amsthm,amsbsy,amsfonts,mathrsfs,enumerate,float,wrapfig,amsmath}
\usepackage[utf8]{inputenc}
\usepackage[outline]{contour}
\usepackage{pdflscape}
\DeclareUnicodeCharacter{2212}{-}
\usepackage{comment}
\newcommand{\M}{\mathcal{M}}
\newcommand{\N}{\mathcal{N}}
\newcommand{\R}{\mathcal{R}}
\newcommand{\nn}{\nonumber}
\newcommand{\be}{\begin{equation}}
\newcommand{\ee}{\end{equation}}
\newcommand{\bea}{\begin{eqnarray}}
\newcommand{\eea}{\end{eqnarray}}
\newcommand{\ba}{\begin{align}}
\newcommand{\ea}{\end{align}}
\newcommand{\F}{\mathcal{F}}
\newcommand{\bF}{\mathbf{F}}
\newcommand{\bS}{\mathbf{S}}
\newcommand{\AS}{\mathbf{AS}}

\usepackage{xcolor}
\newcommand{\MA}[1]{\textcolor{teal}{MA: #1}}

\usepackage{amsmath}
\usepackage{amssymb}
\usepackage{MnSymbol}
\DeclareMathOperator{\PE}{PE}
\DeclareMathOperator{\HWG}{HWG}
\DeclareMathOperator{\HS}{HS}
\DeclareMathOperator{\MQ}{MQ}
\DeclareMathOperator{\EQ}{EQ}

\usepackage{array}
\newcolumntype{L}[1]{>{\raggedright\let\newline\\\arraybackslash\hspace{0pt}}m{#1}}
\newcolumntype{C}[1]{>{\centering\let\newline\\\arraybackslash\hspace{0pt}}m{#1}}
\newcolumntype{R}[1]{>{\raggedleft\let\newline\\\arraybackslash\hspace{0pt}}m{#1}}

\usepackage{tikz}
\usetikzlibrary{decorations.pathmorphing}
\usetikzlibrary{decorations.markings}
\usetikzlibrary{decorations.pathreplacing}
\usetikzlibrary{shapes, shapes.geometric, shapes.symbols, shapes.multipart, shapes.callouts, shapes.misc}
\tikzset{snake it/.style={decorate, decoration=snake}}
\tikzset{7brane/.style={circle, draw=black, fill=black,ultra thick,inner sep=1 pt, minimum size=1 pt,}, c/.default={4pt}}

\tikzset{cross/.style={cross out, draw=black,thick, minimum size=2*(#1-\pgflinewidth), inner sep=0pt, outer sep=0pt}, cross/.default={5pt}}

\tikzset{big7brane/.style={circle, draw=black, fill=black,ultra thick,inner sep=2.5 pt, minimum size=1 pt,}, c/.default={4pt}}
\tikzset{u/.style={circle, draw=black, fill=white,inner sep=2 pt, minimum size=2 pt,},f/.style={square, draw=black, fill=white,ultra thick,inner sep=4 pt, minimum size=2 pt,}}

\tikzset{so/.style={circle, draw=black, fill=red, thick,inner sep=2 pt, minimum size=2 pt,},f/.style={square, draw=black, fill=white,ultra thick,inner sep=4 pt, minimum size=2 pt,}}

\tikzset{sp/.style={circle, draw=black, fill=blue,thick,inner sep=2 pt, minimum size=2 pt,},f/.style={square, draw=black, fill=white,ultra thick,inner sep=4 pt, minimum size=2 pt,}}

\tikzset{uf/.style={rectangle, draw=black, fill=white,inner sep=2.75 pt, minimum size=4 pt,}}

\tikzset{spf/.style={rectangle, draw=black, fill=blue, thick,inner sep=2.5 pt, minimum size=4 pt,}}

\tikzset{sof/.style={rectangle, draw=black, fill=red, thick,inner sep=2.5 pt, minimum size=4 pt,}}
\usetikzlibrary{positioning}

\usepackage{mathtools}
\usepackage{multirow}
\tikzset{hasse/.style={circle, fill,inner sep=2pt}}
\newcommand{\wb}[3]{\overline{\left[\mathcal{W}_{#1} \right]}^{#2}_{#3}}
\usepackage{todonotes}
\usepackage{xstring}
\input{objects}
\usepackage{afterpage}
\usepackage{placeins}
\usepackage{adjustbox}
\usepackage{caption}
\usepackage{subcaption}
\def\UV#1{
    \left[#1\right]_\infty
}

\def\soeq#1#2{
    \raisebox{-0.45\height}{\begin{tikzpicture}
        \node (loc1) {SO$(#1)$};
        \node[above of = loc1] (loc2) {$[#2]$};
        \draw (loc1)--(loc2);
    \end{tikzpicture}}
}

\def\soeqI#1#2{
    \UV{\soeq{#1}{#2}}
}

\def\sueq#1#2{
    \raisebox{-0.45\height}{\begin{tikzpicture}
        \node (loc1) {SU$(#1)$};
        \node[above of = loc1] (loc2) {$[#2]$};
        \node[right of = loc1,label=below:{$S^2$}] (loc3) {};
        \node[right of = loc3] (loc4) {$[1]$};
        \draw (loc1)--(loc2);
        \path[draw,snake it] (loc1)--(loc4);
    \end{tikzpicture}}
}
    
\def\sukeq#1#2#3{
    \raisebox{-0.45\height}{\begin{tikzpicture}
        \node (loc1) {SU$(#1)_{#3}$};
        \node[above of = loc1] (loc2) {$[#2]$};
        \node[right of = loc1,label=below:{$S^2$}] (loc3) {};
        \node[right of = loc3] (loc4) {$[1]$};
        \draw (loc1)--(loc2);
        \path[draw,snake it] (loc1)--(loc4);
    \end{tikzpicture}}
}

\def\sukeqI#1#2#3{
    \UV{\sukeq{#1}{#2}{#3}}
}
  
\def\sukweq#1#2#3{
    \raisebox{-0.45\height}{\begin{tikzpicture}
        \node (loc1) {SU$(#1)_{#3}$};
        \node[above of = loc1] (loc2) {$[#2]$};
        \node[right of = loc1] (locc) {};
        \node[right of = locc,label=below:{$S^2$}] (loc3) {};
        \node[right of = loc3] (loc4) {$[1]$};
        \draw (loc1)--(loc2);
        \path[draw,snake it] (loc1)--(loc4);
    \end{tikzpicture}}
}

\def\suknseq#1#2#3{
    \raisebox{-0.45\height}{\begin{tikzpicture}
        \node (loc1) {SU$(#1)_{#3}$};
        \node[above of = loc1] (loc2) {$[#2]$};
        \draw (loc1)--(loc2);
    \end{tikzpicture}}
}

\def\suknseqI#1#2#3{
    \UV{\suknseq{#1}{#2}{#3}}
}

\title{On brane systems with O${}^+$ planes -- 5d and 6d SCFTs}
\author[a]{Mohammad Akhond,}
\author[b]{Guillermo Arias-Tamargo,}
\author[c,d]{Federico Carta,}
\author[e]{Julius F.\ Grimminger,}
\author[b]{Amihay Hanany}
\affiliation[a]{Department of Physics, Kyoto University, Kyoto 606-8502, Japan}
\affiliation[b]{Theoretical Physics Group, The Blackett Laboratory, Imperial College London, Prince Consort
Road London, SW7 2AZ, UK}
\affiliation[c]{London Institute for Mathematical Sciences, Royal Institution, London W1S 4BS, UK}
\affiliation[d]{Theoretical Particle Physics and Cosmology, King’s College London, Strand, London, WC2R 2LS, United Kingdom}
\affiliation[e]{Mathematical Institute, University of Oxford,\\
Andrew Wiles Building, Woodstock Road, Oxford, OX2 6GG, UK}

\emailAdd{akhond@gauge.scphys.kyoto-u.ac.jp}
\emailAdd{guillermo.arias.tam@gmail.com}
\emailAdd{federico.carta@kcl.ac.uk}
\emailAdd{julius.grimminger@maths.ox.ac.uk}
\emailAdd{a.hanany@imperial.ac.uk}

\abstract{We study Higgs branches of field theories with 8 supercharges in 5 and 6 dimensions, focusing on theories realised on 5-brane webs in Type IIB with an O$7^+$ plane, or a D6-D8-NS5 brane system in Type IIA in the presence of an O$8^+$ plane. We find magnetic quivers for the Higgs branches of these theories. The main consequence of the presence of the orientifold is that it renders the magnetic quiver to be non-simply-laced. We propose a contribution of the O$7^+$ to the usual stable intersection number of 5-branes from tropical geometry, and show that it is consistent with Fayet-Iliopoulos deformations of magnetic quivers which represent mass deformations of 5d SQFTs. From the magnetic quivers, we compute phase diagrams and highest weight generating functions for the Higgs branches, enabling us to identify the global form of the flavour symmetry for several families of 5d SQFTs; among them Bhardwaj's rank-1 theory. For 6d theories realised on a $-4$ curve, we observe the appearance of an additional $D_4$ slice on top of the phase diagram as one goes to the tensionless limit.
}

\begin{document}

\maketitle
\section{Introduction and summary}
\label{sec:intro}
Magnetic quivers (MQ) have proven to be a powerful tool in characterising the Higgs branch of supersymmetric quantum field theories (SQFT) and superconformal field theories (SCFTs) with 8 supercharges in spacetime dimensions $3\leq d\leq 6$. If the SQFT in question is realised on a brane system, then its magnetic quiver can often be derived following a simple procedure, that consists of studying the possible transverse motions of branes along the Higgs branch directions \cite{Cabrera:2018jxt,Cabrera:2019dob,Cabrera:2019izd,Bourget:2019rtl,Bourget:2020asf,Bourget:2020gzi,Bourget:2020mez,Bourget:2021csg,Bourget:2021jwo,Bourget:2021siw,Bourget:2022ehw,Bourget:2022tmw,Bourget:2023uhe,Akhond:2020vhc,Akhond:2021ffo,Akhond:2021knl,Akhond:2022jts}. If we do not have such a system, it is still possible to derive magnetic quivers. For example, many magnetic quivers for $4$d $\mathcal{N}=2$ theories were derived from a geometric engineering picture of Type IIB string theory; and similarly for 5d $\mathcal{N}=1$ from M-theory on a non-compact hypersurface singularity (see for example \cite{Carta:2021whq, Carta:2021dyx, Carta:2022spy, Carta:2022fxc, Closset:2020scj, Closset:2020afy, Giacomelli:2022drw, Giacomelli:2024dbd}). Other ways of finding magnetic quivers include starting with known ones, and tracking how certain operations on the theory of interest (e.g. deformations, discrete gauging, compactification, etc.) translate to the MQ \cite{vanBeest:2020kou, vanBeest:2021xyt, VanBeest:2020kxw, Bourget:2020bxh, Arias-Tamargo:2021ppf, Giacomelli:2024sex, Hanany:2023uzn, Hanany:2023tvn, Mansi:2023faa, Hanany:2022itc, Lawrie:2023uiu, Nawata:2021nse, Bao:2024eoq, Fazzi:2022yca, Fazzi:2023ulb, DelZotto:2023myd, DelZotto:2023nrb, Lawrie:2024zon}. 

In particular, for 5d field theories with 8 supercharges living on brane webs, methods of tropical geometry and brane dynamics have been combined successfully to read magnetic quivers for ordinary brane webs \cite{Cabrera:2018jxt,Bourget:2019rtl,Bourget:2020asf,vanBeest:2020kou,VanBeest:2020kxw,Bourget:2020mez,Bourget:2021csg,Bourget:2021jwo,Bourget:2022ehw,Bourget:2022tmw} (involving only $(p,q)5$-branes and $[p,q]7$-branes), as well as brane webs with O5 planes \cite{Bourget:2020gzi,Akhond:2020vhc,Akhond:2021knl,Akhond:2022jts}. Brane webs with O$7^-$ planes reduce to ordinary brane webs, as the O$7^-$ splits into two $[p,q]7$-branes. Brane webs with O$7^+$ planes, however, are more intricate, and their magnetic quivers were first discussed in \cite{Akhond:2021ffo}. We build on that work in the present paper, and establish a satisfactory set of rules for such cases. The main difference between the magnetic quivers in the present paper and those of \cite{Akhond:2021ffo} is the presence of adjoint hypermultiplets in the former and fundamental, as well as charge 2 hypermultiplets under U(1) gauge nodes in the latter. Since the Coulomb branch of the magnetic quivers considered is insensetive to this choice, one might consider this immaterial and a matter of taste. On the other hand, computation of Hasse diagrams, via the quiver subtraction algorithm is at present only understood for the presentation chosen in the current paper. Moreover this work also establishes a relation between the magnetic quiver for the 5d theories in question and the magnetic quiver for the 6d theories from which they originate via twisted circle compactification, and this perspective favours the choice of adjoint hypermultiplets.

Deriving the magnetic quivers allows us to compute a number of properties of the Higgs branches in question, using $3d$ Coulomb branch technology. The Hasse diagram of symplectic leaves, i.e.\ the phase diagram of the theory under the generalised Higgs mechanism, can be computed via quiver subtraction \cite{Bourget:2019aer, Cabrera:2018ann, Bourget:2019aer, Bourget:2022ehw, Bourget:2022tmw, Bourget:2023dkj, Bourget:2024mgn}.

The Hilbert Series (HS), counting chiral ring operators, can be computed via the monopole formula \cite{Cremonesi:2013lqa}. From this in many cases the Highest Weight Generating function (HWG) \cite{Hanany:2014dia}, encoding all representations of chiral ring operators under the global symmetry algebra, can be obtained for generic parameter families. When this is possible, we are able to identify the global form of the global symmetry group, as this is precisely specified by the set of representations that do or do not show up. To be more precise, this allows us to identify the global form of the isometry group of the Higgs branch. Two issues can occur: 1) Nilpotent operators can show up in the Higgs branch chiral ring (making the Higgs branch a scheme rather than a variety \cite{Bourget:2019rtl}) and may carry global charge not seen by the other operators in the ring \cite{Cremonesi:2015lsa,Bourget:2019rtl}. These nilpotent operators are not captured by magnetic quivers \cite{Bourget:2019rtl}, and hence some global symmetry may be missed. And 2) there are examples where representations of the global symmetry appear in the BPS spectrum in the Coulomb branch that do not appear in the chiral ring \cite{Distler:2019eky}. Keeping these subtleties in mind, in this paper we will read the global form of the Higgs branch symmetry from Hilbert series computations.

In particular, in the first half of the paper we will explain how to extract the magnetic quiver from the brane web, focusing on how the presence of the orientifold modifies the rules. In the second half of the paper we proceed to systematically present the results for several collections of such theories; these include the infinite coupling limit of SO$(K)$ gauge theories with hypermultiplets in the vector representation, SU$(k)$ gauge theories with hypermultiplets in the fundamental and rank two tensor representations, and the recently discovered rank-1 SCFT known as \emph{Bhardwaj's rank-1 theory} \cite{Bhardwaj:2019jtr}, which is not one of Seiberg's $E_n$ rank-1 theories.

The brane webs we study correspond to UV fixed points that flow to 5d gauge theories with SO gauge groups with vector hypermultiplets or SU gauge groups with symmetric, and fundamental matter, and others. These webs have the feature that, at the SCFT point, some of the fivebranes intersect on top of the O7$^+$ plane. It turns out that this makes a big difference when reading the magnetic quiver. As usual, the nodes in the quiver are read from the transverse motion of the possible subwebs. The links between the nodes involve the so-called \emph{stable intersection} between the subwebs,
\begin{align}
    \mathrm{SI}_0 = \sum_{\text{intersections}} \left|\det\left(  \begin{array}{cc}
        p_1 & q_1 \\
        p_2 & q_2
    \end{array}\right)\right|,
\end{align}
where $(p_i,q_i)$ are the RR and NSNS charges of the intersecting fivebranes. Our main result is that, when the intersection occurs on top of the orientifold, the stable intersection needs to be modified schematically as follows (see \eqref{eq:FullyInvariantIntersection} for the fully precise version of this formula):
\begin{align}\label{eq:intersection_modified_O7_intro}
    \mathrm{SI}_{\text{O7}^{+}} = \mathrm{SI}_0 + 2|q_1 q_2|\,. 
\end{align}
Below we argue that this needs to be the case by requiring consistency of various deformations of the theories on the brane web and magnetic quivers. We summarise the magnetic quivers we obtain for the various theories in Tables \ref{tab:5dTheories}, \ref{tab:5d_globalsymm}, \ref{tab:5d_Utheories}, \ref{tab:Bhard}, \ref{tab:6d1} and \ref{tab:6d2}. In some cases, we can use the magnetic quivers to compute the exact HWG's of their Higgs branch chiral rings, and as a result we are able to find which is the global form of the global symmetry group of the SCFTs. In Table \ref{tab:5d_globalsymm} we collect the results of such an analysis for 5d gauge theories with special orthogonal gauge group and in Table \ref{tab:6d2} for the 6d theory. An important remark is that some of the theories under consideration have alternative constructions via webs with O5 planes; they lead to orthosymplectic magnetic quivers, and they can be used to provide highly non-trivial checks that our magnetic quivers are correct (see for instance \cite{Akhond:2021ffo}).

The rest of the paper is organised as follows. In section \ref{sec:intersection_rule} we discuss how to read the magnetic quiver in the presence of the orientifold. Then we proceed to apply these to several different families of theories. In section \ref{sec:orthogonal} we study $5d$ theories with orthogonal gauge groups, in section \ref{sec:specialunitary} theories with special unitary gauge groups, and in section \ref{sec:other_examples} other examples. In section \ref{sec:Hasse_higgsings} we study the Hasse (phase) diagram of these theories, focusing in particular in the differences between finite and infinite coupling.  In section \ref{sec:6d} we show the applicability of our techniques to the $6d$ set-up by discussing theories realised on a $-4$ curve. We discuss a curious pattern of the enhancement of the moduli space by an additional $D_4$ slice as one goes from finite tension to the tensionless limit. Appendices \ref{sec:App_review_monopole} and \ref{app:BraneWebs} provide a quick review of background material, regarding the monopole formula and non-simply laced quivers; as well as brane webs with orientifold planes. In appendix \ref{app:Hasse} we collect a glossary of different quivers that make an appearance throughout the main text and their Hasse diagrams.
\newpage
\begin{landscape}
\begin{table}[]
    \centering
    \begin{tabular}{c|c|c}
        Electric quiver &UV fixed point Magnetic Quiver &$\mathrm{PL}\left[\HWG\right]$  \\\hline\hline
        $ \soeq{K}{K-3}$ & $
    \begin{array}{c}
         \begin{scriptsize}
    \begin{tikzpicture}
    \node[label=below:{$1$}][u](2){};
    \node (dots)[right of=2]{$\cdots$};
    \node[label=below:{$K-3$}][u](2N-5)[right of=dots]{};
    \node[label=below:{$K-2$}][u](2N-4)[right of=2N-5]{};
    \node[label=below:{$2$}][u](2')[right of=2N-4]{};
    \draw (2')to[out=-45,in=45,loop,looseness=10](2');
    \draw(2)--(dots);
    \draw(dots)--(2N-5);
    \draw[ double distance=1.5pt,<-](2N-5)--(2N-4);
    \draw(2N-4)--(2');
    \end{tikzpicture}
    \end{scriptsize}
    \end{array}
$&$
    \sum_{i=1}^{K-2}\mu_i^2t^{2i}+t^{4}+\mu_{K-2} (t^{K-2}+t^{K})-\mu_{K-2}^2t^{2K}\;,
$\\\hline
$ \soeq{K}{K-4}$& $ \begin{array}{c}
         \begin{scriptsize}
    \begin{tikzpicture}
    \node[label=below:{1}][u](2){};
    \node (dots)[right of=2]{$\cdots$};
    \node[label=below:{$K-5$}][u](2N-5)[right of=dots]{};
    \node[label=below:{$K-4$}][u](2N-4)[right of=2N-5]{};
    \node[label=below:{$2$}][u](2')[right of=2N-4]{};
    \node[label=below:{$1$}][u](1')[right of=2']{};
    \draw (2')to[out=45,in=135,loop,looseness=10](2');
    \draw(2)--(dots);
    \draw(dots)--(2N-5);
    \draw[ double distance=1.5pt,<-](2N-5)--(2N-4);
    \draw(2N-4)--(2');
    \draw[double distance=1.5pt,->](2')--(1');
    \end{tikzpicture}
    \end{scriptsize}
    \end{array}$ & $\sum_{i=1}^{K-4}\mu_i^2t^{2i}+\nu^2t^2+t^4+\mu_{K-4}\nu^2\left(t^{K-2}+t^{K}\right)-\mu_{K-4}^2\nu^4t^{2K}$\\\hline
$ \soeq{K}{N}$, $N<K-4$ &$\begin{array}{c}
             \begin{scriptsize}
    \begin{tikzpicture}
    \node[label=below:{$N$}][u](2k-2j-4){};
      \node[label=above:{1}][u][above right of=2k-2j-4](1){};
    \node[label=below:{1}][u](11)[below right of=2k-2j-4]{};
    \node[label=above:{$N-1$}][u](2k-2j-5)[left of=2k-2j-4]{};
    \node(dots)[left of=2k-2j-5]{$\cdots$};
    \node[label=below:{$2$}][u](2)[left of=dots]{};
    \node[label=below:{$1$}][u](111)[left of=2]{};
    \draw(111)--(2);
    \draw(2)--(dots);
    \draw(dots)--(2k-2j-5);
    \draw[<-,double distance=1.5 pt](2k-2j-5)--(2k-2j-4);
    \draw(1)--(2k-2j-4);
    \draw(11)--(2k-2j-4);
    \draw(1)--node[above right]{$K-N-2$}++(11);
    \end{tikzpicture}
    \end{scriptsize}
    \end{array}$ &  $\sum_{i=1}^{N}\mu_i^2t^{2i}+t^2+\left(q+q^{-1}\right)\mu_{N}t^{K-2}-\mu^2_{N}t^{2K-4}$  \end{tabular}
    \caption{Summary of results for Higgs branches of 5d UV fixed point, for theories with special orthogonal gauge group. The first column is the low energy theory in terms of an orthogonal gauge group and matter field transforming in the vector representation, in the last row the number of vectors is restricted to be $N<K-4$. The second column is the magnetic quiver for the Higgs branch at the UV fixed point. The last column is the plethystic logarithm of the highest weight generating function of the chiral ring; the presence or lack thereof of generators transforming in the various representations is what allows us to identify the global form of the global symmetry group reported in table \ref{tab:5d_globalsymm}.}
    \label{tab:5dTheories}
\end{table}
\end{landscape}
\begin{table}
\centering
    \begin{tabular}{c |c| c}
    Electric Quiver  & UV fixed point Magnetic Quiver & Global Symmetry Enhancement  \\\hline\hline
        \electricquiver{1}&\scriptsize\magneticquiver{16}  & $\begin{matrix}
            (\mathrm{Sp}(2k-1)\times \mathrm{SU}(2)_R)/\mathbb{Z}_2\\
            \uparrow\\
            \mathrm{Sp}(2k-2)/\mathbb{Z}_2\times \mathrm{SO}(3)_R \times\mathrm{U}(1)_I
        \end{matrix}$  \\
        \hline
        \electricquiver{2} &\scriptsize\magneticquiver{17} & $\begin{matrix}
            \mathrm{Sp}(2k-2)/\mathbb{Z}_2\times \mathrm{SO}(3)_R\\
            \uparrow\\
            \mathrm{Sp}(2k-3)/\mathbb{Z}_2 \times \mathrm{SO}(3)_R\times\mathrm{U}(1)_I
        \end{matrix}$  \\
        \hline
        \electricquiver{3} &\scriptsize\magneticquiver{18} & $\begin{matrix}
            \left(\mathrm{Sp}(2k-3)\times \mathrm{SU}(2)_R\right)/\mathbb{Z}_2\times \mathrm{SO}(3)_I\\
            \uparrow\\
            \mathrm{Sp}(2k-3)/\mathbb{Z}_2 \times \mathrm{SO}(3)_R \times\mathrm{U}(1)_I
        \end{matrix}$ \\
        \hline
        \electricquiver{4} &\scriptsize\magneticquiver{19} & $\begin{matrix}
            \mathrm{Sp}(2k-4)/\mathbb{Z}_2\times \mathrm{SO}(3)_R\times \mathrm{SO}(3)_I\\
            \uparrow\\
            \mathrm{Sp}(2k-4)/\mathbb{Z}_2 \times \mathrm{SO}(3)_R \times\mathrm{U}(1)_I
        \end{matrix}$ \\
        \hline
        \electricquiver{8}&\scriptsize\magneticquiver{20} &  $\begin{matrix}
            \mathrm{Sp}(2k-2j-4)/\mathbb{Z}_2\times \mathrm{SU}(2)_R \times \mathrm{U}(1)_I\\
            \uparrow\\
            \mathrm{Sp}(2k-2j-4)/\mathbb{Z}_2 \times \mathrm{SO}(3)_R \times\mathrm{U}(1)_I
        \end{matrix}$ \\
        \hline
        \electricquiver{5} &\scriptsize\magneticquiver{21} & $\begin{matrix}
            \mathrm{Sp}(2k-2j-5)\times \mathrm{SO}(3)_R\times \mathrm{U}(1)_I\\
            \uparrow\\
            \mathrm{Sp}(2k-2j-5)/\mathbb{Z}_2 \times \mathrm{SO}(3)_R \times\mathrm{U}(1)_I
        \end{matrix}$  \\
        \hline
        \electricquiver{7}&\scriptsize\magneticquiver{22} &$\begin{matrix}
            ( \mathrm{Sp}(2k-2j-5)\times \mathrm{SU}(2)_R)/\mathbb{Z}_2\times \mathrm{U}(1)_I\\
            \uparrow\\
            \mathrm{Sp}(2k-2j-5)/\mathbb{Z}_2 \times \mathrm{SO}(3)_R \times \mathrm{U}(1)_I
        \end{matrix}$  \\
        \hline
        \electricquiver{6}&\scriptsize\magneticquiver{23} & $\begin{matrix}
           \mathrm{Sp}(2k-2j-6)/\mathbb{Z}_2\times \mathrm{SO}(3)_R\times \mathrm{U}(1)_I\\
           \uparrow\\
           \mathrm{Sp}(2k-2j-6)/\mathbb{Z}_2 \times \mathrm{SO}(3)_R \times\mathrm{U}(1)_I
        \end{matrix}$  \\ \hline
    \end{tabular}
    \caption{Summary of the pattern of enhancement of the global symmetry, for 5d theories with special orthogonal gauge group and hypermultiplets in the vector representation. The first column is the electric theory. The second column is the magnetic quiver at infinite coupling. The last column demonstrates the pattern of the global symmetry enhancement of the SCFT compared with the gauge theory, due to massless instanton operators. In the last four rows, the range of $j$ from top to bottom is respectively $0\leq j\leq k-2$, $0\leq j\leq k-3$, $0\leq j\leq k-3$, $0\leq j\leq k-3$. The global form of the symmetry at the UV fixed point is a consequence of the HWG summarised in Table \ref{tab:5dTheories}, as described in detail in Section \ref{sec:orthogonal}}
    \label{tab:5d_globalsymm}
\end{table}
\begin{table}
\hspace*{-1.95cm}
    \makebox[\textwidth][l]{
    \begin{tabular}{c | c | c }
    Electric Quiver & UV fixed point Magnetic Quiver & Global Symmetry Enhancement \\\hline\hline
            \electricquiver{22} & \scriptsize \magneticquiverU{6} & $\begin{matrix}
        \mathfrak{a}_{K-j-l-5}\oplus \mathfrak{u}_1\oplus \mathfrak{u}_1\oplus\mathfrak{u}_{1}
           \\
           \uparrow\\
  \mathfrak{a}_{K-j-l-5}\oplus \mathfrak{u}_1\oplus \mathfrak{u}_1\oplus \mathfrak{u}_{1,I}
        \end{matrix}$ \\\hline
                \electricquiver{21} & \scriptsize \magneticquiverU{5} & $\begin{matrix}
            \mathfrak{a}_{K-j-5}\oplus \mathfrak{u}_1\oplus \mathfrak{u}_1\oplus\mathfrak{a}_{1}\\
            \uparrow\\
           \mathfrak{a}_{K-j-5}\oplus \mathfrak{u}_1\oplus \mathfrak{u}_1\oplus \mathfrak{u}_{1,I}
        \end{matrix}$\\
        \hline
                \begin{tikzpicture} 
                      \node at (0,0)(gauge){SU$(K)_{(j+1)/2}$};
                      \node (flavour)[above of=gauge]{$[K-j-3]$};
                      \node[label=below:{$S^2$}](empty)[right of=gauge]{};
                      \node(symmetric)[right of=empty]{$\left[1\right]$};
                      \path[draw,snake it](gauge)--(symmetric);
                      \draw(gauge)--(flavour);
         \end{tikzpicture} & \scriptsize 
\begin{tikzpicture}
    \node[label=above:{1}][u](111){};
    \node[label=below:{2}][u](22)[below right of=111]{};
    \node[label=below:{$K-j-2$}][u](k-5)[below left of=111]{};
    \node[label=above:{$K-j-3$}][u](k-6)[left of=k-5]{};
    \node(dots)[left of=k-6]{$\cdots$};
    \node[label=below:{$2$}][u](2)[left of=dots]{};
    \node[label=below:{$1$}][u](1)[left of=2]{};
    \draw(1)--(2);
    \draw(2)--(dots);
    \draw(dots)--(k-6);
    \draw[double distance=1.5pt,<-](k-6)--(k-5);
    \draw(k-5)--(22);
    \draw(k-5)--(111);
    \draw(111)--node[above right]{$j+3$}++(22);
    \draw (22)to[out=45,in=-45,loop,looseness=10](22);
    \end{tikzpicture}\; & $\begin{matrix}
        \mathfrak{a}_{K-j-3}\oplus \mathfrak{u}_1\oplus \mathfrak{u}_1
           \\
           \uparrow\\
  \mathfrak{a}_{K-j-4}\oplus \mathfrak{u}_1\oplus \mathfrak{u}_1\oplus \mathfrak{u}_{1,I}
        \end{matrix}$ \\
        \hline
        \electricquiver{17} & \scriptsize \magneticquiverU{1} & $\begin{matrix}
            \mathfrak{a}_{K-2}\oplus \mathfrak{u}_1\\
            \uparrow\\
            \mathfrak{a}_{K-4}\oplus \mathfrak{u}_1\oplus \mathfrak{u}_1\oplus \mathfrak{u}_{1,I}
        \end{matrix}$ \\
        \hline
        \electricquiver{19} & \scriptsize \magneticquiverU{3} & $\begin{matrix}
         \mathfrak{a}_{K-5}\oplus \mathfrak{u}_1\oplus\mathfrak{a}_{2}\
            \\
            \uparrow\\
\mathfrak{a}_{K-5}\oplus \mathfrak{u}_1\oplus \mathfrak{u}_1\oplus \mathfrak{u}_{1,I}
        \end{matrix}$\\
        \hline
\begin{tikzpicture} 
\node at (0,0)(gauge){SU$(K)_{1}$};
\node (flavour)[above of=gauge]{$[K-4]$};
\node[label=below:{$S^2$}](empty)[right of=gauge]{};
\node(symmetric)[right of=empty]{$\left[1\right]$};
\path[draw,snake it](gauge)--(symmetric);
\draw(gauge)--(flavour);
\end{tikzpicture} & \scriptsize 
         \begin{tikzpicture}
    \node[label=below:{1}][u](2){};
    \node (dots)[right of=2]{$\cdots$};
    \node[label=below:{$K-5$}][u](2N-5)[right of=dots]{};
    \node[label=below:{$K-4$}][u](2N-4)[right of=2N-5]{};
    \node[label=below:{$3$}][u](3')[right of=2N-4]{};
    \node[label=below:{$1$}][u](2')[right of=3']{};
    \draw (3')to[out=45,in=135,loop,looseness=10](3');
    \draw(2)--(dots);
    \draw(dots)--(2N-5);
    \draw[ double distance=1.5pt,<-](2N-5)--(2N-4);
    \draw(2N-4)--(3');
    \draw[double distance=1.5pt,->](3')--(2');
    \end{tikzpicture}\; &$\begin{matrix}
        \mathfrak{a}_{K-4}\oplus \mathfrak{u}_1\oplus \mathfrak{u}_1\\
           \uparrow\\
  \mathfrak{a}_{K-5}\oplus \mathfrak{u}_1\oplus \mathfrak{u}_1\oplus \mathfrak{u}_{1,I}
        \end{matrix}$ \\
        \hline
    \end{tabular}}
    \caption{Summary of results for Higgs branches of 5d UV fixed points, for theories with special unitary gauge group, with $K\ge3$. The case $K=2$ corresponds to Bhardwaj's rank-1 theory, and it is discussed separately in Table \ref{tab:Bhard}. The first column is the low energy theory. The second column is the magnetic quiver for the Higgs branch at the UV fixed point. The third column is the global symmetry algebra of the SCFT.}
    \label{tab:5d_Utheories}
\end{table}

\begin{table}[h]
    \centering
    \begin{tabular}{c|c|c}
    SCFT  & Magnetic Quiver  & Global Symmetry  \\
        \hline\hline
        $Bh(1)$ & \raisebox{-.5\height}{$\begin{tikzpicture}
        \node[label=below:{$3$}][u](1) at (0,0) {};
        \node[label=below:{$1$}][u](2) at (1,0) {};
        \draw (1)to[out=135,in=225,loop,looseness=10](1);
        \draw (1)--(2);
    \end{tikzpicture}$} & $\left(\mathrm{SU}(2)\times\mathrm{SU}(2)_R\right)/\mathbb{Z}_2$\\
    \hline
    $Bh(2)$ &  \raisebox{-.5\height}{$\begin{tikzpicture}
            \node[u,label=below:{3}] (6) at (6,0) {};
            \node[u,label=below:{1}] (7) at (7,0) {};
            \draw (6)to[out=135,in=225,loop,looseness=10](6);
            \draw[->,double distance=1.5pt](6)--(7);
        \end{tikzpicture}$} & $\left(\mathrm{SU}(2)\times\mathrm{SU}(2)_R\right)/\mathbb{Z}_2$\\
    \hline
    $Bh(\alpha)\;,$ $\alpha>2$ & \raisebox{-.5\height}{$\begin{tikzpicture}
            \node[u,label=below:{2}] (6) at (5,0) {};
            \draw (6)to[out=135,in=225,loop,looseness=10](6);
            \node[u,label=below:1] (t) at (6,0) {};
            \draw (t)--(6);
            \node at (5.5,0.3) {$\alpha$};
        \end{tikzpicture}$} & $\begin{cases}
            \mathrm{U}(1)\times\mathrm{SO}(3)_R&\alpha\in2\mathbb{Z}\\
            \left(\mathrm{U}(1)\times\mathrm{SU}(2)_R\right)/\mathbb{Z}_2&\alpha\in2\mathbb{Z}+1
        \end{cases}$
    \end{tabular}
    \caption{Magnetic quivers and global symmetries for Bhardwaj's rank-1 $5d$ $\mathcal{N}=1$ SCFT \cite{Bhardwaj:2019jtr}, which we call $Bh(1)$, and the higher rank generalisations discussed in the paper, which we call $Bh(r)$. The Coulomb branch of the $Bh(1)$ magnetic quiver is Sym${}^3(\mathbb{C}^2)$ which contains a free $\mathbb{C}^2$ factor. The Higgs branch of the 5d SFCT is the singular part of Sym${}^3(\mathbb{C}^2)$ with $\left(\mathrm{SU}(2)\times\mathrm{SU}(2)_R\right)/\mathbb{Z}_2$ symmetry. The Higgs branch of the $Bh(2)$ is the singular part of Sym${}^4(\mathbb{C}^2)$ with $\left(\mathrm{SU}(2)\times\mathrm{SU}(2)_R\right)/\mathbb{Z}_2$ symmetry. The Higgs branch of $Bh(\alpha>2)$ is Sym${}^2(\mathbb{C}^2/\mathbb{Z}_{\alpha})$ whose precise symmetry depends on whether $\alpha$ is even or odd as reported in the table.}
    \label{tab:Bhard}
\end{table}
\begin{landscape}
\begin{table}[h]
    \centering
    \begin{tabular}{c|c|c}
    Electric Theory & 6d SCFT Magnetic Quivers & PL[HWG]  \\
        \hline\hline
  $ \raisebox{-.5\height}{ \begin{tikzpicture}
             \node(gauge){SO$(K)$};
             \node(flavour)[above of=gauge]{$[K-8]$};
             \draw (gauge)--(flavour);
         \end{tikzpicture}}  $  & $  \begin{array}{c}
         \raisebox{-.5\height}{ \begin{scriptsize}
    \begin{tikzpicture}
    \node[label=below:{$1$}][u](1){};
    \node[label=below:{$2$}][u](2)[right of=1]{};
    \node (dots)[right of=2]{$\cdots$};
    \node[label=below:{$K-9$}][u](2N-5)[right of=dots]{};
    \node[label=below:{$K-8$}][u](2N-4)[right of=2N-5]{};
    \node[label=below:{$2$}][u](2')[right of=2N-4]{};
    \node[right of=2']{$3\;\textbf{Adj}$};
    \draw (2')to[out=-45,in=45,loop,looseness=10](2');
    \draw(1)--(2);
    \draw(2)--(dots);
    \draw(dots)--(2N-5);
    \draw[ double distance=1.5pt,<-](2N-5)--(2N-4);
    \draw(2N-4)--(2');
    \end{tikzpicture}
    \end{scriptsize}}    \\
  \raisebox{-.5\height}{ \begin{scriptsize}
             \begin{tikzpicture}
                          \node[label=below:{1}][so](o1){};
                          \node[label=below:{2}][sp](sp1)[right of=o1]{};
                          \node[label=below:{3}][so](so3)[right of=sp1]{};
                          \node(dots)[right of=so3]{$\cdots$};
                          \node[label=below:{$2k-8$}][sp](spk-4)[right of=dots]{};
                          \node[label=below:{$2k-7$}][so](so2k-7)[right of=spk-4]{};
                          \node[label=below:{$2k-8$}][sp](spk-4')[right of=so2k-7]{};
                          \node(dots')[right of=spk-4']{$\cdots$};
                          \node[label=below:{3}][so](so3')[right of=dots']{};
                          \node[label=below:{2}][sp](sp1')[right of=so3']{};
                          \node[label=below:{1}][so](o1')[right of=sp1']{};
                          \node[label=left:{2}][sp](sp11)[above of=so2k-7]{};
                          \node[label=above:{7}][sof](sof)[above of=sp11]{};
                          \draw(o1)--(sp1);
                          \draw(sp1)--(so3);
                          \draw(so3)--(dots);
                          \draw(dots)--(spk-4);
                          \draw(spk-4)--(so2k-7);
                          \draw(spk-4')--(so2k-7);
                          \draw(dots')--(spk-4');
                          \draw(so3')--(dots');
                          \draw(sp1')--(so3');
                          \draw(o1')--(sp1');
                          \draw(so2k-7)--(sp11);
                          \draw(sp11)--(sof);
             \end{tikzpicture}
         \end{scriptsize}  }
    \end{array}  $  & $
        \sum_{i=1}^{K-8}\mu_i^2t^{2i}+t^4
        +\mu_{K-8}\left(t^{K-4}+t^{K-2}\right)
        -\mu_{K-8}^2t^{2K-4}
    $ \\ \hline
    \end{tabular}
    \caption{Magnetic quiver and HWG for $6d$ $\mathcal{N}=(1,0)$ UV fixed point SCFT. The orthosymplectic quiver makes sense only for $K=2k$.}
    \label{tab:6d1}
\end{table}

\begin{table}[h]
    \centering
    \begin{tabular}{c|c|c}
    Electric Theory & 6d SCFT Magnetic Quivers & Global Symmetry Change  \\
        \hline\hline
  $ \raisebox{-.5\height}{ \begin{tikzpicture}
             \node(gauge){SO$(2k)$};
             \node(flavour)[above of=gauge]{$[2k-8]$};
             \draw (gauge)--(flavour);
         \end{tikzpicture}}  $  & $  \begin{array}{c}
         \raisebox{-.5\height}{ \begin{scriptsize}
    \begin{tikzpicture}
    \node[label=below:{$1$}][u](1){};
    \node[label=below:{$2$}][u](2)[right of=1]{};
    \node (dots)[right of=2]{$\cdots$};
    \node[label=below:{$2k-9$}][u](2N-5)[right of=dots]{};
    \node[label=below:{$2k-8$}][u](2N-4)[right of=2N-5]{};
    \node[label=below:{$2$}][u](2')[right of=2N-4]{};
    \node[right of=2']{$3\;\textbf{Adj}$};
    \draw (2')to[out=-45,in=45,loop,looseness=10](2');
    \draw(1)--(2);
    \draw(2)--(dots);
    \draw(dots)--(2N-5);
    \draw[ double distance=1.5pt,<-](2N-5)--(2N-4);
    \draw(2N-4)--(2');
    \end{tikzpicture}
    \end{scriptsize}}    \\
  \raisebox{-.5\height}{ \begin{scriptsize}
             \begin{tikzpicture}
                          \node[label=below:{1}][so](o1){};
                          \node[label=below:{2}][sp](sp1)[right of=o1]{};
                          \node[label=below:{3}][so](so3)[right of=sp1]{};
                          \node(dots)[right of=so3]{$\cdots$};
                          \node[label=below:{$2k-8$}][sp](spk-4)[right of=dots]{};
                          \node[label=below:{$2k-7$}][so](so2k-7)[right of=spk-4]{};
                          \node[label=below:{$2k-8$}][sp](spk-4')[right of=so2k-7]{};
                          \node(dots')[right of=spk-4']{$\cdots$};
                          \node[label=below:{3}][so](so3')[right of=dots']{};
                          \node[label=below:{2}][sp](sp1')[right of=so3']{};
                          \node[label=below:{1}][so](o1')[right of=sp1']{};
                          \node[label=left:{2}][sp](sp11)[above of=so2k-7]{};
                          \node[label=above:{7}][sof](sof)[above of=sp11]{};
                          \draw(o1)--(sp1);
                          \draw(sp1)--(so3);
                          \draw(so3)--(dots);
                          \draw(dots)--(spk-4);
                          \draw(spk-4)--(so2k-7);
                          \draw(spk-4')--(so2k-7);
                          \draw(dots')--(spk-4');
                          \draw(so3')--(dots');
                          \draw(sp1')--(so3');
                          \draw(o1')--(sp1');
                          \draw(so2k-7)--(sp11);
                          \draw(sp11)--(sof);
             \end{tikzpicture}
         \end{scriptsize}  }
    \end{array}  $  & $\begin{matrix}
            \mathrm{Sp}(2k-8)/\mathbb{Z}_2\times \mathrm{SO}(3)_R\\
            \uparrow\\
            \mathrm{Sp}(2k-8)/\mathbb{Z}_2\times \mathrm{SO}(3)_R
        \end{matrix}$ \\ \hline
  $ \raisebox{-.5\height}{ \begin{tikzpicture}
             \node(gauge){SO$(2k+1)$};
             \node(flavour)[above of=gauge]{$[2k-7]$};
             \draw (gauge)--(flavour);
         \end{tikzpicture}}  $  & $  \begin{array}{c}
         \raisebox{-.5\height}{ \begin{scriptsize}
    \begin{tikzpicture}
    \node[label=below:{$1$}][u](1){};
    \node[label=below:{$2$}][u](2)[right of=1]{};
    \node (dots)[right of=2]{$\cdots$};
    \node[label=below:{$2k-8$}][u](2N-5)[right of=dots]{};
    \node[label=below:{$2k-7$}][u](2N-4)[right of=2N-5]{};
    \node[label=below:{$2$}][u](2')[right of=2N-4]{};
    \node[right of=2']{$3\;\textbf{Adj}$};
    \draw (2')to[out=-45,in=45,loop,looseness=10](2');
    \draw(1)--(2);
    \draw(2)--(dots);
    \draw(dots)--(2N-5);
    \draw[ double distance=1.5pt,<-](2N-5)--(2N-4);
    \draw(2N-4)--(2');
    \end{tikzpicture}
    \end{scriptsize} }
    \end{array}  $  & $\begin{matrix}
            (\mathrm{Sp}(2k-7)\times \mathrm{SU}(2)_R)/\mathbb{Z}_2\\
            \uparrow\\
            \mathrm{Sp}(2k-7)/\mathbb{Z}_2\times \mathrm{SO}(3)_R
        \end{matrix}$ \\ \hline
    \end{tabular}
    \caption{Magnetic quiver and global symmetry change for $6d$ $\mathcal{N}=(1,0)$ UV fixed point SCFT. The global symmetry algebra remains the same, however in one case the global form of the group changes due to tensionless strings.}
    \label{tab:6d2}
\end{table}

\end{landscape}

\FloatBarrier

\section{Magnetic quivers and O$7^+$ planes}\label{sec:intersection_rule}

The way to read a magnetic quiver from a brane web was first discussed in \cite{Cabrera:2018jxt} (see also \cite{VanBeest:2020kxw,vanBeest:2020kou}). There are two simple steps: first one identifies the nodes from the different possible subwebs that can move in a Higgs branch direction (the rank of the node comes from the number of equal copies of said subweb that are able to move independently); and second one computes the number of edges linking two nodes from the \emph{stable intersection} between the two subwebs, plus a 7-brane contribution. The stable intersection is a counting of the number of points at which the branes intersect, including a multiplicity coming from the charges of the fivebranes,
\begin{align}
    \mathrm{SI}_0 = \sum_{\text{intersections}} \left|\det\left(  \begin{array}{cc}
        p_1 & q_1 \\
        p_2 & q_2
    \end{array}\right)\right|,
\end{align}
and the second contribution arises when two subwebs end on the same 7-brane: it is positive when they end on different sides of it, and negative otherwise.

When we add orientifolds to the brane web, this story is slightly modified. There are two possibilities compatible with supersymmetry: either to add O5 (or ON, or $(p,q)5$ version) planes parallel to the D5 (or NS5, or $(p,q)5$) branes or O7 planes transverse to the plane of the brane web. The case of O5 planes was studied in \cite{Akhond:2020vhc,Bourget:2020gzi,Akhond:2021knl,Akhond:2022jts}, and leads to orthosymplectic magnetic quivers. The case of an O7$^-$ is particularly simple, because it famously splits into two $[p,q]7$-branes \cite{Sen:1996vd} and one can use the usual rules to derive the O$7^-$ contribution. In what follows, we set out to explore the case of O7$^+$ planes. We will argue that the rules to read the magnetic quiver are the same as for a usual brane web, except that when two subwebs intersect on top of the O7$^+$ plane, the stable intersection gets modified as
\begin{align}\label{eq:intersection_modified_O7}
    \mathrm{SI}_{\text{O7}^{\pm}} = \delta^\pm(\mathrm{SI}_0\pm2|q_1 q_2|)\,, \qquad\textnormal{where}\quad \delta^\pm=\begin{cases}
        1\; \textnormal{for O}7^+\\
        \frac{1}{2}\; \textnormal{for O}7^-
    \end{cases}
\end{align}
involving only the NS charges of the fivebranes. We shall reach this conclusion using the following strategy:

\begin{enumerate}
    \item We will consider a family of six dimensional SCFTs, depending on two parameters $(n,k)$. These are straight-forward systems of D6's, D8's and NS5's, and their magnetic quivers are known.

    \item Upon a twisted circle compactification, we have a 5d KK theory for which we can draw a brane web (even if we cannot bring it to the SCFT point). In this brane web we can easily track various mass deformations, which lead to brane webs which we are able to bring to the SCFT point.

    \item Taking as a starting point the magnetic quiver of the 6d SCFT, we can keep track of the various deformations. Twisted compactification of the theory corresponds to a folding of the magnetic quiver, and mass deformations to Fayet-Iliopoulos (FI) deformations. Moreover, we can track the latter using the algorithm of quiver subtraction. Consistency of these deformations on the web and magnetic quiver sides will lead us to \eqref{eq:intersection_modified_O7}. 
\end{enumerate}

An important remark is that equation \eqref{eq:intersection_modified_O7} is manifestly not $SL(2,\mathbb{Z})$ invariant, due to the special role played by the NS charges. The $SL(2,\mathbb{Z})$ invariant form is discussed at length in subsection \ref{sec:SL2_invariance}.

\subsection{Starting point: 6d SCFT}

Our starting point is the 6d SCFT associated to $2n$ M5 branes next to a $\mathbb{C}^2/Z_{2k+2n}$ singularity in M-theory. Upon circle reduction to Type IIA this becomes a system of $(2k+2n)$ D6 branes intersecting $2n$ NS5 branes
\begin{equation}
    \label{eq:6dSCFTforZ2}
    \begin{tikzpicture}
        \node[7brane] (1) at (1,0) {};
        \node[7brane] (2) at (2,0) {};
        \node (3) at (3,0) {$\cdots$};
        \node[7brane] (4) at (4,0) {};
        \node[7brane] (5) at (5,0) {};
        \draw (0,0)--(1)--(2)--(3)--(4)--(5)--(6,0);
        \node[rotate=90] at (1.5,0.8) {\scriptsize $2k+2n$ D6};
        \draw [decorate,decoration={brace,amplitude=6pt},xshift=0pt,yshift=0pt]
(5.2,-0.2) -- (0.8,-0.2)node [black,midway,xshift=0pt,yshift=-15pt] {
\scriptsize $2n$ NS5};
    \end{tikzpicture}\;,
\end{equation}
with gauge theory description
\begin{equation}
    \begin{tikzpicture}
        \node[uf,label=left:{\scriptsize $2k+2n$}] (0) at (0,0) {};
        \node[u,label=above:{\scriptsize SU$(2k+2n)$}] (1) at (1,0) {};
        \node (2) at (2,0) {$\cdots$};
        \node[u,label=above:{\scriptsize SU$(2k+2n)$}] (3) at (3,0) {};
        \node[uf,label=right:{\scriptsize $2k+2n$}] (4) at (4,0) {};
        \draw (0)--(1)--(2)--(3)--(4);
        \draw [decorate,decoration={brace,amplitude=6pt},xshift=0pt,yshift=0pt]
(3.2,-0.2) -- (0.8,-0.2)node [black,midway,xshift=0pt,yshift=-15pt] {
\scriptsize $2n-1$ nodes};
    \end{tikzpicture}
\end{equation}
on its tensor branch. The Higgs phase is reached in the brane system by making the semi-infinite D6 branes end on D8 branes, and then moving D6 segments between D8 branes as well as moving the NS5 branes
\begin{equation}
    \begin{tikzpicture}
        \draw (0,-1)--(0,1) (1,-1)--(1,1) (2,-1)--(2,1) (4,-1)--(4,1) (5,-1)--(5,1) (7,-1)--(7,1) (8,-1)--(8,1) (10,-1)--(10,1) (11,-1)--(11,1) (12,-1)--(12,1);
        \node[7brane] (1) at (5.4,0.2) {};
        \node[7brane] (2) at (5.8,0.4) {};
        \node[7brane] (3) at (6.2,0.3) {};
        \node[7brane] (4) at (6.6,0.8) {};
        \node at (6,1) {\scriptsize $2n$ NS5};
        \draw (0,0)--(1,0) (1,0.2)--(2,0.2) (1,0.6)--(2,0.6);
        \node at (3,0) {$\cdots$};
        \draw (5,0)--(4,0) (5,-0.6)--(4,-0.6);
        \node at (4.5,-0.3) {$\vdots$};
        \node[rotate=60] at (4.5,-1.3) {\scriptsize $2k+2n-1$};
        \draw (5,-0.1)--(7,-0.1) (5,-0.7)--(7,-0.7);
        \node at (6,-0.4) {$\vdots$};
        \node at (6,-0.9) {\scriptsize $2k+2n$ D6};
        \draw (8,0.7)--(7,0.7) (8,0.1)--(7,0.1);
        \node at (7.5,0.4) {$\vdots$};
        \node[rotate=60] at (7.5,-0.6) {\scriptsize $2k+2n-1$};
        \node at (9,0) {$\cdots$};
        \draw (12,-0.3)--(11,-0.3) (11,0.2)--(10,0.2) (11,0.4)--(10,0.4);
    \end{tikzpicture}\;.
\end{equation}
The Higgs branch emanating from the origin of the tensor branch is captured when all NS5 branes share their $x^6$ position
\begin{equation}
\label{eq:6dSCFTforZ2BW}
    \begin{tikzpicture}
        \draw (0,-1)--(0,1) (1,-1)--(1,1) (2,-1)--(2,1) (4,-1)--(4,1) (5,-1)--(5,1) (7,-1)--(7,1) (8,-1)--(8,1) (10,-1)--(10,1) (11,-1)--(11,1) (12,-1)--(12,1);
        \node[7brane] (1) at (6,0.1) {};
        \node[7brane] (2) at (6,0.4) {};
        \node (3) at (6,0.7) {$\vdots$};
        \node[7brane] (4) at (6,1) {};
        \node at (6,1.3) {\scriptsize $2n$ NS5};
        \draw (0,0)--(1,0) (1,0.2)--(2,0.2) (1,0.6)--(2,0.6);
        \node at (3,0) {$\cdots$};
        \draw (5,0)--(4,0) (5,-0.6)--(4,-0.6);
        \node at (4.5,-0.3) {$\vdots$};
        \node[rotate=60] at (4.5,-1.3) {\scriptsize $2k+2n-1$};
        \draw (5,-0.1)--(7,-0.1) (5,-0.7)--(7,-0.7);
        \node at (6,-0.4) {$\vdots$};
        \node at (6,-0.9) {\scriptsize $2k+2n$ D6};
        \draw (8,0.7)--(7,0.7) (8,0.1)--(7,0.1);
        \node at (7.5,0.4) {$\vdots$};
        \node[rotate=60] at (7.5,-0.6) {\scriptsize $2k+2n-1$};
        \node at (9,0) {$\cdots$};
        \draw (12,-0.3)--(11,-0.3) (11,0.2)--(10,0.2) (11,0.4)--(10,0.4);
    \end{tikzpicture}\;.
\end{equation}

The magnetic quiver for the SCFT Higgs branch can be read from this brane system \cite{Hanany:2018vph, Cabrera:2019izd}, and is given by 
\begin{equation}
    \begin{array}{c}
         \begin{scriptsize}
             \begin{tikzpicture}
                          \node[label=below:{1}][u](1){};
                          \node[label=below:{2}][u](2)[right of=1]{};
                          \node(dots)[right of=2]{$\cdots$};
                          \node[label=above:{$2k+2n-1$}][u](2k-2)[right of=dots]{};
                          \node[label=below:{$2k+2n$}][u](2k-1)[right of=2k-2]{};
                          \node[label=above:{$2k+2n-1$}][u](2k-2')[right of=2k-1]{};
                          \node(dots')[right of=2k-2']{$\cdots$};
                          \node[label=below:{2}][u](2')[right of=dots']{};
                          \node[label=below:{1}][u](1')[right of=2']{};
                          \node[label=left:{$2n$}][u](22)[above of=2k-1]{};
                          \draw(1)--(2);
                          \draw(2)--(dots);
                          \draw(dots)--(2k-2);
                          \draw(2k-2)--(2k-1);
                          \draw(2k-1)--(2k-2');
                          \draw(1')--(2');
                          \draw(2')--(dots');
                          \draw(dots')--(2k-2');
                          \draw(22)--(2k-1);
                          \draw (22)to[out=135,in=45,loop,looseness=10](22);
             \end{tikzpicture}
         \end{scriptsize}
    \end{array}\;.\label{MQ6 SU(2k+2n) quiver}
\end{equation}

\subsection{Twisted compactification and mass deformations}

The $\mathbb{Z}_2$ twisted compactification of the $6d$ theory living on \eqref{eq:6dSCFTforZ2} was studied in \cite{Hayashi:2015vhy}. It can be described by the following 5d brane web 
\begin{align}\label{eq:+KK web opened up}
\begin{array}{c}
     \begin{scriptsize}
         \begin{tikzpicture}[scale=.5]
             \draw(0.5,1)--(-.5,1);
             \draw(.5,1)--(1,.5);
             \draw(-.5,1)--(-1,.5);
             \draw(1,.5)--(-1,.5);
             \draw(1,.5)--(2,0);
             \draw(-1,.5)--(-2,0);
             \draw(-.5,1)--(-.5,7);
             \draw(.5,1)--(.5,7);
             \draw(-1.5,1.5)--(1.5,1.5);
             \draw(-1.5,1.5)--(-1,2);
             \draw(1,2)--(1.5,1.5);
             \draw(1,2)--(-1,2);
             \draw(1,2)--(1,7);
             \draw(-1,7)--(-1,2);
             \draw(1.5,1.5)--(4.5,0);
             \draw(-1.5,1.5)--(-4.5,0);
             \node at (0,3){$\vdots$};
             \node at (5,2){$\udots$};
             \node at (-5,2){$\ddots$};
             \node at (2,5){$\cdots$};
             \node at (-2,5){$\cdots$};
             \draw(-5,4)--(5,4);
             \draw(5,4)--(13,0);
             \draw(-5,4)--(-13,0);
             \draw(5,4)--(4.5,4.5);
             \draw(-5,4)--(-4.5,4.5);
             \draw(4.5,4.5)--(-4.5,4.5);
             \draw(-4.5,7)--(-4.5,4.5);
             \draw(4.5,4.5)--(4.5,7);
             \draw[red](4.5,7)--(5.5,7);
             \draw(4.5,7)--(4,7.5);
             \draw(-4.5,7)--(-5.5,7);
             \draw(-4.5,7)--(-4,7.5);
             \draw[thick](-13,6)--(13,6);
             \node[label=above:{$k$ D5}]at(12,6){};
             \node[label=below:{O7$^+$}][7brane,green]at(0,0){};
             \draw[dash dot,green](-13.5,0)--(13.5,0);
             \draw [thick,decorate,decoration={brace,amplitude=6pt},xshift=0pt,yshift=10pt]
(-5.5,8) -- (5.5,8)node [black,midway,xshift=0pt,yshift=15pt] {
$2n$ NS5};
         \end{tikzpicture}
     \end{scriptsize}
\end{array}
\end{align}
This is a 5d KK theory which has a low energy description in terms of the following quiver,
\begin{equation}\label{+ KK EQ}
    \begin{array}{c}
\begin{tikzpicture}
    \node{${\text{SO}(2k+4n)}-\text{SU$(2k+4n-4)$}-\text{SU$(2k+4n-8)$}-\cdots-\text{SU$(2k+8)$}-{\underset{\underset{\text{\large$\left[(2k+2) \textbf{F}\right]$}}{\textstyle\vert}}{\text{SU}(2k+4)}}$};
    
    \draw [thick,decorate,decoration={brace,amplitude=6pt},xshift=0pt,yshift=10pt]
(-6,0.5) -- (6,0.5)node [black,midway,xshift=0pt,yshift=15pt] {
$n$};
    \end{tikzpicture}
    \end{array}\;.
\end{equation}
We cannot bring this web to the SCFT point, as it cannot be convexified (see Appendix \ref{app:BraneWebs}). Its $5d$ $\mathcal{N}=1$ worldvolume theory is a marginal theory, i.e.\ its UV completion is a $6d$ $\mathcal{N}=(1,0)$ SCFT.

We can do a mass deformation, sending the flavour brane indicated in red in the brane web \eqref{eq:+KK web opened up} vertically towards infinity. In doing this, we reach the first descendant of the marginal theory, which is a 5d theory with a 5d UV completion. The brane web for the SCFT point is
\begin{equation}\label{+nk brane web bis}
   \begin{array}{c}\begin{scriptsize}
    \begin{tikzpicture}[scale=1.2]
   \draw[thick](0,-2)--(1,-1);
   \node[7brane] at (1,-1){};
   \node[label=above:{$2n$}]at(.5,-1.5){};
    \node[label=below right:{O7$^+$}][7brane,green] at (0,-2){};
    \draw[thick](0,-2)--(-2,-2);
    \draw[thick](-5,-2)--(-3,-2);
    \node at (-2.5,-2){$\cdots$};
    \node[7brane] at (-1,-2){};
    \node[7brane] at (-2,-2){};
    \node[7brane] at (-3,-2){};
    \node[7brane] at (-4,-2){};
    \node[7brane] at (-5,-2){};
    \node [label=below:{1}] at (-4.5,-2){};
    \node [label=below:{2}] at (-3.5,-2){};
    \node [label=above:{$2k+2n-1$}] at (-1.5,-2){};
    \node [label=below:{$2k+2n$}] at (-.5,-2){};
    \draw[dash dot,green](-5.5,-2)--(1.5,-2);
    \end{tikzpicture}
    \end{scriptsize}\end{array}\;.
\end{equation}

\paragraph{Magnetic quiver.}
The magnetic quiver for this $5d$ $\mathcal{N}=1$ SCFT -- which is reached by $\mathbb{Z}_2$ twisted compactification of the $6d$ $\mathcal{N}=(1,0)$ SCFT and subsequent mass deformation to reach the first descendant of the marginal theory -- can be obtained from the brane system \eqref{eq:6dSCFTforZ2BW} of the $6d$ $\mathcal{N}=(1,0)$ SCFT by considering the $\mathbb{Z}_2$ invariant moduli.\footnote{This is analogous to reading magnetic quivers of twisted compactifications of $5d$ $\mathcal{N}=1$ theories living on brane webs to $4d$ $\mathcal{N}=2$ theories, studied in \cite{Bourget:2020asf,Bourget:2020mez}.} I.e.\ D6 segments are related to each other by $\mathbb{Z}_2$ reflection,
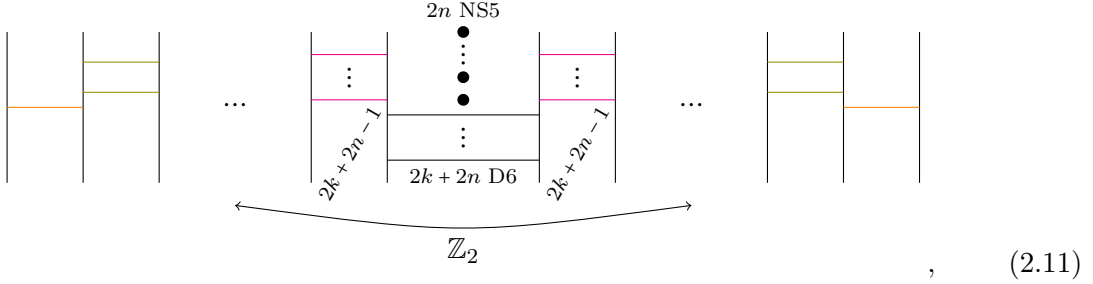
\begin{equation}
    \begin{tikzpicture}
        \draw (0,-1)--(0,1) (1,-1)--(1,1) (2,-1)--(2,1) (4,-1)--(4,1) (5,-1)--(5,1) (7,-1)--(7,1) (8,-1)--(8,1) (10,-1)--(10,1) (11,-1)--(11,1) (12,-1)--(12,1);
        \node[7brane] (1) at (6,0.1) {};
        \node[7brane] (2) at (6,0.4) {};
        \node (3) at (6,0.7) {$\vdots$};
        \node[7brane] (4) at (6,1) {};
        \node at (6,1.3) {\scriptsize $2n$ NS5};
        \draw[orange] (0,0)--(1,0);
        \draw[olive] (1,0.2)--(2,0.2) (1,0.6)--(2,0.6);
        \node at (3,0) {$\cdots$};
        \draw[magenta] (5,0.7)--(4,0.7) (5,0.1)--(4,0.1);
        \node at (4.5,0.4) {$\vdots$};
        \node[rotate=60] at (4.5,-0.6) {\scriptsize $2k+2n-1$};
        \draw (5,-0.1)--(7,-0.1) (5,-0.7)--(7,-0.7);
        \node at (6,-0.4) {$\vdots$};
        \node at (6,-0.9) {\scriptsize $2k+2n$ D6};
        \draw[magenta]  (8,0.7)--(7,0.7) (8,0.1)--(7,0.1);
        \node at (7.5,0.4) {$\vdots$};
        \node[rotate=60] at (7.5,-0.6) {\scriptsize $2k+2n-1$};
        \node at (9,0) {$\cdots$};
        \draw[orange] (12,0)--(11,0);
        \draw[olive] (11,0.2)--(10,0.2) (11,0.6)--(10,0.6);
        \draw[<->] (3,-1.3)  .. controls (6,-1.7) ..  (9,-1.3);
        \node at (6,-1.9) {$\mathbb{Z}_2$};
    \end{tikzpicture}\;,
\end{equation}
with colours indicating $\mathbb{Z}_2$ invariant combinations of D6 branes. The magnetic quiver read from this configuration is
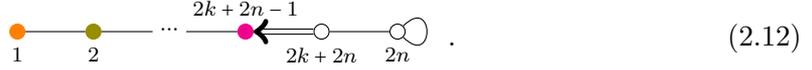
\begin{equation}\label{MQ5 + next to marginal}
    \begin{array}{c}
         \begin{scriptsize}
             \begin{tikzpicture}
                          \node[label=below:{1}][u,orange](1){};
                          \node[label=below:{2}][u,olive](2)[right of=1]{};
                          \node(dots)[right of=2]{$\cdots$};
                          \node[label=above:{$2k+2n-1$}][u,magenta](2k-2)[right of=dots]{};
                          \node[label=below:{$2k+2n$}][u](2k-1)[right of=2k-2]{};
                          \node[label=below:{$2n$}][u](22)[right of=2k-1]{};
                          \draw(1)--(2);
                          \draw(2)--(dots);
                          \draw(dots)--(2k-2);
                          \draw[<-,double distance=1.5pt](2k-2)--(2k-1);
                          \draw(22)--(2k-1);
                          \draw (22)to[out=45,in=-45,loop,looseness=10](22);
             \end{tikzpicture}
         \end{scriptsize}
    \end{array}\;.
\end{equation}
This quiver is the folding \cite{Bourget:2020bxh} of the magnetic quiver \eqref{MQ6 SU(2k+2n) quiver} of the 6d theory.\footnote{Note that the magnetic quiver of a $\mathbb{Z}_k$ twisted compactification is not necessarily a folding of the magnetic quiver of the original theory. See for example the $\mathbb{Z}_4$ twisted compactification discussed in \cite{Bourget:2020asf,Bourget:2020mez}. Considering the $\mathbb{Z}_k$ invariant moduli of the brane system of the original theory is therefore a more reliable method to produce the desired magnetic quiver.} We conclude that \eqref{MQ5 + next to marginal} is the magnetic quiver for the theory living on \eqref{+nk brane web bis}.

\paragraph{Further mass deformations.}
Further mass deformations are most easily seen in the opened up web \eqref{eq:+KK web opened up}. Similarly to the branes in red, we can also move vertically the symmetrical segment on the left side of the web, resulting in

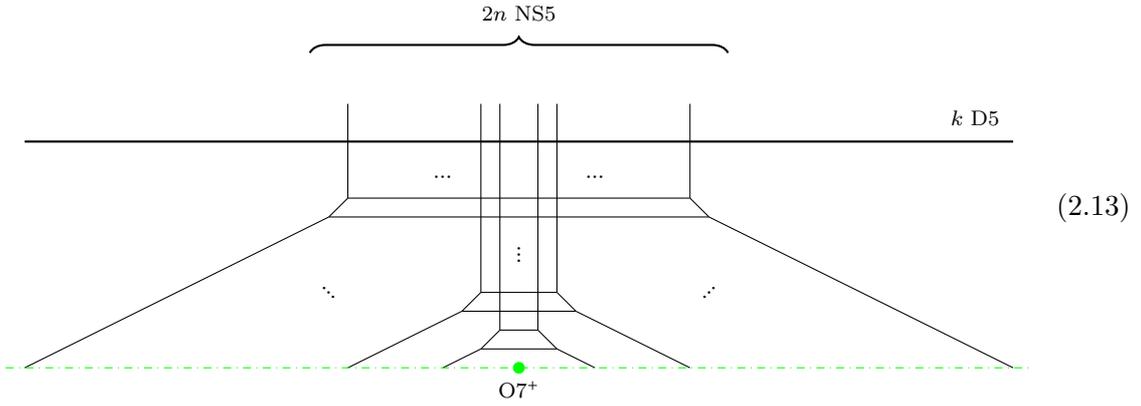
\begin{align}\label{eq:+nk web opened up}
\begin{array}{c}
     \begin{scriptsize}
         \begin{tikzpicture}[scale=.5]
             \draw(0.5,1)--(-.5,1);
             \draw(.5,1)--(1,.5);
             \draw(-.5,1)--(-1,.5);
             \draw(1,.5)--(-1,.5);
             \draw(1,.5)--(2,0);
             \draw(-1,.5)--(-2,0);
             \draw(-.5,1)--(-.5,7);
             \draw(.5,1)--(.5,7);
             \draw(-1.5,1.5)--(1.5,1.5);
             \draw(-1.5,1.5)--(-1,2);
             \draw(1,2)--(1.5,1.5);
             \draw(1,2)--(-1,2);
             \draw(1,2)--(1,7);
             \draw(-1,7)--(-1,2);
             \draw(1.5,1.5)--(4.5,0);
             \draw(-1.5,1.5)--(-4.5,0);
             \node at (0,3){$\vdots$};
             \node at (5,2){$\udots$};
             \node at (-5,2){$\ddots$};
             \node at (2,5){$\cdots$};
             \node at (-2,5){$\cdots$};
             \draw(-5,4)--(5,4);
             \draw(5,4)--(13,0);
             \draw(-5,4)--(-13,0);
             \draw(5,4)--(4.5,4.5);
             \draw(-5,4)--(-4.5,4.5);
             \draw(4.5,4.5)--(-4.5,4.5);
             \draw(-4.5,7)--(-4.5,4.5);
             \draw(4.5,4.5)--(4.5,7);
             \draw[thick](-13,6)--(13,6);
             \node[label=above:{$k$ D5}]at(12,6){};
             \node[label=below:{O7$^+$}][7brane,green]at(0,0){};
             \draw[dash dot,green](-13.5,0)--(13.5,0);
             \draw [thick,decorate,decoration={brace,amplitude=6pt},xshift=0pt,yshift=10pt]
(-5.5,8) -- (5.5,8)node [black,midway,xshift=0pt,yshift=15pt] {
$2n$ NS5};
         \end{tikzpicture}
     \end{scriptsize}
\end{array}
\end{align}
This is a 5d theory with the following quiver gauge theory description
\begin{equation}\label{+ KK EQ_eft}
    \begin{array}{c}
\begin{tikzpicture}
    \node{${\text{SO}(2k+4n)}-\text{SU$(2k+4n-4)$}-\text{SU$(2k+4n-8)$}-\cdots-\text{SU$(2k+8)$}-{\underset{\underset{\text{\large$\left[(2k) \textbf{F}\right]$}}{\textstyle\vert}}{\text{SU}(2k+4)}}$};
    
    \draw [thick,decorate,decoration={brace,amplitude=6pt},xshift=0pt,yshift=10pt]
(-6,0.5) -- (6,0.5)node [black,midway,xshift=0pt,yshift=15pt] {
$n$};
    \end{tikzpicture}
    \end{array}\;.
\end{equation}
Here we have drawn the theory on the Coulomb branch of the gauge theory phase for clarity.  We can also bring the web to the SCFT point,
\begin{equation}\label{+nk brane web 2}
    \begin{array}{c}\begin{scriptsize}
    \begin{tikzpicture}
    \draw[thick](0,1)--(0,3);
    \draw[thick](0,0)--(0,-2);
    \node[7brane] at (0,3){};
    \node[7brane] at (0,2){};
    \node[7brane] at (0,1){};
    \node[7brane] at (0,0){};
    \node[7brane] at (0,-1){};
    \node[label=below right:{O7$^+$}][7brane,green] at (0,-2){};
    \draw[dash dot,green](-5.5,-2)--(1,-2);
    \draw[thick](0,-2)--(-2,-2);
    \draw[thick](-5,-2)--(-3,-2);
    \node at (-2.5,-2){$\cdots$};
    \node[7brane] at (-1,-2){};
    \node[7brane] at (-2,-2){};
    \node[7brane] at (-3,-2){};
    \node[7brane] at (-4,-2){};
    \node[7brane] at (-5,-2){};
    \node [label=below:{1}] at (-4.5,-2){};
    \node [label=below:{2}] at (-3.5,-2){};
    \node [label=below:{$2k-1$}] at (-1.5,-2){};
    \node [label=below:{$2k$}] at (-.5,-2){};
    \node at (0,.5){$\vdots$};
    \node[label=right:{$1$}] at (0,2.5){};
    \node[label=right:{$2$}] at (0,1.5){};
    \node[label=right:{$2n-1$}] at (0,-.5){};
    \node[label=right:{$2n$}] at (0,-1.5){};
    \end{tikzpicture}
    \end{scriptsize}\end{array}\;,
\end{equation}
where we have exploited both reflections in the directions transverse to the orientifold to draw the branes in the upper left quadrant of the total web.

What are the possible mass deformations of this theory? This is most easily seen in the opened up web \eqref{eq:+nk web opened up}. We can take a NS5-brane ending on a (0,1) sevenbrane and a D5-brane ending on a (1,0) sevenbrane from the top right of the web, and move them diagonally,
\begin{align}\label{eq:+nk web opened up bis}
\begin{array}{c}
     \begin{scriptsize}
         \begin{tikzpicture}[scale=.5]
             \draw(0.5,1)--(-.5,1);
             \draw(.5,1)--(1,.5);
             \draw(-.5,1)--(-1,.5);
             \draw(1,.5)--(-1,.5);
             \draw(1,.5)--(2,0);
             \draw(-1,.5)--(-2,0);
             \draw(-.5,1)--(-.5,7);
             \draw(.5,1)--(.5,7);
             \draw(-1.5,1.5)--(1.5,1.5);
             \draw(-1.5,1.5)--(-1,2);
             \draw(1,2)--(1.5,1.5);
             \draw(1,2)--(-1,2);
             \draw(1,2)--(1,7);
             \draw(-1,7)--(-1,2);
             \draw(1.5,1.5)--(4.5,0);
             \draw(-1.5,1.5)--(-4.5,0);
             \node at (0,3){$\vdots$};
             \node at (5,2){$\udots$};
             \node at (-5,2){$\ddots$};
             \node at (2,5){$\cdots$};
             \node at (-2,5){$\cdots$};
             \draw(-5,4)--(5,4);
             \draw(5,4)--(13,0);
             \draw(-5,4)--(-13,0);
             \draw(5,4)--(4.5,4.5);
             \draw(-5,4)--(-4.5,4.5);
             \draw(4.5,4.5)--(-4.5,4.5);
             \draw(-4.5,7)--(-4.5,4.5);
             \draw(4.5,4.5)--(4.5,6);
             \draw[thick](-13,5.5)--(13,5.5);
             \draw(-13,6)--(4.5,6);
             \node[label=above:{$k-1$ D5}]at(12,5.5){};
             \node[label=below:{O7$^+$}][7brane,green]at(0,0){};
             \draw[dash dot,green](-13.5,0)--(13.5,0);
             \draw(4.5,6)--(5.5,7);
             \draw(5.5,7)--(6,7);
             \draw(5.5,7)--(5.5,7.5);
             \draw[thick,->](6.5,6.5)--(7,7);
         \end{tikzpicture}
     \end{scriptsize}
\end{array}
\end{align}
We can take them to infinity, resulting in a web where both the number of NS5 and D5 branes have been reduced by 1, and instead we have a semi-infinite (1,1) fivebrane in the top right. We can do the same with a NS5 and D5 from the top left of the web, and go to the SCFT point,

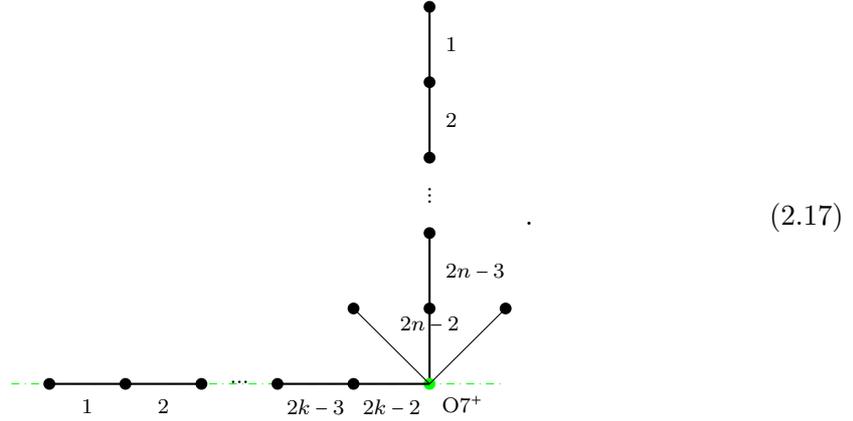
\begin{equation}\label{eq:+nk_1st_deformation}
    \begin{array}{c}\begin{scriptsize}
    \begin{tikzpicture}
    \draw[thick](0,1)--(0,3);
    \draw[thick](0,0)--(0,-2);
    \node[7brane] at (0,3){};
    \node[7brane] at (0,2){};
    \node[7brane] at (0,1){};
    \node[7brane] at (0,0){};
    \node[7brane] at (0,-1){};
    \node[label=below right:{O7$^+$}][7brane,green] at (0,-2){};
    \draw[dash dot,green](-5.5,-2)--(1,-2);
    \draw[thick](0,-2)--(-2,-2);
    \draw[thick](-5,-2)--(-3,-2);
    \node at (-2.5,-2){$\cdots$};
    \node[7brane] at (-1,-2){};
    \node[7brane] at (-2,-2){};
    \node[7brane] at (-3,-2){};
    \node[7brane] at (-4,-2){};
    \node[7brane] at (-5,-2){};
    \node [label=below:{1}] at (-4.5,-2){};
    \node [label=below:{2}] at (-3.5,-2){};
    \node [label=below:{$2k-3$}] at (-1.5,-2){};
    \node [label=below:{$2k-2$}] at (-.5,-2){};
    \node at (0,.5){$\vdots$};
    \node[label=right:{$1$}] at (0,2.5){};
    \node[label=right:{$2$}] at (0,1.5){};
    \node[label=right:{$2n-3$}] at (0,-.5){};
    \node[label=:{$2n-2$}] at (0,-1.5){};
    \draw(0,-2)--(1,-1);
    \draw(0,-2)--(-1,-1);
    \node[7brane]at(-1,-1){};
    \node[7brane]at(1,-1){};
    \end{tikzpicture}
    \end{scriptsize}\end{array}\;.
\end{equation}
where we have made the semi-infinite $(1,\pm1)$ fivebranes end on a corresponding sevenbrane.
Further mass deformations are simple; they correspond to taking an NS5 or D5 brane and move it towards infinity along the left or right diagonal brane. These correspond to introducing a mass and integrating out either a flavour or an instanton particle in the 5d field theory. There are four possibilities in total, each of which will modify the RR or NSNS charge of the left and right diagonal branes according to the number of masses we take to infinity. After an arbitrary number of any such deformation, the brane web will become
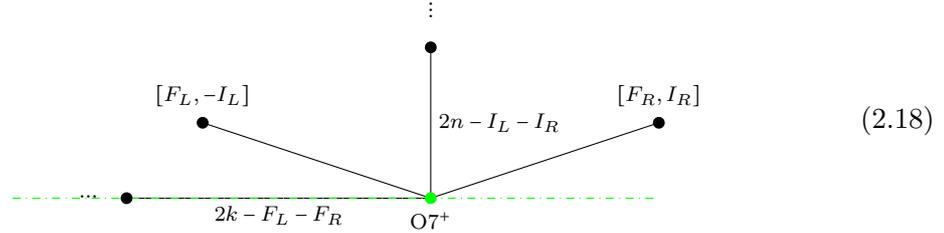
\begin{equation}\label{brane web (F,I) bis}
    \begin{array}{c}
         \begin{scriptsize}
             \begin{tikzpicture}
                 \draw(-4,0)--node[below]{$2k-F_L-F_R$}++(4,0);
                 \draw(0,0)--node[right]{$2n-I_L-I_R$}++(0,2);
                 \draw(0,0)--(3,1);
                 \draw(0,0)--(-3,1);
                 \node[label=above:{{$[F_L,-I_L]$}}] at (-3,1){};
                 \node[label=above:{{$[F_R,I_R]$}}] at (3,1){};
                 \node[7brane,green][label=below:{O7$^+$}]at(0,0){};
                 \draw[dash dot,green](-5.5,0)--(3,0);
                 \node[7brane] at (-4,0){};
                 \node[7brane] at (0,2){};
                 \node[7brane] at (3,1){};
                 \node[7brane] at (-3,1){};
                 \node at (-4.5,0){$\cdots$};
                 \node at (0,2.5){$\vdots$};
             \end{tikzpicture}
         \end{scriptsize}
    \end{array}
\end{equation}
Here, $F_R-1$ is the number of D5 branes we have pulled to the right, $I_L-1$ is the number of NS5 branes we have pulled to the left, and so on.

\subsection{Magnetic quivers and FI deformations}
\label{sec:FIdef}

Mass deformations of the electric theory can be studied via FI deformations of the magnetic quiver. These FI deformations can be realised as \emph{FI quiver subtractions}. We refer the reader to \cite{Bourget:2023uhe} for a pedagogical introduction. This technique has been employed successfully in \cite{Bourget:2020mez,vanBeest:2021xyt,Giacomelli:2022drw} to track a variety of RG flows. In \cite{Bourget:2023uhe} a first step was made to systematically understand the complicated space of FI deformations of simply laced unitary quivers by studying solutions to F-term equations. This work showed that there is no simple rule to translate a given FI deformation to an FI quiver subtraction. In simple cases, however, the FI deformation is reached by subtracting a quiver with trivial Higgs branch from the original quiver.\footnote{Note that this type of `FI quiver subtraction' is very different from the quiver subtraction \cite{Bourget:2019aer} used to compute the Hasse diagram of symplectic leaves.}

In the following, we want to study FI deformations of non-simply laced quivers. In this case, one cannot do an analysis of F-term equations, as the quivers are not Lagrangian (except if nodes connected with non-simply laced edges are U(1) nodes). Inspired by the simply laced case, we will subtract non-simply laced linear quivers (possibly with adjoint loops)
\begin{equation}
    \begin{tikzpicture}
        \node[label=below:{$k$}][u](1){};
        \node(dots)[right of=1]{$\cdots$};
        \node[label=below:{$k$}][u](2)[right of=dots]{};
        \node[label=below:{$k$}][u](3)[right of=2]{};
        \node(dots2)[right of=3]{$\cdots$};
        \node[label=below:{$k$}][u](4)[right of=dots2]{};
        \node[label=below:{$k$}][u](5)[right of=4]{};
        \node(dots3)[right of=5]{$\cdots$};
        \node[label=below:{$k$}][u](6)[right of=dots3]{};
        \draw(1)--(dots);
        \draw(dots)--(2);
        \draw[<-,double distance=1.5 pt](2)--(3);
        \draw(3)--(dots2)--(4);
        \draw[->,double distance=1.5 pt](4)--(5);
        \draw(5)--(dots3)--(6);
        \draw[gray] (4)to[out=45,in=135,loop,looseness=10](4);
    \end{tikzpicture}\;,
\end{equation}
which have trivial Higgs branch if $k=1$, in which case the quiver is Lagrangian. After subtracting such a quiver, we rebalance with a U$(k)$ node (possibly with adjoint), keeping the shortness/length of the nodes the same as in the original quiver by employing non-simply laced edges. That these subtractions actually achieve FI deformations is conjectural and deserves attention in the future.

Based on the operation in the brane web, we propose that the first mass deformation of the first descendant of the marginal theory is realised as the magnetic quiver FI-quiver subtraction
\begin{equation}
    \begin{array}{c}
         \begin{scriptsize}
            \begin{tikzpicture}
                \node at (0,0) {$\begin{tikzpicture}
    \node[label=below:{$1$}][u](1){};
    \node[label=below:{$2$}][u](2)[right of=1]{};
    \node (dots)[right of=2]{$\cdots$};
    \node[label=below:{$2n$}][u](2n)[right of=dots]{};
    \node (dots2)[right of=2n]{$\cdots$};
    \node[label=above:{$2k+2n-1$}][u](2N-5)[right of=dots2]{};
    \node[label=below:{$2k+2n$}][u](2N-4)[right of=2N-5]{};
    \node[label=below:{$2n$}][u](2')[right of=2N-4]{};
    \draw (2')to[out=-45,in=45,loop,looseness=10](2');
    \draw(1)--(2);
    \draw(2)--(dots);
    \draw(dots)--(2n)--(dots2)--(2N-5);
    \draw[ double distance=1.5pt,<-](2N-5)--(2N-4);
    \draw(2N-4)--(2');
    \end{tikzpicture}$};
                \node at (1,-1) {$\begin{tikzpicture}
                          \node(minus){$-$};
                          \node[label=below:{$2n$}][u](1)[right of=minus]{};
                          \node(dots)[right of=1]{$\cdots$};
                          \node[label=below:{$2n$}][u](2)[right of=dots]{};
                          \node[label=below:{$2n$}][u](3)[right of=2]{};
                          \node[label=below:{$2n$}][u](4)[right of=3]{};
                          \draw(1)--(dots);
                          \draw(dots)--(2);
                          \draw[<-,double distance=1.5 pt](2)--(3);
                          \draw(3)--(4);
                        \draw (4)to[out=-45,in=45,loop,looseness=10](4);
             \end{tikzpicture}$};
            \end{tikzpicture}
         \end{scriptsize}
    \end{array}\;.
    \label{FI deformation_1}
\end{equation}
The result of the subtraction before rebalancing is
\begin{equation}
    \begin{scriptsize}
        \begin{tikzpicture}
    \node[label=below:{$1$}][u](1){};
    \node[label=below:{$2$}][u](2)[right of=1]{};
    \node (dots)[right of=2]{$\cdots$};
    \node[label=below:{$2n-1$}][u](2n-1)[right of=dots]{};
    \node[label=below:{$1$}][u](2n)[right of=2n-1]{};
    \node (dots2)[right of=2n]{$\cdots$};
    \node[label=above:{$2k-1$}][u](2N-5)[right of=dots2]{};
    \node[label=below:{$2k$}][u](2N-4)[right of=2N-5]{};
    \draw(1)--(2);
    \draw(2)--(dots);
    \draw(dots)--(2n-1) (2n)--(dots2)--(2N-5);
    \draw[ double distance=1.5pt,<-](2N-5)--(2N-4);
    \end{tikzpicture}
    \end{scriptsize}\;.
\end{equation}
We propose that after subtraction one should rebalance with a U$(2n)$ node with an adjoint, yielding
\begin{equation}\label{+nk MQ bis}
    \begin{array}{c}
         \begin{scriptsize}
    \begin{tikzpicture}
    \node[label=below:{1}][u](2){};
    \node (dots)[right of=2]{$\cdots$};
    \node[label=below:{$2k-1$}][u](2N-5)[right of=dots]{};
    \node[label=below:{$2k$}][u](2N-4)[right of=2N-5]{};
    \node[label=below:{$2n$}][u](3')[right of=2N-4]{};
    \node[label=below:{$2n-1$}][u](2')[right of=3']{};
    \node(dots')[right of=2']{$\cdots$};
    \node[label=below:{$2$}][u](22')[right of=dots']{};
    \node[label=below:{$1$}][u](1')[right of=22']{};
    \draw (3')to[out=45,in=135,loop,looseness=10](3');
    \draw(2)--(dots);
    \draw(dots)--(2N-5);
    \draw[ double distance=1.5pt,<-](2N-5)--(2N-4);
    \draw(2N-4)--(3');
    \draw[double distance=1.5pt,->](3')--(2');
    \draw(dots')--(2');
    \draw(dots')--(22');
    \draw(1')--(22');
    \end{tikzpicture}
    \end{scriptsize}
    \end{array}\;.
\end{equation}
Which we propose to be the magnetic quiver of \eqref{+nk brane web 2}.

We would like to remark that although this derivation of the magnetic quiver has a conjectural component (specially insofar FI quiver subtraction is concerned), for particular cases of low $n$ and $k$ one can test against alternative orthosymplectic magnetic quivers and find perfect agreement \cite{Akhond:2021ffo}; we will comment more on this in following subsections.

We can keep studying the possible mass deformations of the brane webs discussed above in terms of FI deformations and corresponding quiver subtractions in the magnetic quivers. The two deformations leading from \eqref{+nk brane web 2} to the web \eqref{eq:+nk_1st_deformation} correspond to subtracting
\begin{equation}
  2\times  \begin{array}{c}
         \begin{scriptsize}
    \begin{tikzpicture}
    \node[label=below:{1}][u](2){};
    \node (dots)[right of=2]{$\cdots$};
    \node[label=below:{$1$}][u](2N-5)[right of=dots]{};
    \node[label=below:{$1$}][u](2N-4)[right of=2N-5]{};
    \node[label=below:{$1$}][u](3')[right of=2N-4]{};
    \node[label=below:{$1$}][u](2')[right of=3']{};
    \node(dots')[right of=2']{$\cdots$};
    \node[label=below:{$1$}][u](22')[right of=dots']{};
    \node[label=below:{$1$}][u](1')[right of=22']{};
    \draw(2)--(dots);
    \draw(dots)--(2N-5);
    \draw[ double distance=1.5pt,<-](2N-5)--(2N-4);
    \draw(2N-4)--(3');
    \draw[double distance=1.5pt,->](3')--(2');
    \draw(dots')--(2');
    \draw(dots')--(22');
    \draw(1')--(22');
    \end{tikzpicture}
    \end{scriptsize}
    \end{array}\;\label{MQ5 SU(k)+ S+(k-4) F_1}
\end{equation}
from \eqref{+nk MQ bis}.\\

This notation stands for first subtracting the quiver
\begin{equation}
    \begin{array}{c}
         \begin{scriptsize}
    \begin{tikzpicture}
    \node[label=below:{1}][u](2) at (1,0) {};
    \node (dots)[right of=2]{$\cdots$};
    \node[label=below:{$1$}][u](2N-5)[right of=dots]{};
    \node[label=below:{$1$}][u](2N-4)[right of=2N-5]{};
    \node[label=below:{$1$}][u](3')[right of=2N-4]{};
    \node[label=below:{$1$}][u](2')[right of=3']{};
    \node(dots')[right of=2']{$\cdots$};
    \node[label=below:{$1$}][u](22')[right of=dots']{};
    \node[label=below:{$1$}][u](1')[right of=22']{};
    \draw(2)--(dots);
    \draw(dots)--(2N-5);
    \draw[ double distance=1.5pt,<-](2N-5)--(2N-4);
    \draw(2N-4)--(3');
    \draw[double distance=1.5pt,->](3')--(2');
    \draw(dots')--(2');
    \draw(dots')--(22');
    \draw(1')--(22');
    \draw [decorate,decoration={brace,amplitude=6pt},xshift=0pt,yshift=0pt]
(3.2,-0.4) -- (0.8,-0.4)node [black,midway,xshift=0pt,yshift=-15pt] {
\scriptsize $2k-1$};
    \draw [decorate,decoration={brace,amplitude=6pt},xshift=0pt,yshift=0pt]
(9.2,-0.4) -- (5.8,-0.4)node [black,midway,xshift=0pt,yshift=-15pt] {
\scriptsize $2n-1$};
    \end{tikzpicture}
    \end{scriptsize}
    \end{array}\;,
\end{equation}
and rebalancing with a U$(1)$ node, yielding
\begin{equation}\label{eq:FI_Inbetween}
    \begin{array}{c}
         \begin{scriptsize}
    \begin{tikzpicture}
    \node[label=below:{1}][u](2){};
    \node (dots)[right of=2]{$\cdots$};
    \node[label=below:{$2k-2$}][u](2N-5)[right of=dots]{};
    \node[label=below:{$2k-1$}][u](2N-4)[right of=2N-5]{};
    \node[label=below:{$2n-1$}][u](3')[right of=2N-4]{};
    \node[label=below:{$2n-2$}][u](2')[right of=3']{};
    \node(dots')[right of=2']{$\cdots$};
    \node[label=below:{$2$}][u](22')[right of=dots']{};
    \node[label=below:{$1$}][u](1')[right of=22']{};
    \draw (3')to[out=45,in=135,loop,looseness=10](3');
    \draw(2)--(dots);
    \draw(dots)--(2N-5);
    \draw[ double distance=1.5pt,<-](2N-5)--(2N-4);
    \draw(2N-4)--(3');
    \draw[double distance=1.5pt,->](3')--(2');
    \draw(dots')--(2');
    \draw(dots')--(22');
    \draw(1')--(22');
    \node[label=above:{$1$}][u](re) at (3,1) {};
    \draw (2N-4)--(re)--(3');
    \node at (3.5,0.8) {3};
    \end{tikzpicture}
    \end{scriptsize}
    \end{array}\;
\end{equation}
corresponding to turning on an FI deformation on the leftmost and rightmost U$(1)$ node in \eqref{+nk MQ bis}. And then from \eqref{eq:FI_Inbetween} subtracting the quiver
\begin{equation}
    \begin{array}{c}
         \begin{scriptsize}
    \begin{tikzpicture}
    \node[label=below:{1}][u](2) at (1,0) {};
    \node (dots)[right of=2]{$\cdots$};
    \node[label=below:{$1$}][u](2N-5)[right of=dots]{};
    \node[label=below:{$1$}][u](2N-4)[right of=2N-5]{};
    \node[label=below:{$1$}][u](3')[right of=2N-4]{};
    \node[label=below:{$1$}][u](2')[right of=3']{};
    \node(dots')[right of=2']{$\cdots$};
    \node[label=below:{$1$}][u](22')[right of=dots']{};
    \node[label=below:{$1$}][u](1')[right of=22']{};
    \draw(2)--(dots);
    \draw(dots)--(2N-5);
    \draw[ double distance=1.5pt,<-](2N-5)--(2N-4);
    \draw(2N-4)--(3');
    \draw[double distance=1.5pt,->](3')--(2');
    \draw(dots')--(2');
    \draw(dots')--(22');
    \draw(1')--(22');
    \draw [decorate,decoration={brace,amplitude=6pt},xshift=0pt,yshift=0pt]
(3.2,-0.4) -- (0.8,-0.4)node [black,midway,xshift=0pt,yshift=-15pt] {
\scriptsize $2k-2$};
    \draw [decorate,decoration={brace,amplitude=6pt},xshift=0pt,yshift=0pt]
(9.2,-0.4) -- (5.8,-0.4)node [black,midway,xshift=0pt,yshift=-15pt] {
\scriptsize $2n-2$};
    \end{tikzpicture}
    \end{scriptsize}
    \end{array}\;,
\end{equation}
and rebalancing with another U$(1)$ node, corresponding to turning on an FI deformation on the leftmost and rightmost U$(1)$ node in \eqref{eq:FI_Inbetween}.

The result of this double subtraction is
\begin{equation}\label{MQ +nk 1st_def}
    \begin{array}{c}
         \begin{scriptsize}
    \begin{tikzpicture}
    \node[label=below:{1}][u](2){};
    \node (dots)[right of=2]{$\cdots$};
    \node[label=below:{$2k-3$}][u](2N-5)[right of=dots]{};
    \node[label=below:{$2k-2$}][u](2N-4)[right of=2N-5]{};
    \node[label=below:{$2n-2$}][u](3')[right of=2N-4]{};
    \node[label=below:{$2n-3$}][u](2')[right of=3']{};
    \node[label=above:{$1_L$}][u](1L)[above of=2N-5]{};
    \node[label=above:{$1_R$}][u](1R)[above of=2']{};
    \draw(1L)--node[above]{$4$}++(1R);
    \draw(1L)--(2N-4);
    \draw(1R)--(2N-4);
    \draw(1L)--node[above]{3}++(3');
    \draw(1R)--node[right]{3}++(3');
    \node(dots')[right of=2']{$\cdots$};
    \node[label=below:{$2$}][u](22')[right of=dots']{};
    \node[label=below:{$1$}][u](1')[right of=22']{};
    \draw (3')to[out=45,in=135,loop,looseness=10](3');
    \draw(2)--(dots);
    \draw(dots)--(2N-5);
    \draw[ double distance=1.5pt,<-](2N-5)--(2N-4);
    \draw(2N-4)--(3');
    \draw[double distance=1.5pt,->](3')--(2');
    \draw(dots')--(2');
    \draw(dots')--(22');
    \draw(1')--(22');
    \end{tikzpicture}
    \end{scriptsize}
    \end{array}\;,
\end{equation}
with the two rebalancing nodes corresponding to the two new diagonal branes of \eqref{eq:+nk_1st_deformation}.

The rest of mass deformations leading to \eqref{brane web (F,I) bis} also correspond to certain quiver subtractions. 
Integrating out $(F_L-1)$ fundamentals from the left (namely sending D5 branes to infinity along the left diagonal) corresponds to iteratively subtracting
\begin{equation}
    (F_L-1)\times\begin{array}{c}
         \begin{scriptsize}
             \begin{tikzpicture}
                 \node[label=below:{1}][u](1){};
                 \node[label=below:{1}][u](2)[right of=1]{};
                 \node(dots)[right of=2]{$\cdots$};
                 \node[label=below:{1}][u](3)[right of=dots]{};
                 \node[label=below:{1}][u](4)[right of=3]{};
                 \node[label=above:{$1_L$}][u](5)[above of=4]{};
                 \draw(1)--(2);
                 \draw(dots)--(2);
                 \draw(dots)--(3);
                 \draw[double distance=1.5 pt,<-](3)--(4);
                 \draw(4)--(5);
             \end{tikzpicture}
         \end{scriptsize}
    \end{array}
\end{equation}
from \eqref{MQ +nk 1st_def} (with abuse of notation where after each subtraction (killing the $1_R$ node) we rebalance with a U$(1)$ node we then call $1_R$, and the quiver we subtract shortens after each subtraction, similar to the multi-subtraction case discussed just before). 

Similarly integrating out $F_R-1$ fundamental hypermultiplets form the right corresponds to iteratively subtracting
\begin{equation}
    (F_R-1)\times\begin{array}{c}
         \begin{scriptsize}
             \begin{tikzpicture}
                 \node[label=below:{1}][u](1){};
                 \node[label=below:{1}][u](2)[right of=1]{};
                 \node(dots)[right of=2]{$\cdots$};
                 \node[label=below:{1}][u](3)[right of=dots]{};
                 \node[label=below:{1}][u](4)[right of=3]{};
                 \node[label=above:{$1_R$}][u](5)[above right of=4]{};
                 \draw(1)--(2);
                 \draw(dots)--(2);
                 \draw(dots)--(3);
                 \draw[double distance=1.5 pt,<-](3)--(4);
                 \draw(4)--(5);
             \end{tikzpicture}
         \end{scriptsize}
    \end{array}\;.
\end{equation}
On the other hand, sending to infinity $(I_L-1)$ NS5 branes along the left diagonal is identified with the subtraction
\begin{equation}
    (I_L-1)\times\begin{array}{c}
         \begin{scriptsize}
             \begin{tikzpicture}
                 \node[label=below:{1}][u](1){};
                 \node[label=below:{1}][u](2)[left of=1]{};
                 \node(dots)[left of=2]{$\cdots$};
                 \node[label=below:{1}][u](3)[left of=dots]{};
                 \node[label=below:{1}][u](4)[left of=3]{};
                 \node[label=above:{$1_L$}][u](5)[above left of=4]{};
                 \draw(1)--(2);
                 \draw(dots)--(2);
                 \draw(dots)--(3);
                 \draw[double distance=1.5 pt,<-](3)--(4);
                 \draw(4)--(5);
             \end{tikzpicture}
         \end{scriptsize}
    \end{array}\;,
\end{equation}
while doing the same $I_R-1$ times along the right diagonal corresponds to subtracting
\begin{equation}
    (I_R-1)\times\begin{array}{c}
         \begin{scriptsize}
             \begin{tikzpicture}
                 \node[label=below:{1}][u](1){};
                 \node[label=below:{1}][u](2)[left of=1]{};
                 \node(dots)[left of=2]{$\cdots$};
                 \node[label=below:{1}][u](3)[left of=dots]{};
                 \node[label=below:{1}][u](4)[left of=3]{};
                 \node[label=above:{$1_R$}][u](5)[above of=4]{};
                 \draw(1)--(2);
                 \draw(dots)--(2);
                 \draw(dots)--(3);
                 \draw[double distance=1.5 pt,<-](3)--(4);
                 \draw(4)--(5);
             \end{tikzpicture}
         \end{scriptsize}
    \end{array}\;,
\end{equation}
The successive subtractions above lead in the end to the following magnetic quiver
\begin{equation}\label{MQ (F,I)}
    \begin{array}{c}
         \begin{scriptsize}
    \begin{tikzpicture}
    \node[label=below:{1}][u](1){};
    \node[label=below:{2}][u](2)[above right of=1]{};
    \node (dots)[above right of=2]{$\udots$};
    \node[label=left:{$2k-F_L-F_R-1$}][u](2N-5)[above right of=dots]{};
    \node[label=left:{$2k-F_L-F_R$}][u](2N-4)[black,above right of=2N-5]{};
    \node(empty)[right of=2N-4]{};
    \node(empty')[right of=empty]{};
    \node(empty'')[right of=empty']{};
    \node[label=right:{$2n-I_L-I_R$}][u](3')[orange,right of=empty'']{};
    \node[label=above right:{$2n-I_L-I_R-1$}][u](2')[below right of=3']{};
    \node[label=above:{$1_L$}][u](1L)[purple,above of=2N-4]{};
    \node[label=above:{$1_R$}][u](1R)[blue,above of=3']{};
    \draw(1L)--node[above]{$F_LI_R+F_RI_L+2I_LI_R$}++(1R);
    \draw(1L)--node[left]{$I_L$}++(2N-4);
    \draw(1R)--node[above right]{$I_R$}++(2N-4);
    \draw(1L)--node[below right]{$F_L+2I_L$}++(3');
    \draw(1R)--node[right]{$F_R+2I_R$}++(3');
    \node(dots')[below right of=2']{$\ddots$};
    \node[label=below:{$2$}][u](22')[below right of=dots']{};
    \node[label=below:{$1$}][u](1')[below right of=22']{};
    \draw (3')to[out=-90,in=180,loop,looseness=10](3');
    \draw(2)--(dots);
    \draw (1)--(2);
    \draw(dots)--(2N-5);
    \draw[ double distance=1.5pt,<-](2N-5)--(2N-4);
    \draw(2N-4)--(3');
    \draw[double distance=1.5pt,->](3')--(2');
    \draw(dots')--(2');
    \draw(dots')--(22');
    \draw(1')--(22');
    \end{tikzpicture}
    \end{scriptsize}
    \end{array}\;.
\end{equation}
The four nodes in the central part of the quiver correspond to the four branes depicted in \eqref{brane web (F,I) bis}. We have colour coded the nodes in this quiver to make easier the comparison with the web, which we reproduce here once again:

\begin{equation}\label{brane web (F,I) bisbis}
    \begin{array}{c}
         \begin{scriptsize}
             \begin{tikzpicture}
                 \draw[thick,black](-4,0)--node[below]{\textcolor{black}{$2k-F_L-F_R$}}++(4,0);
                 \draw[thick,orange](0,0)--node[right]{\textcolor{black}{$2n-I_L-I_R$}}++(0,2);
                 \draw[thick,blue](0,0)--(3,1);
                 \draw[thick,purple](0,0)--(-3,1);
                 \node[label=above:{{$[F_L,-I_L]$}}] at (-3,1){};
                 \node[label=above:{{$[F_R,I_R]$}}] at (3,1){};
                 \node[7brane,green][label=below:{O7$^+$}]at(0,0){};
                 \draw[dash dot,green](-5.5,0)--(3,0);
                 \node[7brane] at (-4,0){};
                 \node[7brane] at (0,2){};
                 \node[7brane] at (3,1){};
                 \node[7brane] at (-3,1){};
                 \node at (-4.5,0){$\cdots$};
                 \node at (0,2.5){$\vdots$};
             \end{tikzpicture}
         \end{scriptsize}
    \end{array}
\end{equation}

 We see that the number of links between each two nodes corresponds precisely to the modified stable intersection   \eqref{eq:intersection_modified_O7}, as announced! As an example, we can compute the stable intersection between the $1_L$ and $1_R$ nodes, which correspond to the maroon and blue branes,
 \begin{align}
     \textnormal{SI}_0 + 2 |I_R (-I_L)|= \left|\det\left(  \begin{array}{cc}
        F_R & I_R \\
        F_L & -I_L
    \end{array}\right)\right| + 2 I_L I_R= F_L I_R + F_R I_L + 2 I_L I_R\,.
 \end{align}

\subsection{SL$(2,\mathbb{Z})$ invariant intersection number}
\label{sec:SL2_invariance}
The above formula has the obvious problem that it's not SL$(2,\mathbb{Z})$ invariant. In this section, we provide the fully SL$(2,\mathbb{Z})$ invariant form of the intersection number for two fivebranes intersecting on top of an O$7^+$ plane.

\paragraph{Global SL$(2,\mathbb{Z})$.}
The intersection number between two fivebranes of charge $(p_1,q_1)$ and $(p_2,q_2)$ is given by
\begin{equation}
\text{SI}_0\left((p_1,q_1),(p_2,q_2)\right)=\left|\begin{matrix}
        p_1 & q_1\\
        p_2 & q_2
    \end{matrix}\right|=|p_1q_2-p_2q_1|\;.
\end{equation}
Under a global SL$(2,\mathbb{Z})$ transformation
\begin{equation}
\label{eq:SL2MatrixForInvariance}
    M=\begin{pmatrix}
        P & b\\
        Q & d
    \end{pmatrix}\qquad\textnormal{, with}\quad Pd-Qb=1\;,
\end{equation}
we have
\begin{equation}
    \begin{split}
        \text{SI}_0&\left(M.(p_1,q_1),M.(p_2,q_2)\right)=\text{SI}_0\left((Pp_1+bq_1,Qp_1+dq_1),(Pp_2+bq_2,Qp_2+dq_2)\right)\\
        &=|(Pd-Qb)(p_1q_2-p_2q_1)|=|p_1q_2-p_2q_1|=\text{SI}_0\left((p_1,q_1),(p_2,q_2)\right)\;,
    \end{split}
\end{equation}
i.e.\ the intersection number is invariant under global SL$(2,\mathbb{Z})$ transformations.

The modification of the intersection number due to the O$7^+$ plane
\begin{equation}
\label{eq:IntersectionNumberBeforeInvariant}
    \text{SI}_{\mathrm{O}7^+}\left((p_1,q_1),(p_2,q_2)\right)=\left|\begin{matrix}
        p_1 & q_1\\
        p_2 & q_2
    \end{matrix}\right|+2|q_1q_2|
\end{equation}
is certainly not invariant under global SL$(2,\mathbb{Z})$ transformations. This is because the O$7^+$ plane itself is not invariant under global SL$(2,\mathbb{Z})$ transformations. We shall refer to the standard O$7^+$ as O$7^+[1,0]$, and refer to its transformation with \eqref{eq:SL2MatrixForInvariance} as O$7^+[P,Q]$.

Since we can write \eqref{eq:IntersectionNumberBeforeInvariant} as,
\begin{equation}
    \text{SI}_{\mathrm{O}7^+[1,0]}\left((p_1,q_1),(p_2,q_2)\right)=\left|\begin{matrix}
        p_1 & q_1\\
        p_2 & q_2
    \end{matrix}\right|+2\left|\begin{matrix}
        p_1 & q_1\\
        1 & 0
    \end{matrix}\right|\left|\begin{matrix}
        1 & 0\\
        p_2 & q_2
    \end{matrix}\right|
\end{equation}
it is clear how to make the global SL$(2,\mathbb{Z})$ invariance manifest. We can write a general expression
\begin{equation}
    \text{SI}_{\mathrm{O}7^+[u,v]}\left((p_1,q_1),(p_2,q_2)\right)=\left|\begin{matrix}
        p_1 & q_1\\
        p_2 & q_2
    \end{matrix}\right|+2\left|\begin{matrix}
        p_1 & q_1\\
        u & v
    \end{matrix}\right|\left|\begin{matrix}
        u & v\\
        p_2 & q_2
    \end{matrix}\right|\;.
\end{equation}
It's easy to check that now we have global SL$(2,\mathbb{Z})$ invariance:
\begin{equation}
    \begin{split}
        \text{SI}_{\mathrm{O}7^+M.[1,0]}&\left(M.(p_1,q_1),M.(p_2,q_2)\right)=\text{SI}_{\mathrm{O}7^+[P,Q]}\left((Pp_1+bq_1,Qp_1+dq_1),(Pp_2+bq_2,Qp_2+dq_2)\right)\\
        &=\left|\begin{matrix}
            Pp_1+bq_1 & Qp_1+dq_1\\
            Pp_2+bq_2 & Qp_2+dq_2
        \end{matrix}\right|+2\left|\begin{matrix}
            Pp_1+bq_1 & Qp_1+dq_1\\
            P & Q
        \end{matrix}\right|\left|\begin{matrix}
            P & Q\\
            Pp_2+bq_2 & Qp_2+dq_2
        \end{matrix}\right|\\
        &=\left|\begin{matrix}
            p_1 & q_1\\
            p_2 & q_2
        \end{matrix}\right|+2\left|\begin{matrix}
            p_1 & q_1\\
            1 & 0
        \end{matrix}\right|\left|\begin{matrix}
            1 & 0\\
            p_2 & q_2
        \end{matrix}\right|=\text{SI}_{\mathrm{O}7^+[1,0]}\left((p_1,q_1),(p_2,q_2)\right)\;.
    \end{split}
\end{equation}

\paragraph{O$7^+$ SL$(2,\mathbb{Z})$ Monodromy.}
To simplify this discussion, we restrict to talking about the O$7^+[1,0]$ plane for now. All statements can later be generalised to generic O$7^+[u,v]$ planes.

So far we have ignored the monodromy cut of the O$7^+$ plane associated to $M_{\mathrm{O}7^+}=T^4$, and always kept it in the horizontal (i.e.\ in the $(1,0)$ direction away from the O$7^+$). If two fivebranes of charge $(p_1,q_1)$ and $(p_2,q_2)$ intersect on top of the O$7^+$, then by turning the monodromy cut of the O$7^+$ we can act on one of the fivebranes individually, which does not leave the intersection number expressed in its current form invariant.

Without loss of generality we can assume that $q_1,q_2\geq0$ and
\begin{equation}
    \label{eq:BraneOrderCondition}
    p_1q_2-p_2q_1<0\;.
\end{equation}
This means that turning the O$7^+$ monodromy cut counter-clockwise from the $(1,0)$ position, the $(p_2,q_2)$ fivebrane is acted upon first, i.e. this situation
\begin{equation}
\label{eq:BranesCrossingOnO7+1}
    \begin{tikzpicture}
                \node[label=below:{$\mathrm{O}7^+$}][7brane,green]at(0,0){};
                \draw[dash dot,green](-2,0)--(2,0);
                \draw (-2,-1)--(2,1);
                \draw (-2,1)--(2,-1);
                \node at (-1,1) {$(p_1,q_1)$};
                \node at (1,1) {$(p_2,q_2)$};
            \end{tikzpicture}\;.
\end{equation}
After crossing the $(p_2,q_2)$ fivebrane with the O$7^+$ monodromy cut
\begin{equation}
\label{eq:BranesCrossingOnO7+2}
    \begin{tikzpicture}
                \node[label=below:{$\mathrm{O}7^+$}][7brane,green]at(0,0){};
                \draw[dash dot,green](0,-2)--(0,2);
                \draw (-2,-1)--(2,1);
                \draw (-2,1)--(2,-1);
                \node at (-1,1) {$(p_1,q_1)$};
                \node[align=center] at (3.3,1) {$M_{\mathrm{O}7^+}(p_2,q_2)$\\
                $=(p_2+4q_2,q_2)$};
            \end{tikzpicture}\;,
\end{equation}
we would naively obtain
\begin{equation}
    \begin{split}
        \text{SI}_{\mathrm{O}7^+}&\left((p_1,q_1),M_{\mathrm{O}7^+}.(p_2,q_2)\right)=\left|\begin{matrix}
        p_1 & q_1\\
        p_2+4q_2 & q_2
    \end{matrix}\right|+2\left|\begin{matrix}
        p_1 & q_1\\
        1 & 0
    \end{matrix}\right|\left|\begin{matrix}
        1 & 0\\
        p_2+4q_2 & q_2
    \end{matrix}\right|\\
    &=|p_1q_2-p_2q_1-4q_1q_2|+2|q_1q_2|\underset{\eqref{eq:BraneOrderCondition}}{=}|p_1q_2-p_2q_1|-2|q_1q_2|\neq \text{SI}_{\mathrm{O}7^+}\left((p_1,q_1),(p_2,q_2)\right)\;.
    \end{split}
\end{equation}
The naive intersection number is not the same, because when specifying the intersection number of two fivebranes of charge $(p_1,q_1)$ and $(p_2,q_2)$ on top of an O$7^+$, we have to take into account the orientation of the O$7^+$ monodromy cut.
\begin{equation}
    \text{SI}_{\mathrm{O}7^+}\left((p_1,q_1),(p_2,q_2);\epsilon\right)=\left|\begin{matrix}
        p_1 & q_1\\
        p_2 & q_2
    \end{matrix}\right|+(-1)^\epsilon\,2\left|\begin{matrix}
        p_1 & q_1\\
        1 & 0
    \end{matrix}\right|\left|\begin{matrix}
        1 & 0\\
        p_2 & q_2
    \end{matrix}\right|\;,
\end{equation}
where
\begin{equation}
    \epsilon=\left\{\begin{tabular}{ll}
     \multirow{2}{*}{0\textnormal{ , }} & \textnormal{if the O$7^+$ monodromy cut \textbf{can} be turned to the}\\
        & \textnormal{$(1,0)$ direction without crossing a fivebrane.} \\
     \multirow{2}{*}{1\textnormal{ , }} & \textnormal{if the O$7^+$ monodromy cut \textbf{cannot} be turned to the}\\
        & \textnormal{$(1,0)$ direction without crossing a fivebrane.}
\end{tabular}\right.
\end{equation}
We have $\epsilon=0$ in \eqref{eq:BranesCrossingOnO7+1} and $\epsilon=1$ in \eqref{eq:BranesCrossingOnO7+2}.\\

Generalising to O$7^+[u,v]$ planes, we get:\\

\noindent\fbox{\parbox{\textwidth}{%
\paragraph{The fully invariant intersection number:}

The fully invariant intersection number of two fivebranes of charge $(p_1,q_1)$ and $(p_2,q_2)$ on top of an O$7^+[u,v]$ is
\begin{subequations}
\label{eq:FullyInvariantIntersection}
\begin{equation}
    \text{SI}_{\mathrm{O}7^+[u,v]}\left((p_1,q_1),(p_2,q_2);\epsilon\right)=\left|\begin{matrix}
        p_1 & q_1\\
        p_2 & q_2
    \end{matrix}\right|+(-1)^\epsilon\,2\left|\begin{matrix}
        p_1 & q_1\\
        u & v
    \end{matrix}\right|\left|\begin{matrix}
        u & v\\
        p_2 & q_2
    \end{matrix}\right|\;,
\end{equation}
where
\begin{equation}
\label{eq:EpsilonIntersection}
    \epsilon=\left\{\begin{tabular}{ll}
     \multirow{2}{*}{0\textnormal{ , }} & \textnormal{if the O$7^+[u,v]$ monodromy cut \textbf{can} be turned to the}\\
        & \textnormal{$(u,v)$ direction without crossing a fivebrane.} \\
     \multirow{2}{*}{1\textnormal{ , }} & \textnormal{if the O$7^+[u,v]$ monodromy cut \textbf{cannot} be turned to the}\\
        & \textnormal{$(u,v)$ direction without crossing a fivebrane.}
\end{tabular}\right.
\end{equation}
\end{subequations}
}}\\

It is important to note, that two fivebranes of charge $(p_1,q_1)$ and $(p_2,q_2)$ intersecting on top of an O$7^+[u,v]$ with $\epsilon=1$ in \eqref{eq:EpsilonIntersection} may not be consistent, see Appendix \ref{app:BraneWebs} for a detailed explanation. If $\epsilon=0$ however, there is no issue.

\section{Theories with orthogonal gauge group}
\label{sec:orthogonal}

In this section we make use of the methods in Section \ref{sec:intersection_rule} to systematically study the Higgs branch of a collection of theories constructed by brane webs with O7$^+$-planes. In particular, we will focus on the fixed point, as well as the finite coupling limit of SO gauge theories with matter in the vector representation.
Subsequently, we will study special unitary gauge theories with fundamentals, and rank-2 hypers in section \ref{sec:specialunitary}. Further sporadic examples are collected in section \ref{sec:other_examples}. 

An important remark is that several of these theories have alternative constructions as brane setups involving O5 planes. It is by now well understood how to read the magnetic quiver of these systems \cite{Akhond:2020vhc,Bourget:2020gzi,Akhond:2021knl,Akhond:2022jts}, which lead to ortho-symplectic magnetic quivers. They can be used to compute the Higgs branch Hilbert series in an alternative way, and checked against the results from the unitary magnetic quivers presented here (see also \cite{Akhond:2021ffo} for a similar strategy). Therefore, this constitutes a very strong check of our proposal \eqref{eq:FullyInvariantIntersection} for reading the magnetic quivers.

For each of the examples discussed in this section, we follow the same schematic plan: first we present the brane web and the magnetic quiver obtained from it. Some of these have already appeared in the literature, in which case we simply quote the result for ease of the reader. Second, we give our conjectural results for the Highest Weight Generating function of the Higgs branch chiral ring, specifying the HS checks that we perform in each case. They can be used to extract interesting information such as operator spectrum, and the global form of the global symmetry group. 

An important outcome of the analysis is that we can compute the Higgs branch Hasse diagrams, encoding the various phases of the theories. Since these will mix theories with orthogonal and (special) unitary groups, we postpone that analysis to section \ref{sec:Hasse_higgsings}. 

\subsection{The Brane Web}
\label{sec:SOBraneWeb}

The O$7^+$ brane web realisation of $5d$ $\mathcal{N}=1$ special orthogonal gauge theories with vector matter is given in \cite{Bergman:2015dpa}, where their infinite coupling UV fixed points were analysed as well. In this section we briefly present this construction.

\subsubsection{SO(even)}

The brane web for\footnote{Since $\pi_4(SO(K))=0$ for $K>6$, there are no additional discrete choices such as a discrete theta parameter or Chern-Simons level in specifying the theories. This will not be true for $K\leq6$, in which case we only specify this data if it is non-trivial and an unspecidief level is to be understood as 0.}
\begin{equation}
    \soeq{2k}{N=N_L+N_R}
\end{equation}
is \cite{Bergman:2015dpa}
\begin{equation}
    \makebox[\textwidth][c]{\begin{tikzpicture}
        \node[7brane,green] (0) at (0,0) {};
        \draw[dashed,green] (-8,0)--(8,0);
        \draw (-4,0)--(-2,1) (2,1)--(4,0);
        \node at (-4,0.5) {\scriptsize$(2+\alpha,1)$};
        \node at (4,0.5) {\scriptsize$(2-\alpha,-1)$};
        \draw[thick] (-2,1)--(2,1);
        \node at (0,1.3) {\scriptsize$k(1,0)$};
        \node[7brane,blue,label=above:{\scriptsize$[k-2-\alpha-N_L,-1]$}] (L) at (-5,3) {};
        \draw (-2,1)--(-4,2)--(L);
        \node[7brane] (1) at (-9,2) {};
        \node[7brane] (2) at (-8,2) {};
        \node (3) at (-7,2) {$\cdots$};
        \node[7brane] (4) at (-6,2) {};
        \node[7brane] (5) at (-5,2) {};
        \draw (1)--(2);
        \draw[double] (2)--(3);
        \draw[thick] (3)--(4)--(5)--(-4,2);
        \node at (-8.5,2.2) {\scriptsize$1$};
        \node at (-5.5,2.2) {\scriptsize$N_L-1$};
        \node at (-4.5,2.2) {\scriptsize$N_L$};
        \draw [decorate,decoration={brace,amplitude=3pt},xshift=0pt,yshift=0pt](-4.8,1.8) -- (-9.2,1.8) node [midway,xshift=0pt,yshift=-15pt] {\scriptsize $N_L[1,0]$};
        
        \node[7brane,red,label=above:{\scriptsize$[k-2+\alpha-N_R,1]$}] (R) at (5,3) {};
        \draw (2,1)--(4,2)--(R);
        \node[7brane] (10) at (9,2) {};
        \node[7brane] (9) at (8,2) {};
        \node (8) at (7,2) {$\cdots$};
        \node[7brane] (7) at (6,2) {};
        \node[7brane] (6) at (5,2) {};
        \draw (10)--(9);
        \draw[double] (9)--(8);
        \draw[thick] (8)--(7)--(6)--(4,2);
        \node at (8.5,2.2) {\scriptsize$1$};
        \node at (5.5,2.2) {\scriptsize$N_R-1$};
        \node at (4.5,2.2) {\scriptsize$N_R$};
        \draw [decorate,decoration={brace,amplitude=3pt},xshift=0pt,yshift=0pt](9.2,1.8) -- (4.8,1.8) node [midway,xshift=0pt,yshift=-15pt] {\scriptsize $N_R[1,0]$};
    \end{tikzpicture}}
\end{equation}
where $\alpha\in\mathbb{Z}$ as well as the choice of partition $N=N_L+N_R$ do not affect the field theory. Two sevenbranes are coloured red and blue respectively for later reference.

\paragraph{Infinite coupling.} As discussed in \cite{Bergman:2015dpa}: in order to shrink the web to realise the infinite coupling UV fixed point SCFT of the $5d$ $\mathcal{N}=1$ worldvolume gauge theory we need $N\leq2k-3$; furthermore $N=2k-3$ and $N=2k-4$ exhibit special behaviour, while $N<2k-4$ is generic.

\begin{enumerate}
    \item $N<2k-4$: For $N<2k-4$ it is straight forward to shrink the web to the SCFT point
\begin{equation}\label{eq:BW3point3}
    \begin{tikzpicture}
        \node[7brane,green] (0) at (0,0) {};
        \draw[dashed,green] (-6,0)--(2,0);
        \node[7brane] (1) at (-5,0) {};
        \node[7brane] (2) at (-4,0) {};
        \node (3) at (-3,0) {$\cdots$};
        \node[7brane] (4) at (-2,0) {};
        \node[7brane] (5) at (-1,0) {};
        \draw (1)--(2);
        \draw[double] (2)--(3);
        \draw[thick] (3)--(4)--(5)--(0);
        \node at (-4.5,0.2) {\scriptsize$1$};
        \node at (-1.5,0.2) {\scriptsize$N-1$};
        \node at (-0.5,0.2) {\scriptsize$N$};
        \draw [decorate,decoration={brace,amplitude=3pt},xshift=0pt,yshift=0pt](-0.8,-0.2) -- (-5.2,-0.2) node [midway,xshift=0pt,yshift=-15pt] {\scriptsize $N[1,0]$};
        \node[7brane,blue,label=above:{\scriptsize$[k-2-\alpha-N_L,-1]$}] (L) at (-2,2) {};
        \node[7brane,red,label=above:{\scriptsize$[k-2+\alpha-N_R,1]$}] (R) at (2,2) {};
        \draw (L)--(0)--(R);
    \end{tikzpicture}
\end{equation}
\item $N=2k-4$: For $N=2k-4$ we have
\begin{equation}
    k-2-\alpha-N_L=-(k-2+\alpha-N_R)=:-x\;,
\end{equation}
and hence the red and blue sevenbranes at the top of the brane web are of the same kind
\begin{equation}
    \begin{tikzpicture}
        \node[7brane,blue,label=above:{\scriptsize$[x,1]$}] (L) at (-1,2) {};
        \node[7brane,red,label=above:{\scriptsize$[x,1]$}] (R) at (3,2) {};
        \draw (-2,0)--(L) (2,0)--(R);
        \draw[decorate,decoration={coil,aspect=0}] (-3,0)--(3,0);
        \node at (0,-0.5) {rest of brane web};
    \end{tikzpicture}\;.
\end{equation}
Shrinking the brane web to the SCFT point gives
\begin{equation}\label{eq:BW_3point6}
    \begin{tikzpicture}
        \node[7brane,green] (0) at (0,0) {};
        \draw[dashed,green] (-6,0)--(2,0);
        \node[7brane] (1) at (-5,0) {};
        \node[7brane] (2) at (-4,0) {};
        \node (3) at (-3,0) {$\cdots$};
        \node[7brane] (4) at (-2,0) {};
        \node[7brane] (5) at (-1,0) {};
        \draw (1)--(2);
        \draw[double] (2)--(3);
        \draw[thick] (3)--(4)--(5)--(0);
        \node at (-4.5,0.2) {\scriptsize$1$};
        \node at (-1.5,0.2) {\scriptsize$N-1$};
        \node at (-0.5,0.2) {\scriptsize$N$};
        \draw [decorate,decoration={brace,amplitude=3pt},xshift=0pt,yshift=0pt](-0.8,-0.2) -- (-5.2,-0.2) node [midway,xshift=0pt,yshift=-15pt] {\scriptsize $N[1,0]$};
        \node[7brane,blue,label=left:{\scriptsize$[x,1]$}] (R) at (2,2) {};
        \node[7brane,red,label=left:{\scriptsize$[x,1]$}] (L) at (4,4) {};
        \draw[double] (0)--(R);
        \node at (0.5,1.2) {$2$};
        \draw (R)--(L);
        \node at (2.5,3.2) {$1$};
    \end{tikzpicture}\;,
\end{equation}
where the value of $x\in\mathbb{Z}$ does not affect the field theory.
\item $N=2k-3$: For $N=2k-3$ we have
\begin{equation}
    k-2-\alpha-N_L=-(k-2+\alpha-N_R)-1=:-x-1\;,
\end{equation}
and hence the red and blue sevenbranes at the top of the brane web make it non-convex
\begin{equation}
    \begin{tikzpicture}
        \node[7brane,blue,label=above:{\scriptsize$[x+1,1]$}] (L) at (-0,2) {};
        \node[7brane,red,label=above:{\scriptsize$[x,1]$}] (R) at (3,2) {};
        \draw (-2,0)--(L) (2,0)--(R);
        \draw[decorate,decoration={coil,aspect=0}] (-3,0)--(3,0);
        \node at (0,-0.5) {rest of brane web};
    \end{tikzpicture}\;.
\end{equation}
We can use the monodromy
\begin{equation}
    \begin{tikzpicture}
        \node[7brane,blue,label=left:{\scriptsize$[x+1,1]$}] (L) at (1,1) {};
        \draw[dashed] (L)--(3,3);
        \node[7brane,red,label=above:{\scriptsize$[1,0]$}] (R) at (-2,2) {};
        \draw (0,0)--(L) (1,0)--(2,2)--(R);
        \draw[decorate,decoration={coil,aspect=0}] (-3,0)--(3,0);
        \node at (0,-0.5) {rest of brane web};
    \end{tikzpicture}\;,
\end{equation}
perform a Hanany-Witten transition
\begin{equation}
    \begin{tikzpicture}
        \node[7brane,blue,label=left:{\scriptsize$[x+1,1]$}] (L) at (3,3) {};
        \node[7brane,red,label=above:{\scriptsize$[1,0]$}] (R) at (-2,2) {};
        \draw (0,0)--(2,2) (1,0)--(2,2)--(R);
        \draw[double] (2,2)--(L);
        \draw[decorate,decoration={coil,aspect=0}] (-3,0)--(3,0);
        \node at (0,-0.5) {rest of brane web};
    \end{tikzpicture}\;,
\end{equation}
and then shrink the web to
\begin{equation}
    \begin{tikzpicture}
        \node[7brane,green] (0) at (0,0) {};
        \draw[dashed,green] (-6,0)--(2,0);
        \node[7brane,red] (1) at (-5,0) {};
        \node[7brane] (2) at (-4,0) {};
        \node (3) at (-3,0) {$\cdots$};
        \node[7brane] (4) at (-2,0) {};
        \node[7brane] (5) at (-1,0) {};
        \draw (1)--(2);
        \draw[double] (2)--(3);
        \draw[thick] (3)--(4)--(5)--(0);
        \node at (-4.5,0.2) {\scriptsize$1$};
        \node at (-1.5,0.2) {\scriptsize$N$};
        \node at (-0.5,0.2) {\scriptsize$N+1$};
        \draw [decorate,decoration={brace,amplitude=3pt},xshift=0pt,yshift=0pt](-0.8,-0.2) -- (-5.2,-0.2) node [midway,xshift=0pt,yshift=-15pt] {\scriptsize $N+1[1,0]$};
        \node[7brane,blue,label=left:{\scriptsize$[x+1,1]$}] (R) at (2,2) {};
        \draw[double] (0)--(R);
        \node at (0.5,1.2) {$2$};
    \end{tikzpicture}\;,
\end{equation}
where the value of $x\in\mathbb{Z}$ does not affect the field theory.
\end{enumerate}

\subsubsection{SO(odd)}

The brane web for
\begin{equation}
    \soeq{2k+1}{N=N_L+N_R}
\end{equation}
is \cite{Bergman:2015dpa}
\begin{equation}
    \makebox[\textwidth][c]{\begin{tikzpicture}
        \node[7brane,green] (0) at (0,0) {};
        \draw[dashed,green] (-8,0)--(8,0);
        \draw (-4,0)--(-2,1) (2,1)--(4,0) (-4,0)--(4,0);
        \node at (-4,0.5) {\scriptsize$(1+\alpha,1)$};
        \node at (4,0.5) {\scriptsize$(2-\alpha,-1)$};
        \draw[thick] (-2,1)--(2,1);
        \node at (0,1.3) {\scriptsize$k(1,0)$};
        \node[7brane,blue,label=above:{\scriptsize$[k-1-\alpha-N_L,-1]$}] (L) at (-5,3) {};
        \draw (-2,1)--(-4,2)--(L);
        \node[7brane] (1) at (-9,2) {};
        \node[7brane] (2) at (-8,2) {};
        \node (3) at (-7,2) {$\cdots$};
        \node[7brane] (4) at (-6,2) {};
        \node[7brane] (5) at (-5,2) {};
        \draw (1)--(2);
        \draw[double] (2)--(3);
        \draw[thick] (3)--(4)--(5)--(-4,2);
        \node at (-8.5,2.2) {\scriptsize$1$};
        \node at (-5.5,2.2) {\scriptsize$N_L-1$};
        \node at (-4.5,2.2) {\scriptsize$N_L$};
        \draw [decorate,decoration={brace,amplitude=3pt},xshift=0pt,yshift=0pt](-4.8,1.8) -- (-9.2,1.8) node [midway,xshift=0pt,yshift=-15pt] {\scriptsize $N_L[1,0]$};
        
        \node[7brane,red,label=above:{\scriptsize$[k-2+\alpha-N_R,1]$}] (R) at (5,3) {};
        \draw (2,1)--(4,2)--(R);
        \node[7brane] (10) at (9,2) {};
        \node[7brane] (9) at (8,2) {};
        \node (8) at (7,2) {$\cdots$};
        \node[7brane] (7) at (6,2) {};
        \node[7brane] (6) at (5,2) {};
        \draw (10)--(9);
        \draw[double] (9)--(8);
        \draw[thick] (8)--(7)--(6)--(4,2);
        \node at (8.5,2.2) {\scriptsize$1$};
        \node at (5.5,2.2) {\scriptsize$N_R-1$};
        \node at (4.5,2.2) {\scriptsize$N_R$};
        \draw [decorate,decoration={brace,amplitude=3pt},xshift=0pt,yshift=0pt](9.2,1.8) -- (4.8,1.8) node [midway,xshift=0pt,yshift=-15pt] {\scriptsize $N_R[1,0]$};
    \end{tikzpicture}}
\end{equation}
where $\alpha\in\mathbb{Z}$ and the choice of partition $N=N_L+N_R$ do not affect the field theory. Two sevenbranes are coloured red and blue respectively for later reference.

\paragraph{Infinite coupling.} In order to shrink the web to realise the infinite coupling UV fixed point SCFT of the $5d$ $\mathcal{N}=1$ worldvolume gauge theory we need $N\leq2k-2$; furthermore $N=2k-2$ and $N=2k-3$ exhibit special behaviour, while $N<2k-3$ is generic.

The different cases are completely analogous to the SO(even) discussion, and we only present the shrunken webs.

\begin{enumerate}
    \item $N<2k-3$: For $N<2k-3$ the shrunken web realising the SCFT fixed point is
\begin{equation}\label{eq:BW3point14}
    \begin{tikzpicture}
        \node[7brane,green] (0) at (0,0) {};
        \draw[dashed,green] (-6,0)--(2,0);
        \node[7brane] (1) at (-5,0) {};
        \node[7brane] (2) at (-4,0) {};
        \node (3) at (-3,0) {$\cdots$};
        \node[7brane] (4) at (-2,0) {};
        \node[7brane] (5) at (-1,0) {};
        \draw (1)--(2);
        \draw[double] (2)--(3);
        \draw[thick] (3)--(4)--(5)--(0);
        \node at (-4.5,0.2) {\scriptsize$1$};
        \node at (-1.5,0.2) {\scriptsize$N-1$};
        \node at (-0.5,0.2) {\scriptsize$N$};
        \draw [decorate,decoration={brace,amplitude=3pt},xshift=0pt,yshift=0pt](-0.8,-0.2) -- (-5.2,-0.2) node [midway,xshift=0pt,yshift=-15pt] {\scriptsize $N[1,0]$};
        \node[7brane,blue,label=above:{\scriptsize$[k-1-\alpha-N_L,-1]$}] (L) at (-2,2) {};
        \node[7brane,red,label=above:{\scriptsize$[k-2+\alpha-N_R,1]$}] (R) at (2,2) {};
        \draw (L)--(0)--(R);
    \end{tikzpicture}\;.
\end{equation}
\item $N=2k-3$: For $N=2k-3$ the shrunken web realising the SCFT fixed point is
\begin{equation}
    \begin{tikzpicture}
        \node[7brane,green] (0) at (0,0) {};
        \draw[dashed,green] (-6,0)--(2,0);
        \node[7brane] (1) at (-5,0) {};
        \node[7brane] (2) at (-4,0) {};
        \node (3) at (-3,0) {$\cdots$};
        \node[7brane] (4) at (-2,0) {};
        \node[7brane] (5) at (-1,0) {};
        \draw (1)--(2);
        \draw[double] (2)--(3);
        \draw[thick] (3)--(4)--(5)--(0);
        \node at (-4.5,0.2) {\scriptsize$1$};
        \node at (-1.5,0.2) {\scriptsize$N-1$};
        \node at (-0.5,0.2) {\scriptsize$N$};
        \draw [decorate,decoration={brace,amplitude=3pt},xshift=0pt,yshift=0pt](-0.8,-0.2) -- (-5.2,-0.2) node [midway,xshift=0pt,yshift=-15pt] {\scriptsize $N[1,0]$};
        \node[7brane,blue,label=left:{\scriptsize$[x,1]$}] (R) at (2,2) {};
        \node[7brane,red,label=left:{\scriptsize$[x,1]$}] (L) at (4,4) {};
        \draw[double] (0)--(R);
        \node at (0.5,1.2) {$2$};
        \draw (R)--(L);
        \node at (2.5,3.2) {$1$};
    \end{tikzpicture}\;,
\end{equation}
where $x=k-2+\alpha-N_R$ does not affect the field theory.
\item $N=2k-2$: For $N=2k-2$ the shrunken web realising the SCFT fixed point is
\begin{equation}
    \begin{tikzpicture}
        \node[7brane,green] (0) at (0,0) {};
        \draw[dashed,green] (-6,0)--(2,0);
        \node[7brane,red] (1) at (-5,0) {};
        \node[7brane] (2) at (-4,0) {};
        \node (3) at (-3,0) {$\cdots$};
        \node[7brane] (4) at (-2,0) {};
        \node[7brane] (5) at (-1,0) {};
        \draw (1)--(2);
        \draw[double] (2)--(3);
        \draw[thick] (3)--(4)--(5)--(0);
        \node at (-4.5,0.2) {\scriptsize$1$};
        \node at (-1.5,0.2) {\scriptsize$N$};
        \node at (-0.5,0.2) {\scriptsize$N+1$};
        \draw [decorate,decoration={brace,amplitude=3pt},xshift=0pt,yshift=0pt](-0.8,-0.2) -- (-5.2,-0.2) node [midway,xshift=0pt,yshift=-15pt] {\scriptsize $N+1[1,0]$};
        \node[7brane,blue,label=left:{\scriptsize$[x+1,1]$}] (R) at (2,2) {};
        \draw[double] (0)--(R);
        \node at (0.5,1.2) {$2$};
    \end{tikzpicture}\;,
\end{equation}
where $x=k-2+\alpha-N_R$ does not affect the field theory.
\end{enumerate}

\subsection{Finite coupling magnetic quivers}

Let us first focus on the finite coupling Higgs branch, so as to be able to compare with the infinite coupling Higgs branches momentarily. We refer the reader to section \ref{sec:intersection_rule} for the derivation, and only mention the result here.
The brane web for
\begin{equation}
    \soeq{K}{N}
\end{equation}
at finite coupling, at the origin of the moduli space is
\begin{equation}
    \makebox[\textwidth][c]{\begin{tikzpicture}
        \node[7brane,green] (0) at (0,0) {};
        \node[7brane] (1) at (-7,0) {};
        \node[7brane] (2) at (-6,0) {};
        \node (3) at (-5,0) {$\cdots$};
        \node[7brane] (4) at (-4,0) {};
        \node[7brane] (5) at (-3,0) {};
        \draw (1)--(2);
        \draw[double] (2)--(3);
        \draw[thick] (3)--(4)--(5)--(0);
        \node at (-6.5,0.2) {\scriptsize$1$};
        \node at (-3.5,0.2) {\scriptsize$N-1$};
        \node at (-2.5,0.2) {\scriptsize$N$};
        \node at (-0.5,0.2) {\scriptsize$K$};
        \node[7brane,label=above:{$[K-N+y,-1]$}] (l1) at (-2,1) {};
        \node[7brane,label=below:{$[-y-4,1]$}] (l2) at (-2,-1) {};
        \draw (l1)--(-1,0)--(l2);
        \draw[dashed,green] (-8,0)--(2,0);
    \end{tikzpicture}}\;,
\end{equation}
for any $y\in\mathbb{Z}$.

\paragraph{1.} For $N>K$ we read the magnetic quiver
\begin{equation}
    \begin{tikzpicture}
        \node[u,label=below:{\scriptsize$1$}] (1) at (1,0) {};
        \node[u,label=below:{\scriptsize$2$}] (2) at (2,0) {};
        \node (3) at (3,0) {$\cdots$};
        \node[u,label=below:{\scriptsize$K$}] (4) at (4,0) {};
        \node[u,label=above:{\scriptsize$1$}] (4u) at (4,1) {};
        \node[u,label=below:{\scriptsize$K$}] (5) at (5,0) {};
        \node (6) at (6,0) {$\cdots$};
        \node[u,label=below:{\scriptsize$K$}] (7) at (7,0) {};
        \node[u,label=below:{\scriptsize$K$}] (8) at (8,0) {};
        \draw (1)--(2)--(3)--(4)--(5)--(6)--(7);
        \draw[<-,double distance=1.5pt](7)--(8);
        \draw (4)--(4u);
        \draw [decorate,decoration={brace,amplitude=3pt},xshift=0pt,yshift=0pt](8.2,-0.5) -- (0.8,-0.5) node [midway,xshift=0pt,yshift=-15pt] {\scriptsize $N$};
    \end{tikzpicture}\;.
\end{equation}

\paragraph{2.} For $N=K$ we read the magnetic quiver
\begin{equation}
    \begin{tikzpicture}
        \node[u,label=below:{\scriptsize$1$}] (4) at (4,0) {};
        \node[u,label=above:{\scriptsize$1$}] (4u) at (8,1) {};
        \node[u,label=below:{\scriptsize$2$}] (5) at (5,0) {};
        \node (6) at (6,0) {$\cdots$};
        \node[u,label=below:{\scriptsize$N-1$}] (7) at (7,0) {};
        \node[u,label=below:{\scriptsize$N$}] (8) at (8,0) {};
        \draw (4)--(5)--(6)--(7);
        \draw[<-,double distance=1.5pt](7)--(8);
        \draw[->,double distance=1.5pt](8)--(4u);
    \end{tikzpicture}\;.
\end{equation}

\paragraph{3.} For $N<K$ we read the magnetic quiver, 
\begin{equation}
    \begin{tikzpicture}
        \node[u,label=below:{\scriptsize$1$}] (4) at (4,0) {};
        \node[u,label=above:{\scriptsize$1$}] (4u) at (8,1) {};
        \node[u,label=below:{\scriptsize$2$}] (5) at (5,0) {};
        \node (6) at (6,0) {$\cdots$};
        \node[u,label=below:{\scriptsize$N-1$}] (7) at (7,0) {};
        \node[u,label=below:{\scriptsize$N$}] (8) at (8,0) {};
        \draw (4)--(5)--(6)--(7);
        \draw[<-,double distance=1.5pt](7)--(8);
        \draw[double distance=1.5pt](8)--(4u);
    \end{tikzpicture}
\end{equation}
independent of $K$.

The HWG for generic $N<K$ is given by \cite{Hanany:2016gbz,Ferlito:2016grh}
\begin{equation}    \PE\left[\sum_{i=1}^{N}\mu_{i}^2t^{2i}\right]\;,
\end{equation}
From the HWG we notice that the global form of the flavour symmetry is $\mathrm{Sp}(N)/\mathbb{Z}_2 \times \mathrm{SO}(3)_R$.

\subsection{Infinite coupling magnetic quivers}

\subsubsection{SO$(K)$ with $K-3$ vectors}\label{sec:K-3}

The brane web for the infinite coupling limit of SO$(K)$ gauge theory with $K-3$ hypermultiplets in the vector representation is derived in \ref{sec:SOBraneWeb} and also appears in \cite{Akhond:2021ffo}. We present it here again for convenience:
\begin{equation}
    \begin{scriptsize}
    \begin{tikzpicture}
    \draw[thick](1,-1)--(0,-2);
    \node[label=right:{$[1,1]$}][7brane] at (1,-1){};
    \node[7brane,green,label=below right:{O7$^+$}] at (0,-2){};
    \draw[thick](0,-2)--(-2,-2);
    \draw[thick](-5,-2)--(-3,-2);
    \node at (-2.5,-2){$\cdots$};
    \node[7brane] at (-1,-2){};
    \node[7brane] at (-2,-2){};
    \node[7brane] at (-3,-2){};
    \node[7brane] at (-4,-2){};
    \node[7brane] at (-5,-2){};
    \node [label=below:{1}] at (-4.5,-2){};
    \node [label=below:{2}] at (-3.5,-2){};
    \node [label=below:{$K-3$}] at (-1.5,-2){};
    \node [label=below:{$K-2$}] at (-.5,-2){};
    \node[label=right:{2}] at (0,-1.5){};
    \end{tikzpicture}
    \end{scriptsize}\;.
\end{equation}
The magnetic quiver for the infinite gauge coupling limit is easily obtained applying the rules outlined in Section \ref{sec:intersection_rule}, 
\begin{equation}
    \begin{array}{c}
         \begin{scriptsize}
    \begin{tikzpicture}
    \node[label=below:{$1$}][u](2){};
    \node (dots)[right of=2]{$\cdots$};
    \node[label=below:{$K-3$}][u](2N-5)[right of=dots]{};
    \node[label=below:{$K-2$}][u](2N-4)[right of=2N-5]{};
    \node[label=below:{$2$}][u](2')[right of=2N-4]{};
    \draw (2')to[out=-45,in=45,loop,looseness=10](2');
    \draw(2)--(dots);
    \draw(dots)--(2N-5);
    \draw[ double distance=1.5pt,<-](2N-5)--(2N-4);
    \draw(2N-4)--(2');
    \end{tikzpicture}
    \end{scriptsize}
    \end{array}\;.\label{MQ5 SO(2k+1)+(2k-2)}
\end{equation}
The presence of $K-2$ balanced nodes forming the Dynkin diagram of $c_{K-2}$ signal that the enhanced global symmetry at the UV fixed point is $\mathfrak{sp}(K-2)$. We will soon comment on the global structure of this symmetry. The presence of $K-2$ flavours attached to the U(2) gauge node with an adjoint hyper, signals the presence of a monopole operator of scaling dimension $K-2$, which is incidentally the dual coxeter number of the electric gauge group. We therefore see that the 5d theory has a bare instanton operator whose scaling dimension is the dual coxeter number of the gauge algebra, consistent with the observation made in \cite{Zafrir:2015uaa}. The existence of such a bare monopole will be a typical behaviour in all families that we study in subsequent sections.
The proposal of \cite{Akhond:2021ffo} for the magnetic quiver of this theory slightly differs from \eqref{MQ5 SO(2k+1)+(2k-2)}, although they produce the same Hilbert series.

As mentioned in Section \ref{sec:intersection_rule} the 5d UV fixed point of SO$(K)$ gauge theory with $K-3$ hypermultiplets in the vector representation can be obtained by twisted compactification of the 6d (1,0) SU$(K-2)$ gauge theory with $2K-4$ fundamental hypermultiplets \cite{Hayashi:2015vhy}. The magnetic quiver for the aforementioned 6d SCFT is \cite{Hanany:2018vph, Cabrera:2019izd} 
\begin{equation}
    \begin{array}{c}
        \begin{scriptsize}
             \begin{tikzpicture}
                          \node[label=below:{1}][u](1){};
                          \node[label=below:{2}][u](2)[right of=1]{};
                          \node(dots)[right of=2]{$\cdots$};
                           \node[label=below:{$K-3$}][u](2k-2)[right of=dots]{};
                          \node[label=below:{$K-2$}][u](2k-1)[right of=2k-2]{};
                          \node[label=below:{$K-3$}][u](2k-2')[right of=2k-1]{};
                          \node(dots')[right of=2k-2']{$\cdots$};
                      \node[label=below:{2}][u](2')[right of=dots']{};
                          \node[label=below:{1}][u](1')[right of=2']{};
                          \node[label=left:{2}][u](22)[above of=2k-1]{};
                          \draw(1)--(2);
                          \draw(2)--(dots);
                          \draw(dots)--(2k-2);
                          \draw(2k-2)--(2k-1);
                          \draw(2k-1)--(2k-2');
                          \draw(1')--(2');
                          \draw(2')--(dots');
                          \draw(dots')--(2k-2');
                          \draw(22)--(2k-1);
                          \draw (22)to[out=135,in=45,loop,looseness=10](22);
             \end{tikzpicture}
         \end{scriptsize}
    \end{array}\;.\label{MQ6 SU(k)+2k F}
\end{equation}
We now note that performing a $\mathbb{Z}_2$ folding of this quiver, leads us to \eqref{MQ5 SO(2k+1)+(2k-2)}.
In line with the observation \cite{Bourget:2020asf} that the magnetic quiver for the $\mathbb{Z}_l$ twisted compactification of a given theory, is obtained by the $\mathbb{Z}_l$ folding of the magnetic quiver of the original untwisted theory. 

\paragraph{Highest weight generating function:} The HWG of the unfolded quiver \eqref{MQ6 SU(k)+2k F} is \cite{Hanany:2018vph}
\begin{equation}
    \PE\left[\sum_{i=1}^{K-2}\mu_i\mu_{2K-4-i}t^{2i}+t^4+\mu_{K-2}(t^{K-2}+t^{K})-\mu_{K-2}^2t^{2K} \right]\;.
\end{equation}
Here $\mu_{i}$ are $\mathfrak{su}(2K-4)$ highest weight fugacities. The effect of the $\mathbb{Z}_2$ folding at the level of the HWG can be implemented by pairwise identification of the highest weight fugacities
\begin{equation}
    \mu_{i}\sim \mu_{2K-4-i}\;,\qquad i\in\{1,\cdots,K-3\} \:.
\end{equation}
Implementing this identification leads to the HWG for the folded quiver \eqref{MQ5 SO(2k+1)+(2k-2)}
\begin{equation}
    \PE\left[\sum_{i=1}^{K-2}\mu_i^2t^{2i}+t^{4}+\mu_{K-2} (t^{K-2}+t^{K})-\mu_{K-2}^2t^{2K}\right]\;,\label{HWG5 SO(2k+1) + 2k-2 V}
\end{equation}
where the $\mu_i$ are now highest weight fugacities for $\mathfrak{
sp}(K-2)$. Observe that this result encodes the whole spectrum of $\frac{1}{2}$-BPS gauge invariant operators, which allows us to be more precise and extract the global form of the symmetry group as well.
\paragraph{Case $K=2k+1$.} In this case, all the terms in the HWG either transform trivially under the $\mathbb{Z}_2\times\mathbb{Z}_2$ centre of $\mathfrak{sp}(2k-1)\times\mathfrak{su}(2)_R$ algebra, or they transform non-trivially under both factors. This tells us that the global form of the global symmetry group is $\frac{\mathrm{Sp}(2k-1)\times\mathrm{SU}(2)_R}{\mathbb{Z}_2^{\mathrm{diag}}}$.

\paragraph{Case $K=2k$} In this case, all the terms in the HWG transform trivially under the $\mathbb{Z}_2\times\mathbb{Z}_2$ centre of $\mathfrak{sp}(2k)\times\mathfrak{su}(2)_R$ algebra. This tells us that the global form of the global symmetry group is $\frac{\mathrm{Sp}(2k)}{\mathbb{Z}_2}\frac{\times\mathrm{SU}(2)_R}{\mathbb{Z}_2}$.

\paragraph{Special cases}
\begin{enumerate}
    \item When $K=3$ the electric theory corresponds to the UV fixed point associated to pure SO(3) gauge theory. The magnetic quiver \eqref{MQ5 SO(2k+1)+(2k-2)} is identical to its unfolded version \eqref{MQ6 SU(k)+2k F}. The moduli space is the Cartesian product of $(\mathbb{C}^2/\mathbb{Z}_2)\times\mathbb{C}^2$. The reader should note that factorisation of the Hilbert series, does not imply factorisation of the highest weight generating function. The HWG \eqref{HWG5 SO(2k+1) + 2k-2 V} in this case reads 
        \begin{equation}
            \PE\left[\mu^2t^2+t^4+\mu(t+t^3)-\mu^2t^6 \right]\;.
        \end{equation}
        This is the HWG for the product space $(\mathbb{C}^2/\mathbb{Z}_2)\times \mathbb{C}^2$, where we only turn on a highest weight fugacity $\mu$ for the diagonal $\mathfrak{su}(2)$ subalgebra of the full $\mathrm{SO}(3)\times \mathrm{SU}(2)$ global symmetry.
    \item The $K=4$ case is another special instance, which needs some clarification. Firstly, there is an enhancement of $\mathfrak{sp}(2)\subset\mathfrak{su}(4)$ given by the following embedding
            \begin{equation}
                \mu_1^2+\mu_2\rightarrow \nu_1\nu_3\;,\qquad 1+\mu_2+\mu_2^2\rightarrow \nu_2^2\;, \label{embedding sp(2) in su(4)}
            \end{equation}
        where $\nu_i$ are $\mathfrak{su}(4)$ highest weight fugacities. This is in line with the fact that the electric theory SO(4) with one vector, can also be recast as $\text{SU}(2)\times\text{SU}(2)$ quiver theory with a bifundamental. The magnetic quiver for the latter is the mirror of 3d $\mathcal{N}=4$ SQCD with gauge group U(2) and 4 fundamental hypermultiplets. The moduli space is the next to minimal nilpotent orbit closure of $\mathfrak{sl}(4)$. The HWG for this moduli space is
            \begin{equation}
                \PE\left[\nu_1\nu_3t^2+\nu_2^2 t^4\right]\;,
            \end{equation}
        which is consistent with \eqref{HWG5 SO(2k+1) + 2k-2 V}, subject to  the embedding \eqref{embedding sp(2) in su(4)}.
\end{enumerate}

As further non-trivial tests of the proposed HWG \eqref{HWG5 SO(2k+1) + 2k-2 V}, we can explicitly compute the Coulomb branch Hilbert series of \eqref{MQ5 SO(2k+1)+(2k-2)} for $K=5,6,8$. We find a match and report the results in the following. For $K=5$ the HWG can be used to obtain an exact result for the Hilbert series
\begin{equation}
    \HS_{K=5}=\frac{\begin{pmatrix}
        1+ 5 t+ 
 24 t^2 + 87 t^3 + 271 t^4  + 700 t^5+ 1605 t^6+ 3217 t^7 +
 5737 t^8\\+ 9090 t^9 + 13001 t^{10}  + 16715 t^{11} + 19429 t^{12} +20384 t^{13}+\cdots\mathrm{palindrome}\cdots+t^{26}
    \end{pmatrix} }{(1-t^2)^{7} (1+t)^{5}  (1-t^3)^7}\;,
\end{equation}
and we have checked explicitly that the Hilbert series computed using the monopole formula matches this result upto $\mathcal{O}(t^8)$
\begin{equation}
    \HS_{K=5}=1 + 21 t^2 + 14 t^3 + 217 t^4 + 230 t^5 + 1575 t^6 + 1946 t^7 + 8918 t^8 + \mathcal{O}(t^9)\;.
\end{equation}
For $K=6$ one can use the HWG to write down the unrefined Hilbert series as a rational function
\begin{equation}
    \HS_{K=6}=\frac{\begin{pmatrix}
        1 + 25 t^2 + 329 t^4 + 2737 t^6 + 15968 t^8 + 68695 t^{10} + 
   226157 t^{12} + 583363 t^{14}\\ + 1200472 t^{16} + 1993924 t^{18} + 
   2695633 t^{20} + 2978864 t^{22} + \cdots\mathrm{palindrome}\cdots + 
   t^{44}
    \end{pmatrix}}{(1 - t^2)^{11} (1 - t^4)^{11}}\;,
\end{equation}
and we have verified that the Hilbert series computed using the monopole formula matches this result upto $\mathcal{O}(t^{28})$
\begin{equation}
\begin{gathered}
    \HS_{K=6}=1 + 36 t^2 + 681 t^4 + 8688 t^6 + 83376 t^8 + 640695 t^{10} + 
 4110730 t^{12}+ 22694925 t^{14}\\ + 110302751 t^{16} + 480427781 t^{18} + 
 1902140400 t^{20} + 6925794468 t^{22} \\+ 23413973581 t^{24}+
 74087529887 t^{26} + 220916273805 t^{28} + \mathcal{O}(t^{30})\;.
\end{gathered}
\end{equation}
For $K=8$ the HWG can be used to compute the unrefined Hilbert series as a rational function
\begin{equation}
    \frac{\begin{pmatrix}
        1 + 56 t^2 + 1531 t^4 + 27688 t^6 + 373814 t^8 + 4007268 t^{10}+ 
   35410574 t^{12}\\ + 264689513 t^{14} + 1706074927 t^{16} + 
   9625411716 t^{18} + 48111821614 t^{20}\\ + 215194626156 t^{22} + 
   868549278349 t^{24} + 3185862046297 t^{26}\\ + 10684897373331 t^{28} + 
   32938282691854 t^{30} + 93754320533694 t^{32} \\+ 247376164912532 t^{34} + 
   607154752929536 t^{36} + 1390361930656500 t^{38} \\+ 
   2978471467191463 t^{40} + 5982810531136653 t^{42} + 
   11291414657177553 t^{44} \\+ 20058473577431171 t^{46} + 
   33591604012947619 t^{48} + 53104999363156848 t^{50} \\+ 
   79345083211048247 t^{52} + 112155131812219537 t^{54} + 
   150106449173251843 t^{56} \\+ 190354750875572161 t^{58} + 
   228852447460099341 t^{60} + 260952375745699730 t^{62} 
   \\+282301011261103367 t^{64} + 289792728765186536 t^{66} +\cdots\mathrm{palindrome}\cdots+t^{132}
    \end{pmatrix}}{(1 - t^2)^{22}(1 - t^6)^{22}}
\end{equation}
Using the monopole formula, we compute upto $\mathcal{O}(t^{20})$
\begin{equation}
\begin{gathered}
    \HS_{K=8}=1+78 t^2+3016 t^4+77584 t^6+1498003 t^8+23175516 t^{10}+299239381 t^{12}\\+3315821509 t^{14}+32181847958 t^{16}+277906637203 t^{18}+2162236623194 t^{20}+\mathcal{O}(t^{21})\;.
\end{gathered}
\end{equation}

\subsubsection{SO$(K)$ with $K-4$ vectors}\label{SO(K) with K-4 vectors}

The brane web for the infinite coupling limit of SO$(K)$ with $K-4$ hypermultiplets in the vector representation is given by \eqref{eq:BW_3point6}.
We present it here again for convenience:
\begin{equation}
    \begin{scriptsize}
    \begin{tikzpicture}
    \draw[thick](1,-1)--(0,-2);
    \draw[thick](2,0)--(1,-1);
    \node[label=right:{$[1,1]$}][7brane] at (1,-1){};
    \node[label=right:{$[1,1]$}][7brane] at (2,0){};
    \node[label=below right:{O7$^+$}][7brane,green] at (0,-2){};
    \draw[dash dot,green](-5.5,-2)--(2,-2);
    \draw[thick](0,-2)--(-2,-2);
    \draw[thick](-5,-2)--(-3,-2);
    \node at (-2.5,-2){$\cdots$};
    \node[7brane] at (-1,-2){};
    \node[7brane] at (-2,-2){};
    \node[7brane] at (-3,-2){};
    \node[7brane] at (-4,-2){};
    \node[7brane] at (-5,-2){};
    \node [label=below:{1}] at (-4.5,-2){};
    \node [label=below:{2}] at (-3.5,-2){};
    \node [label=below:{$K-5$}] at (-1.5,-2){};
    \node [label=below:{$K-4$}] at (-.5,-2){};
    \node[label=right:{2}] at (0,-1.5){};
    \node[label=right:{1}] at (1,-0.5){};
    \end{tikzpicture}
    \end{scriptsize}\;.
\end{equation}
The magnetic quiver for the 
Higgs branch at infinite coupling can be computed following the rules explained in section \ref{sec:intersection_rule}. We get
\begin{equation}
    \begin{array}{c}
         \begin{scriptsize}
    \begin{tikzpicture}
    \node[label=below:{1}][u](2){};
    \node (dots)[right of=2]{$\cdots$};
    \node[label=below:{$K-5$}][u](2N-5)[right of=dots]{};
    \node[label=below:{$K-4$}][u](2N-4)[right of=2N-5]{};
    \node[label=below:{$2$}][u](2')[right of=2N-4]{};
    \node[label=below:{$1$}][u](1')[right of=2']{};
    \draw (2')to[out=45,in=135,loop,looseness=10](2');
    \draw(2)--(dots);
    \draw(dots)--(2N-5);
    \draw[ double distance=1.5pt,<-](2N-5)--(2N-4);
    \draw(2N-4)--(2');
    \draw[double distance=1.5pt,->](2')--(1');
    \end{tikzpicture}
    \end{scriptsize}
    \end{array}\;.\label{MQ5 SO(2k+1)+ 2k-3 V}
\end{equation}

Alternatively, one can arrive at \eqref{MQ5 SO(2k+1)+ 2k-3 V}, following the discussion in section \ref{sec:FIdef}, via a simple FI-like deformation of \eqref{MQ5 SO(2k+1)+(2k-2)} (this is because in  the electric theory we are doing a mass deformation in order to integrate out one flavour). Following \cite{Bourget:2020mez,vanBeest:2021xyt}, we turn on FI terms at the two U$(2)$ gauge nodes in \eqref{MQ5 SO(2k+1)+(2k-2)}, and subtract
\begin{equation}
    \begin{array}{c}
         \begin{scriptsize}
    \begin{tikzpicture}
    \node[label=below:{$1$}][u](1){};
    \node[label=below:{$2$}][u](2)[right of=1]{};
    \node (dots)[right of=2]{$\cdots$};
    \node[label=below:{$K-3$}][u](2N-5)[right of=dots]{};
    \node[label=below:{$K-2$}][u](2N-4)[right of=2N-5]{};
    \node[label=below:{$2$}][u](2')[right of=2N-4]{};
    \draw (2')to[out=-45,in=45,loop,looseness=10](2');
    \draw(1)--(2);
    \draw(2)--(dots);
    \draw(dots)--(2N-5);
    \draw[ double distance=1.5pt,<-](2N-5)--(2N-4);
    \draw(2N-4)--(2');
    \end{tikzpicture}
    \end{scriptsize}\\
    \begin{scriptsize}
             \begin{tikzpicture}
                          \node(minus){$-$};
                          \node[label=below:{2}][u](1)[right of=minus]{};
                          \node(dots)[right of=1]{$\cdots$};
                          \node[label=below:{2}][u](2)[right of=dots]{};
                          \node[label=below:{2}][u](3)[right of=2]{};
                          \node[label=below:{2}][u](4)[right of=3]{};
                          \draw(1)--(dots);
                          \draw(dots)--(2);
                          \draw[<-,double distance=1.5 pt](2)--(3);
                          \draw(3)--(4);
    \draw (4)to[out=-45,in=45,loop,looseness=10](4);
             \end{tikzpicture}
         \end{scriptsize}
    \end{array}
    \label{FI deformation_2}
\end{equation}
which leads to \eqref{MQ5 SO(2k+1)+ 2k-3 V}, after rebalancing with a U$(2)$ gauge node with an adjoint hypermultiplet.
We propose the following HWG for this magnetic quiver
\begin{equation}
    \PE\left[\sum_{i=1}^{K-4}\mu_i^2t^{2i}+\nu^2t^2+t^4+\mu_{K-4}\nu^2\left(t^{K-2}+t^{K}\right)-\mu_{K-4}^2\nu^4t^{2K} \right]\;,\label{HWG5 SO(2k+1) + 2k-3 V}
\end{equation}
where $\mu_i$, $\nu$ are respectively $\mathfrak{sp}(K-4)$ and $\mathfrak{sp}(1)_I$ highest weight fugacities. Note the enhancement from the classical $\mathfrak{sp}(K-4)\oplus \mathfrak{u}(1)_I$ symmetry. In this case the instantonic symmetry is enhanced to $\mathfrak{sp}(1)_I$, which is naturally acting as an $\mathfrak{su}(2)$ symmetry acting on 2 coincident 5-brane legs in the web diagram. We are now going to distinguish this family into even and odd gauge rank, seeing as the global form of the global symmetry is sensitive to this data.
\paragraph{Case $K=2k+1$.} In this case, the presence of terms $\mu_{K-4}\nu^2(t^{K-2}+t^K)$ signal that the global symmetry is $\frac{\mathrm{Sp}(K-4)\times \mathrm{SU}(2)_R}{\mathbb{Z}_2^{\mathrm{diag}}}\times \mathrm{SO}(3)_I$.
\paragraph{Case $K=2k$.} In this case all terms in the HWG transform trivially with respect to the center of $\mathfrak{sp}(K-4)\oplus\mathfrak{sp}(1)_I\oplus\mathfrak{sp}(1)_R$ symmetry, and hence the global form is given by $\frac{\mathrm{Sp}(K-4)}{\mathbb{Z}_2}\times\mathrm{SO}(3)_I\times\mathrm{SO}(3)_R$.
\paragraph{Special cases}
\begin{enumerate}
    \item When $K=4$, the electric theory in question is the UV SCFT associated with pure SO(4) gauge theory, which is the direct sum of two copies of the $E_1$ SCFT \cite{Akhond:2021knl}. The Higgs branch is therefore given by two copies of $\mathbb{C}^2/\mathbb{Z}_2$. Note, that this is a novel construction of a well known moduli space.
    \item For $K=5$ the Hilbert series was computed in \cite{Akhond:2021ffo}, and the moduli space in question is that of two-SU(2) instantons on $\mathbb{C}^2$.
\end{enumerate}
The HWG can be used to compute the Hilbert series for $K=6$ as a rational function
\begin{equation}
    \HS_{K=6}=\frac{(1+8 t^2+40 t^4+107 t^6+199 t^8+234 t^{10}+\cdots\mathrm{palindrome}\cdots+t^{20})}{(1-t^2)^{5} (1-t^4)^5}\;,
\end{equation}
which is consistent with the perturbative result of the Hilbert series computed using the monopole formula
\begin{equation}
\begin{split}
    \HS_{K=6}&=  1 + 13 t^2 + 100 t^4 + 527 t^6 + 2174 t^8 + 7425 t^{10} + 21997 t^{12} \\&\quad+ 
 58102 t^{14} + 139937 t^{16} + 312042 t^{18} + \mathcal{O}(t^{19})\;,
\end{split}
\end{equation}
while for $K=8$, the HWG predicts the following exact expression for the unrefined Hilbert series:
\begin{equation}
    \HS_{K=8}=\frac{\begin{pmatrix}
        1 + 27 t^2 + 350 t^4 + 3039 t^6 + 19857 t^8 + 102780 t^{10} + 
 436925 t^{12}\\+ 1565790 t^{14} + 4817642 t^{16} + 12914156 t^{18} + 
 30501983 t^{20} + 64031824 t^{22} \\+ 120327288 t^{24} + 203523179 t^{26} + 
 311183551 t^{28} + 431569257 t^{30} \\+ 544202481 t^{32} + 625024568 t^{34} + 
 654489596 t^{36}+\cdots\mathrm{palindrome}\cdots+t^{72}
    \end{pmatrix}}{(1-t^2)^{12}(1-t^6)^{12}}
\end{equation}
which is consistent with the result of the Hilbert series computed using the monopole formula upto $\mathcal{O}(t^{24})$:
\begin{equation}
\begin{split}
    \HS_{K=8}&=1 + 39 t^2 + 752 t^4 + 9721 t^6 + 95286 t^8 + 755753 t^{10} + 
 5049975 t^{12}+ 29237547 t^{14} + 149764183 t^{16} \\&\quad+ 689844328 t^{18} + 
 2894916128 t^{20} + 11186872074 t^{22} + 40164406471 t^{24}+\mathcal{O}(t^{25})\;.
\end{split}
\end{equation}
Both of these results are consistent with the proposed HWG \eqref{HWG5 SO(2k+1) + 2k-3 V}, and serve as additional non-trivial consistency checks. Moreover, they are also consistent with the results obtained from alternative orthosymplectic magnetic quivers (see \cite{Akhond:2021ffo}).

\subsubsection{SO$(K)$ with $N<K-4$ vectors}
Let us now consider the UV fixed point associated with 5d SO$(K)$ gauge theory with $N<K-4$ hypermultiplets in the vector representation. The brane web at the SCFT point is given by \eqref{eq:BW3point3} in the $K$ even case and \eqref{eq:BW3point14} in the $K$ odd one. In either case, using the rules of Section \ref{sec:intersection_rule} leads to the following magnetic quiver for any $K$,
\begin{equation}
      \begin{array}{c}
             \begin{scriptsize}
    \begin{tikzpicture}
    \node[label=below:{$N$}][u](2k-2j-4)[below left of=1]{};
      \node[label=above:{1}][u][above right of=2k-2j-4](1){};
    \node[label=below:{1}][u](11)[below right of=2k-2j-4]{};
    \node[label=above:{$N-1$}][u](2k-2j-5)[left of=2k-2j-4]{};
    \node(dots)[left of=2k-2j-5]{$\cdots$};
    \node[label=below:{$2$}][u](2)[left of=dots]{};
    \node[label=below:{$1$}][u](111)[left of=2]{};
    \draw(111)--(2);
    \draw(2)--(dots);
    \draw(dots)--(2k-2j-5);
    \draw[<-,double distance=1.5 pt](2k-2j-5)--(2k-2j-4);
    \draw(1)--(2k-2j-4);
    \draw(11)--(2k-2j-4);
    \draw(1)--node[above right]{$K-N-2$}++(11);
    \end{tikzpicture}
    \end{scriptsize}
    \end{array}\;.\label{MQ5 SO(2k+1)+ (2k-2j-4)V}
\end{equation}

As in the previous section, and since we can obtain this theory by integrating out $K-4-N$ massive vectors from SO$(K)$ with $K-4$ vectors, we can reach the same magnetic quiver by a FI-like deformation. This amounts to perform quiver subtraction from \eqref{MQ5 SO(2k+1)+ 2k-3 V}. Indeed the following subtraction
\begin{equation}
     \begin{array}{cc}
     &
         \begin{scriptsize}
    \begin{tikzpicture}
    \node[label=below:{1}][u](1){};
    \node[label=below:{2}][u](2)[right of=1]{};
    \node (dots)[right of=2]{$\cdots$};
    \node[label=below:{$K-5$}][u](2N-5)[right of=dots]{};
    \node[label=below:{$K-4$}][u](2N-4)[right of=2N-5]{};
    \node[label=below:{$2$}][u](2')[right of=2N-4]{};
    \node[label=below:{$1$}][u](1')[right of=2']{};
    \draw (2')to[out=45,in=135,loop,looseness=10](2');
    \draw(1)--(2);
    \draw(2)--(dots);
    \draw(dots)--(2N-5);
    \draw[ double distance=1.5pt,<-](2N-5)--(2N-4);
    \draw(2N-4)--(2');
    \draw[double distance=1.5pt,->](2')--(1');
    \end{tikzpicture}
    \end{scriptsize}\\
    \begin{tikzpicture}\begin{scriptsize}
                 \node[label=above:{$- (K-N-4)\times$}]{};
                 \end{scriptsize}
    \end{tikzpicture}&
     \begin{scriptsize}
             \begin{tikzpicture}
                         \node[label=below:{1}][u](1){};
                          \node[label=below:{1}][u](11)[right of=1]{};
                          \node(dots)[right of=11]{$\cdots$};
                          \node[label=below:{1}][u](2)[right of=dots]{};
                          \node[label=below:{1}][u](3)[right of=2]{};
                          \node[label=below:{1}][u](4)[right of=3]{};
                          \node[label=below:{1}][u](5)[right of=4]{};
                          \draw(1)--(11);
                          \draw(11)--(dots);
                          \draw(dots)--(2);
                          \draw[<-,double distance=1.5 pt](2)--(3);
                          \draw(3)--(4);
                          \draw[->,double distance=1.5 pt](4)--(5);
             \end{tikzpicture}
         \end{scriptsize}
    \end{array}\;,
\end{equation}
leads, after appropriate rebalancing, as described in section \ref{sec:FIdef}, to \eqref{MQ5 SO(2k+1)+ (2k-2j-4)V}.
The HWG that we propose is
\begin{equation}
    \PE\left[\sum_{i=1}^{N}\mu_i^2t^{2i}+t^2+\left(q+q^{-1}\right)\mu_{N}t^{K-2}-\mu^2_{N}t^{2K-4}\right]\;,
\end{equation}
where $\mu_i$ are $\mathfrak{sp}(N)$ highest weight fugacities, while $q$ is the fugacity for U(1) charge. This expression contains sufficient information to extract the global form of the global symmetry, and as we shall see the global structure will depend on whether $K$ and $N$ are even or odd.
\paragraph{Case $K=2k$, $N=2n$.} In this case every term in the HWG is in a projective representation of $\mathfrak{sp}(N)$ and $\mathfrak{su}(2)_R$. Therefore the global form of the global symmetry is $\frac{\mathrm{Sp}(N)}{\mathbb{Z}_2}\times \mathrm{U}(1)_I\times \mathrm{SO}(3)_R$.
\paragraph{Case $K=2k+1$, $N=2n$.} In this case only projective representations of $\mathfrak{sp}(N)$ appear, while we have terms in fermionic representations of $\mathfrak{su}(2)_R$. Therefore the global form of the global symmetry is $\frac{\mathrm{Sp}(N)}{\mathbb{Z}_2}\times\mathrm{U(1)_I\times\mathrm{SU}(2)_R}$.
\paragraph{Case $K=2k$, $N=2n+1$.} Here we see that all terms in the HWG are in bosonic representations of $\mathfrak{su}(2)_R$, but there are representations which transform non-trivially under the center of the $\mathfrak{sp}(N)$ algebra and thus the global form is $\mathrm{Sp}(N)\times\mathrm{U}(1)_I\times\mathrm{SO(3)_R}$.
\paragraph{Case $K=2k+1$, $N=2n+1$} Here all terms are either in projective representations of both the $\mathfrak{sp}(N)$ and the $\mathfrak{su}(2)_R$ algebra, or transform trivially with respect to both, and so the global form is $\frac{\mathrm{Sp}(N)\times \mathrm{SU}(2)_R}{\mathbb{Z}_2}\times\mathrm{U}(1)_I$.
\paragraph{Special limits}
\begin{enumerate}
    \item In the extreme case when $N=0$, the electric theory is pure SO$(K)$ whose Higgs branch at the fixed point is $\mathbb{C}^2/\mathbb{Z}_{K-2}$. In this limit, upon setting $\mu_0=1$, the HWG reduces to
   \begin{equation}
    \PE\left[t^2+(q+q^{-1})t^{K-2}-t^{2K-4}\right]\;,
\end{equation}
which is the HS for $\mathbb{C}^2/\mathbb{Z}_{K-2}$. We note that the order of the orbifold $K-2$ is the dual coxeter number of $\mathrm{SO}(K)$.
This is consistent with the results of \cite{Cremonesi:2015lsa}, which found that the Higgs branch of the pure gauge theory at infinite coupling is $\mathbb{C}^2/\mathbb{Z}_{h^\vee}$, where $h^\vee$ is the dual coxeter number of the gauge group.
\item Another special limit is when $N=1$. In this case the quiver becomes simply-laced. The HWG in this limit is
\begin{equation}
\PE\left[\mu^2t^{2}+t^2+\left(q+q^{-1}\right)\mu t^{K-2}-\mu^2 t^{2K-4}\right]\;,
\end{equation}
This is in agreement with the proposed HWG in \cite{Bourget:2019rtl}.
\end{enumerate}
Finally, we provide further consistency checks of the proposed HWG by reporting the results of 
explicit Hilbert series computations for specific values of $K$ and $N$.
For SO$(7)+2\textbf{V}$ the HWG predicts the following exact form for the unrefined Hilbert series
\begin{equation}
    \HS_{SO(7)+2\textbf{V}}=\frac{ \begin{pmatrix}
         1+ 2 t+ 8 t^2+ 14 t^3+ 
 29 t^4+ 50 t^5 + 81 t^6 + 112 t^7 + 153 t^8\\ + 178 t^9 + 207 t^{10}+ 216 t^{11} +\cdots\mathrm{palindrome}\cdots+t^{22} 
    \end{pmatrix}}{(1-t^2)^4(1+t)^2(1-t^5)^4}
\end{equation}
while we find that the Hilbert series computed using the monopole formula is consistent with the above result upto $\mathcal{O}(t^{19})$
\begin{equation}
\begin{gathered}
\HS_{SO(7)+2\textbf{V}}=   1 + 11 t^2 + 60 t^4 + 10 t^5 + 225 t^6 + 80 t^7 + 665 t^8 + 
 350 t^9 + 1694 t^{10} + 1120 t^{11} + 3886 t^{12}\\+ 2940 t^{13}+ 
 8210 t^{14} + 6780 t^{15} + 16195 t^{16}+14228 t^17 + 30125 t^{18}+27730 t^{19} +\mathcal{O}(t^{20})\;.
\end{gathered}
\end{equation}
When $K=8$ and $N=2$, corresponding to the Higgs branch of the fixed point limit of SO(8)+2\textbf{V} the HWG predicts the unrefined Hilbert series to be 
\begin{equation}
    \HS_{SO(8)+2\textbf{V}}=\frac{1 + 7 t^2 + 22 t^4 + 53 t^6 + 94 t^8 + 129 t^{10} + 148 t^{12}+\cdots\mathrm{palindrome}\cdots+t^{24}}{(1-t^2)^4(1-t^6)^4}\;,
\end{equation}
while the computation of the Hilbert series using the monopole formula agrees with the above exact result upto $\mathcal{O}(t^{20})$
\begin{equation}
    \begin{gathered}
    \HS_{SO(8)+2\textbf{V}}= 1 + 11 t^2 + 60 t^4 + 235 t^6 + 745 t^8 + 2016 t^{10} + 4844 t^{12} + 
 10600 t^{14} \\+ 21485 t^{16}+ 40895 t^{18} + 73844 t^{20}+\mathcal{O}(t^{21})\;.
    \end{gathered}
\end{equation}

\section{Theories with special unitary gauge group}
\label{sec:specialunitary}
In this section we further apply our results to 5d field theories, whose low energy regime has an effective field theory description in terms of an $\mathrm{SU}(K)_c$ gauge theory with one hypermultiplet in the 2nd rank symmetric representation of the gauge group, as well as 
$N$ hypermultiplets in the fundamental representation. The electric quiver is given by
\begin{equation}\label{electric quiver SU}
    \sukweq{K}{N}{c}
\end{equation}
There is a consistency condition on the possible allowed values of the number of fundamental flavours $N$, the number of colours $K$, and the CS level $c$. In particular, if $K+N$ is even (resp. odd) then $c$ has to be integer (resp. semi-integer) \cite{Bergman:2015dpa}.

The magnetic quivers typically do not have a polynomial HWG, and so we do not make any attempt to systematically study their Hilbert series. This can be an interesting direction to pursue in future work. 
Instead, we derive the magnetic quivers from the brane webs. It can be easily checked that, for theories related by mass deformations, the magnetic quivers are related by FI deformations, as expected from the general discussion of section \ref{sec:intersection_rule}.

\subsection{The Brane Web}

Brane webs for SU$(K)$ gauge theory with a hypermultiplet in the second rank symmetric, as well as fundamental hypers were constructed in \cite{Bergman:2015dpa, Hayashi:2015vhy}. Here we use these brane webs to extract the magnetic quivers at finite and infinite gauge coupling following the techniques of section \ref{sec:FIdef}. 

For the sake of displaying the brane web, the discussion is divided into the $K$ even and $K$ odd case. However, for derivation of magnetic quivers, a different approach is taken. We will consider the asymptotics of the web, and divide the set of theories into six families. For each family, the procedure of shrinking the web to go to infinite coupling is qualitatively different.

\paragraph{SU(even).}
The brane web is parametrized by four integer numbers: $\alpha$, $N_L$, $N_R$, $k=K/2$. They are respectively the Chern-Simons level in absence of flavour, the number of hypers on the left part of the web, the number of hypers on the right part of the web, and the number of colours. The number of flavours is  $N=N_L+N_R$, and the Chern-Simons level is $c=\alpha+\frac{N_L-N_R}{2}$. This parametersiation is chosen in order to highlight the connection with the brane web, which is given by
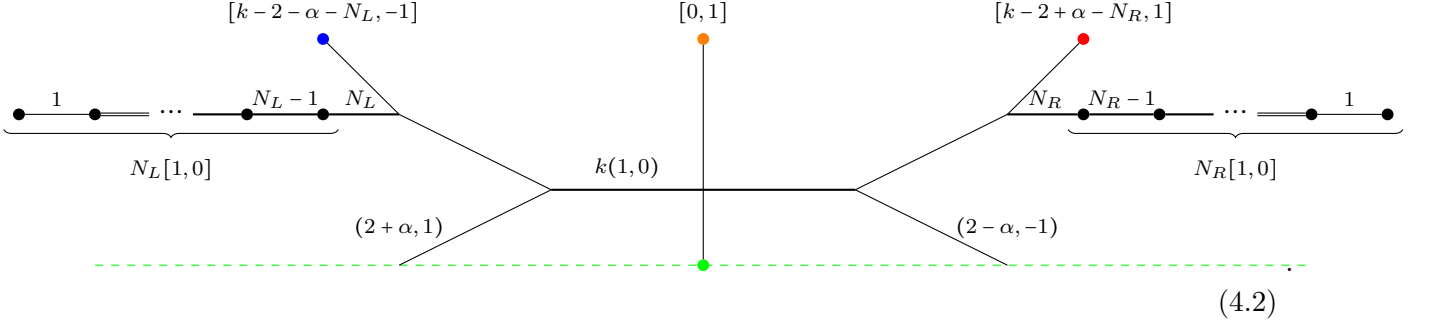
\begin{equation}
    \makebox[\textwidth][c]{\begin{tikzpicture}
        \node[7brane,green] (0) at (0,0) {};
        \draw[dashed,green] (-8,0)--(8,0);
        \draw (-4,0)--(-2,1) (2,1)--(4,0);
        \node at (-4,0.5) {\scriptsize$(2+\alpha,1)$};
        \node at (4,0.5) {\scriptsize$(2-\alpha,-1)$};
        \draw[thick] (-2,1)--(2,1);
        \node at (-1,1.3) {\scriptsize$k(1,0)$};
        \node[7brane,blue,label=above:{\scriptsize$[k-2-\alpha-N_L,-1]$}] (L) at (-5,3) {};
        \draw (-2,1)--(-4,2)--(L);
        \node[7brane] (1) at (-9,2) {};
        \node[7brane] (2) at (-8,2) {};
        \node (3) at (-7,2) {$\cdots$};
        \node[7brane] (4) at (-6,2) {};
        \node[7brane] (5) at (-5,2) {};
        \draw (1)--(2);
        \draw[double] (2)--(3);
        \draw[thick] (3)--(4)--(5)--(-4,2);
        \node at (-8.5,2.2) {\scriptsize$1$};
        \node at (-5.5,2.2) {\scriptsize$N_L-1$};
        \node at (-4.5,2.2) {\scriptsize$N_L$};
        \draw [decorate,decoration={brace,amplitude=3pt},xshift=0pt,yshift=0pt](-4.8,1.8) -- (-9.2,1.8) node [midway,xshift=0pt,yshift=-15pt] {\scriptsize $N_L[1,0]$};
        
        \node[7brane,red,label=above:{\scriptsize$[k-2+\alpha-N_R,1]$}] (R) at (5,3) {};
        \draw (2,1)--(4,2)--(R);
        \node[7brane] (10) at (9,2) {};
        \node[7brane] (9) at (8,2) {};
        \node (8) at (7,2) {$\cdots$};
        \node[7brane] (7) at (6,2) {};
        \node[7brane] (6) at (5,2) {};
        \draw (10)--(9);
        \draw[double] (9)--(8);
        \draw[thick] (8)--(7)--(6)--(4,2);
        \node at (8.5,2.2) {\scriptsize$1$};
        \node at (5.5,2.2) {\scriptsize$N_R-1$};
        \node at (4.5,2.2) {\scriptsize$N_R$};
        \draw [decorate,decoration={brace,amplitude=3pt},xshift=0pt,yshift=0pt](9.2,1.8) -- (4.8,1.8) node [midway,xshift=0pt,yshift=-15pt] {\scriptsize $N_R[1,0]$};

        \node[7brane,orange,label=above:{\scriptsize$[0,1]$}] (M) at (0,3) {};
        \draw (0)--(M);
    \end{tikzpicture}}\;.
\end{equation}
The three 7-branes at the top of the diagram are coloured blue, orange, and red, for later reference. This is particularly useful for keeping track of each 7-brane when performing Hanany-Witten moves.
\paragraph{SU(odd).}
The brane web is parametrized by four integer numbers: $\alpha$, $N_L$, $N_R$, $k=(K-1)/2$. They are respectively the Chern-Simons level in absence of flavour, the number of hypers on the left part of the web, the number of hypers on the right part of the web, and the number of colours. The number of flavours is  $N=N_L+N_R$, and the Chern-Simons level is $c=\alpha-\frac{1}{2}+\frac{N_L-N_R}{2}$.

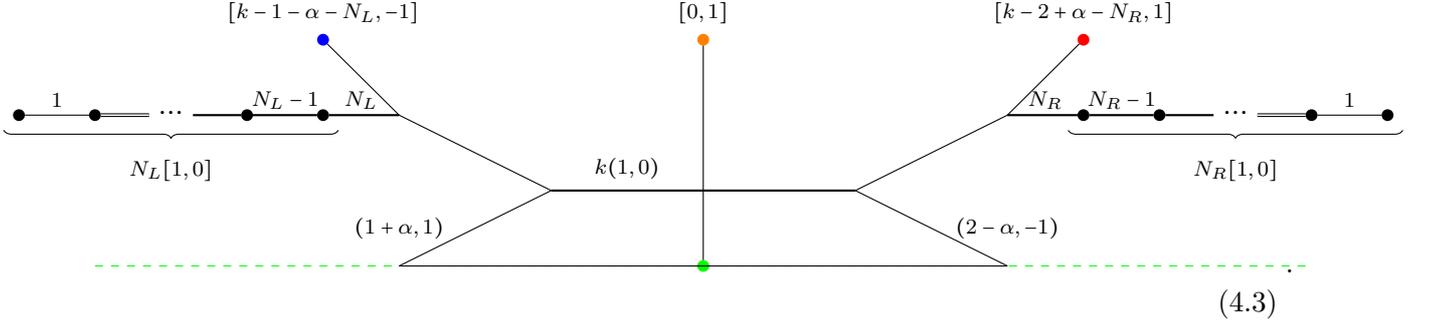
\begin{equation}
    \makebox[\textwidth][c]{\begin{tikzpicture}
        \node[7brane,green] (0) at (0,0) {};
        \draw[dashed,green] (-8,0)--(8,0);
        \draw (-4,0)--(-2,1) (2,1)--(4,0);
        \draw (-4,0)--(4,0);
        \node at (-4,0.5) {\scriptsize$(1+\alpha,1)$};
        \node at (4,0.5) {\scriptsize$(2-\alpha,-1)$};
        \draw[thick] (-2,1)--(2,1);
        \node at (-1,1.3) {\scriptsize$k(1,0)$};
        \node[7brane,blue,label=above:{\scriptsize$[k-1-\alpha-N_L,-1]$}] (L) at (-5,3) {};
        \draw (-2,1)--(-4,2)--(L);
        \node[7brane] (1) at (-9,2) {};
        \node[7brane] (2) at (-8,2) {};
        \node (3) at (-7,2) {$\cdots$};
        \node[7brane] (4) at (-6,2) {};
        \node[7brane] (5) at (-5,2) {};
        \draw (1)--(2);
        \draw[double] (2)--(3);
        \draw[thick] (3)--(4)--(5)--(-4,2);
        \node at (-8.5,2.2) {\scriptsize$1$};
        \node at (-5.5,2.2) {\scriptsize$N_L-1$};
        \node at (-4.5,2.2) {\scriptsize$N_L$};
        \draw [decorate,decoration={brace,amplitude=3pt},xshift=0pt,yshift=0pt](-4.8,1.8) -- (-9.2,1.8) node [midway,xshift=0pt,yshift=-15pt] {\scriptsize $N_L[1,0]$};        
        \node[7brane,red,label=above:{\scriptsize$[k-2+\alpha-N_R,1]$}] (R) at (5,3) {};
        \draw (2,1)--(4,2)--(R);
        \node[7brane] (10) at (9,2) {};
        \node[7brane] (9) at (8,2) {};
        \node (8) at (7,2) {$\cdots$};
        \node[7brane] (7) at (6,2) {};
        \node[7brane] (6) at (5,2) {};
        \draw (10)--(9);
        \draw[double] (9)--(8);
        \draw[thick] (8)--(7)--(6)--(4,2);
        \node at (8.5,2.2) {\scriptsize$1$};
        \node at (5.5,2.2) {\scriptsize$N_R-1$};
        \node at (4.5,2.2) {\scriptsize$N_R$};
        \draw [decorate,decoration={brace,amplitude=3pt},xshift=0pt,yshift=0pt](9.2,1.8) -- (4.8,1.8) node [midway,xshift=0pt,yshift=-15pt] {\scriptsize $N_R[1,0]$};
        \node[7brane,orange,label=above:{\scriptsize$[0,1]$}] (M) at (0,3) {};
        \draw (0)--(M);
    \end{tikzpicture}}\;.
\end{equation}

\paragraph{Infinite coupling.}
We can discuss both cases of SU(odd) and SU(even) together. For the sake of simplifying notation, it is useful to make the following change of parametrization of the theory and brane web,
 \begin{equation}
    \begin{split}
        l &= \left\lceil \frac{K}{2} \right\rceil-2-\alpha-N_L\;,\\
        j &= \left\lfloor \frac{K}{2} \right\rfloor-2+\alpha-N_R\;.
    \end{split}
\end{equation}
where\footnote{Notice that exchanging $j$ and $l$ results in keeping the number of colours fixed, while the level changes sign. The SCFT is invariant upon this exchange, therefore in the following we do not discuss the two symmetric cases independently.}
\begin{align}\label{parameterisation of K,N,c}
  K=N+j+l+4\;,\qquad   N=N_L+N_R\,,\qquad c=\frac{j-l}{2}\;,
\end{align}
The top part of the brane web for the theory looks like
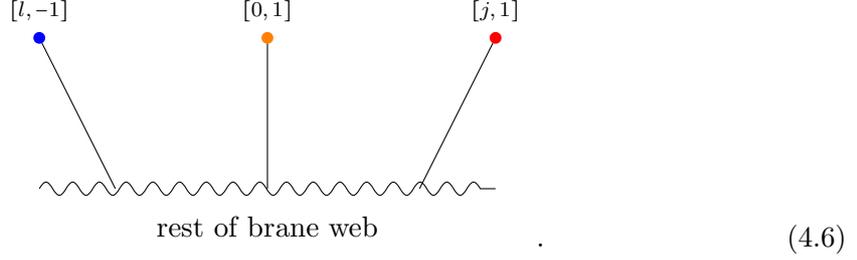
\begin{equation}
    \begin{tikzpicture}
        \node[7brane,blue,label=above:{\scriptsize$[l,-1]$}] (L) at (-3,2) {};
        \node[7brane,red,label=above:{\scriptsize$[j,1]$}] (R) at (3,2) {};
        \node[7brane,orange,label=above:{\scriptsize$[0,1]$}] (M) at (0,2) {};
        \draw (-2,0)--(L) (2,0)--(R) (0,0)--(M);
        \draw[decorate,decoration={coil,aspect=0}] (-3,0)--(3,0);
        \node at (0,-0.5) {rest of brane web};
    \end{tikzpicture}\;.
\end{equation}

The bounds relating $N_L$, $N_R$, $K$ and $\alpha$ \cite{Bergman:2015dpa} such that it is possible to shrink the web to the SCFT point translate to bounds on $j, l$. Moreover, depending on the particular (allowed) values, a different sequence of Hanany-Witten moves is needed to do the shrinking. In what follows, we discuss the various cases:
\begin{enumerate}
    \item $l>0$ and $j>0$: It is straightforward to shrink the web to the SCFT point,
        
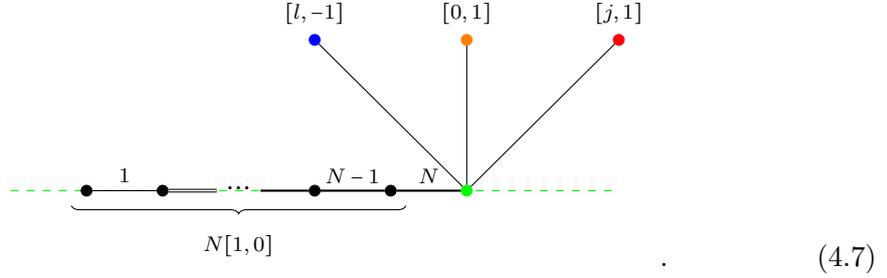
\begin{equation}\label{eq:BW_case1}
            \begin{tikzpicture}
                \node[7brane,green] (0) at (0,0) {};
                \draw[dashed,green] (-6,0)--(2,0);
                \node[7brane] (1) at (-5,0) {};
                \node[7brane] (2) at (-4,0) {};
                \node (3) at (-3,0) {$\cdots$};
                \node[7brane] (4) at (-2,0) {};
                \node[7brane] (5) at (-1,0) {};
                \draw (1)--(2);
                \draw[double] (2)--(3);
                \draw[thick] (3)--(4)--(5)--(0);
                \node at (-4.5,0.2) {\scriptsize$1$};
                \node at (-1.5,0.2) {\scriptsize$N-1$};
                \node at (-0.5,0.2) {\scriptsize$N$};
                \draw [decorate,decoration={brace,amplitude=3pt},xshift=0pt,yshift=0pt](-0.8,-0.2) -- (-5.2,-0.2) node [midway,xshift=0pt,yshift=-15pt] {\scriptsize $N[1,0]$};
                \node[7brane,blue,label=above:{\scriptsize$[l,-1]$}] (L) at (-2,2) {};
                \node[7brane,red,label=above:{\scriptsize$[j,1]$}] (R) at (2,2) {};
                \node[7brane,orange,label=above:{\scriptsize$[0,1]$}] (M) at (0,2) {};
                \draw (L)--(0)--(R) (M)--(0);
            \end{tikzpicture}\;.
        \end{equation}
    \item $l=0$ and $j>0$: Here it is also straight forward to shrink the web to the SCFT point
        
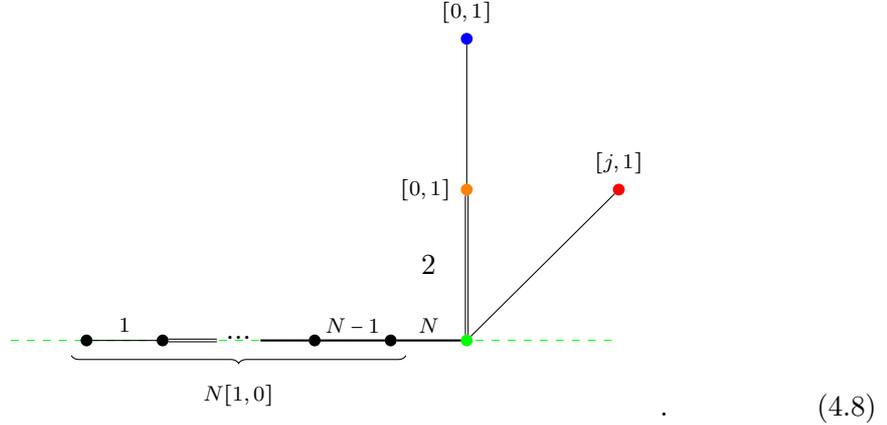
\begin{equation}\label{eq:BW_case2}
            \begin{tikzpicture}
                \node[7brane,green] (0) at (0,0) {};
                \draw[dashed,green] (-6,0)--(2,0);
                \node[7brane] (1) at (-5,0) {};
                \node[7brane] (2) at (-4,0) {};
                \node (3) at (-3,0) {$\cdots$};
                \node[7brane] (4) at (-2,0) {};
                \node[7brane] (5) at (-1,0) {};
                \draw (1)--(2);
                \draw[double] (2)--(3);
                \draw[thick] (3)--(4)--(5)--(0);
                \node at (-4.5,0.2) {\scriptsize$1$};
                \node at (-1.5,0.2) {\scriptsize$N-1$};
                \node at (-0.5,0.2) {\scriptsize$N$};
                \draw [decorate,decoration={brace,amplitude=3pt},xshift=0pt,yshift=0pt](-0.8,-0.2) -- (-5.2,-0.2) node [midway,xshift=0pt,yshift=-15pt] {\scriptsize $N[1,0]$};
                \node[7brane,blue,label=above:{\scriptsize$[0,1]$}] (L) at (0,4) {};
                \node[7brane,red,label=above:{\scriptsize$[j,1]$}] (R) at (2,2) {};
                \node[7brane,orange,label=left:{\scriptsize$[0,1]$}] (M) at (0,2) {};
                \draw (0)--(R) (L)--(M);
                \draw[double] (M)--(0);
                \node at (-0.5,1) {2};
            \end{tikzpicture}\;.
        \end{equation}
    
    \item $l=-1$ and $j>1$: In this case, we need to perform a Hanany-Witten transition in order to convexify the web. First we use the monodromy
        
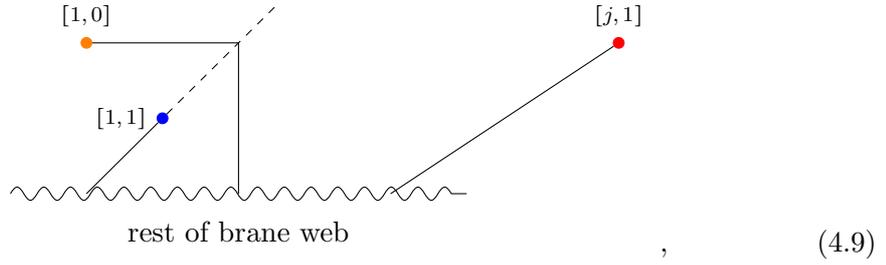
\begin{equation}
            \begin{tikzpicture}
                \node[7brane,blue,label=left:{\scriptsize$[1,1]$}] (L) at (-1,1) {};
                \node[7brane,red,label=above:{\scriptsize$[j,1]$}] (R) at (5,2) {};
                \node[7brane,orange,label=above:{\scriptsize$[1,0]$}] (M) at (-2,2) {};
                \draw[dashed] (L)--(0.5,2.5);
                \draw (-2,0)--(L) (2,0)--(R) (0,0)--(0,2)--(M);
                \draw[decorate,decoration={coil,aspect=0}] (-3,0)--(3,0);
                \node at (0,-0.5) {rest of brane web};
            \end{tikzpicture}\;,
        \end{equation}
        then we perform brane creation
        
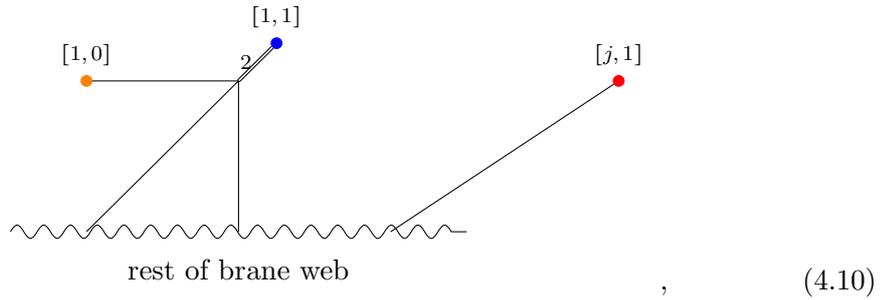
\begin{equation}
            \begin{tikzpicture}
                \node[7brane,blue,label=above:{\scriptsize$[1,1]$}] (L) at (0.5,2.5) {};
                \node[7brane,red,label=above:{\scriptsize$[j,1]$}] (R) at (5,2) {};
                \node[7brane,orange,label=above:{\scriptsize$[1,0]$}] (M) at (-2,2) {};
                \draw[double] (L)--(0,2);
                \node at (0.1,2.25) {\scriptsize$2$};
                \draw (-2,0)--(0,2) (2,0)--(R) (0,0)--(0,2)--(M);
                \draw[decorate,decoration={coil,aspect=0}] (-3,0)--(3,0);
                \node at (0,-0.5) {rest of brane web};
            \end{tikzpicture}\;,
        \end{equation}
        after which the web is convex and we can shrink it to the SCFT point,
        
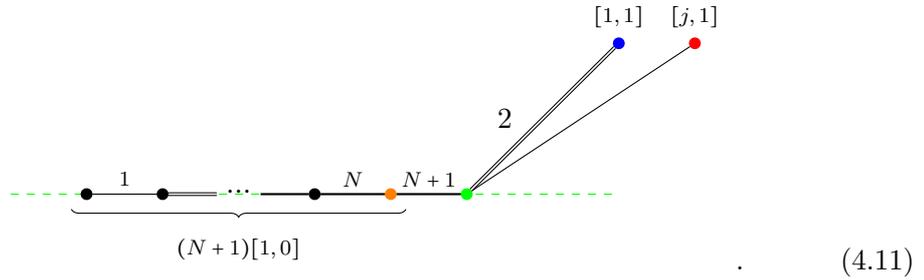
\begin{equation}\label{eq:BW_case3}
            \begin{tikzpicture}
                \node[7brane,green] (0) at (0,0) {};
                \draw[dashed,green] (-6,0)--(2,0);
                \node[7brane] (1) at (-5,0) {};
                \node[7brane] (2) at (-4,0) {};
                \node (3) at (-3,0) {$\cdots$};
                \node[7brane] (4) at (-2,0) {};
                \node[7brane,orange] (5) at (-1,0) {};
                \draw (1)--(2);
                \draw[double] (2)--(3);
                \draw[thick] (3)--(4)--(5)--(0);
                \node at (-4.5,0.2) {\scriptsize$1$};
                \node at (-1.5,0.2) {\scriptsize$N$};
                \node at (-0.5,0.2) {\scriptsize$N+1$};
                \draw [decorate,decoration={brace,amplitude=3pt},xshift=0pt,yshift=0pt](-0.8,-0.2) -- (-5.2,-0.2) node [midway,xshift=0pt,yshift=-15pt] {\scriptsize $(N+1)[1,0]$};
                \node[7brane,blue,label=above:{\scriptsize$[1,1]$}] (L) at (2,2) {};
                \node[7brane,red,label=above:{\scriptsize$[j,1]$}] (R) at (3,2) {};
                \draw (0)--(R);
                \draw[double] (L)--(0);
                \node at (0.5,1) {2};
            \end{tikzpicture}\;.
        \end{equation}

  \item $l=-1$ and $j=0$: Here we also need to perform a Hanany-Witten transition in order to convexify the web. First we use the monodromy
        
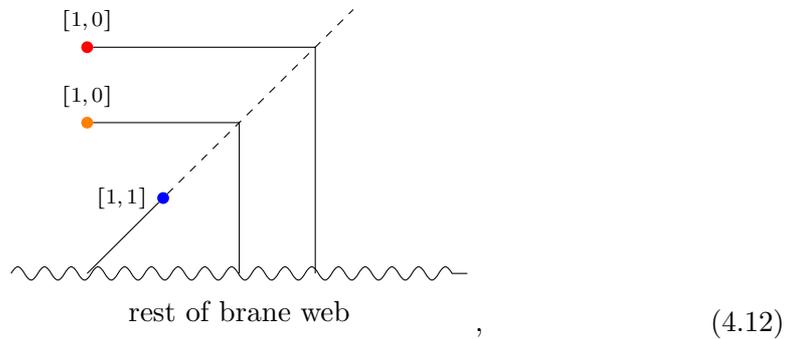
\begin{equation}
            \begin{tikzpicture}
                \node[7brane,blue,label=left:{\scriptsize$[1,1]$}] (L) at (-1,1) {};
                \node[7brane,red,label=above:{\scriptsize$[1,0]$}] (R) at (-2,3) {};
                \node[7brane,orange,label=above:{\scriptsize$[1,0]$}] (M) at (-2,2) {};
                \draw[dashed] (L)--(1.5,3.5);
                \draw (-2,0)--(L) (1,0)--(1,3)--(R) (0,0)--(0,2)--(M);
                \draw[decorate,decoration={coil,aspect=0}] (-3,0)--(3,0);
                \node at (0,-0.5) {rest of brane web};
            \end{tikzpicture}\;,
        \end{equation}
        then we perform brane creation
        
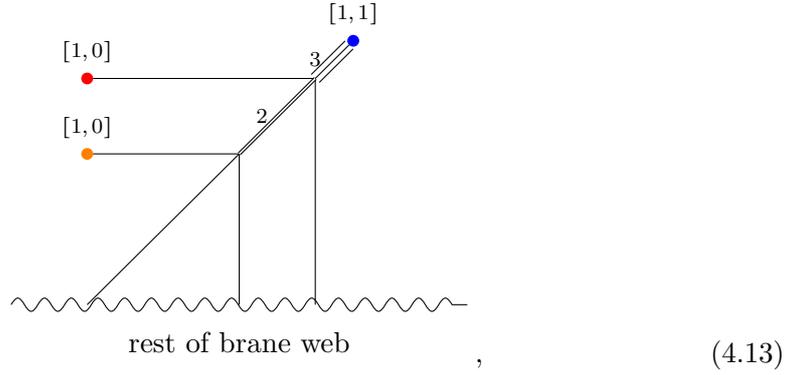
\begin{equation}
            \begin{tikzpicture}
                \node[7brane,blue,label=above:{\scriptsize$[1,1]$}] (L) at (1.5,3.5) {};
                \node[7brane,red,label=above:{\scriptsize$[1,0]$}] (R) at (-2,3) {};
                \node[7brane,orange,label=above:{\scriptsize$[1,0]$}] (M) at (-2,2) {};
                \draw (-2,0)--(0,2) (1,0)--(1,3)--(R) (0,0)--(0,2)--(M);
                \draw[double] (0,2)--(1,3);
                \node at (0.3,2.5) {\scriptsize$2$};
                \draw[transform canvas={xshift=1.5pt,yshift=-1.5pt}] (L)--(1,3);
                \draw (L)--(1,3);
                \draw[transform canvas={xshift=-1.5pt,yshift=1.5pt}] (L)--(1,3);
                \node at (1,3.25) {\scriptsize$3$};
                \draw[decorate,decoration={coil,aspect=0}] (-3,0)--(3,0);
                \node at (0,-0.5) {rest of brane web};
            \end{tikzpicture}\;,
        \end{equation}
        after which we can convexify the web
        
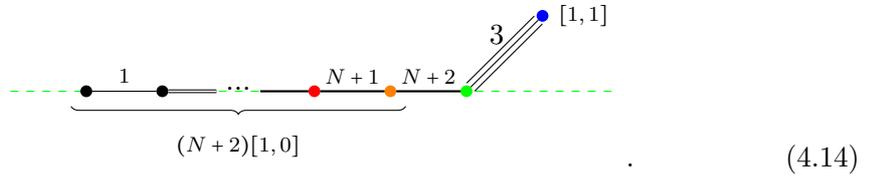
\begin{equation}\label{eq:BW_case4}
            \begin{tikzpicture}
                \node[7brane,green] (0) at (0,0) {};
                \draw[dashed,green] (-6,0)--(2,0);
                \node[7brane] (1) at (-5,0) {};
                \node[7brane] (2) at (-4,0) {};
                \node (3) at (-3,0) {$\cdots$};
                \node[7brane,red] (4) at (-2,0) {};
                \node[7brane,orange] (5) at (-1,0) {};
                \draw (1)--(2);
                \draw[double] (2)--(3);
                \draw[thick] (3)--(4)--(5)--(0);
                \node at (-4.5,0.2) {\scriptsize$1$};
                \node at (-1.5,0.2) {\scriptsize$N+1$};
                \node at (-0.5,0.2) {\scriptsize$N+2$};
                \draw [decorate,decoration={brace,amplitude=3pt},xshift=0pt,yshift=0pt](-0.8,-0.2) -- (-5.2,-0.2) node [midway,xshift=0pt,yshift=-15pt] {\scriptsize $(N+2)[1,0]$};
                \node[7brane,blue,label=right:{\scriptsize$[1,1]$}] (L) at (1,1) {};
                \draw[transform canvas={xshift=1.5pt,yshift=-1.5pt}] (L)--(0);
                \draw (L)--(0);
                \draw[transform canvas={xshift=-1.5pt,yshift=1.5pt}] (L)--(0);
                \node at (0.4,0.75) {3};
            \end{tikzpicture}\;.
        \end{equation}

\item $j=l=0$: In this case, it is again straight forward to shrink the web to the SCFT point
        
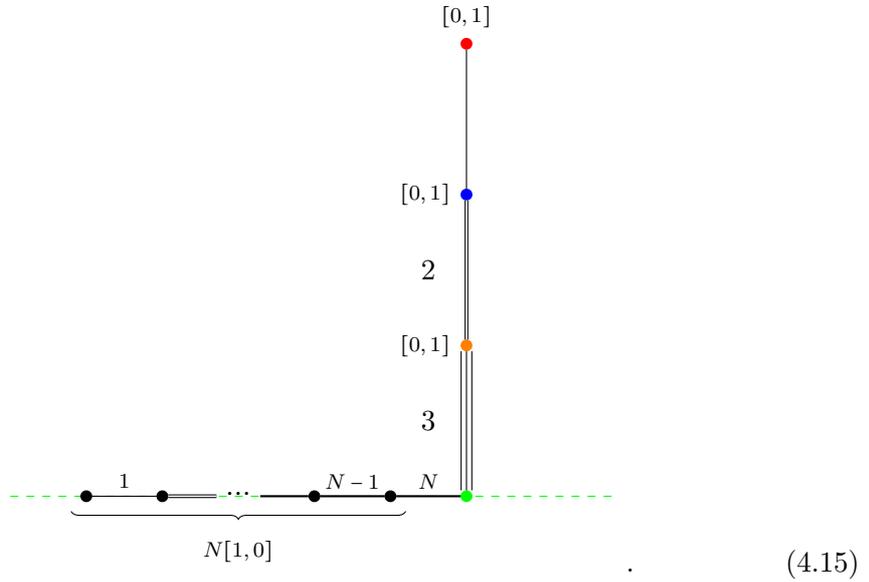
\begin{equation}\label{eq:BW_case5}
            \begin{tikzpicture}
                \node[7brane,green] (0) at (0,0) {};
                \draw[dashed,green] (-6,0)--(2,0);
                \node[7brane] (1) at (-5,0) {};
                \node[7brane] (2) at (-4,0) {};
                \node (3) at (-3,0) {$\cdots$};
                \node[7brane] (4) at (-2,0) {};
                \node[7brane] (5) at (-1,0) {};
                \draw (1)--(2);
                \draw[double] (2)--(3);
                \draw[thick] (3)--(4)--(5)--(0);
                \node at (-4.5,0.2) {\scriptsize$1$};
                \node at (-1.5,0.2) {\scriptsize$N-1$};
                \node at (-0.5,0.2) {\scriptsize$N$};
                \draw [decorate,decoration={brace,amplitude=3pt},xshift=0pt,yshift=0pt](-0.8,-0.2) -- (-5.2,-0.2) node [midway,xshift=0pt,yshift=-15pt] {\scriptsize $N[1,0]$};
                \node[7brane,blue,label=left:{\scriptsize$[0,1]$}] (L) at (0,4) {};
                \node[7brane,red,label=above:{\scriptsize$[0,1]$}] (R) at (0,6) {};
                \node[7brane,orange,label=left:{\scriptsize$[0,1]$}] (M) at (0,2) {};
                \draw (R)--(L);
                \draw[double] (L)--(M);
                \draw[transform canvas={xshift=2pt}] (M)--(0);
                \draw (M)--(0);
                \draw[transform canvas={xshift=-2pt}] (M)--(0);
                \node at (-0.5,3) {2};
                \node at (-0.5,1) {3};
            \end{tikzpicture}\;.
        \end{equation}
        
    \item $l=-1$ and $j=1$: Now we need to perform a Hanany-Witten transition in order to convexify the web. First using the monodromy
        \begin{equation}
            \begin{tikzpicture}
                \node[7brane,blue,label=left:{\scriptsize$[1,1]$}] (L) at (-1,1) {};
                \node[7brane,red,label=above:{\scriptsize$[1,1]$}] (R) at (5,3) {};
                \node[7brane,orange,label=above:{\scriptsize$[1,0]$}] (M) at (-2,2) {};
                \draw[dashed] (L)--(0.5,2.5);
                \draw (-2,0)--(L) (2,0)--(R) (0,0)--(0,2)--(M);
                \draw[decorate,decoration={coil,aspect=0}] (-3,0)--(3,0);
                \node at (0,-0.5) {rest of brane web};
            \end{tikzpicture}\;,
        \end{equation}
        then, performing a brane creation
        \begin{equation}
            \begin{tikzpicture}
                \node[7brane,blue,label=above:{\scriptsize$[1,1]$}] (L) at (0.5,2.5) {};
                \node[7brane,red,label=above:{\scriptsize$[1,1]$}] (R) at (5,3) {};
                \node[7brane,orange,label=above:{\scriptsize$[1,0]$}] (M) at (-2,2) {};
                \draw[double] (L)--(0,2);
                \node at (0.1,2.25) {\scriptsize$2$};
                \draw (-2,0)--(0,2) (2,0)--(R) (0,0)--(0,2)--(M);
                \draw[decorate,decoration={coil,aspect=0}] (-3,0)--(3,0);
                \node at (0,-0.5) {rest of brane web};
            \end{tikzpicture}\;,
        \end{equation}
        after which one can convexify the web
        \begin{equation}\label{eq:BW_case6}
            \begin{tikzpicture}
                \node[7brane,green] (0) at (0,0) {};
                \draw[dashed,green] (-6,0)--(2,0);
                \node[7brane] (1) at (-5,0) {};
                \node[7brane] (2) at (-4,0) {};
                \node (3) at (-3,0) {$\cdots$};
                \node[7brane] (4) at (-2,0) {};
                \node[7brane,orange] (5) at (-1,0) {};
                \draw (1)--(2);
                \draw[double] (2)--(3);
                \draw[thick] (3)--(4)--(5)--(0);
                \node at (-4.5,0.2) {\scriptsize$1$};
                \node at (-1.5,0.2) {\scriptsize$N$};
                \node at (-0.5,0.2) {\scriptsize$N+1$};
                \draw [decorate,decoration={brace,amplitude=3pt},xshift=0pt,yshift=0pt](-0.8,-0.2) -- (-5.2,-0.2) node [midway,xshift=0pt,yshift=-15pt] {\scriptsize $(N+1)[1,0]$};
                \node[7brane,blue,label=right:{\scriptsize$[1,1]$}] (L) at (1,1) {};
                \node[7brane,red,label=above:{\scriptsize$[1,1]$}] (R) at (2,2) {};
                \draw (L)--(R);
                \draw[transform canvas={xshift=1.5pt,yshift=-1.5pt}] (L)--(0);
                \draw (L)--(0);
                \draw[transform canvas={xshift=-1.5pt,yshift=1.5pt}] (L)--(0);
                \node at (0.4,0.75) {3};
            \end{tikzpicture}\;.
        \end{equation}
  
\end{enumerate}

The case for $l=-1$, $j=-1$ UV-completes to a 6d SCFT, and so it is not considered here.

\subsection{Finite coupling magnetic quivers}

The brane web for the electric quiver
\begin{equation}
    \sukweq{K}{N}{c}\,,
\end{equation}
at finite coupling and at the origin of the moduli space is
\begin{equation}
    \makebox[\textwidth][c]{\begin{tikzpicture}
        \node[7brane,green] (0) at (0,0) {};
        \node[7brane] (1) at (-7,0) {};
        \node[7brane] (2) at (-6,0) {};
        \node (3) at (-5,0) {$\cdots$};
        \node[7brane] (4) at (-4,0) {};
        \node[7brane] (5) at (-3,0) {};
        \draw (1)--(2);
        \draw[double] (2)--(3);
        \draw[thick] (3)--(4)--(5)--(0);
        \node at (-6.5,0.2) {\scriptsize$1$};
        \node at (-3.5,0.2) {\scriptsize$N-1$};
        \node at (-2.5,0.2) {\scriptsize$N$};
        \node at (-0.5,0.2) {\scriptsize$K$};
        \node[7brane,label=above:{\scriptsize$[\frac{K}{2}+\frac{\epsilon}{2}-2-\alpha-N_L,-1]$}] (l1) at (-2,1) {};
        \node[7brane,label=below:{\scriptsize$[\frac{K}{2}-\frac{\epsilon}{2}-2+\alpha-N_R,1]$}] (l2) at (-2,-1) {};
        \draw (l1)--(-1,0)--(l2);
        \node[7brane,label=right:{\scriptsize$[0,1]$}] (t) at (0,1) {};
        \draw (t)--(0);
        \draw[dashed,green] (-8,0)--(2,0);
    \end{tikzpicture}}\;,
\end{equation}
where $\epsilon=K \;\mathrm{mod} \;2$.
\paragraph{1.} For $N> K$ we read the magnetic quiver
\begin{equation}
    \begin{tikzpicture}
        \node[u,label=below:{\scriptsize$1$}] (1) at (1,0) {};
        \node[u,label=below:{\scriptsize$2$}] (2) at (2,0) {};
        \node (3) at (3,0) {$\cdots$};
        \node[u,label=below:{\scriptsize$K$}] (4) at (4,0) {};
        \node[u,label=above:{\scriptsize$1$}] (4u) at (4,1) {};
        \node[u,label=below:{\scriptsize$K$}] (5) at (5,0) {};
        \node (6) at (6,0) {$\cdots$};
        \node[u,label=below:{\scriptsize$K$}] (7) at (7,0) {};
        \node[u,label=below:{\scriptsize$K$}] (8) at (8,0) {};
        \node[u,label=above:{\scriptsize$1$}] (8u) at (8,1) {};
        \draw (1)--(2)--(3)--(4)--(5)--(6)--(7);
        \draw[<-,double distance=1.5pt](7)--(8);
        \draw (4)--(4u);
        \draw (8u)--(8);
        \draw [decorate,decoration={brace,amplitude=3pt},xshift=0pt,yshift=0pt](8.2,-0.5) -- (0.8,-0.5) node [midway,xshift=0pt,yshift=-15pt] {\scriptsize $N$};
    \end{tikzpicture}\;.
\end{equation}
The set of balanced nodes imply the symmetry to be $\mathfrak{su}(N)\oplus \mathfrak{u}(1)\oplus\mathfrak{u}(1)$ for $K>2$. When $K=2$, the symmetry is $\mathfrak{so}(2N)\oplus \mathfrak{su}(2)$.
\paragraph{2.} For $N = K$ we read the magnetic quiver 
\begin{equation}
    \begin{tikzpicture}
        \node[u,label=below:{\scriptsize$1$}] (4) at (4,0) {};
        \node[u,label=above:{\scriptsize$1$}] (4u) at (7.3,1) {};
        \node[u,label=below:{\scriptsize$2$}] (5) at (5,0) {};
        \node (6) at (6,0) {$\cdots$};
        \node[u,label=below:{\scriptsize$N-1$}] (7) at (7,0) {};
        \node[u,label=below:{\scriptsize$N$}] (8) at (8,0) {};
        \node[u,label=above:{\scriptsize$1$}] (8u) at (8.7,1) {};
        \draw (4)--(5)--(6)--(7);
        \draw[<-,double distance=1.5pt](7)--(8);
        \draw[->,double distance=1.5pt](8)--(4u);
        \draw (8u)--(8);
    \end{tikzpicture}\;.
\end{equation}
The set of balanced nodes imply the symmetry $\mathfrak{su}(N)\oplus \mathfrak{u}(1)\oplus \mathfrak{u}(1)$.
\paragraph{3.} For $N < K$ we read the magnetic quiver 
\begin{equation}
    \begin{tikzpicture}
        \node[u,label=below:{\scriptsize$1$}] (4) at (4,0) {};
        \node[u,label=above:{\scriptsize$1$}] (4u) at (7.3,1) {};
        \node[u,label=below:{\scriptsize$2$}] (5) at (5,0) {};
        \node (6) at (6,0) {$\cdots$};
        \node[u,label=below:{\scriptsize$N-1$}] (7) at (7,0) {};
        \node[u,label=below:{\scriptsize$N$}] (8) at (8,0) {};
        \node[u,label=above:{\scriptsize$1$}] (8u) at (8.7,1) {};
        \draw (4)--(5)--(6)--(7);
        \draw[<-,double distance=1.5pt](7)--(8);
        \draw[double distance=1.5pt](8)--(4u);
        \draw (8u)--(8);
        \draw[double distance=1.5pt](8u)--(4u);
        \node at (8,1.3) {\scriptsize$K-N$};
    \end{tikzpicture}\;.
\end{equation}
The set of balanced nodes imply the symmetry $\mathfrak{su}(N)\oplus\mathfrak{u}(1)\oplus\mathfrak{u}(1)$.

Note that the magnetic quivers depend on $N$ and $K$ only, and not on the other parameters $\alpha, N_L, N_R$ which do appear in the brane web.

\subsection{Infinite coupling magnetic quivers}
This section contains the infinite coupling magnetic quivers for electric quiver \eqref{electric quiver SU}. The discussion is organised according to the choice of the number of flavours $N$, and the Chern-Simons level $c$. The reader is reminded that these parameters themselves are expressed in terms of $N_L,N_R, j,l$ via \eqref{parameterisation of K,N,c}.
\subsubsection*{1. Case $j>0, \, l>0$} 
The brane web for
\begin{equation}
    N=K-j-l-4\;,\quad c=\frac{j-l}{2}
\end{equation}
corresponding to the UV fixed point that flows to $\text{SU}(K)_\frac{j-l}{2}$ with a symmetric and $K-j-l-4$ fundamentals is \eqref{eq:BW_case1}, which is reproduced here for convenience in the new parametrization,
     \begin{equation}
            \begin{tikzpicture}
                \node[7brane,green] (0) at (0,0) {};
                \node[7brane] (1) at (-5,0) {};
                \node[7brane] (2) at (-4,0) {};
                \node (3) at (-3,0) {$\cdots$};
                \node[7brane] (4) at (-2,0) {};
                \node[7brane] (5) at (-1,0) {};
                \draw[dash dot,green](-5.5,0)--(1.5,0);
                \draw (1)--(2);
                \draw[double] (2)--(3);
                \draw[thick] (3)--(4)--(5)--(0);
                \node at (-4.5,0.2) {\scriptsize$1$};
                \node at (-1.5,0.3) {\scriptsize$K-j-l-5$};
                \node at (-0.5,-0.3) {\scriptsize$K-j-l-4$};
                \node[7brane,label=above:{\scriptsize$[j,-1]$}] (L) at (-2,2) {};
                \node[7brane,label=above:{\scriptsize$[l,1]$}] (R) at (2,2) {};
                \node[7brane,label=above:{\scriptsize$[0,1]$}] (M) at (0,2) {};
                \draw (L)--(0)--(R) (M)--(0);
            \end{tikzpicture}\;.
        \end{equation}

The corresponding magnetic quiver is given by
\begin{equation}
    \begin{array}{c}
         \begin{scriptsize}
             \begin{tikzpicture}
                 \node[label=below:{1}][u](1){};
                 \node[label=below:{2}][u](2)[right of=1]{};
                 \node (dots)[right of=2]{$\cdots$};
                 \node[label=above:{$K-j-l-5$}][u](k-1)[right of=dots]{};
                 \node[label=below:{$K-j-l-4$}][u](k)[right of=k-1]{};
                 \node[label=right:{$1$}][u](1c)[right of=k]{};
                 \node (empty)[right of=1c]{};
                 \node[label=above:{$1$}][u](1r)[above of=empty]{};
                 \node[label=below:{$1$}][u](1l)[below of=empty]{};
                 \draw(1)--(2);
                 \draw(2)--(dots);
                 \draw(dots)--(k-1);
                 \draw[<-,double distance=1.5pt](k-1)--(k);
                 \draw(1l)--node[right]{$j+2$}(1c);
                 \draw(1r)--node[right]{$l+2$}(1c);
                 \draw(k)--(1c);
                 \draw(k)--(1l);
                 \draw(k)--(1r);
                 \draw (1l)to[out=0,in=0]node[right]{$j+l+2$}(1r);
             \end{tikzpicture}
         \end{scriptsize}
    \end{array}
\end{equation}
From the set of balanced nodes, the symmetry is read to be $\mathfrak{su}(K-j-l-4)\oplus \mathfrak{u}(1)\oplus \mathfrak{u}(1)\oplus \mathfrak{u}(1)$, which is the same global symmetry as the classical Higgs branch.
\subsubsection*{2. Case $j> 0, \, l=0$}

Now consider the case 
\begin{equation}
    N=K-j-4\;,\quad c=\frac{j}{2}
\end{equation}
The brane web for this theory is given by \eqref{eq:BW_case2} and repeated here for the reader's convenience
\begin{equation}\label{SU(k)+1S+(k-5)F web_1}
    \begin{array}{c}\begin{scriptsize}
    \begin{tikzpicture}
    \draw[thick](0,0)--(0,-2);
    \node[7brane] at (0,0){};
    \node[7brane] at (0,-1){};
    \draw[dash dot,green](-5.5,-2)--(1.5,-2);
    \draw[thick](0,-2)--(-2,-2);
    \draw[thick](-5,-2)--(-3,-2);
    \node at (-2.5,-2){$\cdots$};
    \node[7brane] at (-1,-2){};
    \node[7brane] at (-2,-2){};
    \node[7brane] at (-3,-2){};
    \node[7brane] at (-4,-2){};
    \node[7brane] at (-5,-2){};
    \node [label=below:{1}] at (-4.5,-2){};
    \node [label=below:{2}] at (-3.5,-2){};
    \node [label=above:{$K-j-5$}] at (-1.5,-2){};
    \node [label=below:{$K-j-4$}] at (-.5,-2){};
    \node[label=left:{1}] at (0,-.5){};
    \node[label=left:{2}] at (0,-1.5){};
    \draw(0,-2)--(3,-1);
    \node[label=above:{$(j,1)$}][7brane]at(3,-1){};
    \node[label=right:{1}] at (0.5,-1.5){};
    \node[label=below right:{O7$^+$}][7brane,green] at (0,-2){};
    \end{tikzpicture}
    \end{scriptsize}\end{array}\;.
\end{equation}
The magnetic quiver can be read from the above web diagram,
\begin{equation}
    \begin{array}{c}
         \begin{scriptsize}
    \begin{tikzpicture}
    \node[label=right:{1}][u](111){};
    \node[label=below:{2}][u](22)[left of=111]{};
    \node[label=below:{1}][u](11)[left of=22]{};
    \node[label=right:{$K-j-4$}][u](k-5)[below of=111]{};
    \node[label=below:{$K-j-5$}][u](k-6)[left of=k-5]{};
    \node(dots)[left of=k-6]{$\cdots$};
    \node[label=below:{$2$}][u](2)[left of=dots]{};
    \node[label=below:{$1$}][u](1)[left of=2]{};
    \draw(1)--(2);
    \draw(2)--(dots);
    \draw(dots)--(k-6);
    \draw[double distance=1.5pt,<-](k-6)--(k-5);
    \draw(k-5)--(22);
    \draw(k-5)--(111);
    \draw(111)--node[above]{$j+2$}++(22);
    \draw[double distance=1.5pt,->](22)--(11);
    \draw (22)to[out=45,in=135,loop,looseness=10](22);
    \end{tikzpicture}
    \end{scriptsize}
    \end{array}\;.\label{MQ5 SU(k)+ S+(k-5) F_1}
\end{equation}

This quiver enjoys an $\mathfrak{su}(K-4-j)\oplus \mathfrak{su}(2)\oplus\mathfrak{u}(1)\oplus\mathfrak{u}(1)$ Coulomb branch isometry, we can interpret the first factor as the flavour symmetry rotating the fundmentals, the second factor as the enhanced instantonic symmetry, while the two abelian factors correspond to the baryonic symmetry and the phase rotation of the symmetric hypermultiplet respectively. 

\subsubsection*{3. Case $j>1,\, l=-1$}
In the case 
\begin{equation}
    N=K-j-3\;,\quad c=\frac{j+1}{2}\;,
\end{equation}
the brane web is given by \eqref{eq:BW_case3}
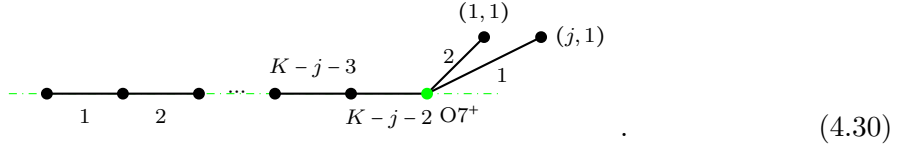
\begin{equation}
    \begin{scriptsize}
    \begin{tikzpicture}
    \draw[thick](.75,-1.25)--(0,-2);
    \draw[thick](1.5,-1.25)--(0,-2);
    \node[label=right:{$(j,1)$}][7brane] at (1.5,-1.25){};
    \node[label=above:{$(1,1)$}][7brane] at (.75,-1.25){};
    \draw[dash dot,green](-5.5,-2)--(1,-2);
    \draw[thick](0,-2)--(-2,-2);
    \draw[thick](-5,-2)--(-3,-2);
    \node at (-2.5,-2){$\cdots$};
    \node[7brane] at (-1,-2){};
    \node[7brane] at (-2,-2){};
    \node[7brane] at (-3,-2){};
    \node[7brane] at (-4,-2){};
    \node[7brane] at (-5,-2){};
    \node [label=below:{1}] at (-4.5,-2){};
    \node [label=below:{2}] at (-3.5,-2){};
    \node [label=above:{\scriptsize{$K-j-3$}}] at (-1.5,-2){};
    \node [label=below:{\scriptsize{$K-j-2$}}] at (-.5,-2){};
    \node[label=below right:{O7$^+$}][7brane,green] at (0,-2){};
    \node[label=right:{2}] at (0,-1.5){};
    \node[label=right:{1}] at (0.7,-1.75){};
    \end{tikzpicture}
    \end{scriptsize}\;.
\end{equation}
The magnetic quiver can be read from the above web diagram,
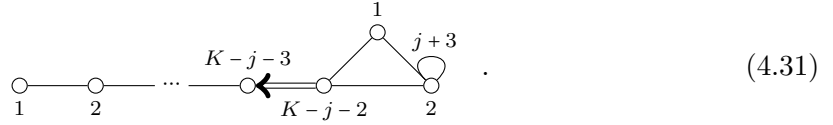
\begin{equation}
    \begin{array}{c}
         \begin{scriptsize}
    \begin{tikzpicture}
    \node[label=above:{1}][u](111){};
    \node[label=below:{2}][u](22)[below right of=111]{};
    \node[label=below:{$K-j-2$}][u](k-5)[below left of=111]{};
    \node[label=above:{$K-j-3$}][u](k-6)[left of=k-5]{};
    \node(dots)[left of=k-6]{$\cdots$};
    \node[label=below:{$2$}][u](2)[left of=dots]{};
    \node[label=below:{$1$}][u](1)[left of=2]{};
    \draw(1)--(2);
    \draw(2)--(dots);
    \draw(dots)--(k-6);
    \draw[double distance=1.5pt,<-](k-6)--(k-5);
    \draw(k-5)--(22);
    \draw(k-5)--(111);
    \draw(111)--node[above right]{$j+3$}++(22);
    \draw (22)to[out=45,in=-45,loop,looseness=10](22);
    \end{tikzpicture}
    \end{scriptsize}
    \end{array}\;.\label{MQ5 SU(k)_3/2+ S+(k-5) F}
\end{equation}
The set of balanced nodes imply the symmetry to be $\mathfrak{su}(K-j-2)\oplus\mathfrak{u}(1)\oplus\mathfrak{u}(1)$.
\subsubsection*{4. Case $j=0,\, l=-1$}
The brane system for
\begin{equation}
    N=K-3\;,\quad c=\frac{1}{2}\;,
\end{equation}
(for $K\ge 3$; note that for $K=2$ we recover Bhardwaj's rank 1 theory which we'll discuss in section \ref{sec:other_examples}) is given by \eqref{eq:BW_case4}
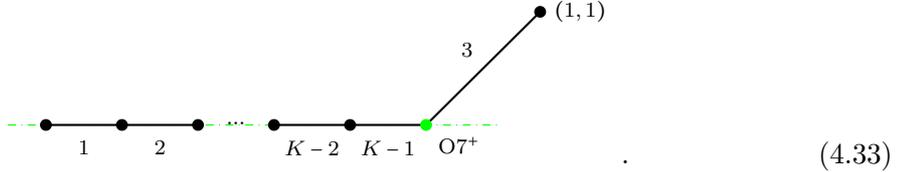
\begin{equation}
\label{eq:suk1/2+1S+(k-3)F}
    \begin{scriptsize}
    \begin{tikzpicture}
    \draw[thick](1.5,-.5)--(0,-2);
    \node[label=right:{$(1,1)$}][7brane] at (1.5,-.5){};
    \draw[dash dot,green](-5.5,-2)--(1,-2);
    \draw[thick](0,-2)--(-2,-2);
    \draw[thick](-5,-2)--(-3,-2);
    \node at (-2.5,-2){$\cdots$};
    \node[7brane] at (-1,-2){};
    \node[7brane] at (-2,-2){};
    \node[7brane] at (-3,-2){};
    \node[7brane] at (-4,-2){};
    \node[7brane] at (-5,-2){};
    \node [label=below:{1}] at (-4.5,-2){};
    \node [label=below:{2}] at (-3.5,-2){};
    \node [label=below:{$K-2$}] at (-1.5,-2){};
    \node [label=below:{$K-1$}] at (-.5,-2){};
    \node[label=below right:{O7$^+$}][7brane,green] at (0,-2){};
    \node[label=right:{3}] at (0.25,-1){};
    \end{tikzpicture}
    \end{scriptsize}\;.
\end{equation}
The corresponding magnetic quiver can be readily obtained 
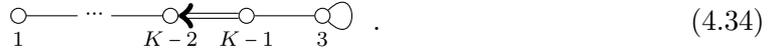
\begin{equation}
    \begin{array}{c}
         \begin{scriptsize}
    \begin{tikzpicture}
    \node[label=below:{$1$}][u](2){};
    \node (dots)[right of=2]{$\cdots$};
    \node[label=below:{$K-2$}][u](2N-5)[right of=dots]{};
    \node[label=below:{$K-1$}][u](2N-4)[right of=2N-5]{};
    \node[label=below:{$3$}][u](2')[right of=2N-4]{};
    \draw (2')to[out=-45,in=45,loop,looseness=10](2');
    \draw(2)--(dots);
    \draw(dots)--(2N-5);
    \draw[ double distance=1.5pt,<-](2N-5)--(2N-4);
    \draw(2N-4)--(2');
    \end{tikzpicture}
    \end{scriptsize}
    \end{array}\;.\label{MQ5 SU(k)+S+(k-3)F}
\end{equation}
The set of balanced nodes imply the symmetry to be $\mathfrak{su}(K-1)\oplus \mathfrak{u}(1)$.

\subsubsection*{5. Case $j=l=0$}
This case corresponds to
\begin{align}
    N=K-4\,\qquad c=0\,.
\end{align} 

The brane web at infinite coupling looks like \eqref{eq:BW_case5}

\begin{equation}
    \begin{array}{c}\begin{scriptsize}
    \begin{tikzpicture}
    \draw[thick](0,1)--(0,-2);
    \node[7brane] at (0,1){};
    \node[7brane] at (0,0){};
    \node[7brane] at (0,-1){};
    \draw[thick](0,-2)--(-2,-2);
    \draw[thick](-5,-2)--(-3,-2);
    \node at (-2.5,-2){$\cdots$};
    \node[7brane] at (-1,-2){};
    \node[7brane] at (-2,-2){};
    \node[7brane] at (-3,-2){};
    \node[7brane] at (-4,-2){};
    \node[7brane] at (-5,-2){};
    \node [label=below:{1}] at (-4.5,-2){};
    \node [label=below:{2}] at (-3.5,-2){};
    \node [label=below:{$K-5$}] at (-1.5,-2){};
    \node [label=below:{$K-4$}] at (-.5,-2){};
    \node[label=right:{1}] at (0,.5){};
    \node[label=right:{2}] at (0,-.5){};
    \node[label=right:{3}] at (0,-1.5){};
    \node[label=below right:{O7$^+$}][7brane,green] at (0,-2){};
    \draw[dash dot,green](-5.5,-2)--(1,-2);
    \end{tikzpicture}
    \end{scriptsize}\end{array}\;.
\end{equation}
We can obtain the magnetic quiver for this brane system using the rules of section \ref{sec:intersection_rule},
\begin{equation}
    \begin{array}{c}
         \begin{scriptsize}
    \begin{tikzpicture}
    \node[label=below:{1}][u](2){};
    \node (dots)[right of=2]{$\cdots$};
    \node[label=below:{$K-5$}][u](2N-5)[right of=dots]{};
    \node[label=below:{$K-4$}][u](2N-4)[right of=2N-5]{};
    \node[label=below:{$3$}][u](3')[right of=2N-4]{};
    \node[label=below:{$2$}][u](2')[right of=3']{};
    \node[label=below:{$1$}][u](1')[right of=2']{};
    \draw (3')to[out=45,in=135,loop,looseness=10](3');
    \draw(2)--(dots);
    \draw(dots)--(2N-5);
    \draw[ double distance=1.5pt,<-](2N-5)--(2N-4);
    \draw(2N-4)--(3');
    \draw[double distance=1.5pt,->](3')--(2');
    \draw(1')--(2');
    \end{tikzpicture}
    \end{scriptsize}
    \end{array}\;.\label{MQ5 SU(k)+ S+(k-4) F}
\end{equation}
For $K>4$, the Coulomb branch symmetry is $\mathfrak{su}(K-4)\oplus\mathfrak{su}(3)\oplus\mathfrak{u}(1)$. 

The case $K=4$ is special, as the moduli space is $d_4/S_4$, the $S_4$ quotient of the closure of the minimal nilpotent orbit of $\mathfrak{so}(8)$, as demonstrated in \cite{Hanany:2023uzn}.

\subsubsection*{6. Case $j=1,\, l=-1$}
The brane web for the UV fixed point of $\mathrm{SU}(K)_1 +1 S^2 + (K-4) F$ is given by \eqref{eq:BW_case6}
\begin{equation}
\label{eq:suk1+S+(k-4)F}
    \begin{scriptsize}
    \begin{tikzpicture}
    \draw[thick](1.5,-.5)--(0,-2);
    \node[label=right:{$(1,1)$}][7brane] at (1.5,-.5){};
    \node[7brane] at (.75,-1.25){};
    \draw[thick](0,-2)--(-2,-2);
    \draw[thick](-5,-2)--(-3,-2);
    \node at (-2.5,-2){$\cdots$};
    \node[7brane] at (-1,-2){};
    \node[7brane] at (-2,-2){};
    \node[7brane] at (-3,-2){};
    \node[7brane] at (-4,-2){};
    \node[7brane] at (-5,-2){};
    \node [label=below:{1}] at (-4.5,-2){};
    \node [label=below:{2}] at (-3.5,-2){};
    \node [label=below:{$K-4$}] at (-1.5,-2){};
    \node [label=below:{$K-3$}] at (-.5,-2){};
    \node[label=right:{1}] at (0.5,-.75){};
    \node[label=right:{3}] at (0,-1.25){};
    \draw[dash dot,green](-5.5,-2)--(1,-2);
    \node[label=below right:{O7$^+$}][7brane,green] at (0,-2){};
    \end{tikzpicture}
    \end{scriptsize}\;.
\end{equation}
From here, one can obtain the magnetic quiver following the rules outlined in section \ref{sec:intersection_rule}. We find
\begin{equation}
    \begin{array}{c}
         \begin{scriptsize}
    \begin{tikzpicture}
    \node[label=below:{1}][u](2){};
    \node (dots)[right of=2]{$\cdots$};
    \node[label=below:{$K-4$}][u](2N-5)[right of=dots]{};
    \node[label=below:{$K-3$}][u](2N-4)[right of=2N-5]{};
    \node[label=below:{$3$}][u](3')[right of=2N-4]{};
    \node[label=below:{$1$}][u](2')[right of=3']{};
    \draw (3')to[out=45,in=135,loop,looseness=10](3');
    \draw(2)--(dots);
    \draw(dots)--(2N-5);
    \draw[ double distance=1.5pt,<-](2N-5)--(2N-4);
    \draw(2N-4)--(3');
    \draw[double distance=1.5pt,->](3')--(2');
    \end{tikzpicture}
    \end{scriptsize}
    \end{array}\;.\label{MQ5 SU(k)_1+ S+(k-4) F}
\end{equation}
From this quiver we read a global symmetry $\mathfrak{su}(K-3)\oplus\mathfrak{u}(1)\oplus\mathfrak{u}(1)$ for $K>3$. The case $K=3$ is special: the moduli space was computed to be $\mathrm{Sym}_0^4\left(\mathbb{C}^2\right) = \mathbb{C}^6/S_4$, the reduced fourth symmetric product of $\mathbb{C}^2$ \cite{Hanany:2023uzn}.

\section{Other examples}\label{sec:other_examples}

In this section, we discuss some sporadic examples. We begin with Bhardwaj's rank 1 theory and discuss higher rank generalisations thereof. Additionally, the isolated example of SU(6) with a single third rank antisymmetric hypermultiplet is considered.

\subsection{Bhardwaj's rank-1 theory and a higher rank generalisation}

In \cite{Bhardwaj:2019jtr} Bhardwaj found a rank-1 $5d$ $\mathcal{N}=1$ SCFT which is not one of Seiberg's $E_N$ rank-1 SCFTs. A braneweb for this theory was first provided in \cite{Kim:2020hhh} (referred to there as `local $\mathbb{P}^2$ plus 1 adjoint'), and the brane web for the SCFT point was provided in \cite{Hayashi:2023boy} and discussed at length in Appendix \ref{app:BraneWebs}
\begin{equation}
\label{eq:bhardwaj_BW}
    \begin{tikzpicture}
        \node[7brane,blue] (1) at (-2,1) {};
        \node[7brane] (2) at (2,0) {};
        \draw[dashed,green] (-3,0)--(3,0);
        \draw[thick] (0,0)--(1);
        \node at (-1,0.8) {3};
        \draw (0,0)--(2);
        \node at (0,-0.5) {O$7^+$};
        \node at (2,0.3) {[1,0]};
        \node[7brane,green] (0) at (0,0) {};
        \node at (-2,1.3) {[2,-1]};
    \end{tikzpicture}\;.
\end{equation}
The magnetic quiver is\footnote{Zhenghao Zhong has pointed out to us another magnetic quiver for Bhardwaj's rank-1 SCFT: $\mathsf{Q}_{ZZ}=\raisebox{-.4\height}{\begin{tikzpicture}
        \node[label=below:{$2$}][u](1) at (0,0) {};
        \node[label=below:{$1$}][u](2) at (1,0) {};
        \draw (1)to[out=135,in=225,loop,looseness=10](1);
        \draw[transform canvas={yshift=2pt}] (1)--(2);
        \draw (1)--(2);
        \draw[transform canvas={yshift=-2pt}] (1)--(2);
        \draw (0.4,0.2)--(0.6,0)--(0.4,-0.2);
    \end{tikzpicture}}$, which has a triply non-simply laced edge (like a $G_2$ Dynkin diagram). We have $\mathcal{C}\left(\raisebox{-.4\height}{\begin{tikzpicture}
        \node[label=below:{$3$}][u](1) at (0,0) {};
        \node[label=below:{$1$}][u](2) at (1,0) {};
        \draw (1)to[out=135,in=225,loop,looseness=10](1);
        \draw (1)--(2);
    \end{tikzpicture}}\right)=\mathcal{C}\left(\mathsf{Q}_{ZZ}\right)\times \mathbb{C}^2$, where the $\mathbb{C}^2$ denotes a free decoupled hypermultiplet. The equality of the two Coulomb branches follows from the low order accidental isomorphism $S_3=D_3$ \cite{Deshuo:2024wip}.
    This raises the question whether $\mathsf{Q}_{ZZ}$ can be derived from another stringy construction of Bhardwaj's rank-1 SCFT, possibly with a $\mathbb{Z}_3$ S-fold \cite{Garcia-Etxebarria:2015wns} given the triple non-simply laced edge (appearing in this context e.g.\ in \cite{Bourget:2020mez}).}
\begin{equation}
\label{eq:bhardwaj_MQ}
    \begin{tikzpicture}
        \node[label=below:{$3$}][u](1) at (0,0) {};
        \node[label=below:{$1$}][u](2) at (1,0) {};
        \draw (1)to[out=135,in=225,loop,looseness=10](1);
        \draw (1)--(2);
    \end{tikzpicture}\;,
\end{equation}
whose Coulomb branch is $\mathrm{Sym}_0^3(\mathbb{C}^2)$, with Hasse diagram
\begin{equation}
    \begin{tikzpicture}
        \node[hasse] (0) at (0,0) {};
        \node[hasse] (1) at (0,-1) {};
        \node[hasse] (2) at (0,-2) {};
        \draw (0)--(1)--(2);
        \node at (-0.3,-0.5) {$A_1$};
        \node at (-0.3,-1.5) {$m$};
    \end{tikzpicture}\;.
\end{equation}
This implies a Higgsing to a theory whose Higgs branch is $A_1$, a natural candidate being the $E_1(=A_1)$ Seiberg SCFT with gauge theory description of pure SU$(2)$. Indeed, the Higgsing is realised in the brane web by sending one of the $(2,-1)$ fivebranes to infinity, leaving behind
\begin{equation}
\label{eq:Bhardwaj5dHiggsingBW}
    \begin{tikzpicture}
        \node[7brane,blue] (1) at (-2,1) {};
        \node[7brane] (2) at (2,0) {};
        \draw[dashed,green] (-3,0)--(3,0);
        \draw[thick] (0,0)--(1);
        \node at (-1,0.8) {2};
        \draw (0,0)--(2);
        \node at (0,-0.5) {O$7^+$};
        \node at (2,0.3) {[1,0]};
        \node[7brane,green] (0) at (0,0) {};
        \node at (-2,1.3) {[2,-1]};
    \end{tikzpicture}\;,
\end{equation}
which represents the SCFT point of the $5d$ $\mathcal{N}=1$ pure SYM theory with gauge algebra so$(3)$, with Higgs branch $A_1$. The reader is referred to section \ref{sec:K-3} for more details.

The generating function for the $n$-th symmetric product of $\mathbb{C}^2$ is 
\begin{equation}
    \PE\left[\frac{\nu}{(1-t x)(1-t x^{-1})}\right]\;,
\end{equation}
that is, the Hilbert series for $\mathrm{Sym}^n(\mathbb{C}^2)$ is given by series expansion in $\nu$ and picking up the $\mathcal{O}(\nu^n)$ coefficient. To obtain the singular component, one has to remove the free $\mathbb{C}^2$ factor leading to the Hilbert series for $\mathrm{Sym}^n_0(\mathbb{C}^2)$. For the case $n=3$ the procedure yields
\begin{equation}
    \HS_{\mathrm{Sym}^3_0(\mathbb{C}^2)}=(1 + [1] (-t-t^5+t^{11}+t^{15}) + t^2  + t^6 + [3] (t^7 -  t^9) - 
   t^{10}  - t^{14}  - 
   t^{16}) \PE\left[ [1] t + [2] t^2 + [3] t^3\right]\;,
\end{equation}
from here it is clear that odd Dynkin labels are correlated with odd powers of $t$ (the fugacity for the R-symmetry). Hence the global form is $\left(\mathrm{SU}(2)\times \mathrm{SU}(2)_R\right)/\mathbb{Z}_2 $.
Recently the $5d$ $\mathcal{N}=1$ superconformal index of Bhardwaj's rank-1 theory has been computed \cite{Kim:2023qwh} upto order 5. The results of this computation are consistent with the moduli space of this theory, in the sense that all operators appearing in the Hilbert series are a subset of those appearing in the index with the same quantum numbers.

\paragraph{6d twisted circle compactification.}

As already pointed out in \cite{Bhardwaj:2019jtr}, Bhardwaj's rank-1 theory can be obtained by a $\mathbb{Z}_2$ twisted compactification of the $6d$ $\mathcal{N}=(1,0)$ SCFT with gauge theory description
\begin{equation}
    \begin{tikzpicture}
        \node[label=below:{1}][uf] (1) at (0,0) {};
        \node[label=below:{SU(1)}][u] (2) at (1,0) {};
        \node[label=below:{SU(1)}][u] (3) at (2,0) {};
        \node[label=below:{1}][uf] (4) at (3,0) {};
        \draw (1)--(2)--(3)--(4);
    \end{tikzpicture}\;.
\end{equation}
The brane system in the tensor branch phase is
\begin{equation}
    \begin{tikzpicture}
        \draw (0,-1)--(0,1) (4,-1)--(4,1);
        \node[7brane] (1) at (1,0) {};
        \node[7brane] (2) at (2,0) {};
        \node[7brane] (3) at (3,0) {};
        \draw (0,0)--(1)--(2)--(3)--(4,0);
    \end{tikzpicture}\;,
\end{equation}
where we can already see the $\mathbb{Z}_2$ symmetry exchanging the left and right, given that the two tensor branch moduli are equal (resulting in a rank-1 theory after twisted compactification). The brane system for the Higgs phase emanating from the origin of the tensor branch is
\begin{equation}
\label{eq:6dN=(2,0)A2BW}
    \begin{tikzpicture}
        \draw (0,-1)--(0,1) (4,-1)--(4,1);
        \node[7brane] (1) at (2,0.3) {};
        \node[7brane] (2) at (2,0.6) {};
        \node[7brane] (3) at (2,0.9) {};
        \draw (0,0)--(4,0);
    \end{tikzpicture}\;,
\end{equation}
with magnetic quiver
\begin{equation}
    \begin{tikzpicture}
        \node[label=left:{$3$}][u](1) at (0,1) {};
        \node[label=left:{$1$}][u](2) at (0,0) {};
        \draw (1)to[out=45,in=135,loop,looseness=10](1);
        \draw (1)--(2);
    \end{tikzpicture}\;.
\end{equation}
Note it is the same as \eqref{eq:bhardwaj_MQ}. Since the magnetic quiver for \eqref{eq:6dN=(2,0)A2BW} is invariant under its $\mathbb{Z}_2$ symmetry, we expect the magnetic quiver for its twisted compactification to be the same. This is further evidence, that the magnetic quiver \eqref{eq:bhardwaj_MQ} derived from \eqref{eq:bhardwaj_BW} is correct.

This is also consistent with Higgsing, as in \eqref{eq:6dN=(2,0)A2BW} one NS5 brane can be sent to infinity. After twisted circle compactification this realises sending a $(2,-1)$ fivebrane in \eqref{eq:bhardwaj_BW} to infinity, reaching \eqref{eq:Bhardwaj5dHiggsingBW}.

\subsubsection{Generalisation to `Bhardwaj rank-$r$ theory'}

Bhardwaj's rank-1 theory, henceforth denoted $Bh(1)$, has no gauge theory phase, similar to Seiberg's $E_0$ theory. As is well known the $E_0$ theory is reached by a mass deformation of the $\tilde{E}_1$ theory, which itself has a gauge theory phase $SU(2)_\pi$. Therefore the $E_0$ theory may in some sense be regarded as
\begin{equation}
    E_0={\suknseqI{2}{-1}{\pi}}\;.
\end{equation}
$Bh(1)$ on the other hand is a descendant of the marginal theory with gauge theory phase $SU(2)_\pi+1 S^2$, where the second rank symmetric is just the adjoint of $SU(2)$. It is therefore regarded as
\begin{equation}
    Bh(1)={\sukeqI{2}{-1}{\pi}}\;.
\end{equation}
A second way to obtain this theory is as part of a family. For $SU(N+1)_{1/2}+(N-2) F+1 S^2$ at infinite coupling we have the brane web \eqref{eq:suk1/2+1S+(k-3)F}
\begin{equation}
\begin{scriptsize}
    \begin{tikzpicture}
    \draw[thick](1.5,-.5)--(0,-2);
    \node[label=right:{$(1,1)$}][7brane] at (1.5,-.5){};
    \draw[thick](0,-2)--(-2,-2);
    \draw[thick](-5,-2)--(-3,-2);
    \node at (-2.5,-2){$\cdots$};
    \node[7brane] at (-1,-2){};
    \node[7brane] at (-2,-2){};
    \node[7brane] at (-3,-2){};
    \node[7brane] at (-4,-2){};
    \node[7brane] at (-5,-2){};
    \node [label=below:{1}] at (-4.5,-2){};
    \node [label=below:{2}] at (-3.5,-2){};
    \node [label=below:{$N-1$}] at (-1.5,-2){};
    \node [label=below:{$N$}] at (-.5,-2){};
    \node[label=right:{3}] at (0.25,-1){};
    \node[label=below right:{O7$^+$}][7brane,green] at (0,-2){};
    \draw[dash dot,green](-5.5,-2)--(1,-2);
    \end{tikzpicture}
    \end{scriptsize}\;.
\end{equation}
For $N=1$ the brane web becomes (SL$(2,\mathbb{Z})$ equivalent) to \eqref{eq:bhardwaj_BW}. The SCFT living on this brane web has no gauge theory phase (since the number of fundamental hypers would be $-1$), and is identified with $Bh(1)$.

This has a natural generalisation to a \emph{Bhardwaj rank-$r$ theory} which may be thought of as
\begin{equation}
    Bh(r)={\sukeqI{r+1}{-1}{r/2}}\;.
\end{equation}

For $SU(N+3)_1+(N-1)F+1 S^2$ the brane web for the infinite coupling fixed point is \eqref{eq:suk1+S+(k-4)F}
\begin{equation}
    \begin{scriptsize}
    \begin{tikzpicture}
    \draw[thick](1.5,-.5)--(0,-2);
    \node[label=right:{$(1,1)$}][7brane] at (1.5,-.5){};
    \node[7brane] at (.75,-1.25){};
    \draw[thick](0,-2)--(-2,-2);
    \draw[thick](-5,-2)--(-3,-2);
    \node at (-2.5,-2){$\cdots$};
    \node[7brane] at (-1,-2){};
    \node[7brane] at (-2,-2){};
    \node[7brane] at (-3,-2){};
    \node[7brane] at (-4,-2){};
    \node[7brane] at (-5,-2){};
    \node [label=below:{1}] at (-4.5,-2){};
    \node [label=below:{2}] at (-3.5,-2){};
    \node [label=below:{$N-1$}] at (-1.5,-2){};
    \node [label=below:{$N$}] at (-.5,-2){};
    \node[label=right:{1}] at (0.5,-.75){};
    \node[label=right:{3}] at (0,-1.25){};
    \node[label=below right:{O7$^+$}][7brane,green] at (0,-2){};
    \draw[dash dot,green](-5.5,-2)--(1,-2);
    \end{tikzpicture}
    \end{scriptsize}\;.
\end{equation}
For $N=0$ the brane web becomes
\begin{equation}
\label{eq:bhardwaj_higher_rank_BW}
    \begin{tikzpicture}
        \node[7brane,label=right:{\scriptsize$[1,1]$}] (1) at (1,1) {};
        \node[7brane,label=right:{\scriptsize$[1,1]$}] (2) at (2,2) {};
        \draw[dashed,green] (-3.5,0)--(3.5,0);
        \draw[thick] (0,0)--(1);
        \node at (0.4,0.6) {\scriptsize$3$};
        \draw (1)--(2);
        \node[7brane,green,label=below:{\scriptsize O$7^+$}] (0) at (0,0) {};
    \end{tikzpicture}\;.
\end{equation}
The SCFT living on this brane web has no gauge theory phase (since the number of fundamental hypers would be $-1$), and is identified with $Bh(2)$. We read the magnetic quiver
\begin{equation}
    \begin{tikzpicture}
            \node[u,label=below:{3}] (6) at (6,0) {};
            \node[u,label=below:{1}] (7) at (7,0) {};
            \draw (6)to[in=45,in=135,loop,looseness=10](6);
            \draw[->,double distance=1.5pt](6)--(7);
        \end{tikzpicture}\;.
\end{equation}
The moduli space is $\mathrm{Sym}_0^4\left(\mathbb{C}^2\right)$ (this can be seen, for example, from the Hasse diagram \eqref{Hasse_equation_6_15_asofnow}). The Hilbert series for this moduli space reads
\begin{equation}
\begin{gathered}
    \HS_{\mathrm{Sym^4_{0}}(\mathbb{C}^2)}=\begin{pmatrix}
    1 + t^2 - t^{10} - t^{14} - t^{20} - t^{24} + t^{32} + t^{34}+[1]( -t + t^9 + 
   t^{25} - t^{33})\\+[2]( -t^6 + 
   t^8 - t^{12} + t^{14} + t^{20} - t^{22} + t^{26} - t^{28})+[4](-t^{12} + t^{16} + t^{18} - t^{22})\\+[5](  t^9 - 2 t^{17} + t^{25})+[6](-t^{12} + t^{16} + t^{18} - t^{22})
\end{pmatrix}\times\\\times\PE\left[[1]t+[2]t^2+[3]t^3+[4]t^4\right]\;,
\end{gathered}
\end{equation}
where we see that odd dynkin labels are correlated with odd powers of $t$ and so the global form is $\left(\mathrm{SU}(2)\times\mathrm{SU}(2)_R\right)/\mathbb{Z}_2$.

For $SU(N+\alpha+1)_{\alpha/2}+(N-1)\mathbf{F}+\mathbf{S}$ (with $\alpha>2$) the brane web for the infinite coupling fixed point is 
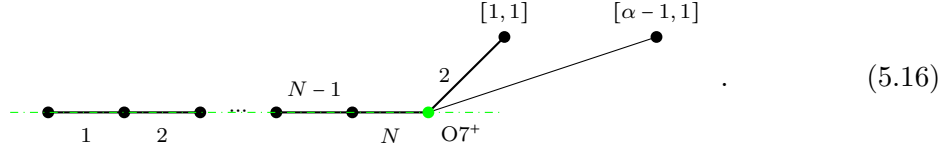
\begin{equation}
    \begin{array}{c}\begin{scriptsize}
    \begin{tikzpicture}
    \draw[thick](1,-1)--(0,-2);
    \node[7brane][label=above:{$[1,1]$}] at (1,-1){};
    \draw[thick](0,-2)--(-2,-2);
    \draw[thick](-5,-2)--(-3,-2);
    \node at (-2.5,-2){$\cdots$};
    \node[7brane] at (-1,-2){};
    \node[7brane] at (-2,-2){};
    \node[7brane] at (-3,-2){};
    \node[7brane] at (-4,-2){};
    \node[7brane] at (-5,-2){};
    \node [label=below:{1}] at (-4.5,-2){};
    \node [label=below:{2}] at (-3.5,-2){};
    \node [label=above:{$N-1$}] at (-1.5,-2){};
    \node [label=below:{$N$}] at (-.5,-2){};
    \node[label=left:{2}] at (0.5,-1.5){};
    \draw(0,-2)--(3,-1);
    \node[label=above:{$[\alpha-1,1]$}][7brane]at(3,-1){};
    \node[label=below right:{O7$^+$}][7brane,green] at (0,-2){};
    \draw[dash dot,green](-5.5,-2)--(1,-2);
    \end{tikzpicture}
    \end{scriptsize}\end{array}\;.
\end{equation}
For $N=0$ the brane web becomes
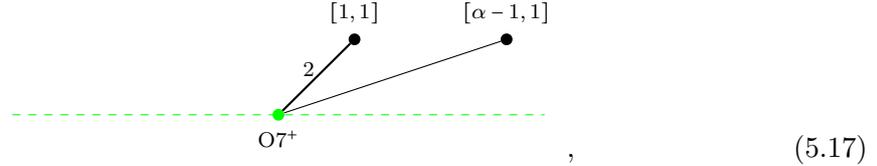
\begin{equation}
    \begin{tikzpicture}
        \node[7brane,label=above:{\scriptsize$[1,1]$}] (1) at (1,1) {};
        \node[7brane,label=above:{\scriptsize$[\alpha-1,1]$}] (2) at (3,1) {};
        \draw[dashed,green] (-3.5,0)--(3.5,0);
        \draw[thick] (0,0)--(1);
        \node at (0.4,0.6) {\scriptsize$2$};
        \draw (0)--(2);
        \node[7brane,green,label=below:{\scriptsize O$7^+$}] (0) at (0,0) {};
    \end{tikzpicture}\;,
\end{equation}
The SCFT living on this brane web has no gauge theory phase (since the number of fundamental hypers would be $-1$), and is identified with $Bh(\alpha)$. We read the magnetic quiver
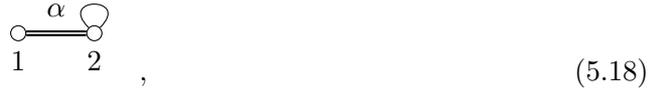
\begin{equation}
    \begin{tikzpicture}
            \node[u,label=below:{2}] (6) at (6,0) {};
            \draw (6)to[out=45,in=-45,loop,looseness=10](6);
            \node[u,label=below:1] (t) at (5,0) {};
            \draw (t)--(6);
            \node at (5.5,0.3) {$\alpha$};
        \end{tikzpicture}\;,
\end{equation}
whose moduli space is Sym$^2\left(\mathbb{C}^2/\mathbb{Z}_{\alpha}\right)$. Note, this is the moduli space of two identical objects moving in a $\mathbb{C}^2/\mathbb{Z}_\alpha$ background. In the brane system the two identical objects are the stack of two (1,1) 5branes, and the $\mathbb{C}^2/\mathbb{Z}_\alpha$ background is the moduli space seen by the $(\alpha-1,1)$ 5-brane.\footnote{We thank Deshuo Liu for a useful discussion on this point.}
In Section \ref{sec:Hasse_higgsings} it is shown that these theories arise naturally on the Higgs branches of 5d infinite coupling UV SCFTs for SU theories with second rank symmetric and fundamental matter.

The generating function for the Hilbert series of $\mathrm{Sym}^n(\mathbb{C}^2/\mathbb{Z}_\alpha)$ is
\begin{equation}
    \PE\left[\frac{\nu(1-t^{2\alpha})}{(1-t^2)(1-xt^\alpha)(1-x^{-1}t^\alpha)}\right]\;,
\end{equation}
where to obtain, say, the expression for $\mathrm{Sym}^2(\mathbb{C}^2/\mathbb{Z}_\alpha)$ one needs to series expand the above expression and extract the $\mathcal{O}(\nu^2)$ coefficient. Even without carrying out this step explicitly, it is apparent that since for $\alpha\in2\mathbb{Z}$ only even powers of $t$ appear the global symmetry is $\mathrm{U}(1)\times\mathrm{SO}(3)_R$, while when $\alpha\in 2\mathbb{Z}+1$, odd Dynkin labels are correlated with odd powers of $t$ and hence the global symmetry is $\left(\mathrm{U}(1)\times\mathrm{SU}(2)_R\right)/\mathbb{Z}_2$
\subsection{SU(6) with third rank antisymmetric}
Consider the 5d SU$(6)_\frac{1}{2}+\frac{1}{2}\Lambda^3+1 S^2$. Let us first take a step back and consider the 6d origin of this theory. It has been conjectured \cite{Hayashi:2019yxj} that the 5d SU$(6)_0+\frac{1}{2}\Lambda^3+1S^2+1F$ theory is the twisted circle compactification of the 6d theory whose electric quiver is
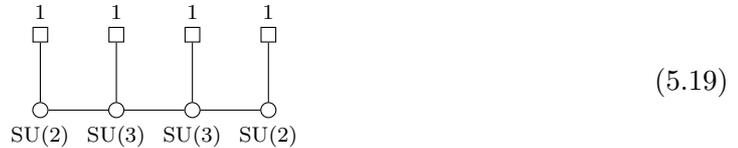
\begin{equation}
    \begin{array}{c}
         \begin{scriptsize}
             \begin{tikzpicture}
                 \node[label=above:{1}][uf](SU1){};
                 \node[label=below:{SU(2)}][u][](SU2)[below of=SU1]{};
                 \node[label=below:{SU(3)}][u][](SU3)[right of=SU2]{};
                 \node[label=below:{SU(3)}][u][](SU3')[right of=SU3]{};
                 \node[label=below:{SU(2)}][u][](SU2')[right of=SU3']{};
                 \node[label=above:{1}][uf][](SU1')[above of=SU2']{};
                \node[label=above:{1}][uf][](SU11)[above of=SU3]{};
                \node[label=above:{1}][uf][](SU11')[above of=SU3']{};
                \draw(SU1)--(SU2);
                \draw(SU3)--(SU2);
                \draw(SU3')--(SU3);
                \draw(SU2')--(SU3');
                \draw(SU1')--(SU2');
                \draw(SU11)--(SU3);
                \draw(SU11')--(SU3');            
             \end{tikzpicture}
         \end{scriptsize}
    \end{array}
\end{equation}
The magnetic quiver for the 6d theory is given by
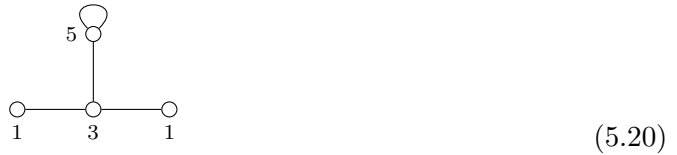
\begin{equation}
    \begin{scriptsize}
        \begin{tikzpicture}
            \node[label=below:{1}][u](1){};
            \node[label=below:{3}][u](3)[right of=1]{};
            \node[label=below:{1}][u](1')[right of=3]{};
            \node[label=left:{5}][u](5)[above of=3]{};
            \draw(1)--(3);
            \draw(1')--(3);
            \draw(5)--(3);
            \draw(5)to[out=45,in=135,loop,looseness=10](5);
        \end{tikzpicture}
    \end{scriptsize}
\end{equation}
The theory of our interest, SU$(6)_0+\frac{1}{2}\Lambda^3+1S^2$ is obtained by giving a large mass to the fundamental hypermultiplet in SU$(6)_0+\frac{1}{2}\Lambda^3+1S^2+1F$, and integrating it out. We therefore expect that the Magnetic quiver for the 5d theory SU$(6)_0+\frac{1}{2}\Lambda^3+1S^2$ is obtained by folding the above magnetic quiver, namely
\begin{equation}
    \begin{scriptsize}
        \begin{tikzpicture}
            \node[label=below:{1}][u](1){};
            \node[label=below:{3}][u](3)[right of=1]{};
            \node[label=right:{5}][u](5)[above of=3]{};
            \draw[<-,double distance=1.5pt](1)--(3);
            \draw(3)--(5);
             \draw(5)to[out=45,in=135,loop,looseness=10](5);
        \end{tikzpicture}
    \end{scriptsize}
\end{equation}
This magnetic quiver is clearly consistent with our conjectured rules for obtaining magnetic quivers from brane webs. The brane web at the fixed point for this theory reads \cite{Hayashi:2019yxj}
\begin{equation}
    \begin{scriptsize}
        \begin{tikzpicture}
            \draw(0,0)--(2,0);
            \node[7brane]at(0,0){};
            \node[7brane]at(1,0){};
            \draw(2,0)--(3,1);
            \node[7brane]at(3,1){};
            \node[label=below:{1}]at(.5,0){};
            \node[label=below:{3}]at(1.5,0){};
            \node[label=below:{5}]at(2.75,.75){};
    \node[label=below right:{O7$^+$}][7brane,green] at (2,0){};
    \draw[dash dot,green](-1,0)--(3,0);
        \end{tikzpicture}
    \end{scriptsize}\;.
\end{equation}

\clearpage

\section{Higgsing Patterns for 5d SCFTs}\label{sec:Hasse_higgsings}
In this section we present the Higgsing patterns of our 5d $\mathcal{N}=1$ SCFTs and compare them to the (classical) Higgsing patterns of their low energy gauge theory phases.\footnote{Note, we only compare to the specific gauge theory phases with special orthogonal or special unitary gauge groups discussed in the paper, and do not ask about any other possible gauge theory phases of our 5d SCFTs.}

\subsection{Classical Higgsings}
Let us first revisit the classical Higgsings of the gauge theories in question, which can be read from the Hasse diagrams provided in Appendix \ref{Hasse diagrams finite}.

\subsubsection{SO($K$)+$N$V}
For $K>2$ and $N\geq1$ there is one minimal partial Higgsing:\footnote{The comparison with the infinite coupling SCFTs will not include $K\leq2$ cases, as those theories do not have a UV completion.}
\begin{equation}
    \begin{tikzpicture}
        \node (1) at (0,0) {$\soeq{K}{N}$};
        \node (2) at (0,3) {$\soeq{K-1}{N-1}$};
        \draw[->,thick] (1)--(2);
        \node[anchor=east] at (0,1.5) {$c_N$};
    \end{tikzpicture}\;,
\end{equation}
where the $c_N$ transition is achieved by giving a VEV to a hyper in the vector representation of SO$(K)$. 

For $N=0$ there is no possible Higgsing.
\subsubsection{SU($K$)+$N$F+S}

For $K>2$ and $N\geq1$ there are two possible minimal Higgsings:
\begin{equation}
\label{eq:SU_Classical_Higgs}
    \begin{tikzpicture}
        \node (1) at (0,0) {$\sueq{K}{N}$};
        \node (2) at (0,4) {$\sueq{K-1}{N-1}$};
        \node (3) at (-5,3) {$\soeq{K}{N}$};
        \draw[->,thick] (1)--(2);
        \draw[->,thick] (1)--(3);
        \node[anchor=west] at (0,2) {$ac_N$};
        \node[anchor=east] at (-2.5,1.5) {$A_{K-1}$};
    \end{tikzpicture}\;,
\end{equation}
where the $ac_N$ transition is achieved by giving a VEV to a hyper in the fundamental representation, and the $A_{K-1}$ transition is achieved by giving a VEV to a hyper in the second rank symmetric representation. The slice $ac_N$ \cite{2003math:5095M} corresponds to the Coulomb branch of either of the following quivers \cite{Bourget:2021siw}
\begin{equation}
    ac_N=\mathcal{C}\left(\begin{array}{c}
         \begin{tikzpicture}
             \node[label=below:{1}][u](1){};
             \node(dots)[right of=1]{$\cdots$}{};
             \node[label=below:{1}][u](1')[right of=dots]{};
             \node[label=below:{1}][u](1'')[right of=1']{};
             \node[label=above:{1}][uf](1f)[above of=1]{};
             \node[label=above:{1}][uf](1f')[above of=1'']{};
             \draw(1)--(1f);
             \draw(1)--(dots);
             \draw(1')--(dots);
             \draw[<-,double distance=2pt](1')--(1'');
             \draw(1'')--(1f');
              \draw [decorate,decoration=
             {brace,amplitude=6pt},xshift=0pt,yshift=0pt]
(3.5,-0.75) -- (-.5,-0.75)node [black,midway,xshift=0pt,yshift=-15pt] {
\scriptsize $N$ nodes};
         \end{tikzpicture}
    \end{array}\right)=\mathcal{C}\left(\begin{array}{c}
         \begin{tikzpicture}
             \node[label=below:{1}][u](1){};
             \node[label=below:{1}][u](11)[right of=1]{};
             \node(dots)[right of=11]{$\cdots$};
             \node[label=below:{1}][u](1')[right of=dots]{};
             \node[label=below:{1}][u](11')[right of=1']{};
             \draw[double distance=2pt,->](1)--(11);
             \draw[double distance=2pt,<-](1')--(11');
             \draw(11)--(dots);
             \draw(1')--(dots);
             \draw(1)to[out=45, in=135](11');
              \draw [decorate,decoration={brace,amplitude=6pt},xshift=0pt,yshift=0pt]
(4,-0.75) -- (0,-0.75)node [black,midway,xshift=0pt,yshift=-15pt] {
\scriptsize $N+1$ nodes};
         \end{tikzpicture}
    \end{array}\right)\;.
\end{equation}
For $N=0$ there is only one possible minimal Higgsing:
\begin{equation}
    \begin{tikzpicture}
        \node (1) at (0,0) {$\sueq{K}{0}$};
        \node (3) at (-5,3) {$\soeq{K}{0}$};
        \draw[->,thick] (1)--(3);
        \node[anchor=east] at (-2.5,1.5) {$A_{K-1}$};
    \end{tikzpicture}\;,
\end{equation}
where the $A_{K-1}$ transition is achieved by giving a VEV to a hyper in the second symmetric representation, as before.

\subsection{Higgsings between SCFTs}
In this section we denote the infinite coupling UV completion of a 5d $\mathcal{N}=1$ gauge theory $T$ as $\UV{T}$. From the Hasse diagrams provided in Appendix \ref{Hasse diagrams SCFTs} we can identify all minimal partial Higgsings of our 5d SCFTs.

\subsubsection{$\UV{SO(K)+(K-3)V}$}
For $K>3$ there is the minimal Higgsing:
\begin{equation}
    \begin{tikzpicture}
        \node (1) at (0,0) {$\soeqI{K}{K-3}$};
        \node (2) at (0,4) {$\soeqI{K-1}{K-4}$};
        \draw[->,thick] (1)--(2);
        \node[anchor=east] at (-0.2,2) {$\left.\begin{matrix}
            c_{K-2} & ,\,K>4\\
            a_{3} & ,\,K=4
        \end{matrix}\;\,\right\}$};
    \end{tikzpicture}\;.
\end{equation}
Note that superficially this is just like the classical Higgsing pattern, however the infinite coupling transverse slice is modified from the classical $c_{K-3}$ to either $c_{K-2}$ or $a_3$ due to the presence of instanton operators.

For $K=3$ we have:
\begin{equation}
    \begin{tikzpicture}
        \node (1) at (0,0) {$\soeqI{3}{0}$};
        \node (2) at (0,4) {trivial};
        \draw[->,thick] (1)--(2);
        \node[anchor=east] at (-0.2,2) {$A_{1}$};
    \end{tikzpicture}\;.
\end{equation}
Note that at finite coupling the pure SO(3) gauge theory has no Higgs branch, while at infinite coupling it has a 1-dimensional Higgs branch due to the presence of the massless instanton operators.

\subsubsection{$\UV{SO(K)+(K-4)V}$}
For $K\geq6$ there are two minimal Higgsings:
\begin{equation}
    \begin{tikzpicture}
        \node (1) at (0,0) {$\soeqI{K}{K-4}$};
        \node (2) at (0,4) {$\soeqI{K-1}{K-5}$};
        \node (3) at (-5,3) {$\soeqI{K-2}{K-5}$};
        \draw[->,thick] (1)--(2);
        \node[anchor=west] at (0.2,2) {$c_{K-4}$};
        \draw[->,thick] (1)--(3);
        \node[anchor=east] at (-2.5,1.3) {$A_{1}$};
    \end{tikzpicture}\;.
\end{equation}
The $c_{K-4}$ transition is the same transition visible classically, however the $A_1$ transition only appears at infinite coupling. The new direction accessible at infinite coupling is similar to the baryonic direction in classical SU$(K)$ SQCD theories with $N\geq K$ flavours \cite{Bourget:2019rtl}, and the symmetric direction in \eqref{eq:SU_Classical_Higgs}, in the sense that it leads to a bifurcation in the Hasse diagram. Physically the bifurcation points out the possibility to perform two independent minimal Higgsings. Note that the $A_1$ transition allows the possibility of connecting the two families of theories SO$(K)+(K-4)V$ and SO$(K)+(K-3)V$ at infinite coupling, which is not possible classically.

For $K=6$ we provide the full phase diagram:
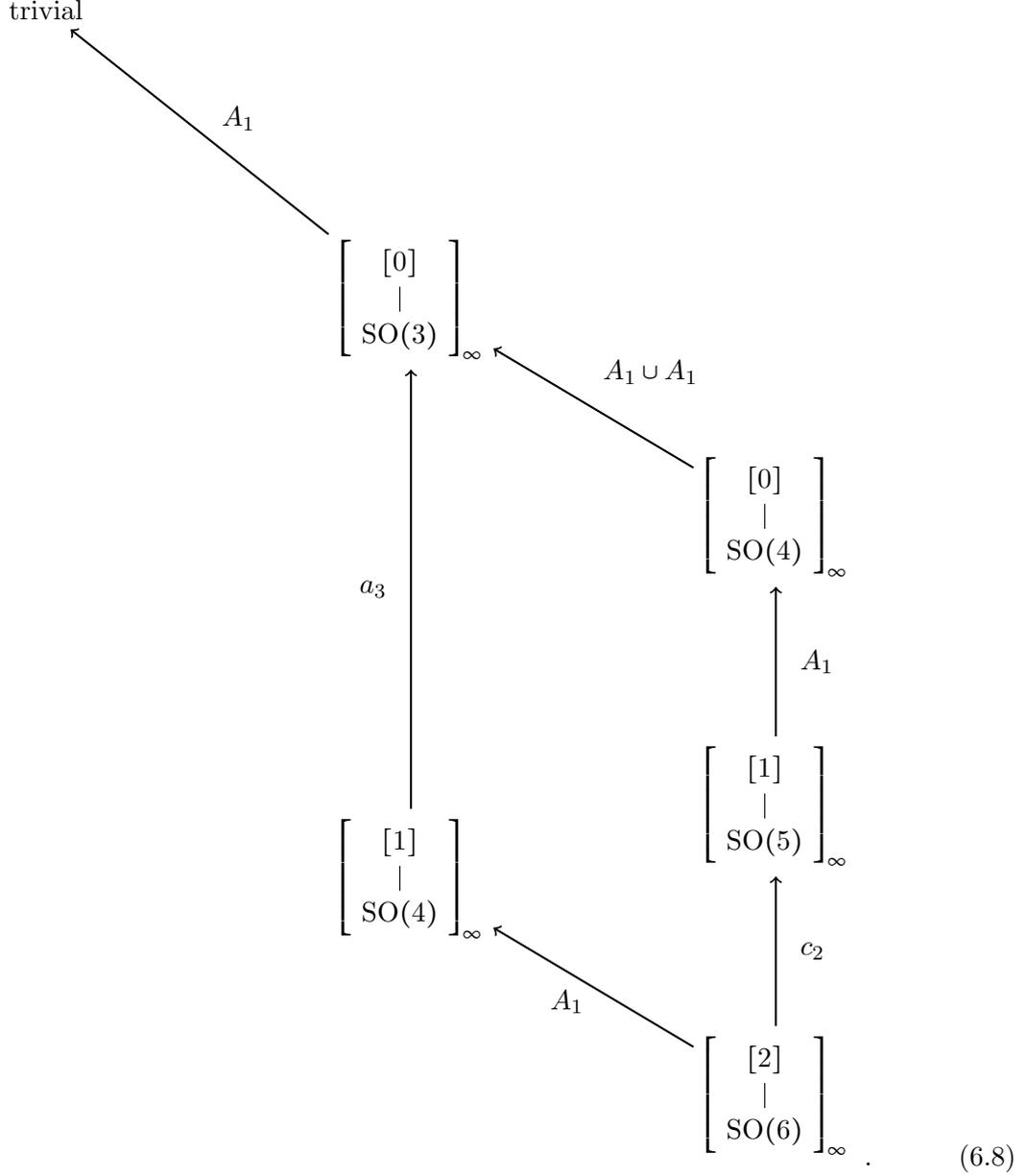
\begin{equation}
    \begin{tikzpicture}
        \node (1) at (0,0) {$\soeqI{6}{2}$};
        \node (2) at (0,4) {$\soeqI{5}{1}$};
        \node (3) at (0,8) {$\soeqI{4}{0}$};
        \node (4) at (-5,3) {$\soeqI{4}{1}$};
        \node (5) at (-5,11) {$\soeqI{3}{0}$};
        \node (6) at (-10,15) {trivial};
        \draw[->,thick] (1)--(2);
        \node[anchor=west] at (0.2,2) {$c_{2}$};
        \draw[->,thick] (2)--(3);
        \node[anchor=west] at (0.2,6) {$A_{1}$};
        \draw[->,thick] (3)--(5);
        \node[anchor=west] at (-2.5,10) {$A_1\cup A_1$};
        \draw[->,thick] (1)--(4);
        \node[anchor=east] at (-2.5,1.3) {$A_{1}$};
        \draw[->,thick] (4)--(5);
        \node[anchor=east] at (-5.2,7) {$a_3$};
        \draw[->,thick] (5)--(6);
        \node[anchor=east] at (-7,13.5) {$A_1$};
    \end{tikzpicture}\;.
\end{equation}
The diagram is depicted such that the diagonal transitions are only possible at infinite coupling. On the other hand, the $c_2$ and $A_1=a_1$ transitions in the right vertical line are already visible in the classical theory. It is worth stressing that the left vertical line is a transition that is present at finite coupling, though at infinite coupling the classical $a_1$ transition is replaced by an $a_3$ transition. Recall from the discussion in section \ref{SO(K) with K-4 vectors}, that the infinite coupling Higgs branch of the SO(5) gauge theory with one vector representation is the 2-instanton moduli space of SU(2). This fact is manifest in this Hasse diagram as the sub-diagram emanating from the infinite coupling limit of the SO(5)+1V theory.

\subsubsection{$\UV{SO(K)+(K-\alpha-2)V}$, $\alpha\geq3$}
For $K>\alpha+2$ there is the minimal Higgsing:
\begin{equation}
    \begin{tikzpicture}
        \node (1) at (0,0) {$\soeqI{K}{K-\alpha-2}$};
        \node (2) at (0,4) {$\soeqI{K-1}{K-\alpha-3}$};
        \draw[->,thick] (1)--(2);
        \node[anchor=west] at (0.2,2) {$c_{K-\alpha-2}$};
    \end{tikzpicture}\;.
\end{equation}
This is visible already at finite coupling.

On the other hand, for $K=\alpha+2$, we have a transition which only appears at infinite coupling:
\begin{equation}
    \begin{tikzpicture}
        \node (1) at (0,0) {$\soeqI{\alpha+2}{0}$};
        \node (2) at (0,4) {trivial};
        \draw[->,thick] (1)--(2);
        \node[anchor=west] at (0.2,2) {$A_{\alpha-1}$};
    \end{tikzpicture}\;,
\end{equation}
where we recall that $\alpha$ is the dual coxeter number for $SO(\alpha+2)$. That is, the infinite coupling Higgs branch of $SO(\alpha+2)$ SYM is the orbifold $\mathbb{C}^2/\mathbb{Z}_\alpha$.

\subsubsection{$\UV{SU(K)_{1/2}+(K-3)F+S}$}
For $K>3$ there are two minimal Higgsings:
\begin{equation}
    \begin{tikzpicture}
        \node (1) at (0,0) {$\sukeqI{K}{K-3}{1/2}$};
        \node (2) at (0,4) {$\sukeqI{K-1}{K-4}{1/2}$};
        \node (3) at (-5,3) {$\soeqI{K+1}{K-2}$};
        \draw[->,thick] (1)--(2);
        \node[anchor=west] at (0.2,2) {$ac_{K-1}$};
        \draw[->,thick] (1)--(3);
        \node[anchor=east] at (-2.5,1.3) {$A_{K-2}$};
    \end{tikzpicture}\;.
\end{equation}
The Higgsing in the vertical direction is visible classically, although the transition is changed from $ac_{K-3}$ to $ac_{K-1}$ at infinite coupling due to instanton operators. The other Higgsing, i.e.\ the $A_{K-2}$ transition, is not visible classically. The Higgsing reached by giving a VEV to a hyper in the second rank symmetric representation of the classical theory (via the $A_{K-1}$ transition) in \eqref{eq:SU_Classical_Higgs} is realised by a further minimal Higgsing of the $\soeqI{K+1}{K-2}$ theory.

In the case $K=3$, the following full phase diagram is
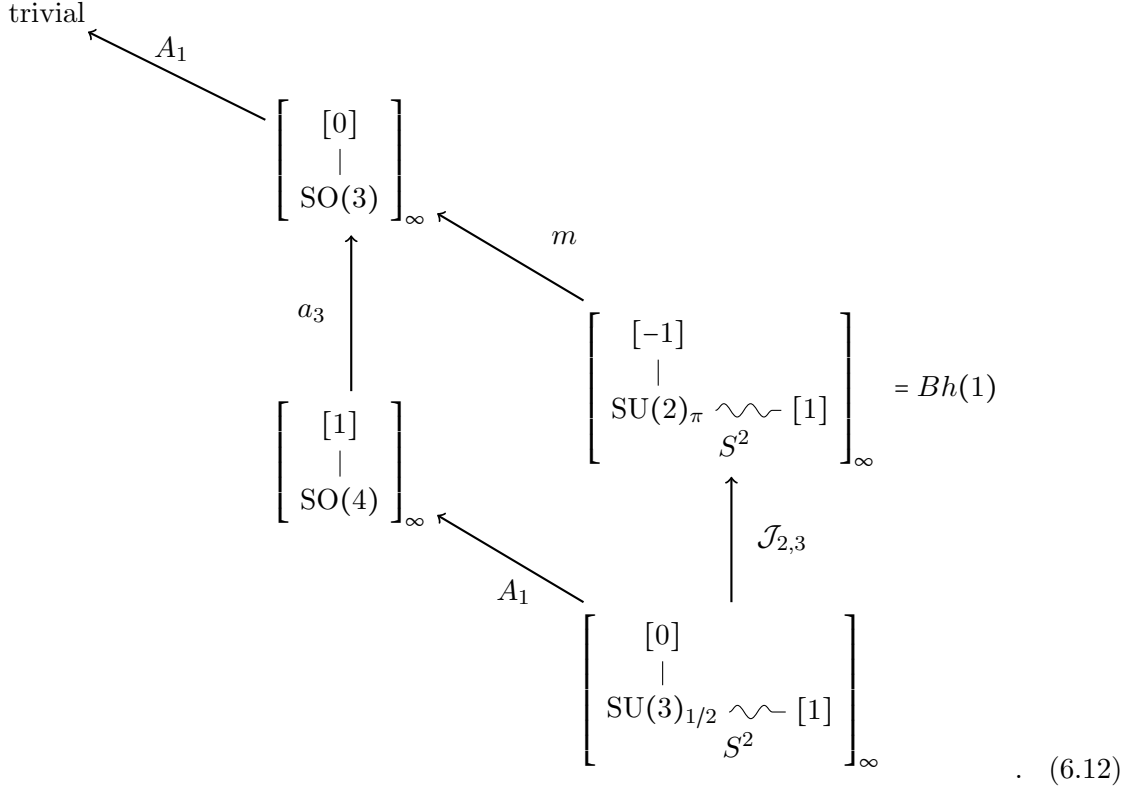
\begin{equation}
    \begin{tikzpicture}
        \node (1) at (0,0) {$\sukeqI{3}{0}{1/2}$};
        \node (3) at (0,4) {$\sukeqI{2}{-1}{\pi}$};
        \node[anchor=west] at (2,4) {$=Bh(1)$};
        \node (4) at (-5,3) {$\soeqI{4}{1}$};
        \node (5) at (-5,7) {$\soeqI{3}{0}$};
        \node (6) at (-9,9) {trivial};
        \draw[->,thick] (1)--(3);
        \node[anchor=west] at (0.2,2) {$\mathcal{J}_{2,3}$};
        \draw[->,thick] (3)--(5);
        \node[anchor=west] at (-2.5,6) {$m$};
        \draw[->,thick] (1)--(4);
        \node[anchor=east] at (-2.5,1.3) {$A_{1}$};
        \draw[->,thick] (4)--(5);
        \node[anchor=east] at (-5.2,5) {$a_3$};
        \draw[->,thick] (5)--(6);
        \node[anchor=east] at (-7,8.5) {$A_1$};
    \end{tikzpicture}\;.
\end{equation}
Here all but one of the transitions are exclusively present at infinite coupling, the exception being the $a_3$ transition taking us from SO(4) with one vector to SO(3), which is enlarged from the classical $a_1$ transition. Note in particular the $m$ transverse slice (introduced in \cite{2015arXiv150205770F}) from the Higgsing of the $Bh(1)$ theory to SO(3), which will reappear in what follows. It corresponds to the variety constructed by removing the two degree one generators from $\mathbb{C}^2$. Alternatively it can be thought of as the singularity that arises when two coincident points on $\mathbb{C}^2$ coincide with a third one.

The second exotic slice in the diagram, $\mathcal{J}_{2,3}$ (first introduced in \cite{Bourget:2022tmw}), corresponds to the Coulomb branch of the following magnetic quiver
\begin{equation}
\mathcal{J}_{2,3}=\mathcal{C}\left(\begin{array}{c}
     \begin{tikzpicture}
        \node[label=below:{1}][u](1){};
        \node[label=below:{2}][u](2)[right of=1]{};
        \node[label=below:{1}][u](1')[right of=2]{};
        \draw[<-,double distance=2pt](1)--(2);
        \draw[->,double distance=3pt](2)--(1');
        \draw(2)--(1');
    \end{tikzpicture}
\end{array}\right)\;.
\end{equation}

\subsubsection{$\UV{SU(K)_{1}+(K-4)F+S}$} For $K>4$ there are three minimal Higgsings:
\begin{equation}
    \scalebox{0.7}{\begin{tikzpicture}
        \node (1) at (0,0) {$\sukeqI{K}{K-4}{1}$};
        \node (2) at (0,4) {$\sukeqI{K-1}{K-5}{1}$};
        \node (3) at (-5,3) {$\soeqI{K+1}{K-3}$};
        \node (4) at (-15,3) {$\sukeqI{K-2}{K-5}{1/2}$};
        \draw[->,thick] (1)--(2);
        \node[anchor=west] at (0.2,2) {$ac_{K-3}$};
        \draw[->,thick] (1)--(3);
        \node[anchor=east] at (-2.5,1.3) {$A_{K-2}$};
        \draw[->,thick] (1)--(4);
        \node[anchor=east] at (-7.5,1.3) {$A_{2}$};
    \end{tikzpicture}}\;.
\end{equation}
The $ac_{K-3}$ transition is the classical direction corresponding to giving a VEV to the fundamental hypermultiplet, while the other two transitions only appear at infinite coupling.

For $K=4$ we have:
\begin{equation}\label{Hasse_equation_6_15_asofnow}
    \makebox[\textwidth][c]{\scalebox{0.9}{\begin{tikzpicture}
        \node (1) at (0,0) {$\sukeqI{4}{0}{1}$};
        \node (3) at (0,4) {$\sukeqI{3}{-1}{1}$};
        \node[anchor=west] at (2,4) {$=$ $Bh(2)$};
        \node (4) at (-5,3) {$\soeqI{5}{1}$};
        \node (5) at (-5,7) {$\soeqI{4}{0}$};
        \node (6) at (-10,11) {$\soeqI{3}{0}$};
        \node (7) at (-14,13) {trivial};
        \node (8) at (-12,7) {$\sukeqI{2}{-1}{\pi}$};
        \node[anchor=east] at (-14,7) {$Bh(1)$ $=$};
        \draw[->,thick] (1)--(3);
        \node[anchor=west] at (0.2,2) {$ac_{1}=A_2$};
        \draw[->,thick] (3)--(5);
        \node[anchor=west] at (-2.7,6) {$A_{1}$};
        \draw[->,thick] (1)--(4);
        \node[anchor=east] at (-2.5,1.3) {$A_{2}$};
        \draw[->,thick] (4)--(5);
        \node[anchor=east] at (-5.1,4.7) {$c_1$};
        \draw[->,thick] (5)--(6);
        \node[anchor=west] at (-7.5,9.5) {$A_1\cup A_1$};
        \draw[->,thick] (6)--(7);
        \node[anchor=east] at (-12,12.5) {$A_{1}$};
        \draw[->,thick] (3)--(8);
        \node[anchor=east] at (-6.5,5.4) {$m$};
        \draw[->,thick] (8)--(6);
        \node[anchor=east] at (-11.2,9) {$m$};
    \end{tikzpicture}}}\;.
\end{equation}
Again, most of these transitions are intrinsic to infinite coupling; the only exception being the $c_1$ slice from SO(5) with one vector to SO(4).

\subsubsection{$\UV{SU(K)_{0}+(K-4)F+S}$}
For $K>4$ there are two minimal Higgsings:
\begin{equation}
    \begin{tikzpicture}
        \node (1) at (0,0) {$\sukeqI{K}{K-4}{0}$};
        \node (2) at (0,4) {$\sukeqI{K-1}{K-5}{0}$};
        \node (3) at (-5,3) {$\sukeqI{K-1}{K-5}{1}$};
        \draw[->,thick] (1)--(2);
        \node[anchor=west] at (0.2,2) {$ac_{K-4}$};
        \draw[->,thick] (1)--(3);
        \node[anchor=east] at (-2.5,1.3) {$a_2$};
    \end{tikzpicture}\;.
\end{equation}
The $a_2$ transition is intrinsic to infinite coupling while the $ac_{K-4}$ transition is present already at the classical level.

For $K=4$ we have only one minimal Higgsing:
\begin{equation}
    \begin{tikzpicture}
        \node (1) at (0,0) {$\sukeqI{4}{0}{0}$};
        \node (3) at (-5,3) {$\sukeqI{3}{-1}{1}$};
        \node[anchor=east] at (-7.3,3) {$Bh(2)$ $=$};
        \draw[->,thick] (1)--(3);
        \node[anchor=east] at (-2.5,1.3) {$a_2$};
    \end{tikzpicture}\;.
\end{equation}
This transition is only present at infinite coupling, and in particular $Bh(2)$ is an SCFT without a non-abelian gauge theory phase.

\subsubsection{$\UV{SU(K)_{\alpha/2}+(K-\alpha-2)F+S}$, $\alpha\geq3$}
For $K>\alpha+2$ there are three minimal Higgsings:
\begin{equation}
    \begin{tikzpicture}
        \node (1) at (0,0) {$\sukeqI{K}{K-\alpha-2}{\alpha/2}$};
        \node (2) at (0,4) {$\sukeqI{K-1}{K-\alpha-3}{\alpha/2}$};
        \node (3) at (5,3) {$\soeqI{K+1}{K-\alpha-1}$};
        \node (4) at (-5,3) {$\soeqI{K-\alpha+1}{K-\alpha-2}$};
        \draw[->,thick] (1)--(2);
        \node[anchor=east] at (-0.2,2) {$ac_{K-\alpha-1}$};
        \draw[->,thick] (1)--(3);
        \node[anchor=west] at (2.5,1.3) {$A_{K-2}$};
        \draw[->,thick] (1)--(4);
        \node[anchor=east] at (-2.5,1.3) {$A_{K+\alpha-2}$};
    \end{tikzpicture}\;.
\end{equation}
The $ac_{K-\alpha-1}$ transition is classical, and the $A_{K-2}$ and $A_{K+\alpha-2}$ transitions happen only at the fixed point.

For $K=\alpha+2$ we have:
\begin{equation}
    \begin{tikzpicture}
        \node (1) at (0,0) {$\sukeqI{\alpha+2}{0}{\alpha/2}$};
        \node (2) at (0,4) {$\sukeqI{\alpha+1}{-1}{\alpha/2}$};
        \node[anchor=east] at (-2.3,4) {$Bh(\alpha)$ $=$};
        \node (3) at (5,4) {$\soeqI{\alpha+3}{1}$};
        \node (4) at (5,8) {$\soeqI{\alpha+2}{0}$};
        \node (5) at (-2,8) {$\soeqI{3}{0}$};
        \node (6) at (1,12) {trivial};
        \draw[->,thick] (1)--(2);
        \node[anchor=east] at (-0.2,2) {$ac_{1}=A_2$};
        \draw[->,thick] (1)--(3);
        \node[anchor=west] at (2.5,1.3) {$A_{\alpha}$};
        \draw[->,thick] (2)--(4);
        \node[anchor=east] at (2.5,6.3) {$A_{\alpha-1}$};
        \draw[->,thick] (3)--(4);
        \node[anchor=west] at (5.2,6) {$c_{1}=A_1$};
        \draw[->,thick] (2)--(5);
        \node[anchor=east] at (-1.2,6) {$A_{\alpha-1}$};
        \draw[->,thick] (5)--(6);
        \node[anchor=east] at (-0.7,10) {$A_{1}$};
        \draw[->,thick] (4)--(6);
        \node[anchor=west] at (3.2,10) {$A_{\alpha-1}$};
    \end{tikzpicture}\;.
\end{equation}
Once more, almost all the transitions are intrinsic to the infinite coupling point, except the transverse slice $c_1$ on the right of the diagram.

\subsubsection{$\UV{SU(K)_{\alpha/2-1}+(K-\alpha-2)F+S}$, $\alpha\geq3$}
For $K>\alpha+2$ there are three minimal Higgsings:
\begin{equation}
    \begin{tikzpicture}
        \node (1) at (0,0) {$\sukeqI{K}{K-\alpha-2}{\alpha/2-1}$};
        \node (2) at (0,4) {$\sukeqI{K-1}{K-\alpha-3}{\alpha/2-1}$};
        \node (3) at (-5,3) {$\soeqI{K-\alpha+2}{K-\alpha-2}$};
        \node (4) at (5,3) {$\sukeqI{K-1}{K-\alpha-3}{\alpha/2}$};
        \draw[->,thick] (1)--(2);
        \node[anchor=west] at (0.2,2) {$ac_{K-\alpha-2}$};
        \draw[->,thick] (1)--(3);
        \node[anchor=east] at (-2.5,1.3) {$A_{K+\alpha-3}$};
        \draw[->,thick] (1)--(4);
        \node[anchor=west] at (2.5,1.3) {$A_{1}$};
    \end{tikzpicture}\;.
\end{equation}
The $ac_{K-\alpha-2}$ is classical, while the other two are at strong coupling.

For $K=\alpha+2$ there are two minimal Higgsings, both of them intrinsic to infinite coupling:
\begin{equation}
    \begin{tikzpicture}
        \node (1) at (0,0) {$\sukeqI{\alpha+2}{0}{\alpha/2-1}$};
        \node (3) at (-5,3) {$\soeqI{4}{0}$};
        \node (4) at (5,3) {$\sukeqI{\alpha+1}{-1}{\alpha/2}$};
        \node[anchor=west] at (7,3) {$=$ $Bh(\alpha)$};
        \draw[->,thick] (1)--(3);
        \node[anchor=east] at (-2.5,1.3) {$A_{2\alpha-1}$};
        \draw[->,thick] (1)--(4);
        \node[anchor=west] at (2.5,1.3) {$A_{1}$};
    \end{tikzpicture}\;.
\end{equation}

\subsubsection{$\UV{SU(K)_{(\alpha-\beta)/2}+(K-\alpha-\beta)F+S}$, $\alpha\geq3$, $\beta\geq3$}
For $K>\alpha+\beta$ there are four minimal Higgsings:
\begin{equation}
    \scalebox{0.7}{\begin{tikzpicture}
        \node (1) at (0,0) {$\sukeqI{K}{K-\alpha-\beta}{\frac{\alpha-\beta}{2}}$};
        \node (2) at (0,4) {$\sukeqI{K-1}{K-\alpha-\beta-1}{\frac{\alpha-\beta}{2}}$};
        \node (3) at (-5,3) {$\soeqI{K}{K-\alpha-\beta}$};
        \node (4) at (-10,3) {$\soeqI{K-\alpha+2}{K-\alpha-\beta}$};
        \node (5) at (-15,3) {$\soeqI{K-\beta+2}{K-\alpha-\beta}$};
        \draw[->,thick] (1)--(2);
        \node[anchor=west] at (0.2,2) {$ac_{K-\alpha-\beta}$};
        \draw[->,thick] (1)--(3);
        \node[anchor=west] at (-2.6,1.8) {$A_{K-1}$};
        \draw[->,thick] (1)--(4);
        \node[anchor=east] at (-5,1.3) {$A_{K+\alpha-3}$};
        \draw[->,thick] (1) .. controls (-12,1) .. (5);
        \node[anchor=north] at (-8,0.7) {$A_{K+\beta-3}$};
    \end{tikzpicture}}\;.
\end{equation}
The slice $ac_{K-\alpha-\beta}$ is classical, while the other three happen only at the fixed point.

For $K=\alpha+\beta$ there are three minimal Higgsings, none of which are visible at the classical level:
\begin{equation}
    \scalebox{0.7}{\begin{tikzpicture}
        \node (1) at (0,0) {$\sukeqI{\alpha+\beta}{0}{\frac{\alpha-\beta}{2}}$};
        \node (3) at (-5,3) {$\soeqI{\alpha+\beta}{0}$};
        \node (4) at (-10,3) {$\soeqI{\beta+2}{0}$};
        \node (5) at (-15,3) {$\soeqI{\alpha+2}{0}$};
        \draw[->,thick] (1)--(3);
        \node[anchor=west] at (-2.6,1.8) {$A_{\alpha+\beta-1}$};
        \draw[->,thick] (1)--(4);
        \node[anchor=east] at (-5,1.3) {$A_{2\alpha+\beta-3}$};
        \draw[->,thick] (1) .. controls (-12,1) .. (5);
        \node[anchor=north] at (-8,0.7) {$A_{\alpha+2\beta-3}$};
    \end{tikzpicture}}\;.
\end{equation}

\clearpage

\section{6d example: SO$(K)$ with $K-8$ vectors}\label{sec:6d}
The techniques presented in this paper are also relevant for studying Higgs branches of 6d $\mathcal{N}=(1,0)$ SQFTs. In this section we demonstrate this by considering a one parameter family of 6d theories with a gauge theory description of SO$(K)$ with $K-8$ vectors.

\subsection{O6 construction}
When $K=2k$, this theory admits a realization using O6-planes, which was considered in \cite{Cabrera:2019dob}.
\begin{equation}
    \begin{array}{c}
         \begin{scriptsize}
             \begin{tikzpicture}
               \draw[dashed](0,0)--(1,0);
               \node[label=below:{O6$^+$}]at(0.5,0){};
               \node[label=below:{O6$^+$}]at(12.5,0){};
               \node[7brane] at (6,0){};
               \node[7brane] at (7,0){};
               \draw(1,0)--(3,0);
               \draw(4,0)--(9,0);
               \draw(1,-1)--(1,1){};
               \draw(2,-1)--(2,1){};
               \draw(3,-1)--(3,1){};
               \node at(3.5,0){$\cdots$};
               \draw(4,-1)--(4,1){};
               \draw(5,-1)--(5,1){};
               \draw(8,-1)--(8,1){};
               \draw(9,-1)--(9,1){};
               \draw(10,-1)--(10,1){};
               \draw(11,-1)--(11,1){};
               \draw(12,-1)--(12,1){};
               \node[label=below:{1}]at(1.5,0){};
               \node[label=below:{1}]at(2.5,0){};
               \node[label=below:{$k-4$}]at(4.5,0){};
               \node[label=below:{$k-4$}]at(5.5,0){};
               \node[label=below:{$k$}]at(6.5,0){};
               \node[label=below:{$k-4$}]at(7.5,0){};
               \node[label=below:{$k-4$}]at(8.5,0){};
               \node at (9.5,0){$\cdots$}{};
               \node[label=below:{$1$}]at(10.5,0){};
               \node[label=below:{$1$}]at(11.5,0){};
               \draw(10,0)--(12,0);
               \draw[dashed](12,0)--(13,0);
             \end{tikzpicture}
         \end{scriptsize}
    \end{array}\;,
\end{equation}
where vertical lines denote D8 branes, horizontal lines denote D6 branes coinciding on O6-planes, while the black dots correspond to NS5 branes. We only specify the charges of the asymptotic O6-planes explicitly. Note that when the O6 plane goes through a D8 brane it switches from $\mathrm{O6}^{\pm}$ to $\widetilde{\mathrm{O6}}^{\pm}$ and vice versa, and when to O6 plane goes through an NS5 brane it switches from $\mathrm{O6}^{+}$ to $\mathrm{O6}^{-}$ and vice versa. The finite coupling magnetic quiver read from this brane system is \cite{Cabrera:2019dob}
\begin{equation}
    \begin{array}{c}
         \begin{scriptsize}
             \begin{tikzpicture}
                          \node[label=below:{1}][so](o1){};
                          \node[label=below:{2}][sp](sp1)[right of=o1]{};
                          \node[label=below:{3}][so](so3)[right of=sp1]{};
                          \node(dots)[right of=so3]{$\cdots$};
                          \node[label=below:{$2k-8$}][sp](spk-4)[right of=dots]{};
                          \node[label=below:{$2k-7$}][so](so2k-7)[right of=spk-4]{};
                          \node[label=below:{$2k-8$}][sp](spk-4')[right of=so2k-7]{};
                          \node(dots')[right of=spk-4']{$\cdots$};
                          \node[label=below:{3}][so](so3')[right of=dots']{};
                          \node[label=below:{2}][sp](sp1')[right of=so3']{};
                          \node[label=below:{1}][so](o1')[right of=sp1']{};
                          \node[label=left:{2}][spf](sp11)[above of=so2k-7]{};
                          \draw(o1)--(sp1);
                          \draw(sp1)--(so3);
                          \draw(so3)--(dots);
                          \draw(dots)--(spk-4);
                          \draw(spk-4)--(so2k-7);
                          \draw(spk-4')--(so2k-7);
                          \draw(dots')--(spk-4');
                          \draw(so3')--(dots');
                          \draw(sp1')--(so3');
                          \draw(o1')--(sp1');
                          \draw(so2k-7)--(sp11);
             \end{tikzpicture}
         \end{scriptsize}
    \end{array}\;.
\end{equation}
The brane system for the Higgs branch emanating from the SCFT point is
\begin{equation}
    \begin{array}{c}
         \begin{scriptsize}
             \begin{tikzpicture}
               \draw[dashed](0,0)--(1,0);
               \node[label=below:{O6$^+$}]at(0.5,0){};
               \node[label=below:{O6$^+$}]at(10.5,0){};
               \node[7brane] at (5.5,.75){};
               \node[7brane] at (5.5,-.75){};
               \draw(1,0)--(3,0);
               \draw(4,0)--(7,0);
               \draw(1,-1)--(1,1){};
               \draw(2,-1)--(2,1){};
               \draw(3,-1)--(3,1){};
               \node at(3.5,0){$\cdots$};
               \draw(4,-1)--(4,1){};
               \draw(5,-1)--(5,1){};
               \draw(6,-1)--(6,1){};
               \draw(7,-1)--(7,1){};
               \draw(8,-1)--(8,1){};
               \draw(9,-1)--(9,1){};
               \draw(10,-1)--(10,1){};
               \node[label=below:{1}]at(1.5,0){};
               \node[label=below:{1}]at(2.5,0){};
               \node[label=below:{$k-4$}]at(4.5,0){};
               \node[label=below:{$k-4$}]at(5.5,0){};
               \node[label=below:{$k-4$}]at(6.5,0){};
               \node at (7.5,0){$\cdots$}{};
               \node[label=below:{$1$}]at(8.5,0){};
               \node[label=below:{$1$}]at(9.5,0){};
               \draw(8,0)--(10,0);
               \draw[dashed](10,0)--(11,0);
             \end{tikzpicture}
         \end{scriptsize}
    \end{array}\;,
\end{equation}
We claim that the magnetic quiver proposed for this brane system in \cite{Cabrera:2019dob} needs a minimal modification of adding flavour to the gauge node in the magnetic quiver corresponding to the NS5 brane moduli, similar to magnetic quivers for 5-brane webs with O5 planes \cite{Bourget:2020gzi,Akhond:2020vhc}. Our proposed magnetic quiver at infinite coupling is
\begin{equation}\label{MQ6 SCFT OSp}
    \begin{array}{c}
         \begin{scriptsize}
             \begin{tikzpicture}
                          \node[label=below:{1}][so](o1){};
                          \node[label=below:{2}][sp](sp1)[right of=o1]{};
                          \node[label=below:{3}][so](so3)[right of=sp1]{};
                          \node(dots)[right of=so3]{$\cdots$};
                          \node[label=below:{$2k-8$}][sp](spk-4)[right of=dots]{};
                          \node[label=below:{$2k-7$}][so](so2k-7)[right of=spk-4]{};
                          \node[label=below:{$2k-8$}][sp](spk-4')[right of=so2k-7]{};
                          \node(dots')[right of=spk-4']{$\cdots$};
                          \node[label=below:{3}][so](so3')[right of=dots']{};
                          \node[label=below:{2}][sp](sp1')[right of=so3']{};
                          \node[label=below:{1}][so](o1')[right of=sp1']{};
                          \node[label=left:{2}][sp](sp11)[above of=so2k-7]{};
                          \node[label=above:{7}][sof](sof)[above of=sp11]{};
                          \draw(o1)--(sp1);
                          \draw(sp1)--(so3);
                          \draw(so3)--(dots);
                          \draw(dots)--(spk-4);
                          \draw(spk-4)--(so2k-7);
                          \draw(spk-4')--(so2k-7);
                          \draw(dots')--(spk-4');
                          \draw(so3')--(dots');
                          \draw(sp1')--(so3');
                          \draw(o1')--(sp1');
                          \draw(so2k-7)--(sp11);
                          \draw(sp11)--(sof);
             \end{tikzpicture}
         \end{scriptsize}
    \end{array}\;.
\end{equation}
The structure of the gauge nodes were already discussed in \cite{Cabrera:2019dob}, the framing with an SO(7) flavour node may be justified as follows. The O6$^+$-plane carries the same charge as an O6$^-$ and 8 $\frac{1}{2}$-D6 branes. A single one of these $\frac{1}{2}$-D6 branes is frozen to give rise to the SO(2k-7) central gauge node, while the other 7 give the SO(7) flavour node in the above magnetic quiver. 

\subsection{O8 construction}
Alternatively we can realise the same 6d theory using the brane system with an O8$^+$-plane \cite{Hanany:1997gh, Apruzzi:2017nck}
\begin{equation}\begin{array}{c}
    \begin{scriptsize}
    \begin{tikzpicture}
              \draw[dashed](0,0)--(0,2);
              \node[label=above:{O8$^+$}] at (0,2){};
              \node[label=above:{$K-8$D8}] at (-1,2){};
              \node[label=above:{$K$D6}] at (-.5,1){};
              \node[label=above:{NS5}][7brane]at(-2,1){};
              \node[7brane]at(2,1){};
              \draw(1,0)--(1,2);
              \draw(-1,0)--(-1,2);
              \draw (-2,1)--(2,1);
    \end{tikzpicture}
    \end{scriptsize}
    \end{array}\;,
\end{equation}
Where $K$ can be odd or even.\footnote{This is in contrast to the O6 construction, where only even $K$ can be realised.}
Before moving on to the infinite coupling case, let us mention the magnetic quiver for finite coupling \cite{Hanany:2016gbz}
\begin{equation}
\begin{array}{c}
     \begin{scriptsize}
    \begin{tikzpicture}
    \node[label=below:{$1$}][u](N){};
    \node[label=below:{$2$}][u](N')[right of=N]{};
    \node(dots')[right of=N']{$\cdots$};
    \node[label=below:{$K-9$}][u](N'')[right of=dots']{};
    \node[label=below:{$K-8$}][u](N''')[right of=N'']{};
    \node[label=above:{$1$}][uf](f)[above of=N''']{};
    \draw(N)--(N');
    \draw(N')--(dots');
    \draw(N'')--(dots');
    \draw[double distance=1.5 pt,<-](N'')--(N''');
    \draw[double distance=1.5 pt,<-](f)--(N''');
    \end{tikzpicture}
    \end{scriptsize}
\end{array}\;.\label{MQ SO(k)+kV}
\end{equation}
The Coulomb branch of this magnetic quiver is the closure of the nilpotent orbit $[2^{K-8}]$ in $\mathfrak{sp}(K-8)$. The HWG is given by
\begin{equation}
    \PE\left[\sum_{i=1}^{K-8}\mu_i^2 t^{2i}\right]\;,
\end{equation}
where $\mu_i$ are $\mathfrak{sp}(K-8)$ highest weight fugacities. Note that since every term in this HWG transforms trivially under the $\mathbb{Z}_2\times\mathbb{Z}_2$ centre of the $\mathfrak{sp}(K-8)\oplus\mathfrak{su}(2)_R$ algebra, the global symmetry that acts faithfully on the chiral ring operators is $\mathrm{Sp}(K-8)/\mathbb{Z}_2\times\mathrm{SO}(3)_R$.

The brane system at infinite coupling SCFT point is
\begin{equation}
    \begin{scriptsize}
        \begin{tikzpicture}
                     \draw(0,0)--(0,2);
                     \draw(1,0)--(1,2);
                     \draw(2,0)--(2,2);
                     \draw(3,0)--(3,2);
                     \draw(4,0)--(4,2);
                     \draw[dashed](5,0)--(5,2);
                     \draw(0,1)--(2,1);
                     \node at (2.5,1){$\cdots$};
                     \draw(3,1)--(5,1);
                     \node[label=below:{$1$}]at(.5,1){};
                     \node[label=below:{$2$}]at(1.5,1){};
                     \node[label=below:{$K-9$}]at(3.5,1){};
                     \node[label=below:{$K-8$}]at(4.5,1){};
                     \node[7brane] at (5,1){};
        \end{tikzpicture}
    \end{scriptsize}\;,
\end{equation}
from which one can read the following magnetic quiver
\begin{equation}
    \begin{array}{c}
       \begin{scriptsize}
    \begin{tikzpicture}
    \node[label=below:{$1$}][u](1){};
    \node[label=below:{$2$}][u](2)[right of=1]{};
    \node (dots)[right of=2]{$\cdots$};
    \node[label=below:{$K-9$}][u](2N-5)[right of=dots]{};
    \node[label=below:{$K-8$}][u](2N-4)[right of=2N-5]{};
    \node[label=below:{$2$}][u](2')[right of=2N-4]{};
    \node[right of=2']{$3\;\textbf{Adj}$};
    \draw (2')to[out=-45,in=45,loop,looseness=10](2');
    \draw(1)--(2);
    \draw(2)--(dots);
    \draw(dots)--(2N-5);
    \draw[ double distance=1.5pt,<-](2N-5)--(2N-4);
    \draw(2N-4)--(2');
    \end{tikzpicture}
    \end{scriptsize}\end{array}\;.
\end{equation}
The first $K-8$ nodes in the above magnetic quiver are obtained by counting the number of D6 brane segments between pairs of D8 branes. The rightmost U(2) node corresponds to the freedom to separate the NS5 brane (and its mirror image) in the transverse directions. The 3 adjoint hypermultiplets are postulated in order to match the Coulomb branch Hilbert series of the orthosymplectic magnetic quiver for this theory for $K=2k$ \eqref{MQ6 SCFT OSp}. It is tempting to conjecture that the self-intersection of the NS5 branes on top of O8$^+$ is 4 -- compared with 0 in the absence of the O8$^+$ -- leading to the adjoint fields in the magnetic quiver \cite{Bourget:2022ehw}. It would be interesting to consider further examples to clarify this point.

We propose the following HWG for the above magnetic quiver
\begin{equation}
    \HWG=\PE\left[\sum_{i=1}^{K-8}\mu_i^2t^{2i}+t^4+\mu_{K-8}\left(t^{K-4}+t^{K-2}\right)-\mu_{K-8}^2t^{2K-4}\right]\;,\label{HWG MQ6d SCFT}
\end{equation}
where $\mu_i$ are $\mathfrak{sp}(K-8)$ highest weight fugacities. The global form of the global symmetries depends on whther $K$ is even or odd:
\paragraph{Case $K=2k$.} In this instance all the terms in the HWG are in projective representations of both $\mathfrak{sp}(2k-8)$ and $\mathfrak{su}(2)_R$ symmetry, as such the faithfully acting global symmetry is $\mathrm{Sp}(2k-8)/\mathbb{Z}_2\times \mathrm{SO}(3)_R$.
\paragraph{Case $K=2k+1$.} In this instance, all the terms in the HWG either transform trivially under $\mathbb{Z}_2\times\mathbb{Z}_2$ centre of the $\mathfrak{sp}(2k-7)\oplus\mathfrak{su}(2)_R$ algebra, or they transform non-trivially under both. Consequently, the global symmetry is $\frac{\mathrm{Sp}(2k-7)\times\mathrm{Su}(2)_R}{\mathbb{Z}_2}$.
As a test of the above conjectured HWG we have performed explicit HS computations for $K=8,10,12$. The results are reported below
\begin{equation}
\begin{split}
    \HS_{K=8}&=\frac{(1-t+t^2)}{(1-t)^2 (1+t)}\\
    \HS_{K=10}&=1 + 10 t^2 + 50 t^4 + 180 t^6 + 530 t^8 + 1351 t^{10} + 3094 t^{12} + 
 6515 t^{14} + 12805 t^{16} + 23775 t^{18} + \mathcal{O}(t^{19})
    \end{split}
\end{equation}
\begin{equation}
    \begin{gathered}
      HS_{K=12}=  1 + 36 t^2 + 639 t^4 + 7491 t^6 + 65598 t^8 + 460062 t^{10} + 
 2703848 t^{12} + 13746645 t^{14} + 61893666 t^{16}\\ + 251253758 t^{18} + 
 932612180 t^{20} + 3201138846 t^{22} + 10254461053 t^{24} + 
 30890710608 t^{26} \\+ 88067690361 t^{28} + 238903842082 t^{30} + 
 619516264054 t^{32} + 1541826036675 t^{34}\\ + 3695516946957 t^{36} + 
 8556356176693 t^{38} + 19188285843951 t^{40} + \mathcal{O}(t^{41})
    \end{gathered}
\end{equation}
We can recognise in the case $K=8$ that this is the Hilbert series for $D_4$ Klein singularity. For $K=9$, the Coulomb branch is a special case of a family studied in \cite{Finkelberg:2019uhm, Hanany:2010qu} and is identified as the intersection of the nilpotent cone of $\mathfrak{sp}(3)$ with the Slodowy slice to the $[4,1^2]$ orbit. Finally let us remark that as a consequence of the results of \cite{Hanany:2023uzn} the Higgs branch of this 6d SCFT is a $\mathbb{Z}_2$ quotient of the symplectic singularity given by the Coulomb branch of
\begin{equation}
    \begin{array}{c}
       \begin{scriptsize}
    \begin{tikzpicture}
    \node[label=below:{$1$}][u](1){};
    \node[label=below:{$2$}][u](2)[right of=1]{};
    \node (dots)[right of=2]{$\cdots$};
    \node[label=below:{$K-9$}][u](2N-5)[right of=dots]{};
    \node[label=below:{$K-8$}][u](2N-4)[right of=2N-5]{};
    \node[label=below:{$1$}][u](2')[below right of=2N-4]{};
    \node[label=above:{$1$}][u](1')[above right of=2N-4]{};
    \draw(1)--(2);
    \draw(2)--(dots);
    \draw(dots)--(2N-5);
    \draw[ double distance=1.5pt,<-](2N-5)--(2N-4);
    \draw(2N-4)--(2');
    \draw(2N-4)--(1');
    \draw(1')--node[right]{$4$}++(2');
    \end{tikzpicture}
    \end{scriptsize}\end{array}\;.\label{MQ SO(2k-2)+2k-8V}
\end{equation}
Curiously, this is the same as the MQ of the 5d SCFT with gauge theory description as $SO(K-2)$ with $(K-8)$ hypermultiplets in the vector representation. The HWG of \eqref{MQ SO(2k-2)+2k-8V} is
\begin{equation}
    \PE\left[\sum_{i=1}^{K-8}\mu_i^2t^{2i}+t^2+\left(q+q^{-1}\right)\mu_{K-8}t^{K-4}-\mu^2_{K-8}t^{2K-8}\right]\;,
\end{equation}
where $q$ is a $U(1)$ fugacity while $\mu_i$ are $\mathfrak{sp}(K-8)$ highest weight fugacities.
The $\mathbb{Z}_2$ quotient removes the $U(1)$ symmetry and so the $t^2$ term in the above HWG picks up a minus sign and the two terms proportional to $\mu_{K-8}$ are exchanged. The invariants of this action are three terms: ${t^4,\mu_{K-8}t^{K-2},\mu_{K-8}^2t^{2K-8}}$, subject to a single relation of the form $\mu_{K-8}^2t^{2K-4}$. Adding all these four terms yields exactly the HWG in \eqref{HWG MQ6d SCFT}. This computation is a consistency test for the conjectured expressions for the HWGs of both quivers.

\subsection{Hasse diagram}
The phase diagram of the Higgs branch of the 6d theory at the finite tension and tensionless points are given by
\begin{equation}
    \begin{tikzpicture}
        \node[hasse] (0l) at (-3,0) {};
        \node[hasse] (1l) at (-3,-1) {};
        \node (2l) at (-3,-2) {$\vdots$};
        \node[hasse] (3l) at (-3,-3) {};
        \node[hasse] (4l) at (-3,-4) {};
        \node[hasse] (5l) at (-3,-5) {};
        \draw (0l)--(1l)--(2l)--(3l)--(4l)--(5l);
        \node at (-4,-0.5) {$c_{1}$};
        \node at (-4,-3.5) {$c_{K-9}$};
        \node at (-4,-4.5) {$c_{K-8}$};
        
        \draw[->] (-2,-2.5)--(-0.5,-2.5);
        \node at (-1.3,-2) {$T\rightarrow 0$};
    
        \node[hasse] (t) at (0,1) {};
        \node[hasse] (0) at (0,0) {};
        \node[hasse] (1) at (0,-1) {};
        \node (2) at (0,-2) {$\vdots$};
        \node[hasse] (3) at (0,-3) {};
        \node[hasse] (4) at (0,-4) {};
        \node[hasse] (5) at (0,-5) {};
        \draw (t)--(0)--(1)--(2)--(3)--(4)--(5);
        \node at (1,0.5) {$D_4$};
        \node at (1,-0.5) {$c_{1}$};
        \node at (1,-3.5) {$c_{K-9}$};
        \node at (1,-4.5) {$c_{K-8}$};
    \end{tikzpicture}
    \label{eq:6dHasse}
\end{equation}
We see that a $D_4$ transition opens up in going from the finite tension $T>0$ to the tensionless limit $T\rightarrow0$. In this case the Higgs branch dimension increases by $1$. 

\subsection{Comment on F-theory constructions with self-intersection $-4$ curves}
We observe that the above discussed theories of $SO(K)$ with $K-8$ vectors can be realized from F-theory on an elliptically fibered non-compact background (Gorenstein singularity), such that the base admits a single rational curve of self-intersection $-4$. 

It was previously discussed in \cite{Bourget:2019aer} that all 6d theories realized by a similar F-theory construction but using a rational curve with self-intersection $-1$, have an $e_8$ transition at the top of the Hasse (phase) diagram as one goes to the tensionless limit. In this case the Higgs branch dimension increases by $29$. Similarly, it was observed in \cite{Hanany:2022itc,Bourget:2022tmw} that all theories realized using a $-2$ curve undergo a $\mathbb{Z}_2$ quotient which replaces the top $d_4$ slice with two slices, $a_1$ on top of $b_3$. In this case the Higgs branch dimension remains constant.

In equation \eqref{eq:6dHasse} we show that all the $SO(K)$ theories realized on a $-4$ curve exhibit a similar behaviour (namely an additional $D_4$ slice at the top). We conjecture that such behaviour happens for all theories on a $-4$ curve, including the gauge groups $F_4$, $E_6$ and $E_7$.

\section{Conclusion and open problems}

This work discusses several aspects of the Higgs branches of 5d and 6d Lagrangian theories, and especially their strongly coupled UV fixed points, constructed with Hanany-Witten brane setups involving O$(\mathrm{d}+2)^+$ orientifold planes. We provide a method to compute their magnetic quivers, and use them to extract physical information such as the precise global symmetry of the theory, $r$-charges of basic operators in the chiral ring, and the Higgs branch Hasse diagram. Additionally, when the computation is feasible, we give the Higgs branch highest weight generating functions and Hilbert series.

One main outcome of this work is that the formula for the intersection number between two sub-webs in a 5-brane web needs to be modified in the presence of an O7${}^+$ orientifold plane. It would be interesting to understand how said contribution arises from string theory in a ``top-down'' way.

A puzzle remains. The Coulomb branches of the magnetic quivers obtained from this modified intersection number are supposed to capture the Higgs branch of the 5d  `electric' theory (at finite or infinite coupling) realised on the brane web. If the 5d theory has a toroidal compactification to a 3d $\mathcal{N}=4$ SCFT with the same Higgs branch (let's call it the 3d electric theory), then one would expect that the magnetic quiver of the 5d theory is the 3d mirror of the 3d electric theory. This in particular would imply that the Higgs branch of the magnetic quiver is isomorphic to the Coulomb branch of the 3d electric theory, which suggests that the dimension of the Higgs branch of the magnetic quiver should match the rank of the 5d electric theory. Most of the magnetic quivers discussed in this paper are non-simply laced quivers, whose Higgs branches (and even their dimensions) are still not known.\footnote{There are few exceptions. If the nodes connected by non-simply laced edges are abelian, the non-simply laced quiver is a standard Lagrangian theory and the Higgs branch can be computed as a hyper-K\"ahler quotient. Furthermore, for some non-simply laced quivers a 3d mirror is known, and the Coulomb branch of the 3d mirror is supposed to match the Higgs branch of the non-simply laced quiver.} However, for low rank cases, some of our magnetic quivers are simply laced and hence their Higgs branches (and Higgs branch dimensions) are easily computed. We find that the Higgs branch dimensions of the magnetic quivers are bigger than the rank of the 5d electric theories. A possible explanation is as follows: The decomposition of a brane web $W$ into subwebs $W_i$ of multiplicity $m_i$,
\begin{equation}
    W=\bigcup_{i=1}^nm_iW_i\;,
\end{equation}
and the intersection numbers of subwebs, $W_i\cdot W_j$, give the adjacency matrix
\begin{equation}
    A_{ij}=W_i\cdot W_j
\end{equation}
and rank vector
\begin{equation}
    v_i=m_i
\end{equation}
of the magnetic quiver (see \cite{Grimminger:2024doq} for a definition of $A$ and $v$ of a quiver). But the decomposition does not give the magnetic quiver itself. The adjacency matrix and rank vector determine the (singular) Coulomb branch of the quiver. However, several quivers with non-identical Higgs branches can have the same adjacency matrix and rank vector\footnote{The unframed unitary quivers for the two 3d $\mathcal{N}=4$ theories: (a) U(1) with 2 hypers of charge 1, (b) U(1) with 1 hyper of charge 2, have the same adjacency matrix and rank vector, and their (singular) Coulomb branches are the same. However their Higgs branches are different. For (a) it is $A_1$ and for (b) it is a point. Both (a) and (b) are magnetic quivers for the $\mathbb{Z}_2$ gauging of a free hyper. But only (b) is a 3d mirror.} (see examples in \cite[Sec. 2.1.2]{Grimminger:2024doq}). So instead of a `magnetic quiver' one may want to use the concept of `magnetic adjacency matrix and rank vector'. This is enough to determine the Higgs branch of the electric theory, but makes no claim about a possible 3d mirror symmetry when the electric theory is compactified to 3 dimensions. This way of thinking is supported by the fact that the magnetic quivers -- or rather magnetic adjacency matrices and rank vectors -- are obtained directly from the brane system of the 5d electric theory, without any reference to a reduction of the electric theory to 3d and subsequent 3d mirror symmetry. If one does want to talk about 3d mirror symmetry, then one has to be careful to identify the right magnetic quiver whose adjacency matrix is obtained from the brane web decomposition. In \cite{Akhond:2021ffo} the magnetic quivers proposed for the infinite coupling SCFT of 5d SO($K$) with $N\leq K-5$ vectors have the same adjacency matrix as the quivers we derive in this paper. But different edges are chosen, such that the Higgs branch dimensions of the magnetic quivers (where computable) match the ranks of the 5d electric theories. It remains an open problem, if (and how) the brane web determines a precise magnetic quiver and not just a magnetic adjacency matrix and rank vector.

Another open problem is related to the physical interpretation of the non-simply laced quivers. It is at the moment unclear if they define some non-Lagrangian 3d $\mathcal{N}=4$ SCFT, or are simply a computational tool useful to study symplectic singularities. As these quivers describe worldvolume theories of brane systems it is highly likely that at least a certain set of non-simply laced quivers define physical theories. It is of great interest to study this problem and settle it. One potentially fruitful avenue would be to use the techniques of \cite{Babinet:2022nxg} to write down the sphere partition functions, or Higgs branch Hilbert series, for non-simply laced quivers.

One interesting output of this study is a complete list of phase diagrams for finite and infinite coupling Higgs branches of the 5d theories under study. While an in depth comparison of these phase diagrams was made, we are yet to identify a general pattern. The search for such a pattern is a very natural direction for future research.

It would be interesting to study the applicability of our methods to brane webs in the presence of $S$-folds \cite{Kim:2021fxx,Acharya:2021jsp}.

While the present work was restricted to brane systems for 5d and 6d field theories, similar methods can be employed to study 3d brane systems. We aim to report the latter analysis in a future publication \cite{D3O3Future}.

\acknowledgments
We thank Lakshya Bhardwaj, Antoine Bourget, Siddharth Dwivedi, Dongwook Ghim, Simone Giacomelli, Hirotaka Hayashi, Rudolphs Kalveks, Deshuo Liu, Bartek Pyszkowski, and Zhenghao Zhong for useful discussions.
MA is supported by a JSPS postdoctoral fellowship grant No. 22F22781.
GAT and AH are supported by the STFC Consolidated Grants ST/T000791/1 and ST/X000575/1
FC is supported by the Leverhulme Research Project Grant 2022 ``Topology from cosmology: axions, astrophysics and machine learning". 
JFG is supported by the EPSRC Open Fellowship (Schafer-Nameki) EP/X01276X/1 and the ``Simons Collaboration on Special Holonomy in Geometry, Analysis and Physics''. 
MA, AH, and JFG are grateful to the Mainz Institute for Theoretical Physics (MITP) and Simons Center for Geometry and Physics (SCGP) for kind hospitality during various stages of this project.
 
\vspace{1cm}
\appendix

\section{Coulomb Branch Hilbert Series}\label{sec:App_review_monopole}

In this appendix, we briefly review the most important results and formulae used in the main text for the computation of Coulomb Branch Hilbert series of a 3d $\mathcal{N}=4$ gauge theory, the so called monopole formula. As this technology is nowadays well established and standard, we will be brief and refer the reader to \cite{Cremonesi:2013lqa} where the monopole formula for gauge theories originates, and to \cite{Cremonesi:2014xha} for non-simply laced quivers.

For Lagrangian theories, the conformal dimension $\Delta(m)$ of a bare monopole operator of magnetic flux $m$ can be computed by 
\begin{equation}\label{conformal dimension}
    \Delta(m)=-\sum_{\alpha\in\Delta_+(\mathfrak{g})}|\alpha(m)|+\frac{1}{2}\sum_{i=1}^{n_H}\sum_{\rho\in \R_i}|\rho(m)|\;,
\end{equation}
where the first sum is over the positive roots of the gauge algebra $\mathfrak{g}$, while the second sum is over the weights of the representation $R_i$ of $\mathfrak{g}$, under which the $i$-th hypermultiplet transforms. The Coulomb branch Hilbert series is  a counting function for the \emph{dressed} monopole operators, grading them by their conformal dimension and charge under the topological symmetry. It can be computed as:
\begin{equation}
    \HS(z,t)=\sum_{m\in\Lambda}z^{J(m)} t^{\Delta(m)}P_G(t;m)\;.
\end{equation}
Here the sum is over the magnetic lattice in which $m$ takes values, $t$ is the fugacity for the $R$-charge of the $\mathcal{N}=2$ subalgebra left unbroken by the monopoles, and $P_G(t;m)$ is a factor taking care of dressing by the scalar component of the adjoint chiral multiplet in the $\mathcal{N}=4$ vector multiplet. Furthermore $z$ is a fugacity for the topological symmetry.

In this paper we also consider magnetic quivers that are non-simply laced, in the sense that among two given nodes there can be a oriented arrow of multiplicity $a$. For example start with considering such line among two abelian nodes. The hypermultiplets contribution to the monopole dimension formula is postulated to change as follows:

\begin{equation}
    \Delta_\text{hyper}\left(\begin{array}{c}
         \begin{scriptsize}\begin{tikzpicture}
         \node[label=above:{$1$}][u](N1){};
         \node[label=above:{$1$}][u](N2)[right of=N1]{};
         \draw[ double distance=1.5pt,->](N1)--(N2);
         \node at (0.2,-0.3) {$a$};
         \end{tikzpicture}  \end{scriptsize}
    \end{array}\right)=|am-n|\;,
    \label{eq:abelian_charges}
\end{equation}
where $m$ denotes the magnetic flux associated with the left U(1) node, while $n$ denotes the magnetic flux for the right U(1) node.
In this case, since the theory is abelian, the multiple line denotes a hypermultiplet with charge $(a,-1)$ under the two U(1) gauge nodes. Similarly, one may consider a hypermultiplet of charge $(a,b)$ under two abelian nodes, the monopole dimension formula will have a contribution
\begin{equation}
    \Delta_\text{hyper}\left(\begin{array}{c}
         \begin{scriptsize}\begin{tikzpicture}
         \node[label=above:{$1$}][u](N1){};
         \node[label=above:{$1$}][u](N2)[right of=N1]{};
         \draw[ double distance=1.5pt,<->](N1)--(N2);
         \node at (0.2,-0.3) {$a$};
         \node at (0.8,-0.3) {$b$};
         \end{tikzpicture}  \end{scriptsize}
\end{array}\right)=|am-bn| \;.
\end{equation}
In the presence of non-abelian nodes, it is no longer possible to interpret the non-simply-laced edge as a hypermultiplet transforming in any representation of the neighbouring gauge nodes. Nevertheless, we will consider magnetic quivers of this type, generalizing what was done above for abelian nodes. The contribution of such lines to the monopole dimension formula is as follows:
\begin{equation}
    \Delta_\text{hyper}\left(\begin{array}{c}
         \begin{scriptsize}\begin{tikzpicture}
         \node[label=above:{$M$}][u](N1){};
         \node[label=above:{$N$}][u](N2)[right of=N1]{};
         \draw[ double distance=1.5pt,<->](N1)--(N2);
         \node at (0.2,-0.3) {$a$};
         \node at (0.8,-0.3) {$b$};
         \end{tikzpicture}  \end{scriptsize}
\end{array}\right)=\sum_{i=1}^M\sum_{j=1}^N|am_i-bn_j|\;.
\end{equation}

This proposal (for $b=1$) has led to numerous non-trivial results for theories with 8 supercharges in 3d \cite{Cremonesi:2014xha}, 4d \cite{Bourget:2020asf}, 5d \cite{Akhond:2021ffo}, and 6d \cite{Bourget:2022tmw} and is by now considered a standard result in the literature about magnetic quivers. For generic $a$ and $b$ Coulomb branches of such quivers have recently been studied in \cite{Grimminger:2024doq}.

\section{Brane Webs, Monodromies, Consistency, and Fixed Points}
\label{app:BraneWebs}

In this section we summarise our conventions for brane webs; we address an important point about the consistency of brane webs, when two fivebranes intersect on top of an O$7^+$ plane; and we illustrate the difficulty in finding the UV fixed point the theory living on a brane web, by `shrinking' the web.

\subsection{Conventions}
We consider webs of $(p,q)$-fivebranes and $[p,q]$-sevenbranes in Type IIB String Theory \cite{Aharony:1997ju,Aharony:1997bh}. The occupied spacetime directions of the branes are summarised in Table \ref{tab:spacetime}.

\begin{table}[h]
\begin{center}
\begin{tabular}{|c|c|c|c|c|c|c|c|c|c|c|}
\hline
Type IIB & $x^0$ & $x^1$ & $x^2$ & $x^3$ & $x^4$ & $x^5$ & $x^6$ & $x^7$ & $x^8$ & $x^9$\\
\hline
$(p,q)5$-brane & $\times$ & $\times$ & $\times$ & $\times$ & $\times$ & \multicolumn{2}{c|}{angle $\alpha$} & & & \\
\hline
$[p,q]7$-Brane & $\times$ & $\times$ & $\times$ & $\times$ & $\times$ & & & $\times$ & $\times$ & $\times$ \\ \hline
\end{tabular}
\caption{Occupation of space-time directions of the $(p,q)$-fivebranes and $[p,q]$-sevenbranes in Type IIB are denoted by $\times$. The angle $\alpha$ depends on the $(p,q)$ charges and the axio-dilaton $\tau$; $\alpha=\arg(p+\tau q)$. We set $\tau=i$ in the rest of the paper, s.t.\ $\tan(\alpha)=q/p$. The 5d $\mathcal{N}=1$ theories exist as effective field theories living on fivebranes suspended between sevenbranes.}
\label{tab:spacetime}
\end{center}
\end{table}

We depict these branewebs by drawing the $(x^5,x^6)$ plane:

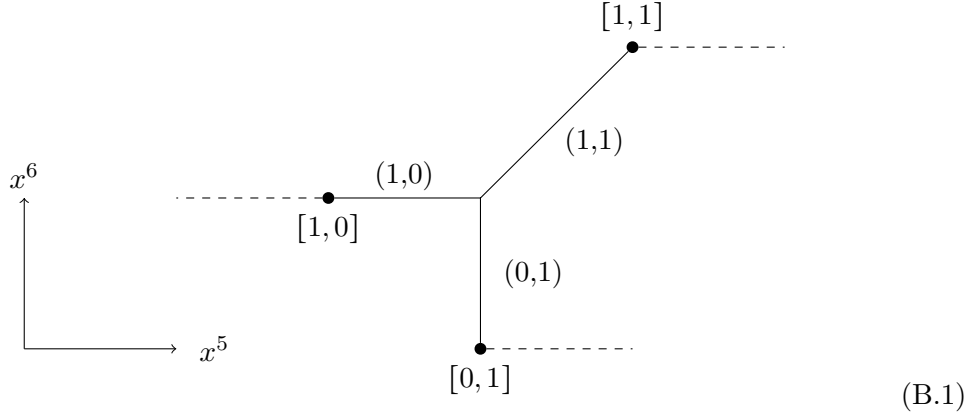
\begin{equation}
    \begin{tikzpicture}
        \node[7brane,label=above:{$[1,1]$}] (1) at (2,2) {};
        \node[7brane,label=below:{$[1,0]$}] (2) at (-2,0) {};
        \node[7brane,label=below:{$[0,1]$}] (3) at (0,-2) {};
        \draw (2)--(0,0)--(3) (0,0)--(1);
        \draw[dashed] (1)--(4,2) (2)--(-4,0) (3)--(2,-2);
        \node at (-1,0.3) {(1,0)};
        \node at (0.7,-1) {(0,1)};
        \node at (1.5,0.75) {(1,1)};

        \draw[->] (-6,-2)--(-4,-2);
        \draw[->] (-6,-2)--(-6,0);
        \node at (-6,0.3) {$x^6$};
        \node at (-3.5,-2) {$x^5$};
    \end{tikzpicture}
\end{equation}

A $(p,q)$ fivebrane ends on a $[p,q]7$-brane, or on a fivebrane vertex which preserves $(p,q)$ charges. Multiple $(p,q)$ fivebranes can end on the same $[p,q]7$-brane as long as the s-rule \cite{Hanany:1996ie,Mikhailov:1998bx,DeWolfe:1998bi,Bergman:1998ej,Benini:2009gi,vanBeest:2020kou,Bergman:2020myx} is not violated.

Each sevenbrane induces an SL$(2,\mathbb{Z})$ monodromy cut, depicted by a dashed line. Our conventions for monodromy matrices associated to monodromy cuts of sevenbranes are summarised in Figure \ref{fig:monodromies}.

\begin{figure}[h]
    \makebox[\textwidth][c]{\begin{tikzpicture}
        \node[anchor=east] at (-6,0) {$M_{[1,0]}=T=\begin{pmatrix}
                1 & 1\\
                0 & 1
            \end{pmatrix}$};
        \node[anchor=east] at (0,0) {$M_{[p,q]}=\begin{pmatrix}
                1-pq & p^2\\
                -q^2 & 1+pq
            \end{pmatrix}$};
        \node[anchor=east] at (-6,-2) {$M_{\mathrm{O}7^+[1,0]}=T^4=\begin{pmatrix}
                1 & 4\\
                0 & 1
            \end{pmatrix}$};
        \node[anchor=east] at (0,-2) {$M_{\mathrm{O}7^+[p,q]}=\begin{pmatrix}
                1-4pq & 4p^2\\
                -4q^2 & 1+4pq
            \end{pmatrix}$};
        \node[anchor=east] at (-6,-4) {$M_{\mathrm{O}7^-[1,0]}=T^{-4}=\begin{pmatrix}
                1 & -4\\
                0 & 1
            \end{pmatrix}$};
        \node[anchor=east] at (0,-4) {$M_{\mathrm{O}7^-[p,q]}=\begin{pmatrix}
                1+4pq & -4p^2\\
                4q^2 & 1-4pq
            \end{pmatrix}$};
        \node at (4,-2) {$\begin{tikzpicture}
            \draw (0,0)--(2,0)--(2,-2);
            \node[7brane,label=below:{$[p,q]$}] (1) at (1,-1) {};
            \draw[dash dot] (1)--(3,1);
            \node at (1,0.5) {$(r,s)$};
            \node at (3,-1) {$M_{[p,q]}.(r,s)$};
            \node at (4,1.3) {\scriptsize $[p,q]$ monodromy cut};
            \draw[dashed] (3.6,1.6)--(4,2);
        \end{tikzpicture}$};
    \end{tikzpicture}}
\caption{Conventions for monodromies of 7-branes and orientifolds. Turning the monodromy cut counter-clockwise leads to an action with the monodromy matrix $M$.}
\label{fig:monodromies}
\end{figure}

\subsection{Web Shrinking and Consistency}

Given a brane web in the Coulomb phase, it is not clear that it can be `shrunk' to a braneweb representing the SCFT point, where all fivebranes meet in a single vertex. This is because not every 5d $\mathcal{N}=1$ effective theory has a UV completion as a 5d $\mathcal{N}=1$ SCFT. Starting from a brane web in the Coulomb phase, it may be possible to perform several Hanany-Witten transitions in order to obtain a `convex' web, which can then be shrunk to realise the SCFT point.

\paragraph{Example 1: SU$(2)$ with 5 fundamental hypers}
Take the brane web for SU$(2)$ with 5 fundamental hypers in a Coulomb phase
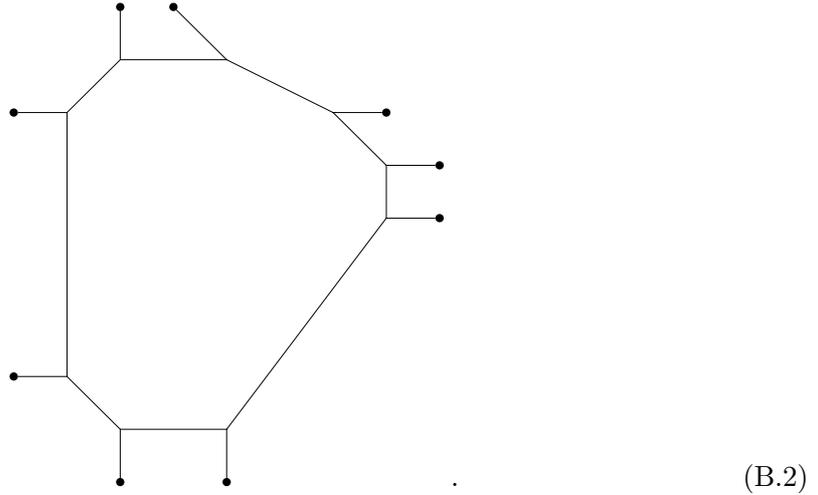
\begin{equation}
    \scalebox{0.7}{\begin{tikzpicture}
        \node[7brane] (1) at (0,1) {};
        \node[7brane] (2) at (-2,-1) {};
        \node[7brane] (3) at (-2,-6) {};
        \node[7brane] (4) at (0,-8) {};
        \node[7brane] (5) at (2,-8) {};
        \node[7brane] (6) at (6,-3) {};
        \node[7brane] (7) at (6,-2) {};
        \node[7brane] (8) at (5,-1) {};
        \node[7brane] (9) at (1,1) {};
        \draw (1)--(0,0)--(-1,-1)--(2) (-1,-1)--(-1,-6)--(3) (-1,-6)--(0,-7)--(4) (0,-7)--(2,-7)--(5) (2,-7)--(5,-3)--(6) (5,-3)--(5,-2)--(7) (5,-2)--(4,-1)--(8) (4,-1)--(2,0)--(9) (2,0)--(0,0);
    \end{tikzpicture}}\;.
\end{equation}
There is no way to directly shrink the web. However, as is well known, one can perform for example the following Hanany-Witten move, employing the monodromy cuts and brane creation:
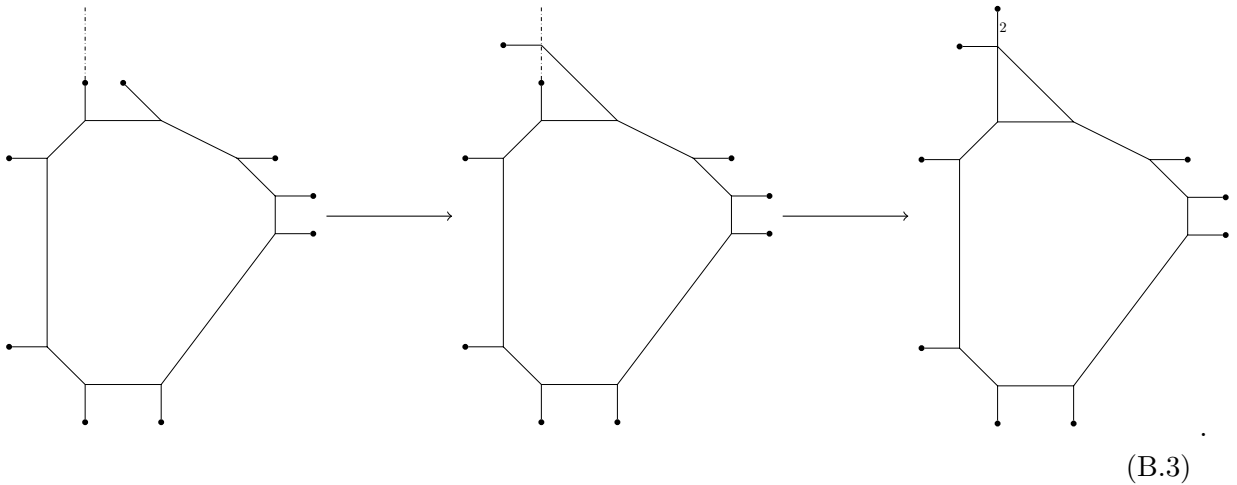
\begin{equation}
    \makebox[\textwidth][c]{\begin{tikzpicture}
        \node (a) at (0,0) {$\scalebox{0.5}{\begin{tikzpicture}
            \node[7brane] (1) at (0,1) {};
            \node[7brane] (2) at (-2,-1) {};
            \node[7brane] (3) at (-2,-6) {};
            \node[7brane] (4) at (0,-8) {};
            \node[7brane] (5) at (2,-8) {};
            \node[7brane] (6) at (6,-3) {};
            \node[7brane] (7) at (6,-2) {};
            \node[7brane] (8) at (5,-1) {};
            \node[7brane] (9) at (1,1) {};
            \draw (1)--(0,0)--(-1,-1)--(2) (-1,-1)--(-1,-6)--(3) (-1,-6)--(0,-7)--(4) (0,-7)--(2,-7)--(5) (2,-7)--(5,-3)--(6) (5,-3)--(5,-2)--(7) (5,-2)--(4,-1)--(8) (4,-1)--(2,0)--(9) (2,0)--(0,0);
            \draw[dash dot] (1)--(0,3);
        \end{tikzpicture}}$};
        \node (b) at (6,0) {$\scalebox{0.5}{\begin{tikzpicture}
            \node[7brane] (1) at (0,1) {};
            \node[7brane] (2) at (-2,-1) {};
            \node[7brane] (3) at (-2,-6) {};
            \node[7brane] (4) at (0,-8) {};
            \node[7brane] (5) at (2,-8) {};
            \node[7brane] (6) at (6,-3) {};
            \node[7brane] (7) at (6,-2) {};
            \node[7brane] (8) at (5,-1) {};
            \node[7brane] (9) at (-1,2) {};
            \draw (1)--(0,0)--(-1,-1)--(2) (-1,-1)--(-1,-6)--(3) (-1,-6)--(0,-7)--(4) (0,-7)--(2,-7)--(5) (2,-7)--(5,-3)--(6) (5,-3)--(5,-2)--(7) (5,-2)--(4,-1)--(8) (4,-1)--(2,0)--(0,2)--(9) (2,0)--(0,0);
            \draw[dash dot] (1)--(0,3);
        \end{tikzpicture}}$};
        \node (c) at (12,0) {$\scalebox{0.5}{\begin{tikzpicture}
            \node[7brane] (1) at (0,3) {};
            \node[7brane] (2) at (-2,-1) {};
            \node[7brane] (3) at (-2,-6) {};
            \node[7brane] (4) at (0,-8) {};
            \node[7brane] (5) at (2,-8) {};
            \node[7brane] (6) at (6,-3) {};
            \node[7brane] (7) at (6,-2) {};
            \node[7brane] (8) at (5,-1) {};
            \node[7brane] (9) at (-1,2) {};
            \draw (0,2)--(0,0)--(-1,-1)--(2) (-1,-1)--(-1,-6)--(3) (-1,-6)--(0,-7)--(4) (0,-7)--(2,-7)--(5) (2,-7)--(5,-3)--(6) (5,-3)--(5,-2)--(7) (5,-2)--(4,-1)--(8) (4,-1)--(2,0)--(0,2)--(9) (2,0)--(0,0);
            \draw[thick] (0,2)--(1);
            \node at (0.15,2.5) {2};
        \end{tikzpicture}}$};
        \draw[->] (a)--(b);
        \draw[->] (b)--(c);
    \end{tikzpicture}}\;.
\end{equation}
The last web can be shrunk to
\begin{equation}
    \scalebox{1}{\begin{tikzpicture}
        \node[7brane] (1) at (-3,0) {};
        \node[7brane] (2) at (-2,0) {};
        \node[7brane] (3) at (-1,0) {};
        \node[7brane] (4) at (0,-1) {};
        \node[7brane] (5) at (0,-2) {};
        \node[7brane] (6) at (1,0) {};
        \node[7brane] (7) at (2,0) {};
        \node[7brane] (8) at (3,0) {};
        \node[7brane] (9) at (0,1) {};
        \draw (1)--(2) (4)--(5) (7)--(8);
        \draw[thick] (2)--(3)--(0,0)--(4) (7)--(6)--(0,0)--(9);
        \node at (-2.5,0.2) {\scriptsize 1};
        \node at (-1.5,0.2) {\scriptsize 2};
        \node at (-0.5,0.2) {\scriptsize 3};
        \node at (1.5,0.2) {\scriptsize 2};
        \node at (2.5,0.2) {\scriptsize 1};
        \node at (-0.2,-0.5) {\scriptsize 2};
        \node at (-0.2,-1.5) {\scriptsize 1};
    \end{tikzpicture}}\;,
\end{equation}
realising the $E_6$ SCFT point.\\

In the presence of an O$7^+$-plane, we have to be more careful. We not only have to find Hanany-Witten moves to convexify the web, but also have to make sure that the shrunk web is consistent under moving the monodromy cut of the O$7^+$. We illustrate this on a simple looking example.

\paragraph{Example 2: Bhardwaj's rank-1 theory.}

If we try to convexify the web for Bhardwaj's rank-1 theory in the same fashion as before, it doesn't work. See Figure \ref{fig:Lakshya_Webs}. However, we can simply turn the monodromy cut of the O$7^+$-plane, and we seemingly have a convex web, see Figure \ref{fig:Lakshya_Webs2} (a), which we can naively shrink reaching the SCFT point, see Figure \ref{fig:Lakshya_Webs2} (b). This is however not a good 'shrinking', as turning the O$7^+$ monodromy cut either direction makes the brane web non-convex, see Figure \ref{fig:Lakshya_Webs2} (c) and (d).

\cite[Figure 18]{Hayashi:2023boy} finds a good convexification of the web, which allows to shrink it, see Figure \ref{fig:Lakshya_Webs3} for a similar convexification and shrinking of the web.

\begin{figure}[h]
    \centering
    \begin{subfigure}[b]{1\textwidth}
        \begin{adjustbox}{center}\scalebox{1}{\begin{tikzpicture}[scale=1]
            \node[7brane,green] (0) at (0,0) {};
            \node[7brane] (1) at (-2,1) {};
            \node[7brane] (2) at (-0,4) {};
            \draw[dashed,green] (-5,0)--(5,0);
            \draw (0,0)--(1,1)--(2,1)--(1,2)--(1,1) (2,1)--(4,0) (-4,0)--(1);
            \draw (1,2)--(2);
            \node at (0,-0.5) {O$7^+$};
            \node at (0.8,0.3) {(1,1)};
            \node at (1.5,0.7) {(1,0)};
            \node at (0.5,1.5) {(0,1)};
            \node at (2.1,1.5) {(1,-1)};
            \node at (3.5,0.7) {(2,-1)};
            \node at (-2,0.5) {[2,1]};
            \node at (0,4.5) {[1,-2]};
        \end{tikzpicture}}
        \end{adjustbox}
        \caption{Original Web, taken from \cite{Kim:2020hhh}}
        \label{fig:Lakshya_Original_Web}
    \end{subfigure}
    \begin{subfigure}[b]{1\textwidth}
        \begin{adjustbox}{center}\scalebox{1}{\begin{tikzpicture}[scale=2]
            \node[7brane,green] (0) at (0,0) {};
            \node[7brane] (2) at (0.9,2.2) {};
            \draw[dashed,green] (-4.5,0)--(4.5,0);
            \draw (0,0)--(1,1)--(2,1)--(1,2)--(1,1) (2,1)--(4,0) (1,2)--(2);
            \draw[dashed] (2)--(0,4);
            \node[7brane] (1) at (4/5+7/20,12/5-9/20) {};
            \draw (-4,0)--(4/5,12/5)--(1);
            \node at (0,-0.2) {O$7^+$};
            \node at (0.8,0.4) {(1,1)};
            \node at (1.5,0.8) {(1,0)};
            \node at (0.7,1.5) {(0,1)};
            \node at (1.4,1.3) {(1,-1)};
            \node at (0.7,2) {[1,-2]};
            \node at (3.2,0.7) {(2,-1)};
            \node at (-2,0.7) {(2,1)};
            \node at (0.3+4/5+7/20,12/5-9/20) {[7,-9]};
            \draw[red] (1)--(4/5+7/5,12/5-9/5);
        \end{tikzpicture}}
        \end{adjustbox}
        \caption{$(2,1)5$-brane affected by monodromies of $[1,-2]7$-brane and O$7^+$-plane. We see, that the $(2,1)5$-brane is affected by the monodromy in such a way, that it collides with itself (as a $(2,-1)5$-brane before O$7^+$ monodromy) as indicated in {\color{red}red}, and we cannot convexify it this way.}
        \label{fig:Lakshya_Monodromies1}
    \end{subfigure}
    \begin{subfigure}[b]{1\textwidth}
        \centering
        \scalebox{0.4}{\begin{tikzpicture}
            \node[7brane,green] (0) at (0,0) {};
            \node[7brane] (1) at (-2,1) {};
            \draw[dashed,green] (-20,0)--(20,0);
            \draw (0,0)--(1,1)--(2,1)--(1,2)--(1,1) (2,1)--(4,0) (-4,0)--(1);
            \draw[dashed] (1)--(20,12);
            \draw (1,2)--(4/5,12/5)--(-8,0);
            \draw (8,0)--(44/7,36/7)--(-13,0);
            \draw (13,0)--(100/9,68/9)--(-88/5,0);
            \node[7brane] (2) at (88/5-1,0+5) {};
            \draw (88/5,0)--(2);
            \node at (0,-0.5) {O$7^+$};
            \node at (0.8,0.3) {(1,1)};
            \node at (1.5,0.7) {(1,0)};
            \node at (0.5,1.5) {(0,1)};
            \node at (2.1,1.5) {(1,-1)};
            \node at (1.5,2.2) {(1,-2)};
            \node at (3.5,0.7) {(2,-1)};
            \node at (-2,0.5) {[2,1]};
            \node at (-2,2) {(11,3)};
            \node at (-2,3.3) {(15,4)};
            \node at (-2,4.5) {(19,5)};
            \node at (8,2) {(3,-1)};
            \node at (12.8,3) {(4,-1)};
            \node at (88/5-1,0+5.5) {[5,-1]};
        \end{tikzpicture}}
        \caption{$(1,-2)5$-brane affected by monodromies of $[2,1]7$-brane and O$7^+$-plane. We see, that the web `continues infinitely' and we cannot convexify it this way.}
        \label{fig:Lakshya_Monodromies2}
    \end{subfigure}
    \caption{Brane web for Bhardwaj's theory, naive attempts to convexify without moving the monodromy cut of the O$7^+$-plane.}
    \label{fig:Lakshya_Webs}
\end{figure}

\begin{figure}[h]
    \centering
    \begin{subfigure}[t]{0.45\textwidth}
        \begin{adjustbox}{center}\scalebox{0.8}{\begin{tikzpicture}[scale=1]
            \node[7brane,green] (0) at (0,0) {};
            \node[7brane] (1) at (4,0) {};
            \node[7brane] (2) at (0,4) {};
            \draw[dashed,green] (-3,3)--(3,-3);
            \draw (0,0)--(1,1)--(2,1)--(1,2)--(1,1) (2,1)--(1);
            \draw (1,2)--(2);
            \node at (0,-0.5) {O$7^+$};
            \node at (0.8,0.3) {(1,1)};
            \node at (1.5,0.7) {(1,0)};
            \node at (0.5,1.5) {(0,1)};
            \node at (2.1,1.5) {(1,-1)};
            \node at (4.5,0) {[2,-1]};
            \node at (0,4.5) {[1,-2]};
        \end{tikzpicture}}
        \end{adjustbox}
        \caption{Original Web, with O$7^+$ monodromy turned.}
        \label{fig:Lakshya_Web_Turned_bis}
    \end{subfigure}
    \hfill
    \begin{subfigure}[t]{0.45\textwidth}
        \begin{adjustbox}{center}\scalebox{0.8}{\begin{tikzpicture}[scale=1]
            \node[7brane,green] (0) at (0,0) {};
            \node[7brane] (1) at (-1,2) {};
            \node[7brane] (2) at (2,-1) {};
            \draw[dashed,green] (-3,3)--(3,-3);
            \draw (1)--(0,0)--(2);
            \node at (0,-0.5) {O$7^+$};
            \node at (2.5,-1) {[2,-1]};
            \node at (-1,2.5) {[1,-2]};
        \end{tikzpicture}}
        \end{adjustbox}
        \caption{Shrunk web.}
        \label{fig:Lakshya_Web_Turned}
    \end{subfigure}
    \begin{subfigure}[t]{0.45\textwidth}
        \begin{adjustbox}{center}\scalebox{0.8}{\begin{tikzpicture}[scale=1]
            \node[7brane,green] (0) at (0,0) {};
            \node[7brane] (1) at (-1,2) {};
            \node[7brane] (2) at (-1,1.5) {};
            \draw (1)--(0,0)--(2,-1);
            \node at (0,-0.5) {O$7^+$};
            \node at (-1.5,1.6) {[2,1]};
            \node at (-1,2.5) {[1,-2]};
            \draw[dashed,green] (0) .. controls (1,-1) .. (2,-1);
            \draw[dashed,green] (0) .. controls (-1,1) .. (-2,1);
            \draw[dashed,green] (2,-1)--(4,-1) (-4,1)--(-2,1);
            \draw (-2,1)--(2);
        \end{tikzpicture}}
        \end{adjustbox}
        \caption{Shrunk web, O$7^+$ monodromy turned counter-clockwise.}
        \label{fig:Lakshya_Web_Turned1}
    \end{subfigure}
    \hfill
    \begin{subfigure}[t]{0.45\textwidth}
        \begin{adjustbox}{center}\scalebox{0.8}{\begin{tikzpicture}[scale=1]
            \node[7brane,green] (0) at (0,0) {};
            \node[7brane] (1) at (1+9/5,-2-2/5) {};
            \node[7brane] (2) at (2,-1) {};
            \draw (-1,2)--(0,0)--(2);
            \node at (0,-0.5) {O$7^+$};
            \node at (2.5,-1) {[2,-1]};
            \node at (0.5+1+9/5,-2-2/5) {[9,-2]};
            \draw[dashed,green] (0) .. controls (1,-1) .. (1,-2);
            \draw[dashed,green] (0) .. controls (-1,1) .. (-1,2);
            \draw[dashed,green] (1,-2)--(1,-3) (-1,3)--(-1,2);
            \draw (1,-2)--(1);
        \end{tikzpicture}}
        \end{adjustbox}
        \caption{Shrunk web, O$7^+$ monodromy turned clockwise.}
        \label{fig:Lakshya_Web_Turned2}
    \end{subfigure}
    \caption{When turning the O$7^+$ monodromy cut in the brane web for Bhardwaj's theory one can shrink the web in a naive attempt to realise the SCFT point. Turning the O$7^+$ monodromy clockwise or counter-clockwise leads to the web becoming non-convex. This signals, that we have not found the shrunken brane web realising the SCFT fixed point.}
    \label{fig:Lakshya_Webs2}
\end{figure}

\begin{figure}[h]
    \makebox[\textwidth][c]{
    \begin{tikzpicture}
        \node (a) at (-4,0) {$\begin{tikzpicture}
            \node[7brane,green] (0) at (0,0) {};
            \node[7brane,blue] (1) at (3,0.5) {};
            \node[7brane] (2) at (0.5,3) {};
            \draw[dashed,green] (-3,0)--(4,0);
            \draw (0,0)--(1,1)--(2,1)--(1,2)--(1,1) (2,1)--(1);
            \draw (1,2)--(2);
            \node at (0,-0.5) {O$7^+$};
            \node at (0.8,0.3) {(1,1)};
            \node at (1.5,0.7) {(1,0)};
            \node at (0.5,1.5) {(0,1)};
            \node at (2.1,1.5) {(1,-1)};
            \node at (3.5,0.5) {[2,-1]};
            \node at (0.5,3.3) {[1,-2]};
        \end{tikzpicture}$};
        \node (b) at (4,0) {$\begin{tikzpicture}
            \node[7brane,green] (0) at (0,0) {};
            \node[7brane,blue] (1) at (1.5,1.25) {};
            \node[7brane] (2) at (0.5,3) {};
            \draw[dashed,green] (-3,0)--(4,0);
            \draw (0,0)--(1,1)--(2,1)--(1,2)--(1,1);
            \draw (1,2)--(2);
            \draw[dash dot] (1)--(3,0.5);
        \end{tikzpicture}$};
        \node (c) at (0,-3) {$\begin{tikzpicture}
            \node[7brane,green] (0) at (0,0) {};
            \node[7brane,blue] (1) at (1.5,1.25) {};
            \node[7brane] (2) at (5.5,0.5) {};
            \draw[dashed,green] (-5,0)--(5,0);
            \draw (0,0)--(1,1)--(3,1)--(1,1.5)--(1,1);
            \draw (3,1)--(2);
            \draw[dash dot] (1)--(0,2);
            \node at (2,1.5) {(4,-1)};
            \node at (5.5,0.8) {[5,-1]};
        \end{tikzpicture}$};
        \node (d) at (0,-6) {$\scalebox{0.7}{\begin{tikzpicture}
            \node[7brane,green] (0) at (0,0) {};
            \node[7brane,blue] (1) at (0,2) {};
            \node[7brane] (2) at (-9,1) {};
            \draw[dashed,green] (-10,0)--(10,0);
            \draw (0,0)--(1,1)--(3,1)--(1,1.5)--(1,1);
            \draw[thick] (1,1.5)--(1);
            \node at (0.5,2) {2};
            \draw (3,1)--(8,0);
            \draw (-8,0)--(2);
            \draw[dash dot] (1)--(-2,3);
            \node at (-9,1.3) {[1,-1]};
        \end{tikzpicture}}$};
        \node (e) at (0,-10) {$\scalebox{0.3}{\begin{tikzpicture}
            \node[7brane,green] (0) at (0,0) {};
            \node[7brane,blue] (1) at (0,2) {};
            \node[7brane] (2) at (2,12) {};
            \draw[dashed,green] (-10,0)--(10,0);
            \draw (0,0)--(1,1)--(3,1)--(1,1.5)--(1,1);
            \draw[thick] (1,1.5)--(1);
            \node at (0.5,2.2) {\Huge 2};
            \draw (3,1)--(8,0);
            \draw (-8,0)--(-20,12)--(2);
            \draw[dash dot] (1)--(-22,13);
            \node at (3,12) {\Huge [1,0]};
        \end{tikzpicture}}$};
        \node (f) at (0,-15) {$\scalebox{0.3}{\begin{tikzpicture}
            \node[7brane,green] (0) at (0,0) {};
            \node[7brane,blue] (1) at (-22,13) {};
            \node[7brane] (2) at (2,12) {};
            \draw[dashed,green] (-10,0)--(10,0);
            \draw (0,0)--(1,1)--(3,1)--(1,1.5)--(1,1);
            \draw[thick] (1,1.5)--(1);
            \node at (-10,7.5) {\Huge 2};
            \node at (-21,13) {\Huge 3};
            \draw (3,1)--(8,0);
            \draw (-8,0)--(-20,12)--(2);
            \draw[dash dot] (1)--(-22,13);
            \node at (3,12) {\Huge [1,0]};
        \end{tikzpicture}}$};
        \node (g) at (0,-19) {$\begin{tikzpicture}
            \node[7brane,green] (0) at (0,0) {};
            \node[7brane,blue] (1) at (-2,1) {};
            \node[7brane] (2) at (2,0) {};
            \draw[dashed,green] (-3,0)--(3,0);
            \draw[thick] (0,0)--(1);
            \node at (-1,0.8) {3};
            \draw (0,0)--(2);
            \node at (0,-0.5) {O$7^+$};
            \node at (2,0.3) {[1,0]};
            \node at (-2,1.3) {[2,-1]};
        \end{tikzpicture}$};
        \draw[->,thick] (a)--(b);
        \draw[->,thick] (b)--(c);
        \draw[->,thick] (c)--(d);
        \draw[->,thick] (d)--(e);
        \draw[->,thick] (e)--(f);
        \draw[->,thick] (f)--(g);
    \end{tikzpicture}}
    \caption{Consistent convexification and shrinking of the brane web for Bhardwaj's rank-1 theory, following moves similar to \cite{Hayashi:2023boy}.}
    \label{fig:Lakshya_Webs3}
\end{figure}

\clearpage

\paragraph{Consistency.}

As we saw in Figure \ref{fig:Lakshya_Webs2} turning the monodromy cut of the O$7^+$ plane in a shrunken web may lead to the web becoming non-convex. In a shrunk web, if the monodromy cut of the O$7^+$ plane is oriented in the $(1,0)$ direction (or oriented in the $(u,v)$ direction for O$7^+[u,v]$), no such issues can occur when turning the monodromy cut. Take the configuration
\begin{equation}
    \begin{tikzpicture}
                \node[label=below:{$\mathrm{O}7^+$}][7brane,green]at(0,0){};
                \draw[dash dot,green](-2,0)--(2,0);
                \draw (-2,-1)--(2,1);
                \draw (-2,1)--(2,-1);
                \node at (-1,1) {$(p_1,q_1)$};
                \node at (1,1) {$(p_2,q_2)$};
            \end{tikzpicture}\;,
\end{equation}
where the O$7^+$ monodromy cut is oriented in the $(1,0)$ direction. Turning it left, first the $(p_2,q_2)$ fivebrane is acted upon, then the $(p_1,q_1)$ fivebrane is acted upon. Without loss of generality let $q_i\geq0$, then the ordering of fivebranes is equivalent to 
\begin{equation}
    p_1q_2-p_2q_1<0\;.
\end{equation}
We have $M_{\mathrm{O}7^+}.(p_2,q_2)=(p_2+4q_2,q_2)$. To check that the web is still convex under turning the O$7^+$ monodromy cut, we have to check that the ordering of the fivebranes is unchanged, which is indeed the case, as
\begin{equation}
    p_1q_2-(p_2+4q_2)q_1=p_1q_2-p_2q_1-4q_2q_1<p_1q_2-p_2q_1<0\;.
\end{equation}

\clearpage

\section{Hasse Diagrams for 5d Magnetic Quivers}
\label{app:Hasse}

In this appendix we summarise the magnetic quivers for the 5d $\mathcal{N}=1$ gauge theories and SCFTs studied in the main text, and provide their Coulomb branch Hasse diagrams.

\subsection{Finite Coupling Gauge Theories}\label{Hasse diagrams finite}
\subsubsection{Summary of Magnetic Quivers}
The relevant magnetic quivers are:
\begin{equation}
    \makebox[\textwidth][c]{\begin{tikzpicture}
        \node at (0,0) {$\begin{split}
    \mathsf{q}_1(K,N)= & \raisebox{-.4\height}{
        \begin{tikzpicture}
            \node at (0.5,0) {};
            \node at (9,0) {};
            \node[u,label=below:{\scriptsize$1$}] (1) at (1,0) {};
            \node[u,label=below:{\scriptsize$2$}] (2) at (2,0) {};
            \node (3) at (3,0) {$\cdots$};
            \node[u,label=below:{\scriptsize$K$}] (4) at (4,0) {};
            \node[u,label=above:{\scriptsize$1$}] (4u) at (4,1) {};
            \node[u,label=below:{\scriptsize$K$}] (5) at (5,0) {};
            \node (6) at (6,0) {$\cdots$};
            \node[u,label=below:{\scriptsize$K$}] (7) at (7,0) {};
            \node[u,label=below:{\scriptsize$K$}] (8) at (8,0) {}; 
            \draw (1)--(2)--(3)--(4)--(5)--(6)--(7);
            \draw[<-,double distance=1.5pt](7)--(8);
            \draw (4)--(4u);
            \draw [decorate,decoration={brace,amplitude=3pt},xshift=0pt,yshift=0pt](8.2,-0.5) -- (0.8,-0.5) node [midway,xshift=0pt,yshift=-15pt] {\scriptsize $N$};
        \end{tikzpicture}
        }=\mathsf{MQ}(SO(K)+N\mathbf{V})\;,\;N>K\\
    \mathsf{q}_2(N)= & \raisebox{-.4\height}{
        \begin{tikzpicture}
            \node at (0.5,0) {};
            \node at (9,0) {};
            \node[u,label=below:{\scriptsize$1$}] (4) at (1,0) {};
            \node[u,label=above:{\scriptsize$1$}] (4u) at (8,1) {};
            \node[u,label=below:{\scriptsize$2$}] (5) at (2,0) {};
            \node (6) at (4.5,0) {$\cdots$};
            \node[u,label=below:{\scriptsize$N-1$}] (7) at (7,0) {};
            \node[u,label=below:{\scriptsize$N$}] (8) at (8,0) {};
            \draw (4)--(5)--(6)--(7);
            \draw[<-,double distance=1.5pt](7)--(8);
            \draw[->,double distance=1.5pt](8)--(4u);
        \end{tikzpicture}
        }=\mathsf{MQ}(SO(K)+N\mathbf{V})\;,\;N=K\\
    \mathsf{q}_3(N)= & \raisebox{-.4\height}{
        \begin{tikzpicture}
            \node at (0.5,0) {};
            \node at (9,0) {};
            \node[u,label=below:{\scriptsize$1$}] (4) at (1,0) {};
            \node[u,label=above:{\scriptsize$1$}] (4u) at (8,1) {};
            \node[u,label=below:{\scriptsize$2$}] (5) at (2,0) {};
            \node (6) at (4.5,0) {$\cdots$};
            \node[u,label=below:{\scriptsize$N-1$}] (7) at (7,0) {};
            \node[u,label=below:{\scriptsize$N$}] (8) at (8,0) {};
            \draw (4)--(5)--(6)--(7);
            \draw[<-,double distance=1.5pt](7)--(8);
            \draw[double distance=1.5pt](8)--(4u);
        \end{tikzpicture}
        }=\mathsf{MQ}(SO(K)+N\mathbf{V})\;,\;N<K\\
    \mathsf{q}_4(K,N)= & \raisebox{-.4\height}{
        \begin{tikzpicture}
            \node at (0.5,0) {};
            \node at (9,0) {};
            \node[u,label=below:{\scriptsize$1$}] (1) at (1,0) {};
            \node[u,label=below:{\scriptsize$2$}] (2) at (2,0) {};
            \node (3) at (3,0) {$\cdots$};
            \node[u,label=below:{\scriptsize$K$}] (4) at (4,0) {};
            \node[u,label=above:{\scriptsize$1$}] (4u) at (4,1) {};
            \node[u,label=below:{\scriptsize$K$}] (5) at (5,0) {};
            \node (6) at (6,0) {$\cdots$};
            \node[u,label=below:{\scriptsize$K$}] (7) at (7,0) {};
            \node[u,label=below:{\scriptsize$K$}] (8) at (8,0) {};
            \node[u,label=above:{\scriptsize$1$}] (8u) at (8,1) {};
            \draw (1)--(2)--(3)--(4)--(5)--(6)--(7);
            \draw[<-,double distance=1.5pt](7)--(8);
            \draw (4)--(4u);
            \draw (8u)--(8);
            \draw [decorate,decoration={brace,amplitude=3pt},xshift=0pt,yshift=0pt](8.2,-0.5) -- (0.8,-0.5) node [midway,xshift=0pt,yshift=-15pt] {\scriptsize $N$};
        \end{tikzpicture}
        }=\mathsf{MQ}(SU(K)+N\mathbf{F}+\mathbf{S})\;,\;N>K\\
    \mathsf{q}_5(N)= & \raisebox{-.4\height}{
        \begin{tikzpicture}
            \node at (0.5,0) {};
            \node at (9,0) {};
            \node[u,label=below:{\scriptsize$1$}] (4) at (1,0) {};
            \node[u,label=above:{\scriptsize$1$}] (4u) at (7.3,1) {};
            \node[u,label=below:{\scriptsize$2$}] (5) at (2,0) {};
            \node (6) at (4.5,0) {$\cdots$};
            \node[u,label=below:{\scriptsize$N-1$}] (7) at (7,0) {};
            \node[u,label=below:{\scriptsize$N$}] (8) at (8,0) {};
            \node[u,label=above:{\scriptsize$1$}] (8u) at (8.7,1) {};
            \draw (4)--(5)--(6)--(7);
            \draw[<-,double distance=1.5pt](7)--(8);
            \draw[->,double distance=1.5pt](8)--(4u);
            \draw (8u)--(8);
        \end{tikzpicture}
        }=\mathsf{MQ}(SU(K)+N\mathbf{F}+\mathbf{S})\;,\;N=K\\
    \mathsf{q}_6(K,N)= & \raisebox{-.4\height}{
        \begin{tikzpicture}
            \node at (0.5,0) {};
            \node at (9,0) {};
            \node[u,label=below:{\scriptsize$1$}] (4) at (1,0) {};
            \node[u,label=above:{\scriptsize$1$}] (4u) at (7.3,1) {};
            \node[u,label=below:{\scriptsize$2$}] (5) at (2,0) {};
            \node (6) at (4.5,0) {$\cdots$};
            \node[u,label=below:{\scriptsize$N-1$}] (7) at (7,0) {};
            \node[u,label=below:{\scriptsize$N$}] (8) at (8,0) {};
            \node[u,label=above:{\scriptsize$1$}] (8u) at (8.7,1) {};
            \draw (4)--(5)--(6)--(7);
            \draw[<-,double distance=1.5pt](7)--(8);
            \draw[double distance=1.5pt](8)--(4u);
            \draw (8u)--(8);
            \draw[double distance=1.5pt](8u)--(4u);
            \node at (8,1.3) {\scriptsize$K-N$};
        \end{tikzpicture}
        }=\mathsf{MQ}(SU(K)+N\mathbf{F}+\mathbf{S})\;,\;N<K
    \end{split}$};
    \end{tikzpicture}}
\end{equation}
where $\mathsf{MQ}(T)$ denotes the magnetic quiver of a classical gauge theory $T$.

\pagebreak

\subsubsection{Hasse Diagrams}

The Coulomb branch Hasse diagrams of the magnetic quivers in question can be computed via quiver subtraction following the rules of \cite{Bourget:2019aer,Bourget:2020mez}.

\paragraph{Notation.} For each node in the Hasse diagram, corresponding to a symplectic leaf, we give the quiver (in orange) whose Coulomb branch is the transverse slice to the symplectic leaf. For each edge in the Hasse diagram we denote the geometry of the corresponding elementary slice.

\paragraph{1}

\begin{subequations}
The Coulomb branch Hasse diagram of
\begin{equation}
    \mathsf{q}_1(K,N)= \raisebox{-.4\height}{
\;,
\end{equation}
where the orange labels denote the magnetic quivers for the transverse slices to each leaf.
\end{subequations}

\pagebreak

\paragraph{2}

\begin{subequations}
The Coulomb branch Hasse diagram of
\begin{equation}
    \mathsf{q}_2(N)= \raisebox{-.4\height}{
        %
\;.
\end{equation}
\end{subequations}

\paragraph{3}

\begin{subequations}
The Coulomb branch Hasse diagram of
\begin{equation}
    \mathsf{q}_3(N)= \raisebox{-.4\height}{
        %
\;.
\end{equation}
\end{subequations}

\paragraph{4}

\begin{subequations}
The Coulomb branch Hasse diagram of
\begin{equation}
    \mathsf{q}_4(K,N)= \raisebox{-.4\height}{
        %
\;.
\end{equation}
\end{subequations}

\pagebreak

\paragraph{5}

\begin{subequations}
The Coulomb branch Hasse diagram of
\begin{equation}
    \mathsf{q}_5(N)=\raisebox{-.4\height}{
        %
\;.
\end{equation}
\end{subequations}

\paragraph{6}

\begin{subequations}
The Coulomb branch Hasse diagram of
\begin{equation}
    \mathsf{q}_6(K,N)= \raisebox{-.4\height}{
        %
}\qquad\textnormal{for }K>N+1\;.
\end{equation}
\end{subequations}

\clearpage

\subsection{SCFTs}\label{Hasse diagrams SCFTs}
\subsubsection{Summary of Magnetic Quivers}

The relevant magnetic quivers are:

\begin{equation}
    \makebox[\textwidth][c]{%
        }=\mathsf{MQ}_{\infty}(SU(N+\alpha+\beta)_{\frac{\alpha-\beta}{2}}+N\mathbf{F}+\mathbf{S})\;,\;N\geq0\;,\;\alpha\geq3\;,\;\beta\geq3
    \end{split}$};
    \end{tikzpicture}}
\end{equation}
where $\mathsf{MQ}_{\infty}(T)$ denotes the magnetic quiver of the (infinite coupling) UV completion of the 5d $\mathcal{N}=1$ gauge theory $T$.

\subsubsection{Hasse Diagrams}

The Coulomb branch Hasse diagrams of the magnetic quivers in question can be computed via quiver subtraction following the rules of \cite{Bourget:2019aer,Bourget:2020mez,Bourget:2022tmw}, and appreciating that:
\begin{enumerate}
    \item 
            \begin{equation}
                \mathcal{C}\left(\raisebox{-.4\height}{
}$ has Coulomb branch Hasse diagram
            \begin{equation}
                %
}\right)/\mathbb{C}^2=\mathrm{Sym}^4\left(\mathbb{C}^2\right)/\mathbb{C}^2=\mathrm{Sym}^4_0(\mathbb{C}^2)
            \end{equation}
            (where the $/\mathbb{C}^2$ denotes the removal of the free $\mathbb{C}^2$ part), and hence (see e.g.\ \cite{Bourget:2022tmw}) $\raisebox{-.4\height}{%
}$ has Coulomb branch Hasse diagram
            \begin{equation}
                %
\;.
            \end{equation}
\end{enumerate}

We now present the Coulomb branch Hasse diagrams of the quivers in question.

\paragraph{Notation.} For each node in the Hasse diagram, corresponding to a symplectic leaf, we give the quiver (in orange) whose Coulomb branch (or rather it's singular part)\footnote{Note that $\mathcal{C}\left(\mathsf{Q}_1(1)=\raisebox{-.4\height}{
        %
        }\right)=\mathrm{Sym}^3(\mathbb{C}^2)$ both have a free $\mathbb{C}^2$ factor, which needs to be removed to obtain the singular part.} is the transverse slice to the symplectic leaf. For each edge in the Hasse diagram we denote the geometry of the corresponding elementary slice.

\clearpage

\paragraph{1}

\begin{subequations}
The Coulomb branch Hasse diagram of
\begin{equation}
    \mathsf{Q}_1(N)= \raisebox{-.4\height}{
        %
\;,
\end{equation}
where the orange labels denote the magnetic quivers for the transverse slices to each leaf.
\end{subequations}

\paragraph{2}

\begin{subequations}
The Coulomb branch Hasse diagram of
\begin{equation}
    \mathsf{Q}_2(N)= \raisebox{-.4\height}{
        %
\;.
\end{equation}
\end{subequations}

\paragraph{3}

\begin{subequations}
The Coulomb branch Hasse diagram of
\begin{equation}
    \mathsf{Q}_3(N,\alpha)= \raisebox{-.4\height}{
        %
\;.
\end{equation}
\end{subequations}

\paragraph{4}

\begin{subequations}
The Coulomb branch Hasse diagram of
\begin{equation}
    \mathsf{Q}_4(N)= \raisebox{-.4\height}{
        %
\;.
\end{equation}
\end{subequations}

\paragraph{5}

\begin{subequations}
The Coulomb branch Hasse diagram of
\begin{equation}
    \mathsf{Q}_5(N)= \raisebox{-.4\height}{
        %
    }\;.
\end{equation}
\end{subequations}

\clearpage

\paragraph{6}

\begin{subequations}
The Coulomb branch Hasse diagram of
\begin{equation}
    \mathsf{Q}_6(N)= \raisebox{-.4\height}{
        %
    }\;.
\end{equation}
\end{subequations}

\clearpage

\paragraph{7}

\begin{subequations}
The Coulomb branch Hasse diagram of
\begin{equation}
    \mathsf{Q}_7(N,\alpha)= \raisebox{-.4\height}{%
    }\;.
\end{equation}
\end{subequations}

\clearpage

\paragraph{8}

\begin{subequations}
The Coulomb branch Hasse diagram of
\begin{equation}
    \mathsf{Q}_8(N,\alpha)= \raisebox{-.4\height}{
        %
    }\;.
\end{equation}
\end{subequations}

\clearpage

\paragraph{9}

\begin{subequations}
The Coulomb branch Hasse diagram of
\begin{equation}
        \mathsf{Q}_9(N,\alpha,\beta)= \raisebox{-.4\height}{
        %
    }\;.
\end{equation}
\end{subequations}

\bibliographystyle{JHEP}

\bibliography{ref}
\end{document}

%% file: objects.tex
\newcommand{\magneticquiver}[1]{%
    \IfEqCase{#1}{%
    {Onkunf}{\raisebox{-.5\height}{       \begin{tikzpicture}
    \node[label=below:{1}][u](1){};
    \node[label=below:{2}][u](2)[right of=1]{};
    \node (dots)[right of=2]{$\cdots$};
    \node[label=below:{$n-1$}][u](2N-5)[right of=dots]{};
    \node[label=below:{$n$}][u](2N-4)[right of=2N-5]{};
    \node[label=above:{2}][u](uf)[above of=2N-4]{};
    \draw(2N-4)--(uf);
    \draw(1)--(2);
    \draw(2)--(dots);
    \draw(dots)--(2N-5);
    \draw[double distance=1.5pt,<-](2N-5)--(2N-4);
    \end{tikzpicture}}}%
{Onk}{\raisebox{-.5\height}{       \begin{tikzpicture}
    \node[label=below:{1}][u](1){};
    \node[label=below:{2}][u](2)[right of=1]{};
    \node (dots)[right of=2]{$\cdots$};
    \node[label=below:{$n-1$}][u](2N-5)[right of=dots]{};
    \node[label=below:{$n$}][u](2N-4)[right of=2N-5]{};
    \node[label=above:{2}][uf](uf)[above of=2N-4]{};
    \draw(2N-4)--(uf);
    \draw(1)--(2);
    \draw(2)--(dots);
    \draw(dots)--(2N-5);
    \draw[double distance=1.5pt,<-](2N-5)--(2N-4);
    \end{tikzpicture}}}%
{1}{\raisebox{-.5\height}{\begin{tikzpicture}  
    \node[label=below:{1}][u](1){};
    \node[label=below:{2}][u](2)[right of=1]{};
    \node(dots)[right of=2]{$\cdots$};
    \node[label=below:{$k-1$}][u](N-1)[right of=dots]{};
    \node[label=below:{$k$}][u](N)[right of=N-1]{};
    \node[label=below:{$k$}][u](N')[right of=N]{};
    \node(dots')[right of=N']{$\cdots$};
    \node[label=below:{$k$}][u](N'')[right of=dots']{};
    \node[label=below:{$k$}][u](N''')[right of=N'']{};
    \node[label=above:{$1$}][uf](f)[above of=N]{};
    \draw(1)--(2);
    \draw(2)--(dots);
    \draw(dots)--(N-1);
    \draw(N-1)--(N);
    \draw(N)--(N');
    \draw(N')--(dots');
    \draw(N'')--(dots');
    \draw[double distance=1.5 pt,<-](N'')--(N''');
    \draw(N)--(f);
    \draw [decorate,decoration={brace,mirror,amplitude=5pt},xshift=0pt,yshift=15pt]
(0,-1.25) -- (8,-1.25)node [black,midway,xshift=0pt,yshift=-15pt] {
$n$};
    \end{tikzpicture}}}%
{1uf}{\raisebox{-.5\height}{\begin{tikzpicture}  
    \node[label=below:{1}][u](1){};
    \node[label=below:{2}][u](2)[right of=1]{};
    \node(dots)[right of=2]{$\cdots$};
    \node[label=below:{$k-1$}][u](N-1)[right of=dots]{};
    \node[label=below:{$k$}][u](N)[right of=N-1]{};
    \node[label=below:{$k$}][u](N')[right of=N]{};
    \node(dots')[right of=N']{$\cdots$};
    \node[label=below:{$k$}][u](N'')[right of=dots']{};
    \node[label=below:{$k$}][u](N''')[right of=N'']{};
    \node[label=above:{$1$}][u](f)[above of=N]{};
    \draw(1)--(2);
    \draw(2)--(dots);
    \draw(dots)--(N-1);
    \draw(N-1)--(N);
    \draw(N)--(N');
    \draw(N')--(dots');
    \draw(N'')--(dots');
    \draw[double distance=1.5 pt,<-](N'')--(N''');
    \draw[double distance=1.5 pt,<-](N)--(f);
    \draw [decorate,decoration={brace,mirror,amplitude=5pt},xshift=0pt,yshift=15pt]
(0,-1.25) -- (8,-1.25)node [black,midway,xshift=0pt,yshift=-15pt] {
$n$};
    \end{tikzpicture}}}%
{2}{\raisebox{-.5\height}{ \begin{tikzpicture} 
    \node[label=below:{1}][u](1){};
    \node[label=below:{2}][u](2)[right of=1]{};
    \node (dots)[right of=2]{$\cdots$};
    \node[label=below:{$k-1$}][u](2N-5)[right of=dots]{};
    \node[label=below:{$k$}][u](2N-4)[right of=2N-5]{};
    \node[label=above:{1}][uf](uf)[above of=2N-4]{};
    \draw[->,double distance=1.5 pt](2N-4)--(uf);
    \draw(1)--(2);
    \draw(2)--(dots);
    \draw(dots)--(2N-5);
    \draw[double distance=1.5pt,<-](2N-5)--(2N-4);
    \end{tikzpicture}}}
{3}{\raisebox{-.5\height}{\begin{tikzpicture}  
    \node[label=below:{1}][u](1){};
    \node[label=below:{2}][u](2)[right of=1]{};
    \node (dots)[right of=2]{$\cdots$};
    \node[label=below:{$n-1$}][u](2N-5)[right of=dots]{};
    \node[label=below:{$n$}][u](2N-4)[right of=2N-5]{};
    \node[label=above:{1}][uf](uf)[above of=2N-4]{};
    \draw[->,double distance=1.5 pt](2N-4)--(uf);
    \draw(1)--(2);
    \draw(2)--(dots);
    \draw(dots)--(2N-5);
    \draw[ double distance=1.5pt,<-](2N-5)--(2N-4);
    \end{tikzpicture}}}
{3uf}{\raisebox{-.5\height}{\begin{tikzpicture}  
    \node[label=below:{1}][u](1){};
    \node[label=below:{2}][u](2)[right of=1]{};
    \node (dots)[right of=2]{$\cdots$};
    \node[label=below:{$n-1$}][u](2N-5)[right of=dots]{};
    \node[label=below:{$n$}][u](2N-4)[right of=2N-5]{};
    \node[label=above:{1}][u](uf)[above of=2N-4]{};
    \draw[<->,double distance=1.5 pt](2N-4)--(uf);
    \draw(1)--(2);
    \draw(2)--(dots);
    \draw(dots)--(2N-5);
    \draw[ double distance=1.5pt,<-](2N-5)--(2N-4);
    \end{tikzpicture}}}
{4}{\raisebox{-.5\height}{\begin{tikzpicture} 
    \node[label=below:{1}][u](1){};
    \node[label=below:{2}][u](2)[right of=1]{};
    \node(dots)[right of=2]{$\cdots$};
    \node[label=below:{$k-1$}][u](N-1)[right of=dots]{};
    \node[label=below:{$k$}][u](N)[right of=N-1]{};
    \node[label=below:{$k$}][u](N')[right of=N]{};
    \node(dots')[right of=N']{$\cdots$};
    \node[label=below:{$k$}][u](N'')[right of=dots']{};
    \node[label=below:{$k$}][u](N''')[right of=N'']{};
    \node[label=above:{$1$}][u](f)[above of=N]{};
    \draw(1)--(2);
    \draw(2)--(dots);
    \draw(dots)--(N-1);
    \draw(N-1)--(N);
    \draw(N)--(N');
    \draw(N')--(dots');
    \draw(N'')--(dots');
    \draw[double distance=1.5 pt,<-](N'')--(N''');
    \draw(N)--(f);
    \draw [decorate,decoration={brace,mirror,amplitude=5pt},xshift=0pt,yshift=15pt]
(0,-1.25) -- (8,-1.25)node [black,midway,xshift=0pt,yshift=-15pt] {
$n$};
    \end{tikzpicture}}}
{5}{\raisebox{-.5\height}{\begin{tikzpicture} 
    \node[label=below:{$1$}][u](N)[right of=N-1]{};
    \node[label=below:{$2$}][u](N')[right of=N]{};
    \node(dots')[right of=N']{$\cdots$};
    \node[label=below:{$n-1$}][u](N'')[right of=dots']{};
    \node[label=below:{$n$}][u](N''')[right of=N'']{};
    \node[label=above:{$1$}][u](f)[above of=N''']{};
    \draw(N)--(N');
    \draw(N')--(dots');
    \draw(N'')--(dots');
    \draw[double distance=1.5 pt,<-](N'')--(N''');
    \draw[double distance=1.5 pt,<-](f)--(N''');
    \end{tikzpicture}}}
{6}{\raisebox{-.5\height}{    \begin{tikzpicture} 
    \node[label=below:{1}][u](2){};
    \node(dots)[right of=2]{$\cdots$};
    \node[label=below:{$k-1$}][u](N-1)[right of=dots]{};
    \node[label=below:{$k$}][u](N)[right of=N-1]{};
    \node[label=below:{$k$}][u](N')[right of=N]{};
    \node(dots')[right of=N']{$\cdots$};
    \node[label=below:{$k$}][u](N'')[right of=dots']{};
    \node[label=below:{$k$}][u](N''')[right of=N'']{};
    \node[label=above:{$1$}][uf](f)[above of=N]{};
    \node[label=above:{$1$}][uf](f')[above of=N''']{};
    \draw(2)--(dots);
    \draw(dots)--(N-1);
    \draw(N-1)--(N);
    \draw(N)--(N');
    \draw(N')--(dots');
    \draw(N'')--(dots');
    \draw[double distance=1.5 pt,<-](N'')--(N''');
    \draw(N)--(f);
    \draw(N''')--(f');
    \draw [decorate,decoration={brace,mirror,amplitude=5pt},xshift=0pt,yshift=15pt]
(0,-1.25) -- (8,-1.25)node [black,midway,xshift=0pt,yshift=-15pt] {
$n$};
    \end{tikzpicture}}}
{6uf}{\raisebox{-.5\height}{    \begin{tikzpicture} 
    \node[label=below:{1}][u](2){};
    \node(dots)[right of=2]{$\cdots$};
    \node[label=below:{$k-1$}][u](N-1)[right of=dots]{};
    \node[label=below:{$k$}][u](N)[right of=N-1]{};
    \node[label=below:{$k$}][u](N')[right of=N]{};
    \node(dots')[right of=N']{$\cdots$};
    \node[label=below:{$k$}][u](N'')[right of=dots']{};
    \node[label=below:{$k$}][u](N''')[right of=N'']{};
    \node[label=above:{$1$}][u](f)[above of=N'']{};
    \draw(2)--(dots);
    \draw(dots)--(N-1);
    \draw(N-1)--(N);
    \draw(N)--(N');
    \draw(N')--(dots');
    \draw(N'')--(dots');
    \draw[double distance=1.5 pt,<-](N'')--(N''');
    \draw[double distance=1.5 pt,<-](N)--(f);
    \draw(N''')--(f);
    \draw [decorate,decoration={brace,mirror,amplitude=5pt},xshift=0pt,yshift=15pt]
(0,-1.25) -- (8,-1.25)node [black,midway,xshift=0pt,yshift=-15pt] {
$n$};
    \end{tikzpicture}}}
{7}{\raisebox{-.5\height}{        \begin{tikzpicture} 
    \node[label=below:{1}][u](1){};
    \node[label=below:{2}][u](2)[right of=1]{};
    \node(dots)[right of=2]{$\cdots$};
    \node[label=below:{$k-1$}][u](N-1)[right of=dots]{};
    \node[label=below:{$k$}][u](N)[right of=N-1]{};
    \node[label=below:{$k$}][u](N')[right of=N]{};
    \node(dots')[right of=N']{$\cdots$};
    \node[label=below:{$k$}][u](N'')[right of=dots']{};
    \node[label=below:{$k$}][u](N''')[right of=N'']{};
    \node[label=above:{$1$}][u](f)[above of=N]{};
    \node[label=above:{$1$}][u](f')[above of=N''']{};
    \draw(1)--(2);
    \draw(2)--(dots);
    \draw(dots)--(N-1);
    \draw(N-1)--(N);
    \draw(N)--(N');
    \draw(N')--(dots');
    \draw(N'')--(dots');
    \draw[double distance=1.5 pt,<-](N'')--(N''');
    \draw(N)--(f);
    \draw(N''')--(f');
    \draw [decorate,decoration={brace,mirror,amplitude=5pt},xshift=0pt,yshift=15pt]
(0,-1.25) -- (8,-1.25)node [black,midway,xshift=0pt,yshift=-15pt] {
$n$};
    \end{tikzpicture}}}
{8}{\raisebox{-.5\height}{\begin{tikzpicture}  
                          \node[label=below:{$2k$}][u](2k){};
                          \node[label=below:{$k$}][u](k)[below left of=2k]{};
                          \node[label=left:{$k$}][u](kk)[above left of=2k]{};
                          \node[label=above:{$1$}][uf](kf)[above of=kk]{};
                          \node[label=below:{$2k$}][u](2k')[right of=2k]{};
                          \node(dots)[right of=2k']{$\cdots$};
                          \node[label=below:{$2k$}][u](2k'')[right of=dots]{};
                          \node[label=below:{$2k$}][u](2k''')[right of=2k'']{};
                          \node[label=above:{$1$}][uf](2kf)[above of=2k''']{};
                          \draw(k)--(2k);
                          \draw(kk)--(2k);
                          \draw(kk)--(kf);
                          \draw(2k)--(2k');
                          \draw(dots)--(2k');
                          \draw(2k'')--(dots);
                          \draw[double distance=1.5 pt,<-](2k'')--(2k''');
                          \draw(2k''')--(2kf);
                          \draw [decorate,decoration={brace,mirror,amplitude=5pt},xshift=0pt,yshift=15pt]
(0,-1.25) -- (4,-1.25)node [black,midway,xshift=0pt,yshift=-15pt] {
$n-1$};
             \end{tikzpicture}}}
{8uf}{\raisebox{-.5\height}{\begin{tikzpicture}  
                          \node[label=below:{$2k$}][u](2k){};
                          \node[label=below:{$k$}][u](k)[below left of=2k]{};
                          \node[label=left:{$k$}][u](kk)[above left of=2k]{};
                          \node[label=below:{$2k$}][u](2k')[right of=2k]{};
                          \node(dots)[right of=2k']{$\cdots$};
                          \node[label=below:{$2k$}][u](2k'')[right of=dots]{};
                          \node[label=below:{$2k$}][u](2k''')[right of=2k'']{};
                          \node[label=above:{$1$}][u](2kf)[above of=2k'']{};
                          \draw(k)--(2k);
                          \draw(kk)--(2k);
                          \draw[double distance=1.5 pt,<-](kk)--(2kf);
                          \draw(2k)--(2k');
                          \draw(dots)--(2k');
                          \draw(2k'')--(dots);
                          \draw[double distance=1.5 pt,<-](2k'')--(2k''');
                          \draw(2k''')--(2kf);
                          \draw [decorate,decoration={brace,mirror,amplitude=5pt},xshift=0pt,yshift=15pt]
(0,-1.25) -- (4,-1.25)node [black,midway,xshift=0pt,yshift=-15pt] {
$n-1$};
             \end{tikzpicture}}}
{9}{\raisebox{-.5\height}{  \begin{tikzpicture}  
                      \node[label=below:{$k$}][u](k){};
                      \node[label=below:{$2k$}][u](2k)[right of=k]{};
                      \node[label=below:{$k$}][u](kk)[right of=2k]{};
                      \node[label=above:{$1$}][uf](kf)[above of=k]{};
                      \node[label=above:{$1$}][uf](2kf)[above of=2k]{};
                      \draw[double distance=1.5 pt,<-](k)--(2k);
                      \draw[double distance=1.5 pt,<-](kk)--(2k);
                      \draw(k)--(kf);
                      \draw(2k)--(2kf);
             \end{tikzpicture}}}
{9uf}{\raisebox{-.5\height}{  \begin{tikzpicture}  
                      \node[label=below:{$k$}][u](k){};
                      \node[label=below:{$2k$}][u](2k)[right of=k]{};
                      \node[label=below:{$k$}][u](kk)[right of=2k]{};
                      \node[label=above:{$1$}][u](2kf)[above of=2k]{};
                      \draw[double distance=1.5 pt,<-](k)--(2k);
                      \draw[double distance=1.5 pt,<-](kk)--(2k);
                      \draw[double distance=1.5 pt,<-](k)--(2kf);
                      \draw(2k)--(2kf);
             \end{tikzpicture}}}
{10}{\raisebox{-.5\height}{     \begin{tikzpicture}  
                          \node[label=below:{$2k$}][u](2k){};
                          \node[label=below:{$k$}][u](k)[below left of=2k]{};
                          \node[label=left:{$k$}][u](kk)[above left of=2k]{};
                          \node[label=above:{$1$}][u](kf)[above of=kk]{};
                          \node[label=below:{$2k$}][u](2k')[right of=2k]{};
                          \node(dots)[right of=2k']{$\cdots$};
                          \node[label=below:{$2k$}][u](2k'')[right of=dots]{};
                          \node[label=below:{$2k$}][u](2k''')[right of=2k'']{};
                          \node[label=above:{$1$}][u](2kf)[above of=2k''']{};
                          \draw(k)--(2k);
                          \draw(kk)--(2k);
                          \draw(kk)--(kf);
                          \draw(2k)--(2k');
                          \draw(dots)--(2k');
                          \draw(2k'')--(dots);
                          \draw[double distance=1.5 pt,<-](2k'')--(2k''');
                          \draw(2k''')--(2kf);
                          \draw [decorate,decoration={brace,mirror,amplitude=5pt},xshift=0pt,yshift=15pt]
(0,-1.25) -- (4,-1.25)node [black,midway,xshift=0pt,yshift=-15pt] {
$n-1$};
             \end{tikzpicture}}}
{11}{\raisebox{-.5\height}{\begin{tikzpicture} 
                      \node[label=below:{$k$}][u](k){};
                      \node[label=below:{$2k$}][u](2k)[right of=k]{};
                      \node[label=below:{$k$}][u](kk)[right of=2k]{};
                      \node[label=above:{$1$}][u](kf)[above of=k]{};
                      \node[label=above:{$1$}][u](2kf)[above of=2k]{};
                      \draw[double distance=1.5 pt,<-](k)--(2k);
                      \draw[double distance=1.5 pt,<-](kk)--(2k);
                      \draw(k)--(kf);
                      \draw(2k)--(2kf);
             \end{tikzpicture}}}
{12}{\raisebox{-.5\height}{\begin{tikzpicture} 
                          \node[label=below:{$k$}][u](2k){};
                          \node[label=above:{$1$}][uf](kf)[above of=2k]{};
                          \node[label=below:{$k$}][u](2k')[right of=2k]{};
                          \node(dots)[right of=2k']{$\cdots$};
                          \node[label=below:{$k$}][u](2k'')[right of=dots]{};
                          \node[label=below:{$k$}][u](2k''')[right of=2k'']{};
                          \node[label=above:{$1$}][uf](2kf)[above of=2k''']{};
                          \draw(2k)--(kf);
                          \draw[double distance=1.5pt,->](2k)--(2k');
                          \draw(dots)--(2k');
                          \draw(2k'')--(dots);
                          \draw[double distance=1.5 pt,<-](2k'')--(2k''');
                          \draw(2k''')--(2kf);
                           \draw [decorate,decoration={brace,mirror,amplitude=5pt},xshift=0pt,yshift=15pt]
(0,-1.25) -- (4,-1.25)node [black,midway,xshift=0pt,yshift=-15pt] {
$n+1$};
             \end{tikzpicture}}}
{12uf}{\raisebox{-.5\height}{\begin{tikzpicture} 
                          \node[label=below:{$k$}][u](2k){};
                          \node[label=below:{$k$}][u](2k')[right of=2k]{};
                          \node(dots)[right of=2k']{$\cdots$};
                          \node[label=below:{$k$}][u](2k'')[right of=dots]{};
                          \node[label=below:{$k$}][u](2k''')[right of=2k'']{};
                          \node[label=above:{$1$}][u](2kf)[above of=dots]{};
                          \draw(2k)--(2kf);
                          \draw[double distance=1.5pt,->](2k)--(2k');
                          \draw(dots)--(2k');
                          \draw(2k'')--(dots);
                          \draw[double distance=1.5 pt,<-](2k'')--(2k''');
                          \draw(2k''')--(2kf);
                           \draw [decorate,decoration={brace,mirror,amplitude=5pt},xshift=0pt,yshift=15pt]
(0,-1.25) -- (4,-1.25)node [black,midway,xshift=0pt,yshift=-15pt] {
$n+1$};
             \end{tikzpicture}}}
{13}{\raisebox{-.5\height}{ \begin{tikzpicture}  
                          \node[label=below:{$k$}][u](k){};
                          \node[label=below:{$k$}][u](kk)[right of=k]{};
                          \node[label=above:{$1$}][uf](f)[above of=k]{};
                          \node[label=above:{$1$}][uf](ff)[above of=kk]{};
                          \draw(k)--(f);
                          \draw(kk)--(ff);
                          \draw[double distance=2pt,<->](k)--(kk);
             \end{tikzpicture}}}
{13uf}{\raisebox{-.5\height}{ \begin{tikzpicture}  
                          \node[label=below:{$k$}][u] at (0,0) (k){};
                          \node[label=below:{$k$}][u](kk)[right of=k]{};
                          \node[label=above:{$1$}][u](f) at (0.5,1) {};
                          \draw(k)--(f);
                          \draw(kk)--(f);
                          \draw[double distance=2pt,<->](k)--(kk);
             \end{tikzpicture}}}
{14}{\raisebox{-.5\height}{  \begin{tikzpicture}  
                          \node[label=below:{$k$}][u](2k){};
                          \node[label=above:{$1$}][u](kf)[above of=2k]{};
                          \node[label=below:{$k$}][u](2k')[right of=2k]{};
                          \node(dots)[right of=2k']{$\cdots$};
                          \node[label=below:{$k$}][u](2k'')[right of=dots]{};
                          \node[label=below:{$k$}][u](2k''')[right of=2k'']{};
                          \node[label=above:{$1$}][u](2kf)[above of=2k''']{};
                          \draw(2k)--(kf);
                          \draw[double distance=1.5pt,->](2k)--(2k');
                          \draw(dots)--(2k');
                          \draw(2k'')--(dots);
                          \draw[double distance=1.5 pt,<-](2k'')--(2k''');
                          \draw(2k''')--(2kf);
                          \draw [decorate,decoration={brace,mirror,amplitude=5pt},xshift=0pt,yshift=15pt]
(0,-1.25) -- (4,-1.25)node [black,midway,xshift=0pt,yshift=-15pt] {
$n+1$};
             \end{tikzpicture}}}
{15}{\raisebox{-.5\height}{\begin{tikzpicture} 
                          \node[label=below:{$k$}][u](k){};
                          \node[label=below:{$k$}][u](kk)[right of=k]{};
                          \node[label=above:{$1$}][u](f)[above of=k]{};
                          \node[label=above:{$1$}][u](ff)[above of=kk]{};
                          \draw(k)--(f);
                          \draw(kk)--(ff);
                          \draw[double distance=2pt,<->](k)--(kk);
             \end{tikzpicture}}}
{16}{\raisebox{-.5\height}{  \begin{tikzpicture} 
    \node[label=below:{$1$}][u](2){};
    \node (dots)[right of=2]{$\cdots$};
    \node[label=below:{$2k-2$}][u](2N-5)[right of=dots]{};
    \node[label=below:{$2k-1$}][u](2N-4)[right of=2N-5]{};
    \node[label=below:{$2$}][u](2')[right of=2N-4]{};
    \draw (2')to[out=-45,in=45,loop,looseness=10](2');
    \draw(2)--(dots);
    \draw(dots)--(2N-5);
    \draw[ double distance=1.5pt,<-](2N-5)--(2N-4);
    \draw(2N-4)--(2');
    \end{tikzpicture} }}
{17}{\raisebox{-.5\height}{     \begin{tikzpicture} 
    \node[label=below:{$1$}][u](2){};
    \node (dots)[right of=2]{$\cdots$};
    \node[label=below:{$2k-3$}][u](2N-5)[right of=dots]{};
    \node[label=below:{$2k-2$}][u](2N-4)[right of=2N-5]{};
    \node[label=below:{$2$}][u](2')[right of=2N-4]{};
    \draw (2')to[out=-45,in=45,loop,looseness=10](2');
    \draw(2)--(dots);
    \draw(dots)--(2N-5);
    \draw[ double distance=1.5pt,<-](2N-5)--(2N-4);
    \draw(2N-4)--(2');
    \end{tikzpicture}  }}
{18}{\raisebox{-.5\height}{     \begin{tikzpicture} 
    \node[label=below:{1}][u](2){};
    \node (dots)[right of=2]{$\cdots$};
    \node[label=below:{$2k-4$}][u](2N-5)[right of=dots]{};
    \node[label=below:{$2k-3$}][u](2N-4)[right of=2N-5]{};
    \node[label=below:{$2$}][u](2')[right of=2N-4]{};
    \node[label=below:{$1$}][u](1')[right of=2']{};
    \draw (2')to[out=45,in=135,loop,looseness=10](2');
    \draw(2)--(dots);
    \draw(dots)--(2N-5);
    \draw[ double distance=1.5pt,<-](2N-5)--(2N-4);
    \draw(2N-4)--(2');
    \draw[double distance=1.5pt,->](2')--(1');
    \end{tikzpicture}   }}
{19}{\raisebox{-.5\height}{   \begin{tikzpicture} 
    \node[label=below:{1}][u](2){};
    \node (dots)[right of=2]{$\cdots$};
    \node[label=below:{$2k-5$}][u](2N-5)[right of=dots]{};
    \node[label=below:{$2k-4$}][u](2N-4)[right of=2N-5]{};
    \node[label=below:{$2$}][u](2')[right of=2N-4]{};
    \node[label=below:{$1$}][u](1')[right of=2']{};
    \draw (2')to[out=45,in=135,loop,looseness=10](2');
    \draw(2)--(dots);
    \draw(dots)--(2N-5);
    \draw[ double distance=1.5pt,<-](2N-5)--(2N-4);
    \draw(2N-4)--(2');
    \draw[double distance=1.5pt,->](2')--(1');
    \end{tikzpicture}  }}
{20}{\raisebox{-.5\height}{  \begin{tikzpicture} 
    \node[label=below:{1}][u](2){};
    \node (dots)[right of=2]{$\cdots$};
    \node[label=above:{$2k-2j-5$}][u](2N-5)[right of=dots]{};
    \node[label=below:{$2k-2j-4$}][u](2N-4)[right of=2N-5]{};
    \node[label=below:{$1$}][u](1')[right of=2N-4]{};
    \node[label=above:{1}][u](uf)[above of=2N-4]{};
    \draw(1')--node[above right]{$3+2j$}(uf);

    \draw(2)--(dots);
    \draw(dots)--(2N-5);
    \draw[ double distance=1.5pt,<-](2N-5)--(2N-4);
    \draw(2N-4)--(1');
    \draw(2N-4)--(uf);
    \end{tikzpicture}   }}
{21}{\raisebox{-.5\height}{ \begin{tikzpicture} 
    \node[label=below:{1}][u](2){};
    \node (dots)[right of=2]{$\cdots$};
    \node[label=above:{$2k-2j-6$}][u](2N-5)[right of=dots]{};
    \node[label=below:{$2k-2j-5$}][u](2N-4)[right of=2N-5]{};
    \node[label=below:{$1$}][u](1')[right of=2N-4]{};
    \node[label=above:{1}][u](uf)[above of=2N-4]{};
    \path [draw](1')--node[above right]{$3+2j$}(uf);
    \draw(2)--(dots);
    \draw(dots)--(2N-5);
    \draw[ double distance=1.5pt,<-](2N-5)--(2N-4);
    \draw(2N-4)--(1');
    \draw(2N-4)--(uf);
    \end{tikzpicture}   }}
{22}{\raisebox{-.5\height}{  \begin{tikzpicture} 
    \node[label=below:{1}][u](2){};
    \node (dots)[right of=2]{$\cdots$};
    \node[label=above:{$2k-2j-6$}][u](2N-5)[right of=dots]{};
    \node[label=below:{$2k-2j-5$}][u](2N-4)[right of=2N-5]{};
    \node[label=below:{$1$}][u](1')[right of=2N-4]{};
    \node[label=above:{1}][u](uf)[above of=2N-4]{};
    \path [draw](1')--node[above right]{$4+2j$}(uf);
    \draw(2)--(dots);
    \draw(dots)--(2N-5);
    \draw[ double distance=1.5pt,<-](2N-5)--(2N-4);
    \draw(2N-4)--(1');
    \draw(2N-4)--(uf);
    \end{tikzpicture}   }}
{23}{\raisebox{-.5\height}{  \begin{tikzpicture} 
    \node[label=below:{1}][u](2){};
    \node (dots)[right of=2]{$\cdots$};
    \node[label=above:{$2k-2j-7$}][u](2N-5)[right of=dots]{};
    \node[label=below:{$2k-2j-6$}][u](2N-4)[right of=2N-5]{};
    \node[label=below:{$1$}][u](1')[right of=2N-4]{};
    \node[label=above:{1}][u](uf)[above of=2N-4]{};
    \path [draw](1')--node[above right]{$4+2j$}(uf);
    \draw(2)--(dots);
    \draw(dots)--(2N-5);
    \draw[ double distance=1.5pt,<-](2N-5)--(2N-4);
    \draw(2N-4)--(1');
    \draw(2N-4)--(uf);
    \end{tikzpicture}   }}
{24}{\raisebox{-.5\height}{  \begin{tikzpicture}   
\node[label=below:{1}][u](1) at (0,0) {};
    \node[label=below:{2}][u](2)[right of=1]{};
    \node (dots)[right of=2]{$\cdots$};
    \node[label=below:{$n-1$}][u](2N-5)[right of=dots]{};
    \node[label=below:{$n$}][u](2N-4)[right of=2N-5]{};
    \node[label=left:{1}][uf](uf) at (3.5,1) {};
    \node[label=right:{1}][uf](uff) at (4.5,1) {};
    \draw(2N-4)--(uff);
    \draw[ double distance=1.5pt,->](2N-4)--(uf);
    \draw(1)--(2);
    \draw(2)--(dots);
    \draw(dots)--(2N-5);
    \draw[double distance=1.5pt,<-](2N-5)--(2N-4);
    \end{tikzpicture}  }}
{24uf}{\raisebox{-.5\height}{  \begin{tikzpicture}   
\node[label=below:{1}][u](1) at (0,0) {};
    \node[label=below:{2}][u](2)[right of=1]{};
    \node (dots)[right of=2]{$\cdots$};
    \node[label=below:{$n-1$}][u](2N-5)[right of=dots]{};
    \node[label=below:{$n$}][u](2N-4)[right of=2N-5]{};
    \node[label=above:{1}][u](uf)[above of=2N-4]{};
    \draw(2N-4) .. controls (4.5,0.5) .. (uf);
    \draw[ double distance=1.5pt,<->](2N-4) .. controls (3.5,0.5) .. (uf);
    \draw(1)--(2);
    \draw(2)--(dots);
    \draw(dots)--(2N-5);
    \draw[double distance=1.5pt,<-](2N-5)--(2N-4);
    \end{tikzpicture}  }}
{25}{\raisebox{-.5\height}{ \begin{tikzpicture}    
    \node[label=below:{1}][u](1){};
    \node[label=below:{2}][u](2)[right of=1]{};
    \node (dots)[right of=2]{$\cdots$};
    \node[label=below:{$n-1$}][u](2N-5)[right of=dots]{};
    \node[label=below:{$n$}][u](2N-4)[right of=2N-5]{};
    \node[label=above:{1}][u](uf)[above of=2N-4]{};
    \node[label=below:{1}][u](uff)[right of=2N-4]{};
    \draw(2N-4)--(uff);
    \draw[ double distance=1.5pt,->](2N-4)--(uf);
    \draw(1)--(2);
    \draw(2)--(dots);
    \draw(dots)--(2N-5);
    \draw[double distance=1.5pt,<-](2N-5)--(2N-4);
    \end{tikzpicture}   }}
 {26}{\raisebox{-.5\height}{     \begin{tikzpicture}  
                          \node[label=below:{$k$}][u](1) at (0,0) {};
                          \node[label=below:{$k$}][u](11) at (1,0) {};
                          \node[label=above:{1}][uf](f) at (0.3,1) {};
                          \node[label=above:{1}][uf](fff) at (-0.3,1) {};
                          \node[label=above:{1}][uf](ff)[above of=11]{};
                          \draw[<->,double distance=1.5pt](1)--(11);
                          \draw(1)--(f);
                          \draw[->,double distance=1.5pt](1)--(fff);
                          \draw(11)--(ff);
             \end{tikzpicture}  }}
 {26uf}{\raisebox{-.5\height}{     \begin{tikzpicture}  
                          \node[label=below:{$k$}][u](1) at (0,0) {};
                          \node[label=below:{$k$}][u](11) at (1,0) {};
                          \node[label=above:{1}][u](f) at (0.5,1) {};
                          \draw[<->,double distance=1.5pt](1)--(11);
                          \draw(1) .. controls (0.5,0.5) .. (f);
                          \draw[<->,double distance=1.5pt](1) .. controls (0,0.5) .. (f);
                          \draw(11)--(f);
             \end{tikzpicture}  }}
 {27}{\raisebox{-.5\height}{  \begin{tikzpicture} 
             \node[label=above:{1}][u](ff){};
                          \node[label=below:{$k$}][u](1)[below left of=ff]{};
                          \node[label=below:{$k$}][u](11)[below right of=ff]{};
                          \node[label=above:{1}][u](f)[above left of=1]{};
                          \draw[<->,double distance=1.5pt](1)--(11);
                          \draw[->,double distance=1.5pt](1)--(f);
                          \draw(1)--(ff);
                          \draw(11)--(ff);
             \end{tikzpicture} }}
{28}{\raisebox{-.5\height}{      \begin{tikzpicture} 
         \node[label=left:{$2k+1$}][u](2k){};
         \node[label=below:{$k+1$}][u](k)[below left of=2k]{};
         \node[label=above:{$k$}][u](kk)[above left of=2k]{};
         \node[label=left:{$1$}][uf](kf)[left of=k]{};
         \node[label=below:{$2k+1$}][u](2k')[right of=2k]{};
         \node(dots)[right of=2k']{$\cdots$};
          \node[label=below:{$2k+1$}][u](2k'')[right of=dots]{};
          \node[label=below:{$2k+1$}][u](2k''')[right of=2k'']{};
          \node[label=above:{$1$}][uf](2kf)[above of=2k''']{};
              \draw(k)--(2k);
              \draw(kk)--(2k);
              \draw(k)--(kf);
                 \draw(2k)--(2k');
               \draw(dots)--(2k');
             \draw(2k'')--(dots);
           \draw[double distance=1.5 pt,<-](2k'')--(2k''');
          \draw(2k''')--(2kf);
          \draw [decorate,decoration={brace,mirror,amplitude=5pt},xshift=0pt,yshift=15pt]
(0,-1.25) -- (4,-1.25)node [black,midway,xshift=0pt,yshift=-15pt] {
$n-1$};
             \end{tikzpicture} }}
{29}{\raisebox{-.5\height}{      \begin{tikzpicture} 
   \node[label=below:{\scriptsize$2k+1$}][u](2k){};
     \node[label=below:{$k+1$}][u](k)[below left of=2k]{};
  \node[label=above:{$k$}][u](kk)[above left of=2k]{};
    \node[label=left:{$1$}][u](kf)[left of=k]{};
      \node[label=below:{$2k+1$}][u](2k')[right of=2k]{};
    \node(dots)[right of=2k']{$\cdots$};
    \node[label=below:{$2k+1$}][u](2k'')[right of=dots]{};
    \node[label=below:{$2k+1$}][u](2k''')[right of=2k'']{};
    \node[label=above:{$1$}][u](2kf)[above of=2k''']{};
            \draw(k)--(2k);
          \draw(kk)--(2k);
              \draw(k)--(kf);
          \draw(2k)--(2k');
         \draw(dots)--(2k');
         \draw(2k'')--(dots);
         \draw[double distance=1.5 pt,<-](2k'')--(2k''');
          \draw(2k''')--(2kf);
     \draw [decorate,decoration={brace,mirror,amplitude=5pt},xshift=0pt,yshift=15pt]
(0,-1.25) -- (4,-1.25)node [black,midway,xshift=0pt,yshift=-15pt] {
$n-1$};
             \end{tikzpicture} }}
}}

\newcommand{\electricquiver}[1]{%
    \IfEqCase{#1}{%
{1}{\raisebox{-.5\height}{ \begin{tikzpicture} 
             \node(gauge){SO$(2k+1)$};
             \node(flavour)[above of=gauge]{$[2k-2]$};
             \draw (gauge)--(flavour);
         \end{tikzpicture}   }}
{2}{\raisebox{-.5\height}{ \begin{tikzpicture} 
             \node(gauge){SO$(2k)$};
             \node(flavour)[above of=gauge]{$[2k-3]$};
             \draw (gauge)--(flavour);
         \end{tikzpicture}   }}
{3}{\raisebox{-.5\height}{ \begin{tikzpicture} 
             \node(gauge){SO$(2k+1)$};
             \node(flavour)[above of=gauge]{$[2k-3]$};
             \draw (gauge)--(flavour);
         \end{tikzpicture}   }}
{4}{\raisebox{-.5\height}{   \begin{tikzpicture} 
             \node(gauge){SO$(2k)$};
             \node(flavour)[above of=gauge]{$[2k-4]$};
             \draw (gauge)--(flavour);
         \end{tikzpicture}   }}
{5}{\raisebox{-.5\height}{  \begin{tikzpicture} 
             \node(gauge){SO$(2k)$};
             \node(flavour)[above of=gauge]{$[2k-2j-5]$};
             \draw (gauge)--(flavour);
         \end{tikzpicture}}}
{6}{\raisebox{-.5\height}{    \begin{tikzpicture} 
             \node(gauge){SO$(2k)$};
             \node(flavour)[above of=gauge]{$[2k-2j-6]$};
             \draw (gauge)--(flavour);
         \end{tikzpicture}     }}
{7}{\raisebox{-.5\height}{   \begin{tikzpicture} 
             \node(gauge){SO$(2k+1)$};
             \node(flavour)[above of=gauge]{$[2k-2j-5]$};
             \draw (gauge)--(flavour);
         \end{tikzpicture}   }}
{8}{\raisebox{-.5\height}{   \begin{tikzpicture} 
             \node(gauge){SO$(2k+1)$};
             \node(flavour)[above of=gauge]{$[2k-2j-4]$};
             \draw (gauge)--(flavour);
         \end{tikzpicture}  }}
{9}{\raisebox{-.5\height}{   \begin{tikzpicture} 
                      \node at (0,0)(gauge){O$(k)$};
                      \node (flavour)[above of=gauge]{$[n]$};
                      \draw(gauge)--(flavour);
         \end{tikzpicture}  }}
{10}{\raisebox{-.5\height}{   \begin{tikzpicture} 
                      \node at (0,0)(gauge){SO$(k)$};
                      \node (flavour)[above of=gauge]{$[n]$};
                      \draw(gauge)--(flavour);
         \end{tikzpicture}  }}
{11}{\raisebox{-.5\height}{ \begin{tikzpicture} 
                      \node at (0,0)(gauge){U$(k)$};
                      \node (flavour)[above of=gauge]{$[n]$};
                      \node[label=below:{$S^2$}](empty)[right of=gauge]{};
                      \node(symmetric)[right of=empty]{$\left[1\right]$};
                      \path[draw,snake it](gauge)--(symmetric);
                      \draw(gauge)--(flavour);
         \end{tikzpicture}   }}
{12}{\raisebox{-.5\height}{ \begin{tikzpicture} 
                      \node at (0,0)(gauge){SU$(k)$};
                      \node (flavour)[above of=gauge]{$[n]$};
                      \node[label=below:{$S^2$}](empty)[right of=gauge]{};
                      \node(symmetric)[right of=empty]{$\left[1\right]$};
                      \path[draw,snake it](gauge)--(symmetric);
                      \draw(gauge)--(flavour);
         \end{tikzpicture}   }}
{13}{\raisebox{-.5\height}{ \begin{tikzpicture} 
                      \node at (0,0)(gauge){U$(2k)$};
                      \node (flavour)[above of=gauge]{$[n]$};
                      \node[label=below:{$S^2$}](empty)[right of=gauge]{};
                      \node[label=below:{$\wedge^2$}](empty2)[left of=gauge]{};
                      \node(symmetric)[right of=empty]{$\left[1\right]$};
                      \node(antisymmetric)[left of=empty2]{$\left[1\right]$};
                      \path[draw,snake it](gauge)--(symmetric);
                      \path[draw,snake it](gauge)--(antisymmetric);
                      \draw(gauge)--(flavour);
         \end{tikzpicture}  }}
{13b}{\raisebox{-.5\height}{ \begin{tikzpicture} 
                      \node at (0,0)(gauge){U$(2k+1)$};
                      \node (flavour)[above of=gauge]{$[n]$};
                      \node[label=below:{$S^2$}](empty)[right of=gauge]{};
                      \node[label=below:{$\wedge^2$}](empty2)[left of=gauge]{};
                      \node(symmetric)[right of=empty]{$\left[1\right]$};
                      \node(antisymmetric)[left of=empty2]{$\left[1\right]$};
                      \path[draw,snake it](gauge)--(symmetric);
                      \path[draw,snake it](gauge)--(antisymmetric);
                      \draw(gauge)--(flavour);
         \end{tikzpicture}  }}
{14}{\raisebox{-.5\height}{ \begin{tikzpicture} 
                      \node at (0,0)(gauge){SU$(2k)$};
                      \node (flavour)[above of=gauge]{$[n]$};
                      \node[label=below:{$S^2$}](empty)[right of=gauge]{};
                      \node[label=below:{$\wedge^2$}](empty2)[left of=gauge]{};
                      \node(symmetric)[right of=empty]{$\left[1\right]$};
                      \node(antisymmetric)[left of=empty2]{$\left[1\right]$};
                      \path[draw,snake it](gauge)--(symmetric);
                      \path[draw,snake it](gauge)--(antisymmetric);
                      \draw(gauge)--(flavour);
         \end{tikzpicture}  }}
{14b}{\raisebox{-.5\height}{ \begin{tikzpicture} 
                      \node at (0,0)(gauge){SU$(2k+1)$};
                      \node (flavour)[above of=gauge]{$[n]$};
                      \node[label=below:{$S^2$}](empty)[right of=gauge]{};
                      \node[label=below:{$\wedge^2$}](empty2)[left of=gauge]{};
                      \node(symmetric)[right of=empty]{$\left[1\right]$};
                      \node(antisymmetric)[left of=empty2]{$\left[1\right]$};
                      \path[draw,snake it](gauge)--(symmetric);
                      \path[draw,snake it](gauge)--(antisymmetric);
                      \draw(gauge)--(flavour);
         \end{tikzpicture}  }}
{15}{\raisebox{-.5\height}{   \begin{tikzpicture} 
                      \node at (0,0)(gauge){U$(k)$};
                      \node (flavour)[above of=gauge]{$[n]$};
                      \node[label=below:{$S^2$}](empty)[right of=gauge]{};
                      \node(symmetric)[right of=empty]{$\left[2\right]$};
                      \path[draw,snake it](gauge)--(symmetric);
                      \draw(gauge)--(flavour);
         \end{tikzpicture} }}
{16}{\raisebox{-.5\height}{   \begin{tikzpicture} 
                      \node at (0,0)(gauge){SU$(k)$};
                      \node (flavour)[above of=gauge]{$[n]$};
                      \node[label=below:{$S^2$}](empty)[right of=gauge]{};
                      \node(symmetric)[right of=empty]{$\left[2\right]$};
                      \path[draw,snake it](gauge)--(symmetric);
                      \draw(gauge)--(flavour);
         \end{tikzpicture} }}
{17}{\raisebox{-.5\height}{   \begin{tikzpicture} 
                      \node at (0,0)(gauge){SU$(K)_{1/2}$};
                      \node (flavour)[above of=gauge]{$[K-3]$};
                      \node[label=below:{$S^2$}](empty)[right of=gauge]{};
                      \node(symmetric)[right of=empty]{$\left[1\right]$};
                      \path[draw,snake it](gauge)--(symmetric);
                      \draw(gauge)--(flavour);
         \end{tikzpicture} }}
{18}{\raisebox{-.5\height}{   \begin{tikzpicture} 
                      \node at (0,0)(gauge){SU$(2k)_{\color{red}1/2}$};
                      \node (flavour)[above of=gauge]{$[2k-3]$};
                      \node[label=below:{$S^2$}](empty)[right of=gauge]{};
                      \node(symmetric)[right of=empty]{$\left[1\right]$};
                      \path[draw,snake it](gauge)--(symmetric);
                      \draw(gauge)--(flavour);
         \end{tikzpicture} }}
{19}{\raisebox{-.5\height}{   \begin{tikzpicture} 
                      \node at (0,0)(gauge){SU$(K)_0$};
                      \node (flavour)[above of=gauge]{$[K-4]$};
                      \node[label=below:{$S^2$}](empty)[right of=gauge]{};
                      \node(symmetric)[right of=empty]{$\left[1\right]$};
                      \path[draw,snake it](gauge)--(symmetric);
                      \draw(gauge)--(flavour);
         \end{tikzpicture} }}
{20}{\raisebox{-.5\height}{   \begin{tikzpicture} 
                      \node at (0,0)(gauge){SU$(K)_{1/2}$};
                      \node (flavour)[above of=gauge]{$[K-5]$};
                      \node[label=below:{$S^2$}](empty)[right of=gauge]{};
                      \node(symmetric)[right of=empty]{$\left[1\right]$};
                      \path[draw,snake it](gauge)--(symmetric);
                      \draw(gauge)--(flavour);
         \end{tikzpicture} }}
{21}{\raisebox{-.5\height}{   \begin{tikzpicture} 
                      \node at (0,0)(gauge){SU$(K)_{j/2}$};
                      \node (flavour)[above of=gauge]{$[K-j-4]$};
                      \node[label=below:{$S^2$}](empty)[right of=gauge]{};
                      \node(symmetric)[right of=empty]{$\left[1\right]$};
                      \path[draw,snake it](gauge)--(symmetric);
                      \draw(gauge)--(flavour);
         \end{tikzpicture} }}
{22}{\raisebox{-.5\height}{   \begin{tikzpicture} 
                      \node at (0,0)(gauge){SU$(K)_{(j-l)/2}$};
                      \node (flavour)[above of=gauge]{$[K-j-l-4]$};
                      \node[label=below:{$S^2$}](empty)[right of=gauge]{};
                      \node(symmetric)[right of=empty]{$\left[1\right]$};
                      \path[draw,snake it](gauge)--(symmetric);
                      \draw(gauge)--(flavour);
         \end{tikzpicture} }}

}}

\newcommand{\plhwg}[1]{%
    \IfEqCase{#1}{%
{1}{\raisebox{-.5\height}{  
    $\sum_{i=1}^{2k-1}\mu_i^2t^{2i}+t^{4}+\mu_{2k-1} (t^{2k-1}+t^{2k+1})-\mu_{2k-1}^2t^{4k+2}$  }}
{2}{\raisebox{-.5\height}{ 
    $\sum_{i=1}^{2k-2}\mu_i^2t^{2i}+t^{4}+\mu_{2k-2} (t^{2k-2}+t^{2k})-\mu_{2k-2}^2t^{4k} $ }}
{3}{\raisebox{-.5\height}{ 
    $\sum_{i=1}^{2k-3}\mu_i^2t^{2i}+\nu^2t^2+t^4+\mu_{2k-3}\nu^2\left(t^{2k-1}+t^{2k+1}\right)-\mu_{2k-3}^2\nu^4t^{4k+2} $ }}
{4}{\raisebox{-.5\height}{ 
    $\sum_{i=1}^{2k-4}\mu_i^2t^{2i}+\nu^2t^2+t^4+\mu_{2k-4}\nu^2\left(t^{2k-2}+t^{2k}\right)-\mu_{2k-4}^2\nu^4t^{4k} $ }}
{5}{\raisebox{-.5\height}{ 
    $\sum_{i=1}^{2k-2j-5}\mu_i^2t^{2i}+t^2+\left(q+q^{-1}\right)\mu_{2k-2j-5}t^{2k-2}-\mu^2_{2k-2j-5}t^{4k-4} $ }}
{6}{\raisebox{-.5\height}{ 
    $\sum_{i=1}^{2k-2j-6}\mu_i^2t^{2i}+t^2+\left(q+q^{-1}\right)\mu_{2k-2j-6}t^{2k-2}-\mu^2_{2k-2j-6}t^{4k-4} $ }}
{7}{\raisebox{-.5\height}{ 
    $\sum_{i=1}^{2k-2j-5}\mu_i^2t^{2i}+t^2+\left(q+q^{-1}\right)\mu_{2k-2j-5}t^{2k-2}-\mu^2_{2k-2j-5}t^{4k-4} $ }}
{8}{\raisebox{-.5\height}{ 
    $\sum_{i=1}^{2k-2j-4}\mu_i^2t^{2i}+t^2+\left(q+q^{-1}\right)\mu_{2k-2j-4}t^{2k-1}-\mu^2_{2k-2j-4}t^{4k-2} $ }}
}}

\newcommand{\magneticquiverU}[1]{%
    \IfEqCase{#1}{%
{1}{\raisebox{-.5\height}{    
                          \begin{tikzpicture}
    \node[label=below:{$1$}][u](2){};
    \node (dots)[right of=2]{$\cdots$};
    \node[label=below:{$K-2$}][u](2N-5)[right of=dots]{};
    \node[label=below:{$K-1$}][u](2N-4)[right of=2N-5]{};
    \node[label=below:{$3$}][u](2')[right of=2N-4]{};
    \draw (2')to[out=-45,in=45,loop,looseness=10](2');
    \draw(2)--(dots);
    \draw(dots)--(2N-5);
    \draw[ double distance=1.5pt,<-](2N-5)--(2N-4);
    \draw(2N-4)--(2');
    \end{tikzpicture} }}
{2}{\raisebox{-.5\height} {
            \begin{tikzpicture}
    \node[label=below:{$1$}][u](2){};
    \node (dots)[right of=2]{$\cdots$};
    \node[label=below:{$2k-2$}][u](2N-5)[right of=dots]{};
    \node[label=below:{$2k-1$}][u](2N-4)[right of=2N-5]{};
    \node[label=below:{$3$}][u](2')[right of=2N-4]{};
    \draw (2')to[out=-45,in=45,loop,looseness=10](2');
    \draw(2)--(dots);
    \draw(dots)--(2N-5);
    \draw[ double distance=1.5pt,<-](2N-5)--(2N-4);
    \draw(2N-4)--(2');
    \end{tikzpicture}  }}
{3}{\raisebox{-.5\height} {
    \begin{tikzpicture}
    \node[label=below:{1}][u](2){};
    \node (dots)[right of=2]{$\cdots$};
    \node[label=below:{$K-5$}][u](2N-5)[right of=dots]{};
    \node[label=below:{$K-4$}][u](2N-4)[right of=2N-5]{};
    \node[label=below:{$3$}][u](3')[right of=2N-4]{};
    \node[label=below:{$2$}][u](2')[right of=3']{};
    \node[label=below:{$1$}][u](1')[right of=2']{};
    \draw (3')to[out=45,in=135,loop,looseness=10](3');
    \draw(2)--(dots);
    \draw(dots)--(2N-5);
    \draw[ double distance=1.5pt,<-](2N-5)--(2N-4);
    \draw(2N-4)--(3');
    \draw[double distance=1.5pt,->](3')--(2');
    \draw(1')--(2');
    \end{tikzpicture}}}
{4}{\raisebox{-.5\height} {
\begin{tikzpicture}
    \node[label=above:{1}][u](111){};
    \node[label=below:{2}][u](22)[below right of=111]{};
    \node[label=below:{1}][u](11)[right of=22]{};
    \node[label=below:{$K-5$}][u](k-5)[below left of=111]{};
    \node[label=below:{$K-6$}][u](k-6)[left of=k-5]{};
    \node(dots)[left of=k-6]{$\cdots$};
    \node[label=below:{$2$}][u](2)[left of=dots]{};
    \node[label=below:{$1$}][u](1)[left of=2]{};
    \draw(1)--(2);
    \draw(2)--(dots);
    \draw(dots)--(k-6);
    \draw[double distance=1.5pt,<-](k-6)--(k-5);
    \draw(k-5)--(22);
    \draw(k-5)--(111);
    \draw(111)--node[left]{$3$}++(22);
    \draw[double distance=1.5pt,->](22)--(11);
    \draw (22)to[out=45,in=135,loop,looseness=10](22);
    \end{tikzpicture}}}
{5}{\raisebox{-.5\height} {
    \begin{tikzpicture}
    \node[label=right:{1}][u](111){};
    \node[label=below:{2}][u](22)[left of=111]{};
    \node[label=below:{1}][u](11)[left of=22]{};
    \node[label=right:{$K-j-4$}][u](k-5)[below of=111]{};
    \node[label=below:{$K-j-5$}][u](k-6)[left of=k-5]{};
    \node(dots)[left of=k-6]{$\cdots$};
    \node[label=below:{$2$}][u](2)[left of=dots]{};
    \node[label=below:{$1$}][u](1)[left of=2]{};
    \draw(1)--(2);
    \draw(2)--(dots);
    \draw(dots)--(k-6);
    \draw[double distance=1.5pt,<-](k-6)--(k-5);
    \draw(k-5)--(22);
    \draw(k-5)--(111);
    \draw(111)--node[above]{$j+2$}++(22);
    \draw[double distance=1.5pt,->](22)--(11);
    \draw (22)to[out=45,in=135,loop,looseness=10](22);
    \end{tikzpicture}}}
{6}{\raisebox{-.5\height}{ 
             \begin{tikzpicture}
                 \node[label=below:{1}][u](1){};
                 \node[label=below:{2}][u](2)[right of=1]{};
                 \node (dots)[right of=2]{$\cdots$};
                 \node[label=above:{$K-j-l-5$}][u](k-1)[right of=dots]{};
                 \node[label=below:{$K-j-l-4$}][u](k)[right of=k-1]{};
                 \node[label=right:{$1$}][u](1c)[right of=k]{};
                 \node (empty)[right of=1c]{};
                 \node[label=above:{$1$}][u](1r)[above of=empty]{};
                 \node[label=below:{$1$}][u](1l)[below of=empty]{};
                 \draw(1)--(2);
                 \draw(2)--(dots);
                 \draw(dots)--(k-1);
                 \draw[<-,double distance=1.5pt](k-1)--(k);
                 \draw(1l)--node[right]{$j+2$}(1c);
                 \draw(1r)--node[right]{$l+2$}(1c);
                 \draw(k)--(1c);
                 \draw(k)--(1l);
                 \draw(k)--(1r);
                 \draw (1l)to[out=0,in=0]node[right]{$j+l+2$}(1r);
             \end{tikzpicture} }}
             }}